\documentclass[aps,prd,nofootinbib,square,sort&compress,10pt,fleqn,showpacs]{revtex4-2}
\usepackage[titletoc]{appendix}
\usepackage[english]{babel}
\usepackage{amssymb}
\usepackage{mathrsfs}
\usepackage{mathalpha}
\usepackage{upgreek}
\usepackage{comment}
\usepackage{yfonts}
\usepackage[fleqn]{amsmath}
\usepackage[T1]{fontenc}
\usepackage[x11names]{xcolor}
\usepackage{empheq}
\usepackage{color}
\allowdisplaybreaks
\usepackage{mathtools}
\usepackage{eucal}
\usepackage{graphicx}
\usepackage{float}
\usepackage[pdfencoding=auto,psdextra]{hyperref}
\usepackage{bookmark}
\usepackage{url}
\usepackage{epsfig}
\usepackage{bm}
\usepackage[utf8x,latin1]{inputenc} 
\usepackage{emerald}

\usepackage{dcolumn}

\usepackage{array}
\let\tempcal \mathcal
\usepackage{dutchcal}

\let\mathcal\tempcal

\makeatletter
\newcommand{\mathleft}{\@fleqntrue\@mathmargin0pt}
\newcommand{\mathcenter}{\@fleqnfalse}

\DeclareMathAlphabet{\mathpzc}{OT1}{pzc}{m}{it}

\def\ds{\displaystyle}

\def\be{\begin{equation}}
\def\ee{\end{equation}}
\def\ba{\begin{eqnarray}}
\def\bal{\begin{eqnarray}\label}
\def\ea{\end{eqnarray}}
\def\bse{\begin{subequations}}
\def\ese{\end{subequations}}

\def\L{{\mathcal L}}

\def\V{{\mathcal V}}
\def\U{{\mathcal U}}

\def\M{{\cal M}}

\def\a{\alpha}
\def\b{\beta}     

\def\d{\delta}

\def\ve{\varepsilon}

\def\pd{\partial}

\def\q{\mathsf{q}}
\def\m{\mathsf{m}}
\def\g{\mathfrak{g}}

\def\x{{\bm x}}
\def\y{{\bm y}}
\def\X{{\bm X}}

\def\MM{{\mathfrak M}}

\def\aap{Astronomy and Astrophysics}
\def\araa{Annual Review of Astronomy and Astrophysics}

\def\apj{The Astrophysical Journal}
\def\apjl{The Astrophysical Journal Letters}

\def\apss{Astrophysics and Space Science}

\def\icarus{Icarus}
\def\mnras{Monthly Notices of the Royal Astronomical Society}
\def\prd{ Physical Review D}
\def\prl{Physics Review Letters}

\def\nat{Nature}
\def\sovast{Soviet Astronomy}

\counterwithin{section}{part}
\counterwithin{equation}{section}

\begin{document}
\title{Lie Group Theory of Multipole Moments and Shape of Stationary Rotating Fluid Bodies}
\author{Sergei M. Kopeikin}
\affiliation{Department of Physics \& Astronomy, University of Missouri, 322 Physics Bldg., Columbia, Missouri 65211, USA}
\email{E-mail: kopeikins@missouri.edu}
\begin{abstract}
We present a novel and rigorous framework for determining the equilibrium configurations of uniformly rotating, self-gravitating fluid bodies. This work addresses the longstanding challenge of accurately modeling the rotational deformation of celestial objects such as stars and planets. By integrating classical Newtonian potential theory with modern mathematical techniques, we develop a unified formalism that significantly enhances both the precision and generality of shape modeling in astrophysical contexts.
Our methodology employs Lie group theory and exponential mapping to characterize vector flows associated with rotational deformations. We derive functional equations governing perturbations in density and gravitational potential, which are analytically resolved using the shift operator and Neumann series summation. This approach extends Clairaut's classical linear perturbation theory into the nonlinear regime.
The resulting formulation yields an exact nonlinear differential equation for the shape function, describing hydrostatic equilibrium under rotation without relying on the assumption of slow rotation. This equation generalizes Clairaut's classical result -- which applies only in the linear, slow-rotation regime -- by incorporating nonlinear effects and accommodating sufficiently large rotational speeds, the upper limit of which remains subject to further investigation. We refer to this as the generalized Clairaut equation to emphasize its role as a natural extension of Clairaut's original theory. We validate this formulation by deriving exact solutions, including the Maclaurin spheroid, Jacobi ellipsoid, and the unit-index polytrope.
Furthermore, we introduce spectral decomposition techniques to analyze the radial harmonics of the shape function and associated gravitational perturbations. Utilizing Wigner's formalism for angular momentum addition, we compute higher-order spectral corrections, thereby enhancing both the accuracy and computational efficiency of existing models. Our framework also derives rigorously defined boundary conditions for the spectral radial harmonics, enabling the systematic computation of Love numbers and gravitational multipole moments with high accuracy. This work represents a substantial advancement in the mathematical theory of rotating fluid figures, offering a comprehensive, non-perturbative methodology for modeling rotational deformations in astrophysical and planetary systems.
\end{abstract}
\maketitle
\newpage
\tableofcontents
\newpage

\newcounter{equationschapter}[section]
\section{Introduction}\label{sec1}
\stepcounter{equationschapter}
\setcounter{equationschapter}{1}
\renewcommand{\theequation}{1.\arabic{equation}}

\subsection{Problem Statement and Its Relevance}

Determining the equilibrium shape of an isolated, massive fluid body (such as a star or planet) that is rotating rigidly about its axis is a challenging mathematical problem that remains unsolved despite 250 years of effort by theoretical physicists and mathematicians. The problem's statement is deceptively simple: Assume the fluid body is initially static, spherically symmetric, and non-rotating with a fixed radius $r=a_0$ in spherical coordinates $(r, \theta, \varphi)$. The body is then brought into a state of uniform rotation with angular velocity $\omega$ directed along the $z$ axis. The goal is to determine the shape of the body's surface as a function of the rotation rate, internal density distribution, and gravitational potential. An additional goal is to determine the multipole moments of the external gravitational field of the rotating body and their exact analytic dependence on the angular speed of the body's rotation $\omega$. Measuring the shape and multipole moments parameters allows us to extract valuable information about the internal structure of the body and the fluid's equation of state. 

To solve the problem, the body is assumed to be in hydrostatic equilibrium, meaning the forces due to gravity, pressure, and centrifugal effects are balanced at every point within the fluid. The problem is governed by several basic equations:
\begin{enumerate}
\item Equation of hydrostatic equilibrium
\bal{001}
\pd_i p&=&-\mu\pd_iV\;,
\ea
where $\pd_i=\pd/\pd x^i$, $\mu=\mu(\x)$ is the fluid's density, $p=p(\x)$ is the pressure, 
\bal{h002}
V&=&\Phi+W\;,
\ea 
is the total gravitational potential of a rotating body consisting of the gravitational potential $\Phi=\Phi(\x)$ of the fluid itself and the centrifugal potential $W=W(\x)$.
\item Poisson's equation for gravitational potentials $\Phi$:
\bal{002}
\Delta \Phi&=&-4\pi G\mu\;,
\ea
where $G$ is the universal gravitational constant, and $\Delta=\equiv \delta^{ij}\frac{\pd^2}{\pd x^i\pd x^j}$ is the Laplacian.
\item Poisson's equation for centrifugal potential $W$:
\bal{a005}
\Delta W&=&2\omega^2\;,
\ea
where $\omega$ is the constant angular velocity of body's rotation.
\item Equation of state, 
\bal{003}
p&=&p(\mu)\;.
\ea
\end{enumerate}
The boundary of the fluid body is a level surface of equal gravitational potential, $V=V(\x)$. In the case of hydrodynamic equilibrium, the equipotential surface also coincides with the surfaces of equal density, $\mu$, and pressure, $p$ \citep{MacMillan1930}. The body's boundary is defined by the condition that the pressure, $p$, is zero at the surface: $p(\x)=0$. Before imposing the rotational perturbation, the equipotential surfaces are spheres, and the boundary of the body is a sphere with a radius $r=a_0$. Rotation deforms the equipotential surfaces and distorts the shape of the body's boundary. Several theoretical approaches have been proposed to determine the rotational distortion of the body's boundary based on various mathematical techniques applied to solve the system of equations \eqref{001}--\eqref{003}.

Despite their seemingly straightforward appearance, these equations represent a system of nonlinear, coupled differential equations, posing significant mathematical challenges. A key difficulty arises from the interdependence between the gravitational potential, $\Phi$, and the shape of the rotating body. Specifically, determining the body's equilibrium configuration requires knowledge of the perturbed gravitational potential. However, this potential is defined by the integral:
\bal{004} 
\Phi(\x)&=&\int\limits_{\cal D}\frac{\mu(\x')d^3x'}{|\x-\x'|}\;, \ea 
evaluated over the volume ${\cal D}$ occupied by the fluid comprising the rotating body. Crucially, the domain ${\cal D}$ is not known a priori and must be determined as part of the solution. This creates a classical coupling problem: the gravitational potential depends on the body's shape, yet the shape itself is governed by the potential. 

To address this fundamental interdependence, earlier methodologies frequently employed a Laplace expansion of the Green's function $|\x - \x'|^{-1}$ associated with the Laplace operator, as shown in Eq. \eqref{004}, expressed as a series of Legendre polynomials centered at the interior point $\x$ within the integration domain ${\cal D}$. However, this expansion is known to diverge within the bounded region between two concentric surfaces -- referred to as the interior and exterior Brillouin spheres -- both centered at the body's center of mass. The interior Brillouin sphere is defined by the radius $r_{\rm i}$, which is the shortest distance from the center of mass to any point on the body's surface. In contrast, the exterior Brillouin sphere has a radius $r_{\rm e}$, representing the greatest such distance, and it fully encloses the entire mass of the body. While all matter lies within the exterior sphere, some of it extends beyond the interior one. The divergence of the Laplace expansion within the region between these two spheres has raised significant concerns regarding the mathematical rigor and physical reliability of this method for determining the equilibrium configuration of rotating fluid bodies. For a more detailed discussion, see \citep{Zharkov_1978book,Hubbard_2014}.

A further major challenge in solving the system of non-linear equations \eqref{001}--\eqref{003} was to suggest a self-consistent method for calculating the non-linear corrections to the linearized theory of rotational deformations of fluid bodies proposed by A. Clairaut \citep{Clairaut-book-1743} in 1743. It took over 150 years to develop equations describing the second-order corrections to Clairaut's theory by G.~H. Darwin \citep{Darwin_1899} and W. de Sitter \citep{deSitter_1924BAN}, and another 50-100 years to develop a systematic approach to derive corrections to Clairaut's theory up to the 7-th order inclusively \citep{Zharkov_1978book,Kopal_1960book,Lanzano_1982,Nettelmann_2021}. Despite these advances, no existing theoretical framework has yet provided a general, closed-form methodology for computing the boundary shape and gravitational field of a rotating fluid body at arbitrary orders of perturbation theory. This longstanding obstacle has challenged some of the most prominent thinkers for centuries, with progress historically slow and a complete, general solution remaining elusive. A detailed account of the key historical developments and foundational contributions to this problem is provided in Section \ref{section1}. In the subsequent sections, we introduce a novel and systematic approach for determining the equilibrium shape of a rotating fluid body, along with its associated gravitational potential, Love numbers, and multipole moments. This framework is grounded in nonlinear perturbation theory and the theory of Lie groups of diffeomorphisms. By leveraging these mathematical tools, we derive a general nonlinear differential equation governing the shape of the rotating body and formulate a unified method for constructing successive approximations to its solution. This approach facilitates the computation of nonlinear corrections to Clairaut's classical theory to arbitrary orders of approximation, thereby representing a substantial advancement in the mathematical theory of figures of rotating fluid bodies.

A key advantage of the theoretical approach developed in this manuscript is its ability to yield an exact, self-consistent solution for the equilibrium shape of a rigidly rotating, self-gravitating fluid body.  This marks a significant improvement over approximate or perturbative methods used by the previous researchers. Our approach enables more accurate and comprehensive modeling across a wide range of astrophysical and planetary contexts. In planetary science, precise modeling of gas giants such as Jupiter and Saturn -- whose interiors are predominantly fluid -- is essential for probing their internal structure, rotation profiles, and equations of state. The observed oblateness of these planets encodes critical information about internal mass distribution and angular momentum that low-order approximations fail to capture. The shape of a planet directly influences its gravitational field, which in turn affects the dynamics of surrounding systems such as planetary rings, satellite orbits, and tidal interactions. For instance, Saturn's oblateness governs the structure and stability of its ring system, while tidal deformations play a key role in the thermal and orbital evolution of its moons. In stellar astrophysics, many stars rotate rapidly, resulting in substantial deviations from spherical symmetry. Therefore, accurately determining their equilibrium shapes is crucial for modeling internal processes such as differential rotation, meridional circulation, and chemical mixing, all of which influence stellar evolution and nucleosynthesis. In systems involving accretion disks -- such as those surrounding black holes, neutron stars, or young stellar objects -- the shape and dynamics of the rotating fluid play a pivotal role in jet formation and angular momentum transport, and an exact treatment enhances our ability to model these complex environments. Finally, gravitational wave astronomy can greatly benefits from the methodology present in this manuscript, as rapidly rotating neutron stars (pulsars) can exhibit measurable oblateness that affects the amplitude and frequency of the gravitational waves they emit, making accurate shape modeling essential for interpreting signals detected by observatories such as LIGO and Virgo. In all these contexts, an exact, general method for determining the figure of rotating fluid bodies -- such as the one developed in this manuscript -- offers a powerful and versatile tool for advancing both theoretical understanding and observational interpretation across multiple domains of astrophysics and planetary science.

\subsection{Historical Background}\label{section1}
The Newtonian theory of rotating celestial bodies \citep{Jardetzky_2005} is crucial in both geophysical and astrophysical studies. It enhances our understanding of the gravitational fields of planets \citep{Zharkov_1978book,Hubbard_2013ApJ} and stars \citep{Chandra_1939book,Tassoul_1978book,horedt_2004book}, and contributes to the improvement of geodetic models \citep{Moritz_1990,Arora2011}. The primary objective is to determine the shape distortion of a rapidly rotating fluid body as a function of its rotation speed ($\omega$) and the body's equation of state. This distortion influences the symmetry of the gravitational field and is characterized by the body's multipole moments. By measuring these multipole moments, researchers can gain insights into the internal structure of the celestial body and the equation of state of matter deep within its interior. This method provides the only means to understand the conditions under which matter exists at high density and pressure in the central regions of astrophysical objects \citep{Wahl_2017,Hubbard_2024}.

The Newtonian theory of figures of astronomical bodies has been extended to general relativity to better understand the nuclear equations of state in the cores of compact astrophysical objects, such as neutron stars \citep{Hinderer-2008ApJ,Poisson_2009PhRvD,Damour-2009PhRvD}. This extension gained significant attention following the detection of gravitational waves from neutron-star binaries \citep{Abbott-2017PhRvL,Chia-2024PhRvD}, highlighting the necessity for detailed studies on how relativistic stars respond to external gravitational forces caused by tides.

This paper primarily focuses on the study of rotational deformations of astronomical bodies within the framework of Newtonian theory. These deformations are driven by radial and quadrupole harmonics of the centrifugal rotation potential. Due to the highly nonlinear nature of deformation theory, analyzing rotational deformations aids in understanding nonlinear effects that arise solely from the self-interaction of perturbation harmonics related to the body's internal structure. The topic of tidal effects will be addressed in a separate paper.

This section provides a historical overview of the mathematical methods previously employed by researchers to determine the shape and gravitational field of rotating celestial bodies in Newtonian theory.

\subsubsection{Homogeneous Bodies: From Maclaurin to Chandrasekhar and Lebovitz}
The simplest case in this theory involves determining the shape and potential of a rotating fluid body with constant density, $\rho$. Maclaurin first solved this problem in 1742, confirming Newton's assertion that a rotating fluid forms an oblate ellipsoid of revolution. Researchers such as Jacobi, Dedekind, Dirichlet, Riemann, Poincar\'e, and Cartan extended this work by discovering more general spheroidal configurations of rotating homogeneous bodies. 

These results are concisely summarized and independently corroborated in Chandrasekhar's seminal monograph Ellipsoidal Figures of Equilibrium \citep{chandr87}. The book systematically applies the virial tensor theorem methodology, originally developed for homogeneous rotating fluid bodies and strongly championed by N. Lebovitz in a series of foundational works \citep{Lebovitz_1961,Lebovitz_1967,Lebovitz_1996}. 

This direction in the development of the theory has practical limitations, as real astronomical bodies do not maintain a constant density profile across their entire structure. A more general theory must consider the heterogeneous density distribution of stars and giant planets, which are composed of matter subject to specific equations of state.

\subsubsection{Non-Homogeneous Bodies: From Clairaut and Darwin-de Sitter to Higher-Order Theories}
Clairaut established the foundational elements of the theory of rotating inhomogeneous bodies in 1743 \citep{Clairaut-book-1743}. Clairaut's theory assumes that the rotating body is in hydrostatic equilibrium and has the shape of an oblate ellipsoid, which is flattened at the poles due to rotation. The theory formulates a linear differential equation for the oblateness of the body. Clairaut went further and established a theorem relating the surface gravity at any point on the rotating ellipsoid to its position, allowing for the calculation of the body's ellipticity from gravity measurements at different latitudes. This theorem is considered as a basis of geometric geodesy.

Airy \citep{Airy_1826} improved Clairaut's linear theory by considering second-order effects in the body's oblateness. He developed integral equations for the level surface of rotating fluids, although he couldn't convert them to differential equations. Airy concluded that the Earth's surface must be depressed below the level of the true ellipsoid in middle latitudes, although he did not provide a numerical estimate of this depression. His work laid the groundwork for understanding the more complex shape of the Earth due to its rotation and varying density distribution.

Darwin achieved more comprehensive results about 70 years later \citep{Darwin_1899}, building on earlier work by Helmert \citep{Helmert_1880}, Callandreau \citep{Callandreau_1889,Callandreau_1897}, and Wiechert \citep{Wiechert_1897}. Darwin extended the theory by successfully converting the integral equations into differential equations, allowing for more precise calculations of the rotating body's shape. He also expanded the gravitational potential of the body into a power series using Legendre polynomials, which provided a more accurate spectral representation of the gravitational field. Darwin's contributions significantly advanced the understanding of the equilibrium figures of rotating fluid masses and the effects of rotation on their shapes. 

Willem de Sitter revisited the problem of the Earth's figure in the early 20th century \citep{deSitter_1924BAN}, building on Darwin's second-order theory \citep{Darwin_1899}. He introduced several refinements, notably the use of the mean radius as the independent variable, which simplifies the mathematical formulation. De Sitter also incorporated geodetic concepts such as the geoid and the normal surface, accounting for isostatic compensation and deviations from hydrostatic equilibrium near the surface. He recalculated second-order corrections using improved observational data and corrected a numerical error in Darwin's evaluation of the mean density integral. Furthermore, de Sitter evaluated Wiechert's two-layer Earth model \citep{Wiechert_1897} and demonstrated its consistency with observational constraints, in contrast to Darwin's use of Roche's model. He also derived small but non-negligible corrections to the Earth's ellipticity and moment of inertia arising from internal structural deviations.

Extension to Clairaut and Darwin-de Sitter approximations \citep{Darwin_1899,deSitter_1924BAN} was proposed by Kopal \citep{Kopal_1960book} which laid the groundwork for analyzing the equilibrium shapes of rotating fluid bodies using perturbative expansions of the Clairaut equation. Lanzano \citep{Lanzano1962,Lanzano_1974, Lanzano_1982} extended Kopal's work by developing a third-order correction to Clairaut equation for bodies with arbitrary internal density distributions. These equations have been used in geophysical study of the shape, gravity, and moment of inertia for highly flattened celestial bodies by Rambaux et al \citep{Rambaux2015}.

Expanding the body's gravitational potential into a power series using Legendre polynomials has a serious disadvantage: near the body's surface, the series may converge slowly or not at all, reducing computational accuracy. This issue caused doubts about the validity of applying the Legendre polynomial series for calculating the oblateness of the rotating body until the work of Hubbard \citep{Hubbard_2014}, who showed that the series converge on level surfaces if rotational distortion stays below a critical value.

\subsubsection{Lyapunov's Examination of Rotating Fluid Bodies}

Due to convergence issues inherent in the Legendre polynomial series expansion of the internal gravitational potential in rotating self-gravitating bodies, Lyapunov \citep{Lyapunov_1906} developed an alternative approach that avoids these limitations. His method introduces a system of integro-differential equations capable of describing the body's figure with arbitrary precision. However, the analytical complexity of these equations -- especially at higher orders -- has posed significant challenges for practical implementation.

Lyapunov's approach represented a significant advancement over earlier methods, offering a more rigorous and versatile framework for analyzing the equilibrium figures of rotating bodies. His theory established sufficient conditions for the existence and uniqueness of solutions to the governing integro-differential equations. Furthermore, Lyapunov addressed the stability of classical configurations such as the Maclaurin and Jacobi ellipsoids, extending his analysis to encompass more general equilibrium figures that bifurcate from these canonical shapes.

One of Lyapunov's key contributions was the development of the Lyapunov series -- a power series expansion in terms of a small parameter $\m$, representing the ratio of centrifugal to gravitational forces. This series enabled the systematic analysis of slowly rotating, inhomogeneous fluid bodies, offering valuable insights into their equilibrium configurations and stability properties. Importantly, Lyapunov also estimated the radius of convergence of the series, thereby delineating the regime within which the expansion remains valid and physically meaningful under rotational distortion. Sretenskii provided an in-depth examination of Lyapunov's findings in his review paper \citep{sret_1938}.

Recent developments in Lyapunov's theory have been advanced by Kholshevnikov and collaborators \citep{Khol_1994, Khol_2003det, Khol_2003zb, Kholshev_2007}, who have revisited and extended Lyapunov's original ideas. Their work addresses previously unresolved problems and explores new applications within celestial mechanics and astrophysical fluid dynamics. Kholshevnikov's contributions have deepened the theoretical understanding of the stability and equilibrium figures of rotating celestial bodies, reaffirming the continued relevance and applicability of Lyapunov's methods in contemporary research.

\subsubsection{Chandrasekhar's Theory on the Deformation of Rotating Polytropic Stars}

The methods for calculating the shape of rotating celestial bodies, developed from Clairaut to Lyapunov, initially focused on determining the Earth's figure. By the early 20-th century, these methods needed to address astrophysical problems, such as star formation and internal structure, considering rotational deformation.

Astrophysicists either overlooked geodesists' theoretical developments or found them unsuitable for astrophysical problems. Consequently, Chandrasekhar \citep{Chandrasekhar_1933MNRAS} developed his own theory of rotational distortion of stars, limited to the first approximation with respect to the rotational parameter $\m$ and mainly applicable to stars with a polytropic equation of state. However, he did not address boundary conditions, complicating its application and drawing criticism \citep{Jardetzky_2005}. Kroghdal \citep{Kroghdal_1942ApJ} later discussed these boundary conditions in detail, providing a more comprehensive framework for their application in astrophysical contexts. Additionally, Chandrasekhar and Krogdahl \citep{Chandrasekhar_1942ApJ} showed that Chandrasekhar's theory is equivalent to Clairaut's but in different variables.

Extensions of Chandrasekhar's theory to the second order in the small parameter $\m$ were made by Occhionero \citep{Occhionero_1967}, Anand \citep{Anand_1968}, and Aikawa \citep{Aikawa_1971}. Occhionero's work focused on refining the perturbation techniques to better model the rotational deformation of stars, while Anand's research extended the theory to include more complex degenerate configurations. Aikawa further developed these extensions, providing a more detailed analysis of the second-order effects. However, no comparison was made with Darwin's second-order theory, leaving a gap in the comprehensive understanding of rotational deformation in astrophysical bodies. A comprehensive review of the works of these authors, along with more recent results, is discussed in the monograph by Horedt \citep{horedt_2004book}, which provides an extensive overview of polytropic models and their applications in astrophysics.

\subsubsection{The Zharkov-Trubitsyn Framework for Rotating Giant Planets}
Advancements in computer technology and symbolic computations at the end of the 20th century significantly enhanced the study of rotating fluids and celestial bodies. These advancements enabled researchers to extend classical theories, such as those by Clairaut and Darwin, to more complex and accurate models. The increase in computational power allowed for the handling of more complex equations and simulations, which was crucial for studying the intricate dynamics of rotating fluids and celestial bodies. Symbolic computation tools enabled the manipulation of mathematical expressions and the derivation of higher-order approximations, improving the precision of models used to describe the shapes and gravitational fields of rotating bodies.

Zharkov and Trubitsyn \citep{Zharkov_1970,zharkov_1976SvA,Zharkov_1978book,Efimov_1978,Zharkov-book-1986} developed a comprehensive theory that provided a robust framework for studying rotating celestial bodies. Their work allowed for more accurate modeling of the internal structure and gravitational potential of these bodies in hydrostatic equilibrium. This theory enhanced earlier methods by integrating higher-order approximations, tackling the complexities of non-uniform density distributions, and considering the differential rotation of astronomical bodies. The improved models have been applied to various celestial bodies, including planets and satellites, enhancing our understanding of their shapes, internal structures, and gravitational fields. These advancements have had a profound impact on fields such as geophysics, astrophysics, and planetary science, enabling more precise and comprehensive studies of rotating fluids within celestial bodies \citep{Denis_1998GeoJI,Nettelmann_2021,Nettelmann_2017}.

The Zharkov-Trubitsyn theory utilizes the total gravitational potential  $V=V(\x)$, introduced in Eq. \eqref{h002}. The level surface corresponding to this potential is defined by the condition:
\bal{xa1}
V({\bm y})&:=&\Phi({\bm y})+W({\bm y})=\frac{4\pi G}{3}r^2A_0(r)\;.
\ea
In this expression, $\Phi$ denotes the gravitational potential of the fluid, and $W$ represents the centrifugal potential, $A_0(r)$ is yet unknown function which has a constant value at each fixed value of radial coordinate $r$. The level surface is described by the relation: ${\bm y}={\bm x}+\X$, where ${\bm x}$ corresponds to a point on the unperturbed (reference) surface, and $\X={\bm n}X(\x)$ is the displacement vector indicating the radial deviation of the perturbed surface from its reference configuration. Here, ${\bm n} = {\bm x}/r$, is the unit radial vector, and $r=|\x|$ is the radial distance from the origin.

In the case of a stationary, rotating fluid body with its axis of rotation aligned along the $z$-axis, the resulting deformation of the body is axisymmetric \citep{Tassoul_1978book}. The solution to the condition \eqref{xa1} is sought through an expansion of ${\bm y}={\bm n}y(\x)$ in terms of the even-order Legendre polynomials $P_{2n}(\cos\theta)$:
\bal{xa2}
y&=&r\left[1+\sum_{n=0}^\infty s_{2n}(r)P_{2n}(\cos\theta)\right]\;,
\ea
where $r$ denotes the mean radius of the undisturbed level surface. The function $s_0(r)$ represents the residual radial deformation associated with the monopole component of the centrifugal potential, while the coefficients $s_{2n}(r)$ for $n \ge 1$ characterize the spectral radial harmonics that describe the deviation of the perturbed level surface from spherical symmetry. The principal objective of Zharkov-Trubitsyn theory is to determine these coefficients $s_{2n}(r)$, which quantify the rotationally induced deformation of the level surfaces as governed by the body's internal structure and equation of state.

This is achieved by substituting the expansion \eqref{xa2} into Eq. \eqref{xa1}, resulting in various infinite products and power series in $P_n(\cos\theta)$. Traditional recurrence relations and orthogonality properties of the Legendre polynomials are used to transform the products into a sum, presenting the total potential $V$ as:
\bal{xa3}
V({\bm y})&=&\frac{4\pi Gr^2}{3}\sum_{n=0}A_{2n}(r)P_{2n}(\cos\theta)\;.
\ea   
Here, the coefficients $A_{2n}(r)$ represent an algebraic sum of integrals involving the fluid density $\rho=\rho(r)$, the radial harmonics $s_{2n}$, and their multiple products.

The condition \eqref{xa1} requires that $A_{2n}(r)=0$ for all $n\ge 1$. This results in a set of integro-algebraic equations for the radial harmonics $s_{2n}(r)$. The coefficient $A_0(r)$ fixes the value of the potential $V(\y)$ in the left-hand side of Eq. \eqref{xa1} by setting, $A_0(r):=\ds\frac{3}{4\pi Gr^2}V(r)$ on the level surface of the body. This provides a supplementary condition for determining the radial distribution of density $\rho=\rho(r)$ for a given equation of state $p=p(\rho)$ by solving the equation of hydrostatic equilibrium \eqref{001} on the level surfaces in terms of the radial harmonics $s_{2n}(r)$.  Substituting the resulting density $\rho(r)$ into the system of integro-algebraic equations and solving it, the radial harmonics $s_{2n}(r)$ can be determined. The values of the harmonics $s_{2n}(r)$ at the surface of the body are crucial to determine theoretical values of the zonal harmonic coefficients $J_{2n}$ in the multipolar expansion of the body's gravitational field. Comparison of the theoretical and experimental values of $J_{2n}$ allows to analyze the interior structure of the celestial body \citep{Zharkov_1978book,Zharkov-book-1986,Hubbard-book}.

The Zharkov-Trubitsyn theory, while providing a robust framework for studying rotating astronomical bodies, faces several shortcomings:
\begin{enumerate}
\item[--] The Legendre polynomial series used to expand the total potential $V$ inside the body diverges in a small shell between the spheres with the minimal and maximal radii of the level surface. Although this divergence typically does not affect practical computations, it raises concerns about the theoretical robustness of the method.
\item[--] The Zharkov-Trubitsyn approach is computationally expensive due to the lack of a general iterative form for the perturbation equations. Each approximation's equations must be calculated independently, requiring extensive computational algebra.  
\item[--] The accuracy of the theory tends to degrade as higher-order approximations are introduced. This limitation may stem from its reliance on integro-differential equations, which inherently involve non-local dependencies-integrals that require knowledge of the solution over extended spatial domains. Such non-locality increases computational complexity and reduces numerical stability, as the solution at any given point is influenced by values across the entire domain. For example, the seventh-order expansion in the Zharkov-Trubitsyn theory exhibits noticeable discrepancies when compared with more accurate numerical solutions derived from purely differential formulations \citep{Nettelmann_2021}.
\item[--]  The theory is most effective for polytropic models, which may not accurately capture the complex internal structures of celestial bodies. Polytropic models are simplified representations that may not fully account for the variations in density and composition found in real celestial bodies.
\end{enumerate}

\subsubsection{Hubbard's Concentric Maclaurin Spheroid Method}

To address the issues with the Zharkov-Trubitsyn theory, Hubbard \citep{Hubbard_2012ApJ,Hubbard_2013ApJ} proposed the Concentric Maclaurin Spheroid (CMS) method. This method belongs to the class of the non-perturbative treatments of hydrostatic equilibrium in rotating fluid bodies. It approximates the real density distribution within a body using $N$ concentric, constant-density spheroids, allowing for exact calculation of each spheroid's potential. By applying the principle of superposition for the Newtonian gravitational potential, a self-consistent solution for the shapes of the interfaces between spheroids and the interior gravitational potential can be found iteratively. Hubbard \citep{Hubbard_2013ApJ} claims this method is simpler and more precise than perturbation methods like the Zharkov-Trubitsyn theory.

Increasing the number of concentric Maclaurin layers allows the CMS method to achieve any desired level of accuracy, as demonstrated by the \textit{Juno} mission to Jupiter \citep{Wahl_2017}. Several refinements have been made to the CMS model. For instance, the method has been extended to handle differential rotation on cylinders, which is crucial for accurately modeling the interior dynamics of gas giants like Jupiter and Saturn  \citep{Militzer_2019}. This extension allows for a more precise calculation of gravitational moments and better alignment with observational data from missions like Juno and Cassini. 

Additionally, the CMS method has been adapted to include tidal potentials from satellites, enhancing its applicability to studying the tidal responses of giant planets. This adaptation has revealed significant changes in calculated tidal Love numbers, which are essential for interpreting high-precision measurements of planetary gravitational fields \citep{Wahl_2017Icar}.

Despite these advancements, the CMS method still faces challenges. The high computational cost remains a significant limitation, although recent acceleration techniques have been developed to reduce this burden \citep{Militzer_2019,Nettelmann_2021}. These techniques allow for efficient optimization of model parameters and increase the precision of calculated gravitational harmonics. Debras and Chabrier \citep{Debras_2018AA} pointed out that the treatment of the outermost layers in the framework of the CMS method leads to irreducible errors in the calculation of the gravitational moments and thus on the inferred physical quantities for the giant planets. They have quantified these errors and evaluated the maximum precision that can be reached with the CMS method in the present and future exploitation of Juno's data.

Wisdom \citep{Wisdom_1996} proposed an alternative non-perturbative treatment of finding the radial harmonics $s_{2n}(r)$ and zonal gravitational harmonics $J_{2n}$ of rapidly rotating giant planets of the solar system like Saturn and Jupiter which he calls the Centrifugal Liquid Core (CLC) method. Wisdom's method relies upon iterative adjustment of the shape, gravitational potential, density, and pressure inside the planet starting from some, properly chosen approximation to the real solution. The CLC model specifically focuses on how the liquid core of a rotating planet responds to the centrifugal force. Recently, Wisdom and Hubbard \citep{Wisdom_2016Icar} compared the CMS and CLC methods and found that the two methods are in remarkable agreement.

\subsubsection{Molodensky's Theory}

Molodensky's theory of figures of celestial bodies \citep{molodensky1953} is primarily applicable to studies of the Earth's figure. Molodensky introduces a new formulation of the equations describing the rotational and tidal deformation of the Earth's figure, significantly advancing the determination of the Earth's geoid. His approach involves solving the boundary value problem for the Earth's gravitational field, taking into account the effects of rotation and tidal forces. This method allows for a more accurate representation of the Earth's shape, particularly the geoid, which is the hypothetical sea level surface under the influence of Earth's gravity and rotation. Similar ideas were proposed and developed by Harold Jeffreys \citep{bolt_1993GMS}.

Together, the works of Molodensky and Jeffreys have significantly advanced the field of geodesy, offering robust methods for determining the shape and gravitational field of rotating celestial bodies. Their theories continue to be fundamental in modern geophysical and astronomical studies, aiding in the precise measurement and modeling of the Earth's figure and its variations over time.

\subsection{Notations}

Throughout this paper, we adopt the following notational conventions:
\begin{itemize}
\item[--] Roman indices $i,j,k,...$ range over the set $\{1,2,3\}$ and are used to label spatial components in Euclidean 3-space. 
\item[--] Euclidean vectors are denoted by boldface letters, for example, ${\bm a}=(a^i)=(a^1,a^2,a^3)$.
\item[--] The symbol $\delta_{ij}=\delta^{ij}={\rm diag}(1,1,1)$ denotes the unit matrix, also known as the Kronecker delta. In Cartesian coordinates, it serves as the metric tensor, enabling the raising and lowering of indices and defining the Euclidean inner product structure of the space.
\item[--] The Euclidean dot product of two vectors, such as ${\bm a}$ and ${\bm b}$, is denoted by ${\bm a} \cdot {\bm b} \equiv a^i b^i=\delta_{ij}a^ia^j$. Throughout this paper, we adopt the Einstein summation convention, whereby any repeated index implies summation over the range from 1 to 3, regardless of the specific letter used.  For example, the expression $a_i b_i$ corresponds explicitly to $a_1 b_1 + a_2 b_2 + a_3 b_3$.
\item[--] Parentheses around a pair of vector indices denote symmetrization. For example, $a^{(i}b^{j)}=\ds\frac12\left(a^ib^j+a^jb^i\right)$.
\item[--] Square brackets around a pair of vector indices denote antisymmetrization. For example, $a^{[i}b^{j]}=\ds\frac12\left(a^ib^j-a^jb^i\right)$.
\item[--] The Cartesian coordinates of the reference frame associated with a rigidly rotating fluid body are denoted by ${\bm x}=(x^i)=(x^1,x^2,x^3)$.
\item[--] The standard spherical coordinates $(r,\theta,\varphi)$ are related to the Cartesian coordinates $x^i$ by the transformation equations
$$
x^1=r\sin\theta\cos\varphi\qquad,\qquad  x^2=r\sin\theta\sin\varphi\qquad,\qquad x^3=r\cos\theta\;,
$$
where $r$ denotes the radial distance, $\theta$ is the polar angle measured from the positive $x^3$-axis,  and $\varphi$ is the azimuthal angle measured in the $x^1x^2$-plane from the positive $x^1$-axis.
\item[--] The symbol $\omega$ denotes the constant magnitude of the angular velocity of the rigid rotation of the fluid body.
\item[--] The angular velocity vector of the rigidly rotating fluid body is denoted by ${\bm\omega}:=(\omega^i)=(0,0,\omega)$, where $\omega$ is the constant magnitude of the angular velocity. The vector ${\bm\omega}$ is directed along the $x^3$-axis of the Cartesian coordinate system.
\item[--] The symbol $\mathfrak{M}$ denotes the base manifold of the Lie group of diffeomorphisms. Geometrically, $\mathfrak{M}$ represents the spherically-symmetric volume $\V$ corresponding to the reference configuration of the rotating fluid body.
\item[--] The vector ${\bm\xi}:=(\xi^i)=(\xi^1,\xi^2,\xi^3)$ is the generator of the vector flow that describes the infinitesimal displacement of a fluid element from its unperturbed to perturbed position. It is defined over the base manifold $\mathfrak{M}$ and depends on the coordinates $\x$ on the manifold.
\item[--] The vector ${\bm\zeta}:=(\zeta^i)=(\zeta^1,\zeta^2,\zeta^3)$ is the generator of the vector flow that describes the infinitesimal displacement of a level surface from its unperturbed to perturbed position. The vector ${\bm\zeta}$ is defined over the base manifold $\mathfrak{M}$ and depends on the coordinates $\x$ on the manifold.
\item[--] The vector ${\bm X}:=(X^i)=(X^1,X^2,X^3)$ represents the finite displacement describing the radial translation of the perturbed level surface relative to the unperturbed one. The vector $\X$ is defined over the base manifold $\mathfrak{M}$ and depends on the coordinates $\x$ on the manifold.
\item[--] The scalar quantity $X:=|{\bm X}|$ denotes the radial height function, where $|\X|$ represents the Euclidean norm of the displacement vector $\X$, i.e., $|\X|=\left({\bm X}\cdot{\bm X}\right)^{1/2}$. 
\item[--] The scalar function $f:=X/r$ is referred to as the shape function.
\item[--] The operator $\pd_i:=\pd/\pd x^i$ denotes the partial derivative with respect to the $i$-th Cartesian coordinate. Occasionally, we will use the notation ${\bm\nabla} := (\pd_i)$ to represent the gradient operator in the coordinate system $\x$.
\item[--] The operator $\delta_{\bm\xi}$ denotes the infinitesimal Eulerian variation induced by the vector field ${\bm\xi}$.
\item[--] The operator $L_{\bm\xi}:=\xi^i\pd_i$ denotes the linear directional derivative along the vector field ${\bm\xi}$.
\item[--] The operator $\pounds_{\bm\xi}$ denotes the Lie derivative along the congruence of the integral curves generated by the vector field ${\bm\xi}$. 
\item[--] The symbol $\hat{\mathsf T}_{\bm X}=\sum_{k=1}^\infty X^{i_1}\ldots X^{i_k}\pd_{i_1}\ldots\pd_{i_k}$ is the shift operator associated with the displacement vector ${\bm X}=(X^i)$. Note that the Einstein summation convention is used: repeated indices are implicitly summed over the range 1 to 3.
\item[--] The symbol $\Delta$ denotes the Laplace operator defined with respect to the coordinate system introduced on the base manifold $\mathfrak{M}$. By definition, it is given by $\Delta := g^{ij}\pd_i\pd_j$ where $g^{ij}$ is the contravariant metric tensor of the chosen coordinate system.
\item[--] The symbol $a$ denotes the radius of the spherical volume occupied by the unperturbed reference configuration of the fluid body. In the case of a Jacobi ellipsoid, this radius is denoted by $r_0$ to avoid ambiguity with the conventional notation $a$ used for the semi-major axis of the ellipsoid.
\item[--] The scalar function $\rho:=\rho(r)$ denotes the unperturbed fluid density in the reference configuration and depends solely on the radial coordinate, defined as $r = |\x|$.
\item[--] The function $\bar\rho(r)$ denotes the average unperturbed fluid density within a spherical volume of radius $r$. Similar to $\rho(r)$, it depends exclusively on the radial coordinate, defined as $r = |\x|$.
\item[--] The function $\mu(\x)$ denotes the perturbed fluid density within the rotating fluid body and depends on all three spatial coordinates.
\item[--] The symbol $\M$ denotes the constant total mass of the fluid body.
\item[--] The notation $\bar\rho(a)$ refers to the constant average density of the entire fluid body, defined by $\bar{\rho}(a)= \ds\frac{3\M}{4\pi a^3}$, where $\M$ is the total mass and $a$ is the radius of the spherical volume $\V$ in the reference configuration. 
\end{itemize}
Other notations are introduced and explained in the text as they appear.

\subsection{Structure and Organization of the Paper}\label{org_pap}

The main results of the paper are organized into eight sections, each addressing different aspects of the development and application of an advanced non-linear theory of finite rotational deformations in fluid heterogeneous bodies, such as planets and stars. Details of supplementary mathematical calculations are placed in the appendix, which also consists of several sections.

Section \ref{basman1} defines the unperturbed state of the rotating fluid body. The non-rotating body is considered spherically symmetric, with all functions defining hydrostatic equilibrium -- such as density, pressure, and gravitational potential -- depending solely on the radial coordinate $r$. The only external perturbation considered in this paper is the centrifugal potential, which consists of two separate harmonics: monopole $W_R$ and quadrupole $W_Q$. The monopole component $W_R$ represents purely radial perturbation, leading to merely radial deformation of the rotating body. It is reasonable to combine it with the unperturbed state of the non-rotating body. Section \ref{basman1} provides a precise mathematical definition of this reference configuration and formulates the equations of its hydrostatic equilibrium.

Section \ref{sec2} discusses the linear theory of infinitesimal perturbations of the rotating fluid caused by the quadrupole component $W_Q$ of the centrifugal potential. The theory operates on the base manifold $\mathfrak{M}\in\mathbb{R}^3$, which consists of all internal points $\x$ of the spherically symmetric volume $\V$ of the body reference configuration, including its boundary $\pd\V$. Section \ref{sec2} deals with the Eulerian variations of the fluid density $\d_{\bm\xi}\rho$, pressure $\d_{\bm\xi}p$, and gravitational potential $\delta_{\bm\xi}{\cal U}$. These variations are induced by the infinitesimal translation $\xi^i=\xi^i(\x)$ -- known as a vector diffeomorphism -- of a fluid element from its undisturbed position $\x\in\mathfrak{M}$ in the reference configuration to the displaced point $\x+{\bm\xi}\in\mathfrak{M}^\dagger$. Here, $\mathfrak{M}^\dagger$ denotes the deformed base manifold, corresponding to the three-dimensional volume $\V^\dagger$ occupied by the fluid after the external perturbation $W_Q$ has been applied. Mathematical consideration of the Eulerian variations is most effectively done within the framework of the Lie algebra $\g$ of diffeomorphisms ${\bm\xi}$, which operate in the tangent space $T\mathfrak{M}$ of the base manifold $\mathfrak{M}$. The Lie algebra $\g$ is equipped with the Lie bracket $[{\bm\xi},{\bm\eta}]$ of the vector fields ${\bm\xi}\in\g$ and ${\bm\eta}\in\g$, which naturally coincides with the Lie derivative $\pounds_{\bm\xi}{\bm\eta}=[{\bm\xi},{\bm\eta}]$ of one vector field with respect to another. The Lie derivative defines the infinitesimal Eulerian variation of density, $\d_{\bm\xi}\rho\equiv\pounds_{\bm\xi}\rho$, and pressure, $\d_{\bm\xi}p\equiv\pounds_{\bm\xi}p$, and is also used for the definition of the infinitesimal variation $\d_{\bm\xi}\U$ of the gravitational potential $\U$. We demonstrate that the density variation $\delta_{\bm\xi}\rho$ couples linearly with the perturbation of the gravitational field $\delta_{\bm\xi}\mathcal{U}$. This linear coupling plays a central role in deriving the Helmholtz equation for the total gravitational field perturbation, defined as ${\cal K} \equiv \delta_{\bm\xi}\mathcal{U} + W_Q$. This relation is also known as Molodensky's equation \citep{molodensky1953}. We further show how the gravitational field perturbation ${\cal K}$ is connected to the geometry of the level surfaces, and we employ Molodensky's equation to derive the classical Clairaut equation, which describes the figure of a slowly rotating fluid body.

Section \ref{sec3} develops the mathematical framework necessary to analyze the non-linear deformation of a rotating fluid body under the influence of the quadrupole centrifugal potential $W_Q$, which is parameterized as $W_Q \to \tau W_Q$ by the homotopy parameter $\tau \in [0,1]$. When $\tau = 0$, the rotational perturbation is absent; when $\tau = 1$, it reaches its full strength as defined by Eq. \eqref{bw5aa}.
This perturbation causes each fluid element to move from its original position $\x \in \mathfrak{M}$ to a new position $\x_\tau \in \mathfrak{M}_\tau$ in the perturbed base manifold $\mathfrak{M}_\tau$. The resulting finite displacement, represented by the vector ${\pmb{\mathscr X}}_\tau = \x_\tau - \x$, defines a diffeomorphism directed along the integral curve generated by the infinitesimal generator ${\bm\xi}\in\g$, and connecting the points $\x$ and $\x_\tau$. The magnitude of this displacement is determined by the parameter $\tau$, which is measured along the integral curve and denoted as a subscript in the associated functions.
Rotation transforms the initially spherically-symmetric volume $\V$ of the reference configuration into the perturbed volume $\V_\tau$. Since the fluid is compressible, we allow $\V_\tau \neq \V$. The finite diffeomorphism ${\pmb{\mathscr X}}_\tau$ is constructed using the exponential operator of the directional derivative, $L_{\bm\xi} \equiv \xi^i \partial_i$, applied to the coordinates of the point $\x$. These diffeomorphisms ${\pmb{\mathscr X}}_\tau \in \mathfrak{G}$ form a Lie group of diffeomorphisms, $\mathfrak{G} = \mathrm{Diff}(\mathfrak{M})$.
The exponential map is also employed to define finite Eulerian perturbations of fluid density and pressure. We examine the gauge freedom inherent in the diffeomorphism group $\mathfrak{G}$ when defining hydrostatic perturbations of fluid density. It is shown that each vector element ${\pmb{\mathscr X}}_\tau \in \mathfrak{G}$ possesses two arbitrary degrees of freedom. This freedom permits the elimination of the two tangential components of ${\pmb{\mathscr X}}_\tau$ on the unit sphere, retaining only the radial component ${\pmb{\mathscr X}}_\tau = {\mathscr X}_\tau {\bm n}$, where ${\bm n} = \x/r$ is the unit radial vector, and the scalar ${\mathscr X}_\tau={\mathscr X}_\tau(\x)$ is a function of the coordinates $\x\in\mathfrak{M}$. Adopting this radial gauge significantly simplifies the subsequent non-linear analysis of fluid and gravitational field perturbations, which is developed in the following sections.

Section \ref{gvzc34} explores the correspondence between diffeomorphisms that transport fluid elements and those that deform level surfaces, thereby altering the body's shape. It is the deformation of these level surfaces that perturbs the internal gravitational field of the rotating fluid body, transforming its initially spherically-symmetric reference configuration into a more complex geometry with volume $\V_\tau$.
A key difficulty in this analysis is that the perturbed volume $\V_\tau$ must be determined simultaneously with the perturbed density $\rho_\tau$ and gravitational field. Since the gravitational field perturbation depends on integrating $\rho_\tau$ over the unknown domain $\V_\tau$, this creates an apparent circularity. This issue is resolved by pulling back $\V_\tau$ to the known unperturbed volume $\V$ using the inverse diffeomorphism -- a technique originally introduced by Chandrasekhar in his foundational work on the equilibrium of rotating bodies \citep{chandr87}. This transformation reformulates the gravitational potential perturbation as the sum of two mathematical integrals: a volume integral over $\V$ involving $\rho_\tau$, and a surface integral over the boundary $\pd\V$ of the reference configuration, derived from the fluid's surface density. The surface integral corresponds to a solution of the homogeneous Laplace equation and must be included in the analysis.
The section further establishes a functional relationship between the finite gravitational field perturbation $K_\tau$ and the corresponding density perturbation $\varrho_\tau$. By exploiting the non-linear coupling between fluid density and gravitational potential, a closed-form, non-linear partial differential equation for $K_\tau$ is derived. While a linearized version of this equation was originally introduced by Molodensky \citep{molodensky1953}, the present work extends the formulation to the non-linear regime, allowing for the treatment of strong perturbations beyond the scope of the Clairaut linear theory.

Section \ref{sec4} presents a comprehensive nonlinear analysis of rotational deformations in fluid bodies, with a particular focus on the characteristics and significance of level surfaces of gravitational potential, density, and pressure. The section begins by examining finite (nonlinear) distortions in the geometry of these level surfaces, defined via the height function, which quantifies the radial displacement $X_\tau = X_\tau(\x)$ of a perturbed level surface from its unperturbed counterpart of radius $r$ in the reference configuration. It is demonstrated that, in the general case of a compressible fluid, $X_\tau$ does not coincide with the radial displacement ${\pmb{\mathscr X}}_\tau$ of a fluid element. To derive the governing functional equation for the height function $X_\tau$, the Lie group method is employed. From this point forward -- both within this section and throughout the remainder of the paper -- we focus on the maximal radial deformations of the level surfaces, denoted $X \equiv X_1$, corresponding to the vector flow of the Lie group of diffeomorphisms evaluated at the parameter value $\tau = 1$. The functional equation for the vector field $\X = {\bm n}X(\x)$ is solved using the mathematical formalism of the shift operator $\hat{\mathsf{T}}_\X$. The resulting solution expresses the gravitational field and density perturbations as an infinite Neumann series involving the operator $\hat{\mathsf{T}}_\X$ acting on the unperturbed gravitational potential $U(r)$ or density $\rho(r)$ of the reference configuration. By applying the Lagrange inversion theorem, the height function $X$ is further expressed as a power series in the gravitational field perturbation $K$. This formulation enables the determination of rotational deformations to arbitrary precision, provided that an analytical or numerical solution of the governing equations for $K$, as derived in the preceding section, is available. The latter part of Section \ref{sec4} introduces an alternative methodology for determining rotational deformations that circumvents the direct solution of the equation for gravitational perturbation. This approach constitutes a nonlinear extension of Clairaut's theory, advancing beyond the quadratic approximations previously developed by Darwin and de Sitter. It leverages the recursive structure of the partial derivatives of the shift operator $\hat{\mathsf{T}}_\X$, which allows any order derivative to be expressed in terms of the deformation gradient matrix $A_{ij} \equiv \pd_i X^j$, where $X^i = n^i X(\x)$ denotes the components of the vector-valued height function. By exploiting this recursive property, along with the differential equations governing the unperturbed and perturbed gravitational potentials $U$ and $K$, a master equation for the height function $X = X(\x)$ is derived. This nonlinear partial differential equation is applicable to rapidly rotating fluid bodies, where large deformations render the Clairaut and Darwin-de Sitter approximations inadequate. Additionally, the section derives the corresponding nonlinear master equation for the shape function $f := X/r$ of the rotating body.

Section \ref{sec5} addresses the decomposition of the master equations governing the radial spectral harmonics of the height and shape functions, denoted by $X_l = X_l(r)$ and $f_l = f_l(r)$, respectively. These functions depend solely on the radial coordinate $r$ and serve as coefficients in the Legendre expansions of the height, $X(r,\theta)$, and shape, $f(r,\theta)$, functions, each associated with the Legendre polynomial $P_l(\cos\theta)$ of order $l$. Due to the nonlinear nature of the equations, we apply the Wigner technique \citep{gelfand_1963} to perform spectral decomposition of nonlinear terms -- such as double, triple, and higher-order products of the height and shape functions -- into their corresponding spectral harmonics. The axial symmetry of the rotating fluid body ensures that the master equation decouples into a system of equations for harmonics with even indices, $l = 0, 2, 4, \ldots$, which can be solved iteratively. The Wigner technique, along with its associated notational framework, significantly simplifies the algebraic structure of the equations, yielding more compact and tractable forms compared to alternative approaches, such as the Zharkov-Trubitsyn theory \citep{Zharkov_1978book}. To demonstrate the efficacy of this method, we explicitly present the equations resulting from the first two iterations, which directly reproduce the classical Clairaut and Darwin-de Sitter theories. Additionally, this section includes a spectral analysis of the gravitational perturbation equations, both within the interior of the fluid body and in the surrounding vacuum domain.

Section \ref{sec6} presents the boundary conditions for the gravitational potential perturbation $K$ and the height function $X$ at the center ($r = 0$) and the outer boundary ($r = a$) of the fluid reference configuration. These conditions are expressed as series expansions in terms of Legendre polynomials. The coefficients of these expansions correspond to the matching conditions for the spectral radial harmonics of the height and shape functions. Moreover, they characterize the integrated response of fluid bodies to a specific external perturbation harmonic of degree $l$. The amplitude of this response is described by the Love number $k_l$. Each Love number $k_l$ quantifies the multipole moment (zonal harmonic $J_l$) of degree $l$ in the external gravitational potential of the rotating body. These multipole moments can also be derived independently by expanding the body's external potential in a Legendre polynomial series. A comparison between the multipole moments obtained through both approaches serves to validate the internal consistency of the theory and its agreement with alternative formulations.

Finally, Section \ref{appmac} provides an analysis of exact solutions to the nonlinear master equation governing the height and shape functions. We show that, in the case of a fluid with constant density, the solution for the shape function reproduces the classical Maclaurin and Jacobi ellipsoids. In addition, we present a second exact solution corresponding to a polytropic fluid with index unity. Our analysis confirms that the solution for the height function is consistent with the findings of previous researchers.

In summary, the proposed theory underscores the power of integrating the mathematical frameworks of Lie groups and Neumann series to both formulate and solve operator equations arising in the nonlinear analysis of rotational deformations in fluid bodies. This unified approach offers a novel and rigorous reinterpretation of Newtonian gravitational potential theory, effectively bridging classical results with modern mathematical techniques. By combining historical insights with advanced tools from functional analysis and differential geometry, the suggested theory of the figure of rotating planets and stars significantly expands our capacity to analyze their internal structure and equation of state.
This methodology not only improves the precision of theoretical predictions but also enhances our ability to interpret observational data related to planetary flattening, internal mass distribution, and stellar morphology. As a result, it opens new pathways for modeling astrophysical bodies under strong rotational influence, with promising applications in planetary science, stellar evolution, and gravitational wave astronomy.

\section{Base Manifold and Reference Configuration of a Rotating Fluid Body}\label{basman1}
\stepcounter{equationschapter}
\setcounter{equationschapter}{2}
\renewcommand{\theequation}{2.\arabic{equation}}
The challenge of determining the shape of a rigidly rotating fluid body begins with defining the underlying base manifold, which serves as the geometrical foundation for the mathematical development of the theory. The most straightforward choice for such a background manifold is a spherically-symmetric volume of a non-rotating fluid body. Indeed, a bulk of fluid with density $\rho$ and pressure $p$, influenced by its own gravity and situated in empty space, adopts a spherically symmetric shape to minimize its total energy. In this ground-energy state, the density $\rho$ and pressure $p$ depend solely on the radial coordinate $r$, and their radial profiles can be derived by solving the equations of hydrostatic equilibrium \citep{Tassoul_1978book,horedt_2004book}.

The perturbation caused by the rigid rotation of the fluid is considered an external disturbance, the magnitude of which is determined by the constant rotational velocity $\omega$. We assume that the vector ${\bm\omega}$ of the angular velocity is directed along the $z$ axis. The rotation distorts the spherically-symmetric shape of the undisturbed fluid, and the objective is to theoretically evaluate the distortion of the body's shape. It turns out that the rotational perturbation includes a spherically-symmetric (monopole) component, which does not alter the spherical symmetry of the undisturbed fluid configuration. The fluid responds to this monopole component of the perturbation by adjusting the radius of its spherically symmetric volume and redistributing the radial profiles of density and pressure, but nothing else. Therefore, it is necessary to decide whether to interpret the spherical monopole component of the rotational perturbation as part of the undisturbed body's formation.

In the course of this study, we have determined that, from the mathematical standpoint of perturbation theory, it is most optimal to consider the monopole component of the rotational perturbation as an integral part of the undisturbed configuration of the body, which we refer to as the reference configuration. The reference configuration has a fixed radius $r=a$, which depends on the constant angular velocity $\omega$ as a parameter. Non-radial distortions of the body are calculated relative to this reference configuration. This section outlines the equations used in modeling the reference configuration. We adopt a Cartesian coordinate system $\x:=(x^i)=(x^1,x^2,x^3)$, fixed to the rigidly rotating reference configuration of the fluid body. The $x^3=z$ axis is aligned with the angular velocity vector $\omega^i=(0,0,\omega)$, where $\omega$ denotes the constant angular speed of rotation.

\subsection{Density and Gravitational Potential of a Self-Gravitating Fluid Body}

We consider a massive, self-gravitating body composed of a compressible fluid with a spatially varying density distribution, denoted by $\rho=\rho(\x)$, defined within a compact domain $\V$ bounded by the surface of the body. The density function is assumed to be continuous and differentiable to the degree required for solving the governing field equations, such as Poisson's equation \eqref{tyc438} and the equation of hydrostatic equilibrium \eqref{9ab}. In the development of the perturbative formalism presented in subsequent sections, it is further assumed that $\rho(\x)$  is analytic, i.e., $\rho(\x)\in\mathbb{C}^\infty(\V)$ including up to the boundary $\pd\V$. 

While this analyticity assumption is satisfied by many physically realistic models of astrophysical bodies (excluding compact objects such as black holes), it is known that certain idealized models -- most notably polytropic stars -- exhibit singular behavior in the derivatives of the density near the surface. Specifically, for polytropic indices $n > 1$ (see Section \ref{rhe4} for definition) the first or higher-order radial derivatives of $\rho(\x)$ diverge as the surface is approached, even though the density itself vanishes smoothly at a finite radius \citep{Chandra_1939book,horedt_2004book}. This behavior raises a legitimate concern regarding the extent to which the analyticity condition constrains the applicability of the perturbation theory developed in this study.

A rigorous analysis of the convergence properties of the infinite series employed throughout the present formulation is necessary to fully address this issue.  Preliminary observations, however, suggest that the theory may remain valid even in the presence of such singularities. Notably, we have demonstrated that all surface terms involving derivatives of the density cancel identically in the expressions for physically observable quantities, such as the total mass, gravitational multipole moments, and Love numbers (see Sections~\ref{sec6} and~\ref{lov45} for details). This cancellation appears to be a structural feature of the formalism and is conjectured to persist at higher orders of approximation. If confirmed, this would imply that the requirement of finite surface derivatives of the density may be relaxed without compromising the physical validity of the results.

A more comprehensive mathematical investigation of this conjecture is currently underway. It is anticipated that treating the density as a distribution (also referred to as a generalized function \citep{Gelfand_1964,shilov_1968}), rather than as a classical function, may offer a natural extension of the present formalism to a broader class of astrophysical models, including those exhibiting non-analytic or weakly singular behavior at the boundary. Preliminary considerations regarding the application of distribution theory to a uniformly dense body are presented in Sec.~\ref{subD1}, specifically in the paragraph following Eq.~\eqref{b6v5}.

The undisturbed gravitational potential $\U=\U(\x)$ of the fluid body satisfies the Poisson equation:
\bal{tyc438}
\Delta\U=-4\pi G\rho\;,
\ea
where $\Delta=\delta^{ij}\pd_{ij}$ is the Laplace operator. A particular solution to the Poisson equation \eqref{tyc438}, which is regular at infinity, is given by
\bal{vt3s}
\U(\x)&=&G\int_\V\frac{\rho(\x')d^3x'}{|\x-\x'|}\;,
\ea 
where $\V$ is a spherical volume of the reference configuration with radius $r=a$, occupied by the fluid. 

The undisturbed density $\rho(\x)$ and potential $\U(\x)$ are spherically symmetric, meaning they depend solely on the radial coordinate $r$. This allows us to reformulate the solution \eqref{vt3s} in terms of radial integrals: 
\bal{6bb}
\U(r)&=&4\pi G\int\limits_r^{a}\rho(s)sds+\frac{G\M(r)}{r}\;,
\ea
where
\bal{pm6ad}
\M(r)&:=&4\pi\int_0^r\rho(s)s^2 ds\;.
\ea
is the fluid's mass contained within a spherical volume of radius $r$. 

An important variable defining the magnitude of the fluid's response to rotational perturbation is the ratio of the density $\rho(r)$ to the average density $\bar\rho(r)$ within a volume of radius $r$. This ratio is denoted by
\bal{vxz4}
{\upalpha}(r):=\frac{\rho(r)}{\bar\rho(r)}\;,
\ea
where the average density is
\bal{jky7}
\bar\rho(r)&:=&\frac{M(r)}{{\cal V}(r)}=\frac3{r^3}\int_0^r\rho(s)s^2ds\;.
\ea
The variable ${\upalpha}$ allows us to express the ratio of the density $\rho$ to the gravitational acceleration $\U':=d\U/dr$, as follows:
\bal{bct6}
4\pi G\frac{\rho}{\U'}=-\frac{3\upalpha}{r}=-\frac{\bar\rho'}{\bar\rho}-\frac{3}{r}=-\frac{d\ln\M(r)}{dr}\;.
\ea 
The variable $\upalpha=\upalpha(r)$ appears in the Clairaut equation \citep{Zharkov-book-1986,Jardetzky_2005}, which defines the shape of the rotating body as a function of the rotational speed $\omega$ through appropriate boundary conditions. 

\subsection{Centrifugal Potential and Rotational Parameter}

This paper considers a single type of external perturbation that deforms the volume of the fluid body: the rigid rotation of the fluid with a constant angular velocity ${\bm\omega}$ directed along the $z$ axis.  
The centrifugal potential generated by this rotation satisfies the Poisson equation:
\bal{3ab}
\Delta W=2\omega^2\;,
\ea
where $\omega^2 ={\bm\omega\cdot\bm\omega}$, and the vector ${\bm\omega}=(\omega^i)=(0,0,\omega)$ represents the angular velocity. A particular solution to \eqref{3ab} is \citep{Zharkov_1978book}:
\bal{nuc2}
W&=&\frac12\omega^2r^2\sin^2\theta\;.
\ea
The centrifugal potential \eqref{nuc2} can be decomposed in the sum of monopole, $W_R=W_R(r)$, and quadrupole, $W_Q=W_Q(r,\theta)$, components:
\bal{3ac}
W=W_R+W_Q\;.
\ea
Here, 
\bal{iii7}
W_R=\frac13\omega^2r^2\qquad\quad &,&\quad\qquad W_Q=-\frac13\omega^2r^2 P_2(\cos\theta)\;,
\ea
where $P_2(\cos\theta)$ is the Legendre polynomial of the second order. 

The monopole term, $W_R$, is solely a function of the radial coordinate $r$. Therefore, this term can induce only a uniform, spherically-symmetric deformation in the shape of the fluid body. The centrifugal monopole represents a particular solution to the Poisson equation \eqref{3ab}:
\bal{3aba}
\Delta W_R=2\omega^2\;.
\ea
The quadrupole component, $W_Q$, of the centrifugal perturbing potential satisfies the homogeneous Laplace equation,
\bal{3afk}
\Delta W_Q=0\;,
\ea
and induces non-radial deformations in the fluid body, altering its shape from a sphere to a spheroidal configuration.

It has been proposed \citep{Zharkov_1978book, Hubbard_1975ApJ} and is now widely accepted to characterize the strength of deformation induced by the centrifugal potential using a dimensionless rotational parameter: 
\bal{tttr4} 
\m&=&\frac{\omega^2a^3}{G\M}=\frac{3\omega^2}{4\pi G\bar\rho(a)}\;, 
\ea 
where $\bar\rho(a)$ is the mean density, $\M=\M(a)$ is the total mass of the body, and $a$ denotes the mean radius of the rotating fluid body, identified with the radius of the undisturbed reference configuration.
The parameter $\m$ is particularly useful because it is expressed entirely in terms of observable quantities: the angular velocity $\omega$, the mass $\M$, and the mean radius $a$. The angular velocity can be determined from the observed rotational period and radius. The mass, a key quantity in Keplerian dynamics, can be accurately inferred from the orbital motion of satellites or spacecraft. The mean radius is typically obtained through satellite altimetry, radar ranging, or stellar occultation measurements.

The rotational parameter characterizes the ratio of centrifugal acceleration at the equator, $\omega^2 a$, to the dominant gravitational acceleration at the surface, $G\M/a^2$. It can also be interpreted as an approximation of the ratio between the rotational kinetic energy and the gravitational potential energy of the body \citep{Tassoul_1978book}. In perturbative theories describing the equilibrium shape of rotating celestial bodies, successive approximations are employed to determine both the body's figure and its gravitational field. An approximation of order $n$ includes all terms proportional to $\m^{n}$ in the perturbative expansion of the governing equations used to compute the shape and gravitational multipole moments. The centrifugal potential, expressed in terms of the rotational parameter $\m$, consists of two components:
\bal{bw5}
W_R&=&\frac{\m}{3}\frac{G\M}{a}\left(\frac{r}{a}\right)^2\;,\\\label{bw5aa}
W_Q&=&-\frac{\m}{3}\frac{G\M}{a}\left(\frac{r}{a}\right)^2P_2(\cos\theta)\;.
\ea

A rotating body becomes gravitationally unbound when the centrifugal acceleration at the equator, $\omega^2 a$, equals the gravitational acceleration, $G\M/a^2$. To ensure structural stability and prevent rotational disruption, the following condition must be satisfied: $\omega^2 a\le Gm/a^2$, which limits the parameter $\m$ to: 
\bal{qwer12}
0\le\m\le 1\;.
\ea
This constraint is more stringent than the classical upper bound, $\m<1.5$, derived by Poincar\'e \citep{Poincare_1902}. A tighter limit was later proposed by Quilghini \citep{Quilghini_1959a}, who found:
\bal{poin23}
0\le\m\le 0.75\;.
\ea
In practice, the value of $\m$ is typically small for most astrophysical objects. For example, the Sun has $\m\simeq 2.2\times 10^{-5}$, reflecting its relatively slow rotation. The parameter increases for more compact and rapidly rotating objects. A millisecond pulsar with a rotational period of 1 ms has $\m=0.31$, indicating significant rotational effects. Among planets, Earth has $\m\simeq 3.5\times 10^{-3}$, while Jupiter, due to its rapid rotation and large size, reaches  $\m\simeq 0.083$. 

We can use the values of the parameter $\m$ to estimate the number of approximations needed to calculate the shape of a celestial body, ensuring consistency with the uncertainty $\delta J_2$ in the measurement of the body's second zonal harmonic (quadrupole moment) $J_2$. It is well-known \citep{Zharkov_1978book} that $J_2 \sim \m$, and its uncertainty can be estimated as $\delta J_2 \simeq \m^\alpha$, where $\alpha$ is a number less than or equal to the number of approximations required to calculate $J_2$ with accuracy compatible with the uncertainty $\delta J_2$. For example, the measurement of the solar quadrupole moment \citep{J2-Sun-2023} gives $J^{\rm Sun}_2 = (2.2 \pm 0.4) \times 10^{-7}$. This corresponds to $\delta J^{\rm Sun}_2\simeq \m_{\rm Sun}^{1.6}$, indicating that the theoretical calculation of the Sun's shape must include all terms up to the second order ($\sim \m^2$) to reach an adequate interpretation of the measurement result. Recent measurements of Jupiter's gravitational field by the Juno spacecraft \citep{J2-Jupiter_2018} showed that the quadrupole moment of Jupiter is $J_2^{\rm Jupiter} = (1.4697 \pm 0.0001) \times 10^{-2}$. The uncertainty in the measurement of $J_2$ for Jupiter is approximately $10^{-6}$, which is comparable to the value $\m_{\rm Jupiter}^{6.5}$. This suggests that the theory of the shape of the rotating Jupiter must be developed, at least, up to the seventh order in the parameter $\m$ \citep{Nettelmann_2017, Nettelmann_2021}.

It is important to note that planetary scientists often use a slightly different definition of the rotational parameter. This parameter is defined as follows:
\bal{ttr5} 
\q &= &\frac{\omega^2 R_{\rm e}^3}{G \M} \;, 
\ea
where $R_{\rm e}$ is the equatorial radius of the rotating body. Comparing the two definitions \eqref{tttr4} and \eqref{ttr5} reveals that these parameters are related by the formula:
\bal{hjuv} 
\m &= &\q \left( \frac{a}{R_{\rm e}} \right)^3 \;. 
\ea
The parameter $\m$ is more useful for performing perturbative analysis of the rotating fluid configuration, while the parameter $\q$ is used in presenting the external gravitational field of the body and its multipole moments induced by rotation \citep{Zharkov_1978book}.

\subsection{Reference Configuration of Rotating Fluid Body}

The primary objective of our study is to investigate the non-spherical deformations of the body's shape induced by the quadrupole potential $W_Q$. Accordingly, we incorporate the monopole component $W_R$ of the centrifugal potential into the set of quantities that define the spherically symmetric reference configuration of the fluid. By definition, this reference configuration occupies the volume $\V$, bounded by the surface $\pd\V$, with a radial extent $r=a$. All points $\x \in \mathcal{V} \cup \partial \mathcal{V}$ constitute the base manifold $\mathfrak{M}$ associated with the Lie group of diffeomorphisms, which will be discussed in detail later. The center of this reference configuration coincides with the body's center of mass, which also serves as the origin of the coordinate system, where the radial coordinate satisfies $r=0$.

The functions defining the reference configuration of the fluid on the base manifold $\mathfrak{M}$ depend exclusively on the radial coordinate $r$. These functions include the fluid density $\rho=\rho(r)$, pressure $p=p(r)$, the fluid's gravitational potential $\U=\U(r)$, and the monopole component $W_R=W_R(r)$ of the centrifugal potential. These two potentials are combined through a linear superposition:
\bal{bs4x} 
U(r)&=&\U(r)+W_R(r)\;, 
\ea
which defines the effective gravitational potential associated with the reference configuration of the fluid body.

In the analysis that follows, it will be convenient in several instances to introduce an auxiliary (effective) density function $\sigma=\sigma(r)$, defined by the relation:
\bal{hunb6}
\sigma(r)&:=&\rho(r)-\frac{\omega^2}{2\pi G}=\rho(r)-\frac{2\m}{3}\bar\rho(a)\;.
\ea
The effective density $\sigma(r)$ is defined only within the domain corresponding to the fluid's reference configuration, specifically for $r\le a$.
In terms of the density $\sigma$, the equation governing the effective potential $U$ becomes: 
\bal{mku6}
\Delta U&=&-4\pi G\sigma\;,
\ea
where $\Delta\equiv \delta^{ij}\frac{\pd^2}{\pd x^i\pd x^j}$ is the Laplace operator in the coordinate system $\x\in\MM$.

The solution to this Poisson equation is spherically symmetric and can be expressed explicitly using Eqs. \eqref{bs4x} and \eqref{vt3s} as:
\bal{om5} 
U(r)&=&4\pi G\int\limits_r^{a}\rho(s)sds+\frac{G\M(r)}{r}+\frac13\omega^2 r^2\;, 
\ea 
or, equivalently, in terms of the auxiliary density $\sigma$: 
\bal{7bb}
U(r)&=&4\pi G\int\limits_r^{a}\sigma(s)sds+\frac{GM(r)}{r}+\m\frac{G\M(a)}{a}\;,
\ea
where $M(r)$ denotes the mass enclosed within radius $r$, defined in terms of $\sigma$ as: 
\bal{mk8v}
M(r):=4\pi\int_0^r\sigma(s)s^2ds=\M(r)-\frac{2\omega^2}{3G} r^3=\M(r)-\frac{2\m}{3}\frac{r^3}{a^3}\M(a)\;,
\ea
Expression \eqref{7bb} for the effective potential $U$ closely resembles Eq. \eqref{6bb}, which describes the gravitational potential of the fluid.

Derivatives of the effective gravitational potential $U$ can be readily computed using Eq. \eqref{7bb}. It is particularly convenient to express these derivatives in terms of the first derivative $U'(r)$, the auxiliary density $\sigma(r)$, and the radial derivatives of the physical density $\rho(r)$. The first derivative is given by: 
\bal{yvc5} U'(r)&=&-\frac{GM(r)}{r^2}\;, 
\ea 
while the second and third derivatives take the form:
\ba\label{klm8h} 
U''(r)&=&-\frac{2U'(r)}{r}-4\pi G\sigma(r)\;,\\ 
\label{ov3as} 
U'''(r)&=&\frac{6U'(r)}{r^2}+\frac{8\pi G\sigma(r)}{r}-4\pi G\rho'(r)\;. 
\ea 
For higher-order derivatives with $n \ge 4$, the general expression is:
\bal{jn7bv}
U^{(n)}(r)&=&(-1)^{n+1}n!\Biggl\{\frac{U'(r)}{r^{n-1}}+\frac{2\pi G}{3}\left[\frac{2\sigma(r)}{r^{n-2}}+\left(1-\frac{6}{n}\right)\frac{\rho'(r)}{r^{n-3}}+\left(1-\frac{3}{n}\right)\frac{\rho''(r)}{r^{n-4}}\right]\\\nonumber
&&+4\pi G\sum_{k=1}^{n-4}\frac{(-1)^{k}}{(k+3)!}\left(1-\frac{k+3}{n}\right)\left[\frac{\rho^{(k+2)}(r)}{r^{n-k-4}}+2(k+2)\frac{\rho^{(k+1)}(r)}{r^{n-k-3}}+(k+1)(k+2)\frac{\rho^{(k)}(r)}{r^{n-k-2}}\right]\Biggr\}\;,
\ea
where $\rho^{(k)}(r):=\pd^k_r\rho(r)$ denotes the $k$-th derivative of the density function $\rho(r)$ with respect to the radial coordinate $r$.
The summation terms involving higher-order derivatives of the density function $\rho$ are independent of the angular velocity $\omega$. The dependence on $\omega$ arises solely through the terms containing $U'(r)$ and $\sigma(r)$.

We consider the case in which the fluid density at the surface of the body, denoted by $\rho(a)$, as well as its first and higher-order derivatives, remain finite and non-vanishing. Under this assumption, the values of the effective gravitational potential $U$ and its derivatives evaluated at the surface $r = a$ are determined as follows: 
\bal{oku8}
U(a)&=&\phantom{+}\frac{G\M}{a}\left(1+\frac{\m}{3}\right)\;,\\\label{cok7}
U'(a)&=&\phantom{+}\U'(a)\left(1-\frac{2\m}{3}\right)\;,\\\label{cok8}
U''(a)&=&-\frac{2\U'(a)}{a}\left(1+\frac{\m}{3}\right)-4\pi G\rho(a)\;,\\\label{cok9}
U'''(a)&=&\phantom{+}\frac{6\U'(a)}{a^2}+\frac{8\pi G\rho(a)}{a}-4\pi G\rho'(a)\;,
\ea
where $\M:=\M(a)$ is the total mass of the fluid body, which remains constant. Higher-order derivatives of the potential $U$ at the surface $r=a$ can be computed by using Eq. \eqref{jn7bv}. Notably, the higher-order surface derivatives for $n\ge 3$ are independent of the parameter $\m$ -- only the surface values of the potential $U$, its first derivative $U'$, and second derivative $U''$ exhibit explicit dependence on $\m$.

To characterize the fluid's response to rotational perturbations, we introduce a variable on the base manifold $\mathfrak{M}$, analogous to Eq. \eqref{vxz4}, but defined in terms of the auxiliary density $\sigma$ of the reference configuration. This variable is denoted by: 
\bal{v4}
{\upbeta}(r):=\frac{\sigma(r)}{\bar\sigma(r)}\;,
\ea
where the average value of $\sigma$ over the volume of radius $r$ is given by: 
\bal{y7}
\bar\sigma(r)&:=&\frac3{r^3}\int_0^r\sigma(s)s^2ds=\bar\rho(r)-\frac{2\m}{3}\bar\rho(a)\;.
\ea  
The first derivative $U'(r)$ of the effective gravitational potential $U$, the auxiliary density $\sigma(r)$, and the variable $\upbeta(r)$ are related through the identity: 
\bal{bni5}
4\pi G\frac{\sigma}{U'}=-\frac{3\upbeta}{r}\;,
\ea
which mirrors the structure of Eq. \eqref{bct6}.
Furthermore, the variables $\upalpha$, defined in Eq. \eqref{vxz4}, and $\upbeta(r)$ are interconnected via the series expansion:
\bal{uu2}
\upbeta&=&\upalpha+(\upalpha -1)\sum_{n=1}^\infty\left[\frac{2\m}{3}\frac{\bar\rho(a)}{\bar\rho(r)}\right]^n\;.
\ea
The variable $\upbeta$ plays a dual role in the analysis of rotational deformations of fluid bodies. It serves both as a physical parameter quantifying the fluid's response to rotational perturbations and as a mathematical construct that simplifies the governing equations. Unlike earlier formulations that relied on the parameter $\upalpha$, which required explicit retention of all terms in the expansion series (as shown in Eq. \eqref{uu2}), the use of $\upbeta$ significantly reduces algebraic complexity and enhances the tractability of higher-order perturbative calculations.

\subsection{Mathematical Formulation of Hydrostatic Equilibrium in the Reference Configuration}\label{rhe4}

The reference configuration of a rotating fluid plays a central role in the computation of rotational deformations via perturbation theory. The defining parameters of this configuration -- such as the radius, internal density distribution, and gravitational field -- are fundamentally governed by the fluid's equation of state, which dictates the balance between pressure and gravity under hydrostatic equilibrium. In general, the equation of state expresses pressure as a function of density $\rho$, temperature $T$, and other thermodynamic variables \citep{Landau1987}. However, in this analysis, we restrict our attention to a barotropic equation of state, wherein pressure depends solely on density, i.e., $p = p(\rho)$. This assumption is consistent with the condition of rigid rotation in the fluid \citep{Tassoul_1978book}.

We further assume that the fluid density $\rho$ in the reference configuration is inhomogeneous and varies only with the radial coordinate, such that $\rho := \rho(r)$. Given the barotropic assumption, the pressure likewise becomes a function of the radial coordinate alone, $p = p(r)$. The density is permitted to vary continuously throughout the fluid interior and is assumed to possess non-zero derivatives at the surface, reflecting a physically realistic boundary behavior consistent with the equation of state. The unperturbed reference state is characterized by an isotropic stress tensor, 
\bal{1}
t_{ij}&=&-p\delta_{ij} \;,
\ea
where $\delta_{ij}={\rm diag}(1,1,1)$ denotes the identity matrix and $p$ is the pressure. 

The fluid in the reference configuration satisfies the equation of hydrostatic equilibrium \citep{Zharkov_1978book}:
\bal{9ab}
\rho\pd_i U+\pd_jt^{ij}&=&0\;,
\ea
where $U$ denotes the effective gravitational potential, as defined in Eqs. \eqref{bs4x}--\eqref{7bb}. Since all quantities in Eq. \eqref{9ab} depend solely on the radial coordinate $r$, the equation simplifies to: 
\bal{w9ab}
\rho U'-p'&=&0\;,
\ea
where, here and throughout the paper, a prime denotes differentiation with respect to $r$, e.g., $U' = dU/dr$.

For a fluid obeying a polytropic equation of state, $p = \mathsf{K}_0 \rho^{1 + 1/n}$, where $\mathsf{K}_0$ and $n$ (the polytropic index) are constants, Eq. \eqref{w9ab} is typically reformulated into the Lane-Emden equation \citep{lemden}. This is achieved by introducing a dimensionless function $\Theta = \Theta(r)$ such that $\rho = \rho_0 \Theta^n$, where $\rho_0$ is the central density. Substituting this into the polytropic equation of state yields:
\bal{eqst39}
p&=&p_0\Theta^{n+1}\;,
\ea
with $p_0 = \rho_0 \mathsf{K}_0$. Combining Eqs. \eqref{mku6}, \eqref{w9ab}, and \eqref{eqst39} leads to a second-order ordinary differential equation for $\Theta$, known as the Lane-Emden equation \citep{Tassoul_1978book, horedt_2004book}. This equation is solved subject to the boundary conditions $\Theta(0) = 1$ and $\Theta'(0) = 0$. An additional boundary condition, $\Theta(a) = 0$, ensures that the density vanishes at the surface $r = a$, thereby defining the radius of the fluid body. The resulting solutions for $\Theta$ provide the radial profiles of pressure and density and are referred to as polytropes of index $n$ \citep{Tassoul_1978book, horedt_2004book}.

In the subsequent analysis, we assume that the solution to the hydrostatic equilibrium equations for the reference configuration -- describing the radial profiles $\rho = \rho(r)$, $p = p(r)$, and $U = U(r)$ -- is known exactly, either through analytical expressions or high-precision numerical integration. Accordingly, all required radial derivatives of these functions, which are essential for the perturbative treatment developed in later sections, are also assumed to be accurately determined.

\section{Linear Perturbations of the Base Manifold}\label{sec2}
\stepcounter{equationschapter}
\setcounter{equationschapter}{3}
\renewcommand{\theequation}{3.\arabic{equation}}

A distinctive feature of this study is the application of the formalism of the Lie group of fluid diffeomorphisms to develop a nonlinear extension of Clairaut's theory of the figure for rotating fluid bodies. The Lie group of fluid diffeomorphisms acts on the base manifold $\mathfrak{M}$, which represents the reference configuration of the fluid. Rotation induces displacements of fluid elements from their positions in this reference configuration, resulting in a deformation of the base manifold. These diffeomorphisms encapsulate both the magnitude and direction of the deformation, making them a powerful tool for analyzing changes in the shape of the rotating body through Lie group techniques. This approach, inspired by the foundational ideas of V. Arnold \citep{arnold, khesin}, is systematically developed in the subsequent sections of the paper.

\subsection{Fluid Diffeomorphisms and Their Lie Algebra}

We consider a stationary, rotating fluid body described in a rigidly rotating Cartesian coordinate system $\x = (x^i)$ that is comoving with the fluid. In this frame, the fluid velocity vanishes. The unperturbed state of the fluid is defined by incorporating the radial component \eqref{bw5} of the centrifugal potential $W_R$ into the reference configuration, which occupies a spherically symmetric volume $\mathcal{V}$ of radius $a$. The union of all interior and boundary points of this volume constitutes the base manifold $\mathfrak{M}$: $\x \in \mathcal{V} \cup \partial \mathcal{V}$.

The non-radial, quadrupole component \eqref{bw5aa} of the centrifugal potential $W_Q$ induces a physical deformation of the volume $\mathcal{V}$, transforming it into a spheroidal configuration. This chapter focuses on the regime of slow rotation, where the deformation remains small and can be treated within the linearized framework of Clairaut's theory, which neglects nonlinear effects. The perturbations considered here arise from the stationary rotation of an astrophysical fluid body (e.g., a gaseous planet or star) and are therefore time-independent. Consequently, all functions in this and subsequent sections depend solely on the spatial coordinates $\x$, and time derivatives are excluded from the analysis.

The perturbing potential $W_Q$ displaces each fluid element (or fluid parcel \citep{fparcel}) from its equilibrium position $\x = x^i$ to a perturbed position described by
\bal{4} w^i = x^i + \xi^i\;, 
\ea 
where $x^i$ are the Lagrangian coordinates of the fluid parcel \citep{frefframe}, $w^i$ are the corresponding Eulerian coordinates, and $\xi^i = \xi^i(\x)$ is the infinitesimal Lagrangian displacement vector -- also referred to as an infinitesimal fluid diffeomorphism \citep{chandr87, Tassoul_1978book}. The displacement magnitude is proportional to the perturbation $W_Q$, i.e., $|{\bm \xi}| \sim \m$, where $\m$ is the rotational parameter defined in \eqref{tttr4}. The precise magnitude and direction of the displacement vector ${\bm \xi} = (\xi^i)$ are not known a priori and must be determined by solving the governing differential equations, which will be addressed in later sections. The collection of all such vectors ${\bm \xi}$ at each point $\x \in \mathfrak{M}$ forms the tangent space $T_{\x} \mathfrak{M}$ to the base manifold.

This tangent space can be endowed with the structure of a Lie algebra  $\g_\x$ of infinitesimal diffeomorphisms, which provides a geometric framework for analyzing perturbations in the fluid and gravitational fields. The Lie algebra $\g_\x$ is defined by equipping the tangent space with a Lie bracket $[{\bm \xi}, {\bm \eta}]$ for any two vector fields ${\bm \xi} \in \g_\x$ and ${\bm \eta} \in \g_\x$, given by the Lie derivative: $[{\bm \xi}, {\bm \eta}] := \L_{\bm \xi} {\bm \eta}=-\L_{\bm\eta} {\bm \xi}$. This bracket satisfies the Jacobi identity, a fundamental property of Lie algebras \citep{arnold}. The collection of all tangent spaces $T_\x \mathfrak{M}$ forms the tangent (vector) bundle $T \mathfrak{M}$.
  
The corresponding collection of Lie algebras $\g_\x$ over the manifold defines a (tangent) Lie algebroid $\g \rightarrow \mathfrak{M}$ \citep{Mackenzie_2005}. The Lie bracket structure on $\mathfrak{M}$ is preserved via the anchor map, which, in the case of the tangent algebroid, is defined by the Leibniz rule \citep{Mackenzie_2005}: 
\bal{anch2} 
[{\bm \xi}, f {\bm \eta}] &= {\bm \xi}(f) {\bm \eta} + f [{\bm \xi}, {\bm \eta}]\;, 
\ea 
where $f \in \mathbb{C}^1$ is a smooth scalar function, and ${\bm \xi}(f) = \xi^i \pd_i f$ denotes the action of the vector field ${\bm \xi}$ on $f$. 

\subsection{Level Surfaces: Key to the Geometry of Rotating Fluid Bodies}\label{levs8}

In a rotating fluid body, three distinct types of three-dimensional surfaces are commonly recognized. Surfaces on which the total gravitational potential $V$ remains constant are termed equipotential or level surfaces. Surfaces of constant fluid density $\rho$ are referred to as {\it isopycnal} surfaces, while those of constant pressure $p$ are known as {\it isobaric} surfaces. It is well established that, under hydrostatic equilibrium and assuming the angular velocity of the fluid's steady rotation is independent of the vertical coordinate $z$, these three surfaces coincide. In such a configuration, each surface simultaneously maintains constant values of gravitational potential, density, and pressure \citep{Tassoul_1978book}. For brevity, we shall refer to these coinciding surfaces collectively as the {\it level surface}.

Level surfaces are fundamental to the analysis of both the internal structure and the overall shape of a rotating fluid body \citep{Zharkov_1978book, Vanicek_book}. The collection of these surfaces exhibits several important geometric and physical characteristics that are essential for understanding the equilibrium configuration of such bodies \citep{MacMillan1930}:
\begin{itemize}
\item[]\textbf{Continuity} -- Level surfaces are continuous and free of any breaks or discontinuities.
\item[]\textbf{Convexity} --  Each level surface is convex at every point\footnote{This condition may be violated in configurations involving rapid rotation. In the present analysis, we assume that the angular velocity is sufficiently small to preserve the convexity of the level surfaces.}. That is, in a neighborhood of any point on the surface, the local curvature is non-negative, implying that the surface either curves outward or remains locally planar.
\item[]\textbf{Non-intersection} -- Level surfaces do not intersect and are nested concentrically, resembling the pattern of tree rings in a cross-section of a trunk. 
\item[]\textbf{Smooth curvature} -- The local curvature of the surfaces varies smoothly, except possibly at the outer boundary where abrupt changes in matter density may occur.
\item[]\textbf{Centering} -- The geometric center of the entire family of level surfaces coincides with the center of mass of the fluid body.
\end{itemize}
The boundary of a fluid body is defined by the condition of vanishing pressure: $p({\bm x})\left|_{\rm surface}\right.=0$. In fluid bodies lacking a boundary layer, this condition also implies that the surface density vanishes \citep{Tassoul_1978book}: $\rho({\bm x})\left|_{\rm surface}\right.=0$. For a fluid in hydrostatic equilibrium, the surfaces of constant pressure and density coincide with equipotential surfaces of the gravitational field. As a result, the boundary of the fluid body also corresponds to an equipotential surface. Therefore, the shape of a fluid body in hydrostatic equilibrium is governed by the structure of the gravitational field's level surfaces within the interior, along with the boundary conditions imposed on the gravitational potential and its first derivatives at the surface of the rotating body  \citep{Zharkov_1978book}.

The central objective of the theory of hydrostatic perturbations in rotating fluid bodies is to establish the governing principles and equations that describe the geometry of perturbed level surfaces within the body's interior. These perturbations arise in response to the centrifugal force and depend on the fluid density $\rho$, the equation of state $p = p(\rho)$, and the angular velocity of rotation $\omega$.
Before delving into the details of this perturbation theory, it is essential to clarify the relationship between fluid diffeomorphisms --representing the displacement of individual fluid parcels -- and the diffeomorphisms associated with the deformation of level surfaces induced by rotation. In the reference configuration of the base manifold $\MM$, each level surface comprises fluid parcels sharing identical values of density, pressure, and gravitational potential. The quadrupole component of the centrifugal potential, $W_Q$, displaces a fluid parcel from its original position $\x$ to a new location ${\bm w}$, as described by Eq. \eqref{4}. However, the surface formed by the new positions of these displaced parcels may not constitute a level surface, since the density may vary differently in different radial directions. This implies that the fluid diffeomorphism $\xi^i$ alone cannot, in general, capture the deformation of level surfaces in a compressible fluid.

To address this issue, we introduce a new infinitesimal (geometric) diffeomorphism, denoted by $\zeta^i = \zeta^i(\x)$, which represents the displacement of a level surface element initially located at the point $\x \in \MM$ to a new position:
\bal{qno2}
y^i&=&x^i+\zeta^i\;.
\ea
In the linearized framework, this expression characterizes the geometry of the perturbed level surfaces, thereby describing the deformation of the fluid body relative to the spherical level surfaces in the reference configuration. The alteration in the geometry of a level surface is captured by the vector field $\zeta^i(\x)$, whose magnitude is proportional to the rotational parameter $\m$.

As will be demonstrated below, the fluid diffeomorphism $\xi^i$ is related to the geometric diffeomorphism $\zeta^i$ via a first-order partial differential equation -- see Eq. \eqref{wcx41}. Importantly, the identity $\zeta^i = \xi^i$ holds only in the special case of an incompressible fluid, where the displacement field is divergence-free, i.e., $\theta := \partial_i \xi^i = 0$. Both types of diffeomorphisms play fundamental roles in perturbation theory and will be employed as appropriate, depending on the specific characteristics of the problem under investigation.

\subsection{Hydrostatic Perturbations}\label{hdr77}
The external perturbation $W_Q$ induces variations in all relevant physical quantities, including the density, pressure, and gravitational potential of the body. These variations can be categorized into two types, corresponding to the fluid diffeomorphism \eqref{4}. 

The {\it Lagrangian} perturbation of the density, denoted $\Delta_{\bm \xi} \rho$, characterizes the change in density experienced by a fluid element as it is displaced from its original position $\xi^i$ to a new position $w^i = x^i + \xi^i$:
\bal{hb7g} \Delta_{\bm \xi} \rho &:=& \rho^\dagger({\bm w}) - \rho({\bm x})\;, \ea 
where $\rho^\dagger({\bm w})$ and $\rho({\bm x})$ represent the perturbed and unperturbed densities of the same fluid element, evaluated at positions ${\bm w} = {\bm x} + {\bm \xi}$ and ${\bm x}$, respectively. In contrast, the {\it Eulerian} perturbation, $\delta_{\bm \xi} \rho$, describes the change in density at a fixed point on the base manifold $\MM$: 
\bal{5} \delta_{\bm \xi} \rho &:=& \rho^\dagger({\bm x}) - \rho({\bm x}) \;, \ea 
where both $\rho^\dagger({\bm x})$ and $\rho({\bm x})$ are evaluated at the same spatial location ${\bm x} \in \MM$. The underlying principle of both Lagrangian and Eulerian perturbations is that the variation in density, governed by the infinitesimal displacement vector ${\bm \xi}$, must satisfy the fluid continuity equation. This equation can be expressed in either Lagrangian coordinates $x^i$, which track the position of a fluid parcel in the unperturbed configuration, or Eulerian coordinates $w^i$, which describe the perturbed configuration of the base manifold \citep{Tassoul_1978book}. In the subsequent analysis, we adopt the Lagrangian coordinate framework $x^i$ for our mathematical formulation. This choice aligns with the methodology of Lie groups and offers distinct advantages in the treatment of nonlinear perturbations, which will be discussed in detail in the following sections.

We consider a barotropic fluid, in which the pressure $p$ depends solely on the density $\rho$ via an equation of state of the form $p = p(\rho)$. By applying the chain rule to this functional relationship, the infinitesimal Eulerian variation of pressure is given by: 
\ba\label{5a} \delta_{\bm \xi} p &:=& p^\dagger({\bm x}) - p({\bm x}) = \frac{\pd p}{\pd \rho} \delta_{\bm \xi} \rho \;, \ea 
where $p^\dagger({\bm x}) := p\left[\rho^\dagger({\bm x})\right]$ denotes the perturbed pressure evaluated at the same spatial point ${\bm x} \in \MM$ in the fluid's reference configuration, and $p({\bm x}) := p\left[\rho({\bm x})\right]$ is the corresponding unperturbed value.

The infinitesimal variations of density and pressure are closely associated with the Lie derivative \citep{kopeikin_2011book, Petrov_2017book} along the vector field ${\bm \xi} \in \g_\x$, and are expressed as: 
\bal{6} 
\delta_{\bm \xi} \rho &:= &\pounds_{\bm \xi} \rho \;, \\ 
\label{ligt5v} \delta_{\bm \xi} p &:=& \frac{\pd p}{\pd \rho} \pounds_{\bm \xi} \rho \;, \ea 
where $\pd_i := \pd / \pd x^i$ denotes partial differentiation. Here, the Lie derivative of the fluid density is defined as \citep{Petrov_2017book}:
\bal{lid4} 
\pounds_{\bm \xi} \rho &=& -\pd_i \left( \xi^i \rho \right) = -\xi^i \pd_i \rho - \rho \pd_i \xi^i \;. 
\ea 
The expression for the Lie derivative of the fluid density reflects the fact that $\rho$ is not a conventional scalar field, as it incorporates the infinitesimal volume element of a fluid parcel, which depends on the divergence of the vector field ${\bm \xi}$, defined as $\theta := \partial_i \xi^i$. Consequently, as the parcel is advected along ${\bm \xi}$, both its particle content and volume vary. The Lie derivative of $\rho$ thus captures both the directional derivative of the density and the divergence of ${\bm \xi}$. This formulation is consistent with the fluid continuity equation \citep{chandr87, Tassoul_1978book}. Notably, the present analysis assumes a compressible fluid, implying $\theta \neq 0$.

The variation of pressure, $\delta_{\bm \xi} p$, can be reformulated by introducing the adiabatic bulk modulus \citep{Landau1959}, defined as: 
\bal{t5}
\uplambda&:=&\rho\left(\frac{\pd p}{\pd\rho}\right)_S\;,
\ea
where the partial derivative $\left( \pd p / \pd \rho \right)_S$ is evaluated at constant entropy $S$. Since we consider a stationary rotational configuration without heat exchange, the entropy remains constant throughout the fluid. Consequently, the subscript $S$ may be omitted in all subsequent thermodynamic derivatives.
Utilizing Eqs. \eqref{6} and \eqref{t5}, the Eulerian variation of pressure becomes: \bal{t21} \delta_{\bm \xi} p = -\xi^i \pd_i p - \uplambda\, \pd_i \xi^i \;. \ea

The deformation of the stress tensor \eqref{1}, in the most general case, receives contributions from variations in pressure $p$, the metric tensor $\delta_{ij}$, and changes in the thermodynamic state of matter due to the displacement of material elements from their equilibrium positions \citep{Landau1959}:  
\bal{ta5x3}
\delta_{\bm\xi}t_{ij}&:=&t^\dagger_{ij}({\bm x})-t_{ij}({\bm x})=\left(-\xi^p\pd_k p
-\uplambda u_{kk}\right)\delta_{ij}-2(\upmu-p) u_{ij}\;,
\ea
where $\upmu$ is the shear modulus, and the strain tensor is defined as
\bal{7a}
u_{ij}:=\frac{1}{2}\left(\pd_i\xi_j+\pd_j\xi_i\right)\;.
\ea
Here, $\uplambda$ denotes the bulk modulus, characterizing resistance to compression, while $\upmu$ quantifies resistance to shear deformation. In the general theory of elasticity \citep{Landau1959, Audoly_2010}, the strain tensor may include nonlinear terms quadratic in the derivatives of the displacement vector $\xi^i$. However, within the linearized regime, such terms are neglected. This linearization effectively combines the contributions of $p$ and $\upmu$ in the stress tensor, leading to the definition of an {\it effective shear modulus}: 
\bal{axe3z}
\upmu_{\rm eff}&=&\upmu-p\;,
\ea
which vanishes in the case of a static fluid, i.e., $\upmu_{\rm eff} = 0$. Consequently, for a rigidly rotating fluid, the deformation of the stress tensor \eqref{1} reduces to a purely volumetric (bulk) response:
\bal{7}
\delta_{\bm\xi}t_{ij}&:=&
\left(-\xi^p\pd_k p-\uplambda u_{kk}\right)\delta_{ij}=
\delta_{\bm\xi}p\,\delta_{ij}\;,
\ea
which is directly proportional to the Lie derivative of the fluid pressure.

It is important to emphasize that the variation of the stress tensor in a solid body or a viscous fluid does not, in general, coincide with the Lie derivative of the unperturbed stress tensor, as expressed in Eq. \eqref{7}. This discrepancy arises because, in these more general media, the stress tensor depends explicitly on the effective shear modulus ($\upmu_{\rm eff} \neq 0$) or on the viscosity coefficient \citep{Landau1959, Tassoul_1978book}. While this issue is absent in the case of an ideal (inviscid) fluid, it can manifest in viscous fluids, particularly under differential rotation.

However, when a viscous fluid undergoes rigid-body rotation, the viscosity-dependent terms vanish, and the variation of the stress tensor remains consistent with Eq. \eqref{7} \citep{Tassoul_1978book}. Therefore, the formalism developed here is applicable not only to ideal fluids but also to viscous fluids, provided the rotation is rigid. In such cases, contributions from viscosity and effective shear modulus are absent, regardless of the fluid's internal constitution.

\subsection{Gravitational Perturbations}\label{uuu23}

The unperturbed gravitational potential $\U := \U({\bm x})$ of the fluid body satisfies the Poisson equation \eqref{tyc438}, whose solution is given by Eq. \eqref{vt3s} as an integral over the volume $\V$ of the base manifold $\MM$. In the perturbed configuration, the gravitational potential $\U^\dagger := \U^\dagger({\bm x})$ satisfies the modified Poisson equation: 
\bal{rv6x}
\Delta \U^\dagger=-4\pi G\rho^\dagger\;,
\ea
where $\rho^\dagger$ denotes the perturbed density of the fluid, now occupying a deformed volume $\V^\dagger$ due to the influence of the quadrupole component of the centrifugal potential $W_Q$. A particular solution to Eq. \eqref{rv6x} is given by: 
\bal{8a}
\U^\dagger({\bm x})&=&G\int\limits_{\V^\dagger}\frac{\rho^\dagger({\bm y}')d^3y'}{|{\bm x}-{\bm y}'|}\;,
\ea
where the integration is performed over coordinates ${\bm y}' \in \mathbb{R}^3$. Although the perturbed potential $\U^\dagger$ is evaluated at the same spatial point ${\bm x} \in \mathbb{R}^3$ as the unperturbed potential $\U$, a direct comparison between Eqs. \eqref{vt3s} and \eqref{8a} is not immediately possible due to the difference in integration domains. To facilitate this comparison, the integration volume $\V^\dagger$ in Eq. \eqref{8a} must be mapped back to the unperturbed spherical volume $\V$ of the base manifold $\MM$ using the pullback diffeomorphism.

Recall that a barotropic, rotating fluid in hydrostatic equilibrium occupies a volume whose boundary coincides with a level surface of constant density, pressure, and gravitational potential \citep{MacMillan1930, Tassoul_1978book}. In the presence of rotational perturbations, the level surfaces of the gravitational potential are described by the position vector ${\bm y} = (y^i)$, defined by Eq. \eqref{qno2} as $\y = \x + {\bm\zeta}$, where ${\bm x}$ denotes a point on the unperturbed level surface, and ${\bm \zeta} = (\zeta^i)$ is an infinitesimal displacement vector belonging to the Lie algebra $\mathfrak{g}_{\bm x}$. The vector ${\bm \zeta}$ can be determined, for example, by requiring that the total potential remains constant on the perturbed surface: $U^\dagger({\bm x} + {\bm\zeta}) + W_Q({\bm x} + {\bm\zeta}) = {\rm const}$.

This approach is employed in the Zharkov-Trubitsyn theory of level surfaces in rotating bodies \citep{Zharkov_1978book}, which assumes that ${\bm\zeta}$ has only a radial component, i.e., ${\bm \zeta} = \zeta {\bm n}$, where ${\bm n}$ is a unit radial vector and $\zeta = \zeta(\x)$ is a scalar function defined on $\x \in \MM$. This assumption arises from interpreting Eq. \eqref{qno2} in spherical coordinates $(r, \theta, \varphi)$ as an infinitesimal transformation of a level surface from a sphere of radius $r$ to a more general surface $R = r + \zeta(r, \theta, \varphi)$, clearly indicating that ${\bm\zeta}$ is purely radial. In contrast, we adopt a more general method for computing the components of the deformation vector $\zeta^i$, which is outlined in detail in Section \ref{popo9}.

At this stage, it is useful to recall that the vector field ${\bm\zeta} \in \g_\x$ introduced in Eq. \eqref{qno2} is distinct from the vector field ${\bm\xi} \in \g_\x$ defined in Eq. \eqref{4}. The vector ${\bm\xi}$ represents the displacement of a fluid parcel, described by the transformation $\x \to {\bm w} = \x + {\bm\xi}$, whereas ${\bm\zeta}$ characterizes the displacement of a point on a level surface, given by $\x \to \y = \x + {\bm\zeta}$. By construction, the displacement ${\bm\xi}$ ensures mass conservation for the fluid element under external perturbations. In contrast, ${\bm\zeta}$ preserves the values of fluid density, pressure, and gravitational potential on the perturbed level surface, maintaining them equal to those in the reference configuration -- see Section \ref{popo9} for further discussion. These two vector fields coincide only under the assumptions of fluid incompressibility and volume conservation, which are not adopted in the present analysis.

Building on these considerations, we adopt the pullback transformation of the level surfaces, $\y' = \x' + {\bm\zeta}'$, as defined in Eq. \eqref{qno2} and applied at the point $\x'$. In our formulation, the infinitesimal displacement vector ${\bm\zeta}'$ is not restricted to being radial; rather, it may be any smooth vector field that transforms the level surfaces. This leads to the following expression for the perturbed potential:
\bal{8aqw} 
\U^\dagger({\bm x})&=&G\int\limits_{\V}\frac{\rho^\dagger({\bm x}'+{\bm\zeta}')}{|{\bm x}-({\bm x}'+{\bm\zeta}')|}\left(1+\frac{\pd\zeta'^i}{\pd x'^i}\right)d^3x'\;, 
\ea 
where $\zeta'^i := \zeta^i(\x')$, and all terms quadratic in ${\bm\zeta}$  have been omitted. 

All subsequent calculations are performed to linear order in the displacement vector ${\bm\zeta}$. Expanding the integrand in Eq. \eqref{8aqw} in a Taylor series about the point ${\bm x}'$ and retaining only terms linear in $\zeta^i$, we obtain: 
\bal{8juy} 
\U^\dagger({\bm x})&=&G\int\limits_{\V}\frac{\rho^\dagger({\bm x}')d^3x'}{|{\bm x}-{\bm x}'|}+G\int\limits_{\V}\frac{\pd}{\pd x'^i}\left[\frac{\rho({\bm x}')\zeta^i(x')}{|{\bm x}-{\bm x}'|}\right]d^3x'\;, \ea 
where, in the second integral, the difference between the perturbed and unperturbed fluid densities has been neglected, as it is of order ${\cal O}(\zeta)$ and the integrand already contains the infinitesimal displacement ${\bm\zeta}$. The divergence in the second term on the right-hand side of Eq. \eqref{8juy} can be converted into a surface integral via Gauss's theorem, yielding:
\bal{a8juy} 
\U^\dagger({\bm x})&=&G\int\limits_{\V}\frac{\rho^\dagger({\bm x}')d^3x'}{|{\bm x}-{\bm x}'|}+G\oint\limits_{\mathbb{S}^2}\frac{\rho({\bm a})\zeta^i({\bm a})}{|{\bm x}-{\bm a}|}d^2S_i\;, 
\ea 
where $d^2S_i := d^2S_i({\bm a})$ denotes the oriented surface element at the point $\x' = {\bm a}$, and $\mathbb{S}^2$ indicates that the integration is performed over the boundary $\partial \V$ of the volume $\V$ occupied by the fluid in the unperturbed reference configuration, which is a sphere $\mathbb{S}^2$ of radius $a = |{\bm a}|$.

Expression \eqref{a8juy} represents the same perturbed potential $\U^\dagger$ as given in Eq. \eqref{rv6x}, but now expressed as the sum of a volume integral and a surface integral. The volume integral in Eq. \eqref{a8juy} is evaluated over the unperturbed, spherically symmetric volume $\V$ of the reference configuration, where all integration points lie on the base manifold, i.e., $\x' \in \MM$. This allows for a direct comparison with the unperturbed potential $\U$, since both integrals are now computed over the same spatial domain. The surface integral in Eq. \eqref{a8juy} corresponds to a solution of the Laplace equation, with the fluid density $\rho({\bm a})$ evaluated on the spherical boundary $\partial \V=\mathbb{S}^2$ of the unperturbed configuration. If the surface density vanishes -- as is the case for certain idealized models such as polytropic stars -- this surface term becomes zero. However, in more general scenarios where the fluid possesses a non-zero surface density layer, the surface integral cannot be neglected and must be included in the analysis. The evaluation of this surface contribution can be deferred to the solution of the boundary conditions that match the interior and exterior solutions of the perturbed gravitational potential $\U^\dagger$ at the spherical boundary $r = a$ of the reference configuration.

In general, the variation of the gravitational potential, $\delta_{\bm\xi} \U$, induced by the external perturbation of the centrifugal potential $W_Q$, can be defined as the difference between the perturbed and unperturbed volume integrals: 
\bal{8c} 
\delta_{\bm\xi} \U := G\int\limits_{\V}\frac{\rho^\dagger({\bm x}')d^3x'}{|{\bm x}-{\bm x}'|}-G\int\limits_{\V}\frac{\rho({\bm x}')d^3x'}{|{\bm x}-{\bm x}'|} = G\int\limits_{\V} \frac{\rho^\dagger({\bm x}') - \rho({\bm x}')}{|{\bm x} - {\bm x}'|} d^3x' = G\int\limits_{\V} \frac{\delta_{{\bm\xi}} \rho({\bm x}')}{|{\bm x} - {\bm x}'|} d^3x'\;, 
\ea
demonstrating that the variation $\delta_{\bm\xi} \U$ is a functional of the Eulerian density perturbation $\delta_{\bm\xi} \rho$, as introduced in Eq. \eqref{5}.
The definition in Eq. \eqref{8c} is fully consistent with the expression in Eq. \eqref{a8juy}. Specifically, the final integral in Eq. \eqref{8c} represents a particular solution to the Poisson equation:
\bal{9}
\Delta\left(\delta_{\bm\xi}\U\right)&=&-4\pi G\delta_{\bm\xi}\rho\;,
\ea
where $\Delta := \delta^{ij}\pd_i\pd_j$ denotes the Laplace operator. The general solution to Eq. \eqref{9} is defined up to a solution of the homogeneous Laplace equation, which corresponds to the surface integral in Eq. \eqref{a8juy}. The value of this surface term will be determined later by applying the appropriate boundary conditions that match the interior and exterior solutions of the perturbed gravitational potential $\U^\dagger$ at the spherical boundary $r = a$ of the reference configuration.

It is important to emphasize that the diffeomorphism defined in Eq. \eqref{4} is induced by the external centrifugal potential $W_Q$ and applies exclusively to the coordinates of the fluid parcels. Consequently, the Eulerian variation operator $\delta_{\bm\xi}$ does not act on the external perturbation itself \citep{chandr87}. This implies that the Eulerian variations of the external potentials vanish: 
\bal{nbvrt7} 
\delta_{\bm\xi} W_R &=& 0 \qquad,\qquad \delta_{\bm\xi} W_Q = 0\;. 
\ea 
The physical interpretation of Eq. \eqref{nbvrt7} is that the fluid's perturbations do not exert any back-reaction on the external potential $W=W_R+W_Q$. As a result, the Eulerian variation of the effective potential $U$, as defined in Eq. \eqref{bs4x}, coincides with the variation of the gravitational potential $\U$: 
\bal{va5c} 
\delta_{\bm\xi} U &=& \delta_{\bm\xi} \U\;. 
\ea
This property will be utilized later in the development of the nonlinear perturbation theory describing the shape of a rotating fluid body.

\subsection{Coupling of Density Variation with Gravitational Perturbation}\label{sec23}

It is essential to elucidate the relationship between the Eulerian variation of density, $\delta_{\bm\xi}\rho$, and the corresponding variation in the effective gravitational potential, $\delta_{\bm\xi} U$. This relationship has already been partially established through the integral equation \eqref{8c}, which is mathematically equivalent to the Poisson differential equation \eqref{9}. In the context of a barotropic equation of state, the coupling between the density and gravitational field perturbations becomes more direct, governed by a linear algebraic relation. This relation represents the first integral of the perturbed hydrostatic equilibrium equation for a rotating fluid. We derive this fundamental equation in this section.

To this end, we consider the linearized perturbation of the hydrostatic equilibrium equation \eqref{9ab}, induced by the quadrupole component $W_Q$ of the centrifugal potential. This perturbation is obtained by taking the Eulerian variation of the equilibrium equation and equating it to the external force arising from $W_Q$:
\bal{10}
\delta_{\bm\xi}\left(\rho\pd_i U+\pd_jt^{ij}\right)=-\rho\pd_i W_Q\;.
\ea
The variation of the product $\rho \pd_i U$ follows the Leibniz rule.  When applying this rule, it is important to note that the operations of taking the Eulerian variation and partial differentiation commute, regardless of the geometric nature of the quantity involved \citep{Petrov_2017book}:
\bal{lie765}
\pd_i\delta_{\bm\xi}&=&\delta_{\bm\xi}\pd_i\;.
\ea
Applying the Leibniz rule and incorporating Eqs. \eqref{t21}, \eqref{7}, \eqref{nbvrt7}, and \eqref{lie765}, we recast Eq. \eqref{10} into the following form:
\bal{11}
\delta_{\bm\xi}\rho\pd_i U+\rho\pd_i(\delta_{\bm\xi}U+W_Q)+\pd_i(\xi^j\pd_j p)+\pd_i\left(\uplambda\,\pd_j\xi^j\right)=0\;,
\ea
where the gradients of $\delta_{\bm\xi}U$ and $W_Q$ have been combined into a single term for clarity.

We now take into account that the functions $\rho = \rho(r)$, $p = p(r)$, and $U = U(r)$ in Eq. \eqref{11} correspond to the unperturbed configuration of the base manifold $\MM$ and depend solely on the radial coordinate $r$. This allows us to express all partial derivatives of these functions in terms of derivatives with respect to $r$. Specifically, we use the identity $\pd_i U = U' n^i$, where the prime denotes differentiation with respect to the radial coordinate, and $n^i = x^i / r$ is the radial unit vector. With this substitution, the terms in Eq. \eqref{11} can be grouped accordingly, and the equation takes the following form:
\bal{11yu}
\left(U'\delta_{\bm\xi}\rho-\rho'{\cal K}\right)n^i
+\pd_i\left(\rho{\cal K}+\xi^j\pd_j p+\uplambda\,\pd_j\xi^j\right)=0\;,
\ea 
where, once again, the prime denotes differentiation with respect to $r$. We have also introduced the notation
\bal{za46}
{\cal K}(\x)&:=&\delta_{\bm\xi}U(\x)+W_Q(\x)\;,
\ea
to represent the total gravitational perturbation at a point $\x$ on the base manifold $\MM$.

It is noteworthy that the terms involving pressure and the bulk modulus in Eq. \eqref{11} can be expressed in terms of the Eulerian variation of density. Specifically, we have:
\bal{rt74}
\xi^j\pd_j p+\uplambda\,\pd_j\xi^j=\frac{\pd p}{\pd\rho}\xi^j\pd_j\rho+\uplambda\,\pd_j\xi^j=\frac{\uplambda}{\rho}\left[\xi^j\pd_j\rho+\rho\pd_j\xi^j \right]=-\frac{\uplambda}{\rho}\delta_{\bm\xi}\rho\;.
\ea
The right-hand side of Eq. \eqref{rt74} can be further simplified using the Adams-Williamson equation, which describes hydrostatic equilibrium in spherically symmetric configurations \citep{Zharkov_1978book}:
\bal{ko9h}
\rho U'-\frac{\uplambda}{\rho}\rho'=0\;.
\ea
This equation is equivalent, in the case of a barotropic fluid, to the standard hydrostatic equilibrium condition \eqref{w9ab}. This equivalence can be demonstrated by invoking Eq. \eqref{t5}, which yields:
\bal{t6}
\frac{\uplambda}{\rho}\rho'=\frac{\pd p}{\pd\rho}\rho'=p'\;,
\ea
reducing Eq. \eqref{ko9h} to $\rho U' - p' = 0$, which is precisely the form of Eq. \eqref{w9ab} for a self-gravitating fluid sphere in hydrostatic equilibrium.

By incorporating Eq. \eqref{ko9h}, we can rewrite Eq. \eqref{rt74} in the following form:
\bal{mket67}
\xi^j\pd_j p+\uplambda\,\pd_j\xi^j=-\frac{\rho}{\rho'}\,U'\,\delta_{\bm\xi}\rho\;.
\ea
Substituting this expression into Eq. \eqref{11yu} simplifies the equation to:
\bal{11ws}
\left[\frac{U'}{\rho'}\delta_{\bm\xi}\rho-{\cal K}\right]\rho'n^i
-\pd_i\left[\rho\left(\frac{U'}{\rho'}\delta_{\bm\xi}\rho-{\cal K}\right)\right]=0\;.
\ea
Taking the partial derivative in the second term and canceling like terms yields:
\bal{11pp}
\rho\pd_i\left[{\cal K}-\frac{U'}{\rho'}\delta_{\bm\xi}\rho\right]=0\;,
\ea
which represents the gradient form of the perturbed hydrostatic equilibrium equation \eqref{11}, describing the interplay between variations in matter variables and the gravitational field. This equation can be integrated, leading to the first integral:
\bal{io9n6v}
{\cal K}-\frac{U'}{\rho'}\delta_{\bm\xi}\rho&=&{\rm const.}
\ea
The constant of integration must be zero, since in the absence of the external perturbation $W_Q$, the left-hand side of Eq. \eqref{io9n6v} vanishes identically.

Thus, the first integral \eqref{io9n6v} of the variational form of the hydrostatic equilibrium equation \eqref{10} establishes a direct relationship between the first Eulerian variation of fluid density and the Eulerian variation of the gravitational potential:
\bal{za44}
\delta_{\bm\xi}\rho=\frac{\rho'}{U'}{\cal K}\;.
\ea
This fundamental result was originally derived by M.S. Molodensky \citep{molodensky1953}. It is important to emphasize that Eq. \eqref{za44} is valid exclusively for bodies composed of stationary, rotating fluids without differential rotation. This restriction arises from the absence of an effective shear modulus in fluids, i.e., $\upmu_{\rm eff} = 0$. In cases involving differentially rotating viscous fluids or solid bodies, Eq. \eqref{za44} must be appropriately modified to incorporate additional terms accounting for bulk viscosity or shear elasticity.

\subsection{Molodensky's Equation for Gravitational Perturbation}\label{y1ub23}

We have established that the variation in the gravitational field satisfies the Poisson equation \eqref{9}, where the variation in density appears on the right-hand side as the source term for the perturbation in the gravitational potential. Additionally, Eq. \eqref{za44} provides a linear relationship between the density variation and the gravitational field, enabling us to reformulate Eq. \eqref{9} into a Helmholtz-type equation that governs only the variation in the gravitational field.

To proceed, we observe that $W_Q$ satisfies the Laplace equation \eqref{3afk}, allowing Eq. \eqref{9} to be equivalently expressed as: 
\bal{xv56}
\Delta\left(\delta_{\bm\xi}U\right)&=&-4\pi G\delta_{\bm\xi}\rho-\Delta W_Q\;.
\ea
By combining the Laplacian terms and applying the definition \eqref{za46} for the total perturbation ${\cal K}$ of the gravitational field induced by the body's rotation, we recast Eq. \eqref{xv56} into the form:
\bal{him4x3}
\Delta{\cal K}+4\pi G\delta_{\bm\xi}\rho&=&0\;.
\ea
This equation can be further simplified by substituting the expression for the density perturbation $\delta_{\bm\xi} \rho$, given in terms of ${\cal K}$ via Eq. \eqref{za44}. This substitution transforms Eq. \eqref{him4x3} into the homogeneous Helmholtz equation for the gravitational field perturbation ${\cal K} = {\cal K}(\x)$: 
\bal{za45}
\Delta{\cal K}+\kappa^2 {\cal K}&=&0\;,
\ea
where the coefficient 
\bal{mu53x}
\kappa&:=&\sqrt{\frac{4\pi G\rho'}{U'}}\;,
\ea 
is a function of the radial coordinate, $\kappa=\kappa(r)$, defined by the internal density distribution $\rho = \rho(r)$ and the effective gravitational potential $U = U(r)$. Here, the prime denotes differentiation with respect to the radial coordinate. The dimensionality of $\kappa$ is $[\kappa] = (\rm{length})^{-1}$.

Equation \eqref{za45}, known as the Molodensky equation for gravitational field perturbations \citep{molodensky1953}, forms the foundation of Molodensky's theory of the Earth's figure. This theory aims to determine the Earth's shape and gravity field based on measurements of gravitational anomalies at the topographic surface \citep{Molodensky_1958}. It is widely applied in geodesy for computing gravitational anomalies and determining heights above the (quasi)geoid \citep{Guo2023}, and its relevance has recently been explored in planetary science \citep{Tenzer_2018}.

Although it is possible to analyze the boundary conditions associated with the linearized theory to fully resolve the perturbations in density, pressure, and gravitational field, this approach proves insufficient for modeling the shape and gravitational field of rapidly rotating massive planets and stars \citep{Zharkov_1978book,Zharkov-book-1986,Hubbard-book}. These celestial bodies require the inclusion of non-linear perturbations due to their high rotational velocities \citep{Zharkov_1970,zharkov_1976SvA,Nettelmann_2021}. Therefore, we will not pursue the technical elaboration of the linearized theory further. Instead, we will focus on the fundamental principles and key results of the non-linear theory, which constitutes the primary objective of this study.

\subsection{Exploring Infinitesimal Distortions of Level Surfaces: The Clairaut Equation}\label{popo9}

The level surface of the effective gravitational potential, denoted by $U = U({\bm x})$, is defined as the set of points where the potential assumes a constant value. Owing to the spherical symmetry of the unperturbed potential, $U({\bm x}) = U(r)$, these level surfaces form a family of concentric spheres. Each sphere can be uniquely identified by a continuous parameter, most naturally chosen as the radial coordinate $r$, which varies from $r = 0$ at the center of the body to $r = a$ at the spherical boundary of the reference configuration.

The introduction of the rotational quadrupole perturbation $W_Q$ modifies the gravitational potential from $U$ to 
\bal{jh68}
U^\dagger({\bm x})&:=&U({\bm x})+{\cal K}({\bm x})\;,
\ea
where $U({\bm x})$ and ${\cal K}({\bm x})$ are defined in Eqs. \eqref{bs4x} and \eqref{za46}, respectively. This modification manifests in two distinct ways: first, through the direct addition of the perturbation potential $W_Q$ to the effective gravitational potential $U$ at each point ${\bm x}$; and second, through the induced variation in the body's potential, $\delta_{\bm \xi} U$, resulting from the density perturbation $\delta_{\bm \xi} \rho$ caused by the centrifugal effects associated with $W_Q$.

The coordinates of points on the level surfaces of the perturbed gravitational potential $U^\dagger$ are denoted by the vector $\y = (y^i)$, as defined in Eq. \eqref{qno2}. To determine the translation vector ${\bm \zeta}$ appearing in this expression, we derive the functional equation governing the deformed level surface. This equation follows from Eq. \eqref{za44}, which encapsulates the condition of hydrostatic equilibrium under perturbations and relates the variation ${\cal K}$ in the gravitational field to the perturbation in the fluid's density.

This equation can be recast in the following form: 
\bal{ash9b}
{\cal K}({\bm x})&=&-\zeta(\x) U'(r)=-\zeta^i(\x)\pd_iU(r)\;,
\ea
where the displacement vector
\bal{un7v5}
\zeta^i(\x)&=&n^i\zeta(\x)+\zeta^i_\perp\;,
\ea
 is expressed as the sum of two components. The first term, $n^i\zeta(\x)$ is purely radial, while the second term, $\zeta^i_\perp$, is tangential to the radial direction and satisfies $n^i\zeta^i_\perp=0$. The scalar function $\zeta$ is given by:
\bal{wcx41}
\zeta&=&-\frac{\delta_{\bm\xi}\rho}{\rho}=\xi+\frac{\rho}{\rho'}\theta\;,
\ea
where $\theta = \pd_i \xi^i$ denotes the divergence of the fluid diffeomorphism vector field $\xi^i \in \g_\x$. 

The equation of hydrostatic equilibrium \eqref{ash9b} imposes no constraint on the tangential component $\zeta^i_\perp$ which remains undetermined. This reflects a gauge freedom in the choice of the vector field $\zeta^i$. It is convenient to fix this gauge by setting $\zeta^i_\perp=0$, corresponding to the adoption of the radial gauge, which is also used to restrict the freedom in the fluid diffeomorphism ${\bm\xi}$. Since the magnitude of the fluid diffeomorphism scales with the rotation parameter, $\xi \sim \m$, the corresponding magnitude of $\zeta$ also scales as $\zeta \sim \m$ as follows from Eq. \eqref{wcx41}. Here, $\m$ denotes the parameter characterizing the rotational speed of the body, as defined in Eq. \eqref{tttr4}.

By substituting Eq. \eqref{ash9b} into Eq. \eqref{jh68} and applying the Taylor expansion of the potential $U(\x)$, Eq. \eqref{jh68} can be reformulated to express the perturbed potential in terms of a shifted argument of the effective gravitational potential:  
\bal{mn6vc3}
U({\bm x})+{\cal K}({\bm x})&=&U(\x)-\zeta^i\pd_iU(r)=U(\x-{\bm\zeta})+{\cal O}\left(\zeta^2\right)\;,
\ea
where ${\bm \zeta} = (\zeta^i)$ and $\zeta = \zeta({\bm x})$ represents the radial displacement of the original (spherical) level surface due to the perturbation induced by the gravitational field ${\cal K}$. This displacement vector is the same infinitesimal translation introduced earlier in Eq. \eqref{qno2}. Its connection to the generator ${\bm \xi}$ is established in Eq. \eqref{wcx41}, which reveals that the difference between ${\bm \zeta}$ and ${\bm \xi}$ arises from the divergence $\theta = \pd_i \xi^i$ of the vector field ${\bm \xi}$.

Equation \eqref{mn6vc3} can be equivalently expressed in terms of a push-forward transformation as: 
\bal{6vc4}
U(\x+{\bm\zeta})+{\cal K}(\x+{\bm\zeta})&=&U(\x)\;,
\ea
where we have used the fact that, under the linear approximation, ${\cal K}(\x) = {\cal K}(\x + {\bm \zeta})$. This formulation replaces, in the linear regime, the condition given in Eq. \eqref{xa1}, which was originally employed in the Zharkov-Trubitsyn theory to determine the distortion ${\bm \zeta}$ of the level surfaces. 

The infinitesimal radial displacement $\zeta$ of the level surface in our framework can be determined via two distinct approaches. The first method utilizes Eq. \eqref{ash9b}, which yields the direct expression:
\bal{bruns}
\zeta&=&-\frac{{\cal K}}{U'}\;.
\ea
This formulation for $\zeta = \zeta(\x)$ is applicable when the effective potential $U = U(r)$ is known from the solution of the reference configuration, and the perturbing potential ${\cal K} = {\cal K}(\x)$ is available from the solution of the Molodensky equation \eqref{za45}. In the context of geodesy, Eq. \eqref{bruns} is recognized as Bruns' theorem \cite{Vanicek_book, torge2012}.

In cases where the perturbing potential ${\cal K}$ is not explicitly known, an alternative approach can be employed to determine the radial shift $\zeta$ without relying on Eq. \eqref{bruns}. This method involves deriving a differential equation for $\zeta$ in which ${\cal K}$ does not appear explicitly. Such an equation can be obtained directly from the Molodensky equation \eqref{za45}, which governs the gravitational perturbation ${\cal K}$.

By substituting the expression for ${\cal K}$ from Eq. \eqref{ash9b} into the Molodensky equation \eqref{za45} and carrying out the differentiation, we obtain the following differential equation for the radial displacement $\zeta$: 
\bal{mn5c32}
\Delta\zeta-\frac{4\zeta'}{r}-8\pi G\sigma\frac{\zeta'}{U'}+\frac{2\zeta}{r^2}&=&0\;,
\ea
where Eqs. \eqref{klm8h} and \eqref{ov3as} have been used to eliminate the second and third derivatives of the potential $U$, respectively. Eq. \eqref{mn5c32} has been also derived by S. Molodensky \citep{molodensky1953}.

This equation can be further simplified by introducing the shape function $s = \zeta / r$, which admits a decomposition into a series of Legendre polynomials:
\bal{io8n6}
s&=&\sum_{l=0}^\infty s_l(r)P_l(\cos\theta)\;.
\ea 
Substituting this expansion into Eq. \eqref{mn5c32} yields the governing equation for the spectral harmonics $s_l$ of the shape function: 
\bal{vf5x2}
s_l''+\frac{6\upbeta}{r}\left(s'_l+\frac{s_l}{r}\right)-\frac{l(l+1)s_l}{r^2}&=&0\;,
\ea
where Eq. \eqref{bni5} has been applied, and the function $\upbeta = \upbeta(r)$ is defined in Eq. \eqref{v4}.

Equation \eqref{vf5x2} is, in fact, the classical Clairaut equation \citep{Jardetzky_2005, Zharkov_1978book} governing the spectral harmonics $s_l$ ($l \ge 2$) of the shape function $s = s(r, \theta)$. Solving this equation with appropriate boundary conditions on $s_l$ and its radial derivative $s'_l = ds_l/dr$ determines the geometric configuration of the level surfaces. The derivation of the Clairaut equation \eqref{vf5x2} within the linearized approximation -- based on the small parameter $\m$ characterizing the body's rotational speed -- is attributed to S.M. Molodensky \citep{molodensky1953}.

The next step in our analysis is to extend this linearized Clairaut theory to higher-order approximations and ultimately to formulate a fully non-linear theory describing the figure of a rotating fluid body.

\section{Nonlinear Hydrostatic Perturbations}\label{sec3}
\stepcounter{equationschapter}
\setcounter{equationschapter}{4}
\renewcommand{\theequation}{4.\arabic{equation}}

The previous section presented a linearized theory of rotational perturbations in a fluid body, formulated in terms of infinitesimal diffeomorphisms of fluid parcels and level surfaces, along with first-order Eulerian variations in fluid density, pressure, and the gravitational field. These quantities were defined on the tangent bundle $T\MM$ of the base manifold $\MM$. While this linear theory is adequate for modeling the rotational deformation of slowly rotating celestial bodies -- where perturbations are of the same order as the small parameter $\m$ -- it becomes insufficient for rapidly rotating bodies such as Jupiter, where the deformation reaches a significant magnitude, approximately 10\% of the planetary radius \citep{J2-Jupiter_2018, Nettelmann_2021}. In such cases, the perturbation theory must incorporate higher-order effects in $\m$ to accurately describe small-scale variations in the gravitational field and to compute large (finite rather than infinitesimal) deformations of the body's shape.

This section develops a non-linear theory of hydrostatic perturbations in the density distribution of a rotating fluid body. It begins with a formulation of finite fluid diffeomorphisms and their associated Lie group structure, and proceeds to derive the finite, non-linear Eulerian perturbations in density, pressure, and the gravitational field.

\subsection{Fluid Diffeomorphisms: From Lie Algebra to Lie Group}\label{fdx45}

The advanced theory of non-linear hydrostatic perturbations necessitates a generalization of the infinitesimal Eulerian variations in a fluid's density, pressure, and gravitational field to their finite counterparts. This extension enables a more comprehensive analytical treatment of finite deformations within the fluid body, accommodating large-scale perturbations and yielding a more accurate mathematical representation of fluid behavior under the extreme conditions characteristic of planetary and stellar interiors. By incorporating finite variations, the theoretical framework can effectively capture the influence of external forces on the intrinsic displacements of fluid parcels. This leads to the development of enhanced analytical models and more precise numerical simulations, particularly in the context of hydrostatic studies concerning the internal structure, equilibrium shape, and gravitational field of rotating planets and stars.

The vector flow $\x_\tau$ is generated by the vector field ${\bm\xi}$. This flow induces a finite diffeomorphism, denoted by $\phi_\tau$, which maps the unperturbed base manifold $\MM$ to the perturbed manifold $\MM_\tau$, i.e., $\phi_\tau: \MM \to \MM_\tau$. The base manifold $\MM$ is compact and has a volume $\V = {\rm Vol}(\MM)$, including its boundary. Correspondingly, the perturbed manifold $\MM_\tau$ has volume ${\rm Vol}(\MM_\tau)$. It is important to note that, in the general case of a compressible fluid, the volume ${\rm Vol}(\MM)$ of the base manifold may differ from the volume ${\rm Vol}(\MM_\tau)$ of the perturbed manifold due to the deformation induced by the flow, such as rotational motion.

The parameter $\tau$ in the vector flow parameterizes the strength of the external centrifugal perturbation $W_Q$ through a linear mapping: $W_Q \to \tau W_Q$. This is equivalent to introducing a $\tau$-dependent rotational parameter: $\m \to \tau \m$, where $\m$ is fixed and defined in Eq. \eqref{tttr4}. When $\tau = 0$, the quadrupole centrifugal perturbation is absent, corresponding to the undisturbed reference configuration. When $\tau = 1$, the full quadrupole perturbation of interest, $W_Q$, is present. Intermediate values $0 < \tau < 1$ allow us to consider a continuous family of quadrupole centrifugal perturbations, interpolating from $W_Q = 0$ (at $\tau = 0$) to the full perturbation $W_Q$ (at $\tau = 1$).

Additionally, $\tau$ parameterizes the rotationally induced perturbation of the gravitational field of the reference configuration, denoted by $U_\tau = U_\tau(\x)$ (see Eq. \eqref{b5z3}). This perturbation includes both linear and non-linear terms in $\tau$ and, as shown later in Eq. \eqref{cc87}, is expressed using an exponential mapping.

It is important to emphasize that $\tau$ should not be interpreted as time. Instead, we work within the Lagrangian coordinate system $\x$, which rigidly co-rotates with the fluid. In this frame, all functions defined on the base manifold $\MM$ depend solely on the spatial coordinates $\x$, and thus all time derivatives vanish identically. Within this framework, $\tau$ serves as a homotopy parameter \citep{homotopy}, interpolating between two continuous functions: the unperturbed gravitational potential $U(\x)$ of the reference configuration (see Section \ref{basman1}), and the fully perturbed gravitational potential $V(\x) := U(\x) + K(\x)$. Here, $K(\x)$ encapsulates the total perturbation, including both the gravitational field variation and the quadrupole centrifugal potential, as detailed later in Eq. \eqref{gt6vc}. This homotopy defines a smooth transition from the unperturbed base manifold $\MM = \MM_0$ to the perturbed manifold $\MM_\tau$. It is represented by a continuous function $H(\tau, \x): \MM \times [0,1] \to \MM_\tau$, mapping the product space $\MM \times [0,1]$ into $\MM_\tau$, such that $H(0, \x) = U(\x)$ and $H(1, \x) = V(\x)$ for all $\x \in \MM$.

The parameter $\tau$ thus serves as a valuable organizational tool in the perturbation analysis. In the previous section, we implicitly set $\tau = 1$, omitting it from the notation, as we were working within the framework of linear perturbation theory -- where distinguishing between different orders of $\tau$ is unnecessary. However, this distinction becomes crucial in the context of non-linear theory, where higher-order contributions must be systematically tracked and analyzed.

Each vector flow is a smooth map: ${\bm x}\rightarrow{\bm x}_\tau=\phi_\tau{\bm x}$, generated by a vector field $\xi^i = \xi^i(\x)$, which is tangent to the integral curve of the flow at the point $\x \in \MM$. This vector field defines the entire congruence associated with the deformation of the fluid body. The flow map is governed by the ordinary differential equation: 
\bal{f1}
\frac{dx^i_\tau}{d\tau}=\xi^i\left(\x_\tau\right)\;,
\ea
with the initial condition $\x_0 = \x$. Under standard regularity assumptions, this initial value problem admits a unique solution. Assuming the vector field is analytic over the entire base manifold $\MM$, Eq. \eqref{f1} can be solved locally around the initial point $\x = \x_0$ using a Taylor expansion: 
\bal{jknt5c}
x^i_\tau=x^i+\tau{\dot x}^i+\frac12\tau^2\ddot{x}^i+\ldots\;,
\ea
where the derivatives are computed using Eq. \eqref{f1}, yielding $\dot{x}^i = \xi^i(\x)$, $\ddot{x}^i = \xi^j(\x) \partial_j \xi^i(\x)$, and so on. This leads to the exponential map: 
\bal{f2}
x^i_\tau &=&\exp\left(\tau L_{\bm\xi}\right)x^i=\sum_{n=0}^\infty\frac{\tau^n}{n!}L_{\bm\xi}^nx^i\;,
\ea
where $L_{\bm\xi} := \xi^i \partial_i$ is the directional derivative along the tangent vector ${\bm\xi} = (\xi^i)$ taken at the intiial point $\x=\x_0$. All evaluations of $\xi^i$ and its derivatives are taken at the initial point $x^i = x^i_0$. The flow map ${\bm x} \mapsto {\bm x}_\tau=\phi_\tau\x$ thus extends the elements ${\bm\xi}$ of the Lie algebra $\g_x$ to a one-parameter Lie group ${\mathfrak G}_\tau$. The collection of all such one-parameter subgroups over the base manifold forms the full Lie group ${\mathfrak G}=\cup{\mathfrak G}_\tau$ of diffeomorphisms on $\MM$.

Equation \eqref{f2} can be interpreted as describing a finite translation of a fluid parcel from its unperturbed position ${\bm x} \in {\mathbb R}^3$ to a new position ${\bm w} := \x_\tau \in {\mathbb R}^3$:  
\bal{f3v}
w^i=x^i+{\mathscr X}_\tau^i(\x)\;,
\ea
where the vector ${\pmb{\mathscr X}}_\tau=({\mathscr X}_\tau^i)$ represents the finite displacement of the fluid parcel and is given by the exponential map:
\bal{f3}
{\mathscr X}_\tau^i(\x)=\left[\exp\left(\tau L_{\bm\xi}\right)-1\right]x^i=\sum_{n=1}^\infty\frac{\tau^n}{n!}L_{\bm\xi}^nx^i\;.
\ea
This expression can be recast using the identity $\xi^i = L_{\bm\xi} x^i$, yielding: 
\bal{4c5}
{\mathscr X}_\tau^i(\x)&=&\sum_{n=0}^\infty\frac{\tau^{n+1}}{(n+1)!}L_{\bm\xi}^n\xi^i\;.
\ea
The vector ${\mathscr X}_\tau^i\in{\mathfrak G}_\tau$ generalizes the infinitesimal translation $\xi^i\in\g_\x$ introduced in Eq. \eqref{4} to the domain of a finite size. Here, the coordinates $w^i\in\MM_\tau$ represent the Eulerian position of the fluid element, while $x^i \in \MM$ are its Lagrangian coordinates on the base manifold $\MM$. 

The collection of all independent vector fields ${\bm\xi} = (\xi^i)$ forms a Lie algebra $\mathfrak{g}$ associated with the diffeomorphism group on $\MM$. The structure of this Lie algebra is fully characterized by the Lie derivative operator $\pounds_{\bm\xi}$, which defines the Lie bracket between any two vector fields. The vector field ${\pmb{\mathscr X}}_\tau$ is an element of the Lie group associated with the Lie algebra of vector fields ${\bm\xi}$ via the exponential map \eqref{f3} \citep{khesin}. This exponential map \eqref{f2} extends the Lie bracket structure across the entire base manifold.

\subsection{Hydrostatic Perturbations as an Exponential Map}\label{hhh89}

The finite Lagrangian variation of the fluid density, denoted by $\rho$, is defined as
\bal{yu36}
\Delta_\tau\rho:=\rho_\tau\left(\x_\tau\right)-\rho(\x)\;,
\ea
where $\rho(\x)$ represents the unperturbed density of a fluid parcel located at the point $\x$ on the reference manifold $\MM$, and $\rho_\tau(\x_\tau)$ denotes the perturbed density of the same fluid parcel, now located at the point $\x_\tau$ on the deformed manifold $\MM_\tau$ due to a perturbation.

In general, $\rho_\tau(\x_\tau) \neq \rho(\x_\tau)$, as the fluid is compressible and its density may vary along the flow induced by the perturbation. The definition in \eqref{yu36} inherently involves quantities defined on two distinct manifolds, $\MM$ and $\MM_\tau$, which complicates analytical treatment -- particularly because the geometric structure of the deformed manifold $\MM_\tau$ is typically unknown a priori. In light of these complications, Lagrangian variations such as \eqref{yu36} are not utilized in this manuscript in the formulation of the theory of figures for rotating fluid bodies.

From a mathematical standpoint, it is more convenient to work with the finite Eulerian variation of the density, denoted by 
\bal{b6vc}
\varrho_\tau &:=&\rho_\tau(\x)-\rho(\x)\;,
\ea
where 
\bal{xcw70}
\rho_\tau(\x)&=&\rho_\tau\left(\phi^{-1}_\tau\x_\tau\right)\;,
\ea
represents the pullback of the perturbed density $\rho_\tau(\x_\tau)$ from the point $\x_\tau \in \MM_\tau$ to the corresponding point $\x \in \MM$. This pullback is defined via the inverse diffeomorphism $\x_\tau \to \x=\phi^{-1}_\tau\x_\tau$, which is obtained from Eq. \eqref{f2} as $\x = \exp(-\tau L_{\bm\xi}) \x_\tau$. The Eulerian variation \eqref{b6vc} is thus entirely defined on the reference manifold $\MM$, thereby circumventing the need to work with coordinates $\x_\tau$ on the deformed manifold $\MM_\tau$. This makes it more suitable for analytical developments, particularly in contexts where the geometry of $\MM_\tau$ is not explicitly known.

A finite Eulerian perturbation of a fluid attribute (such as density or pressure) is defined by extending the concept of the linear Eulerian variation, $\delta_{\bm\xi}$, into the nonlinear regime induced by the finite displacement vector field given in Eq. \eqref{f3}. According to Eq. \eqref{6}, the linearized variation of the density, $\delta_{\bm\xi} \rho$, corresponds to the Lie derivative of $\rho$ along the vector field ${\bm\xi}$, thereby revealing the underlying structure of the Lie algebra generated by ${\bm\xi}$.

To construct a nonlinear extension, the Lie algebra is promoted to its associated Lie group, $\mathfrak{g} \mapsto \mathfrak{G}$, via the exponential map \citep{Petrov_2017book}, defined by
\bal{41bq}
\exp\left(\tau{\delta_{\bm\xi}}\right)&:=&\sum_{n=0}^\infty\frac{\tau^n}{n!}\delta^n_{\bm\xi}=1+\tau\delta_{\bm\xi} +\frac{\tau^2}{2!}\delta^2_{\bm\xi} +\frac{\tau^3}{3!}\delta^3_{\bm\xi} +...\;.
\ea
Since the first-order Eulerian variation commutes with partial derivatives with respect to spatial coordinates, as shown in Eq. \eqref{lie765}, the exponential map inherits this commutation property: 
\bal{f4}
\exp\left({\delta_{\bm\xi}}\right)\pd_i &=&\pd_i\exp\left({\delta_{\bm\xi}}\right)\;.
\ea
Using the exponential map, the pullback of the density $\rho$ to the reference configuration is defined as
\bal{mnk8g}
\rho_\tau(\x)&=&\exp\left(\tau{\delta_{\bm\xi}}\right)\rho(\x)=\sum_{n=0}^\infty\frac{\tau^n}{n!}\delta^n_{\bm\xi}\rho(\x)\;.
\ea
This definition ensures that the perturbed density satisfies the continuity equation along the integral curve $\x_\tau$:
\bal{b2tv5}
\frac{\pd\rho_\tau(\x)}{\pd\tau}&=&\delta_{\bm\xi}\rho_\tau(\x)\;,
\ea
where the variational operator $\delta_{\bm\xi} \rho_\tau$ is interpreted as the Lie derivative, $\delta_{\bm\xi} \rho_\tau = \pounds_{\bm\xi} \rho_\tau = -\partial_i(\xi^i \rho_\tau)$, with the minus sign reflecting the pullback transformation. Equation \eqref{b2tv5} expresses the conservation of mass for a fluid parcel as it is transported along the flow field.

The finite Eulerian variation of the density, as defined in Eq. \eqref{b6vc}, can thus be expressed via the exponential map as
\ba\label{ty74x}
\varrho_\tau&:=&\bigl[\exp\left(\tau{\delta_{\bm\xi}}\right)-1\bigr]\rho=\sum_{n=1}^\infty\frac{\tau^n}{n!}\delta^n_{\bm\xi}\rho\;.
\ea
This formulation generalizes the linear perturbation theory expressed in Eq. \eqref{6} to the nonlinear regime. In particular, the linear Eulerian variation is recovered as the leading-order term: $\delta_{\bm\xi}\rho=\lim\limits_{\tau\to 0}\pd_\tau\varrho_\tau$. 

The finite Eulerian perturbation of pressure is defined via the equation of state as
\bal{oa9}
p_\tau(\x)&:=&p(\rho_\tau(\x))=p\left(e^{\tau\delta_{\bm\xi}}\rho(\x)\right)\;.
\ea
By the equivariance property of the Lie group of diffeomorphisms (as discussed in Appendix \ref{appA1}), the composition of the equation of state with the exponential map satisfies: $p\left(e^{\tau\delta_{\bm\xi}}\rho(\x)\right)=e^{\tau\delta_{\bm\xi}}p(\rho(\x))$. This implies that the finite perturbation of pressure is also governed by the exponential map:
\bal{oa4c}
p_\tau(\x)&=&\exp\left(\tau{\delta_{\bm\xi}}\right)p(\x)=\sum_{n=0}^\infty\frac{\tau^n}{n!}\delta^n_{\bm\xi}p(\x)\;.
\ea
The higher-order variations $\delta_{\bm\xi}^n p$ can be computed recursively using the equation of state $p = p(\rho)$ and successive applications of the chain rule for composite functions. For instance: 
\bal{pon4}
\delta_{\bm\xi}p&=&\frac{\pd p}{\pd\rho}\delta_{\bm\xi}\rho\;,\\
\delta^2_{\bm\xi}p&=&\frac{\pd^2 p}{\pd\rho^2}\left(\delta_{\bm\xi}\rho\right)^2+\frac{\pd p}{\pd\rho}\delta^2_{\bm\xi}\rho\;,
\ea
and so on. For general higher-order variations appearing in the expansion \eqref{oa4c}, Fa\`a di Bruno's formula \citep{deBruno} is employed to systematically compute the derivatives of composite functions. This procedure is further elaborated in Section \ref{nco51}.

\subsection{Gauge Freedom in Fluid Diffeomorphisms}\label{kji6v5}

\subsubsection{Gauge Freedom in Lie Algebra Generators}\label{bu2y}

The perturbation in density governs corresponding perturbations in both pressure and the gravitational field of the fluid body. Typically, an infinitesimal density perturbation of the fluid, denoted by $\delta_{\bm{\xi}}\rho$, is attributed to an infinitesimal displacement ${\bm{\xi}} \in \mathcal{G}$ of the fluid element, which transports it from the point $\x$ to the point ${\bm w}$, as described in Eq. \eqref{4}.

However, this interpretation warrants careful consideration. In the macroscopic framework of continuum mechanics, where the fluid is modeled as a continuous medium, it is possible for the same physical density perturbation, $\delta_{\bm{\xi}}\rho$, to arise from different infinitesimal displacements of fluid elements. This phenomenon, known as {\it gauge freedom}, is well recognized in cosmological studies of the large-scale structure of the universe \citep{Mukhanov_1992PhR,Kopeikin_2013PhRvD}, and was also identified in the Newtonian perturbation theory of rotating fluid bodies by Friedman and Schutz \citep{Friedman_1978ApJ}.

Gauge freedom is characterized by a transformation of the generator of the Lie algebra:
\bal{nbv4c}
{\bm\xi}\mapsto{\bm\chi}={\bm\xi}+{\bm\eta}\;,
\ea 
where the gauge vector field ${\bm\eta} = (\eta^i)$ is chosen to preserve the density perturbation:
\bal{gag32}
\delta_{\bm\chi}\rho=\delta_{{\bm\xi}+{\bm\eta}}\rho=\delta_{\bm\xi}\rho+\delta_{\bm\eta}\rho=\delta_{\bm\xi}\rho\;.
\ea
Since the density perturbation is defined as the Lie derivative, the gauge invariance condition expressed in Eq. \eqref{gag32} requires:
\bal{gf340}
\delta_{\bm\eta}\rho=\pounds_{\bm\eta}\rho = -{\bm\nabla}\cdot(\rho{\bm\eta})=0\;,
\ea
where ${\bm\nabla} = (\partial_i)$ denotes the gradient operator.

This differential condition can be addressed using Helmholtz's theorem, which states that any vector field $\rho{\bm\eta}$ satisfying Eq. \eqref{gf340} can be decomposed into the sum of a curl-free component, ${\bm\nabla}\Phi$, and a divergence-free component, ${\bm\nabla} \times {\bm A}$ \citep{Arfken-Weber}:
\bal{g1}
\rho{\bm\eta}&=&{\bm\nabla}\Phi+{\bm\nabla}\times{\bm A}\;,
\ea
where ${\bm A} = {\bm A}(\x)$ is an arbitrary smooth vector field, and $\Phi = \Phi(\x)$ is a non-singular harmonic function satisfying the Laplace equation, $\Delta \Phi = 0$.

Friedman and Schutz \citep{Friedman_1978ApJ} referred to the gauge transformation in Eq. \eqref{nbv4c}, with the gauge field given by Eq. \eqref{g1}, as ``trivial''. However, it is important to note that the original formulation of the ``trivial'' gauge transformation by Friedman and Schutz omits the gauge degree of freedom associated with the curl-free scalar field $\Phi$ in Eq. \eqref{g1}. This degree of freedom presents only in case of hydrostatic perturbations.

An illustrative example of a gauge diffeomorphism is an infinitesimal rigid rotation of the fluid's reference configuration by a constant angle ${\bm\alpha}$ about a fixed axis. The corresponding gauge vector field is given by:
\bal{rot71}
\eta^i&=&\left({\bm\alpha}\times\x\right)^i\;.
\ea
Evaluating the density variation along this vector field yields:
\bal{rot72}
\delta_{\bm\eta}\rho&=&\pounds_{\bm\eta}\rho = -{\bm\nabla}\cdot\left[\rho\left({\bm\alpha}\times\x\right)\right]=-{\bm\nabla}\rho\cdot\left({\bm\alpha}\times\x\right)+\rho{\bm\alpha}\cdot\left({\bm\nabla}\times\x\right)\;.
\ea
Assuming the density depends solely on the radial coordinate, i.e., $\rho = \rho(r)$, we have ${\bm\nabla}\rho = \rho' {\bm n}$, where the prime denotes differentiation with respect to $r$, and ${\bm n} = \x / r$ is the unit radial vector. Under this assumption, the first term on the right-hand side vanishes: ${\bm\nabla}\rho\cdot\left({\bm\alpha}\times\x\right)=0$, since the radial gradient is orthogonal to the tangential vector ${\bm\alpha} \times \x$. The second term also vanishes because ${\bm\nabla} \times \x = 0$. This result implies that a rigid rotation of a fluid sphere about a fixed axis in space does not induce any perturbation in the fluid density. Physically, this reflects the invariance of the density distribution under such a symmetry transformation.

\subsubsection{Gauge Freedom in Exponential Map of a Lie Group}

The gauge freedom in choosing generators of the Lie algebra naturally extends to gauge invariance of the Lie group of diffeomorphisms on the base manifold $\MM$. The exponential map generates the total variation of density $\varrho_\tau$, as given in Eq. \eqref{ty74x}. Let us consider a more general gauge transformation: 
\bal{b5c3}
{\bm\xi}\mapsto{\bm\chi}={\bm\xi}+{\bm\upsilon}\;,
\ea
where ${\bm\upsilon}$ must ensure the gauge invariance of the finite Eulerian vartiation of density, $\varrho_\tau$. This condition is satisfied if and only if:
\bal{mb6c8}
\exp\left(\tau\pounds_{{\bm\xi}+{\bm\upsilon}}\right)\rho(\x)&=&\exp\left(\tau\pounds_{\bm\xi}\right)\rho(\x)\;,
\ea
where ${\bm\upsilon}$ has yet to be determined. It is always possible (in the region of existence of the exponential map) to choose the gauge vector field ${\bm\upsilon}$ in a form suitable for the application of the Baker-Campbell-Hausdorff (BCH) formula \citep{enwiki_BCH}:
\bal{crd5}
{\bm\upsilon}&=&{\bm\eta}+\frac{\tau}{2}\pounds_{\bm\xi}{\bm\eta}+\frac{\tau^2}{12}\pounds^2_{\bm\xi}{\bm\eta}+\frac{\tau^2}{12}\pounds^2_{\bm\eta}{\bm\xi}-\frac{\tau^3   }{24}\pounds_{\bm\eta}\pounds^2_{\bm\xi}{\bm\eta}+\ldots\;,
\ea
where the residual terms are well-known and consist of the commutators (Lie derivatives) of the vector fields ${\bm\eta}$ and ${\bm\xi}$ \citep{enwiki_BCH}.

Using the BCH formula \citep{enwiki_BCH} for the exponential function of the sum of two non-commuting vector fields, we can rewrite the left-hand side of Eq. \eqref{mb6c8} as follows: 
\bal{iuz4e}
\exp\left(\tau\pounds_{{\bm\xi}+{\bm\upsilon}}\right)\rho(\x)&=&
\exp\left(\tau\pounds_{\bm\xi}\right)\exp\left(\tau\pounds_{\bm\eta}\right)\rho(\x)\;.
\ea
The product of the two exponents on the right-hand side of Eq. \eqref{iuz4e} must be equal to the exponent on the right-hand side of Eq. \eqref{mb6c8}, which imposes the following condition on the gauge field ${\bm\eta}$:
\bal{tr4c4}
\exp\left(\tau\pounds_{\bm\eta}\right)\rho(\x)&=&\rho(\x)\;.
\ea
This condition aligns with the gauge requirement $\pounds_{\bm\eta}\rho(\x)=0$, imposed on the vector field ${\bm\eta}$ within the Lie algebra space. It provides the same solution for the gauge vector field ${\bm\eta}$ as shown in Eq. \eqref{g1}. With this understanding, and knowing the generator ${\bm\xi}$, we can determine the generalized gauge vector field ${\bm\upsilon}$ using Eq. \eqref{crd5}.

\subsubsection{The Radial Gauge}

The gauge freedom for the generators ${\bm\xi}$ of the Lie group ${\mathfrak G}$ is constrained by the condition $\pounds_{\bm\eta}\rho=0$, which is imposed on the three otherwise free components of the gauge vector field ${\bm\eta}$. This implies that any two components of the gauge vector ${\bm\eta}$ can be chosen arbitrarily. Consequently, any linear diffeomorphism ${\bm\xi}$ in a stationary rotating fluid possesses two nonphysical degrees of freedom associated with the Lie-invariance of the fluid density $\rho$ along the gauge vector field ${\bm\eta}$, as defined by Eq. \eqref{g1}. This freedom is particularly advantageous for theoretical calculations, as it permits the elimination of two coordinate components of the diffeomorphism ${\bm\xi}$ that lack physical significance. For example, the vector field ${\bm\xi}$ can be chosen coinciding with coordinate $x=x^1$ of the rigidly rotating coordinate system: ${\bm\xi}=(\xi,0,0)$, where $\xi=\xi(\x)$. This choice will be implemented in Section \ref{ee31}.

A particularly advantageous choice for our analysis is the radial gauge, in which the generator of the Lie algebra is constrained to be purely radial, taking the form ${\bm\xi} = {\bm n}\xi(\x)$ or ${\bm\zeta} = {\bm n}\zeta(\x)$, where ${\bm n}$ denotes the unit radial vector. This gauge naturally extends the local generators of the Lie algebra $\mathfrak{g}$ to the global vector flows of the Lie group ${\mathfrak G}$ across the entire base manifold $\MM$, as illustrated previously via the exponential map. The radial gauge proves especially useful in deriving the differential equations governing gravitational perturbations and the equilibrium shape of a rotating fluid body, as will be discussed in subsequent sections.

The presence of gauge freedom implies that, in the context of hydrostatic perturbations, an identical variation in fluid density within the perturbed configuration can be produced through multiple distinct permutations of fluid parcels from their original positions in the unperturbed reference state. These permutations are governed by the gauge vector field ${\bm\eta}$, as defined in Eq. \eqref{g1}. Although such rearrangements can be implemented in various ways, they do not alter the geometry of the level surfaces and can always be interpreted as resulting from displacements of fluid parcels exclusively along radial directions from their initial positions.

\section{Nonlinear Perturbations of Gravitational Field}\label{gvzc34}
\stepcounter{equationschapter}
\setcounter{equationschapter}{5}
\renewcommand{\theequation}{5.\arabic{equation}}

\subsection{Gravitational Potential and Level Surfaces}

The quadrupole centrifugal potential $\tau W_Q$ generates a vector flow $\phi_\tau$, which perturbs the fluid density by an amount $\varrho_\tau$, defined via the exponential map \eqref{ty74x}. This perturbation modifies the total density on the base manifold, such that $\rho(\x) \to \rho_\tau(\x)=\rho(\x)+\varrho_\tau(\x)$, with $\rho_\tau(\x)$ given explicitly in Eq. \eqref{mnk8g}. The gravitational potential $\U_\tau(\x)$ corresponding to the perturbed density $\rho_\tau$ satisfies the Poisson equation: 
\bal{pois5}
\Delta \U_\tau(\x)&=&-4\pi G\rho_\tau(\x)\;.
\ea
The rotation-induced change in density at each point on the base manifold $\MM$ is accompanied by a geometric deformation of the manifold itself, $\MM \to \MM_\tau$, and a rearrangement of fluid parcels. This rearrangement alters both the shape of the reference configuration and, in the case of a compressible fluid, its volume $\V$. Importantly, the vector flow generator $\xi^i$, which defines the fluid displacement, does not fully determine the deformation of the base manifold $\MM$. Likewise, the finite diffeomorphism ${\mathscr X}_\tau^i$ alone does not uniquely define the shape of the rotating body.

The infinitesimal change in the volume occupied by a fluid parcel is characterized by the divergence of the vector flow, $\theta = \pd_i \xi^i$, which represents the algebraic part of the operator for the infinitesimal Eulerian variation of density, $\delta_{\bm\xi} \rho = \pounds_{\bm\xi} \rho$. Since the fluid is assumed to be compressible, we have $\theta \neq 0$. Therefore, the rearrangement of fluid parcels due to rotational perturbation must account for both the geometric deformation of the fluid volume and the change in its numerical value.

The physical deformation of the base manifold is thus associated with a reconfiguration of level surfaces -- those of constant density, pressure, and gravitational potential (see Section \ref{levs8}). To describe the infinitesimal deformation of these surfaces, we introduce a generator ${\bm\zeta} = (\zeta^i)$ and adopt the radial gauge $\zeta^i = n^i \zeta(\x)$, as detailed in Section \ref{levs8}. In the reference configuration, each level surface on the base manifold $\MM$ is spherically symmetric and labeled by the radial distance $r \in [0, a]$ measured from the center of mass which coincides with the origin of the coordinate system $\x$.

Under a large rotational deformation induced by the potential $\tau W_Q$, each point $\x$ on an unperturbed spherical level surface is displaced to a new position with coordinates: 
\bal{k9x}
y^i_\tau=x^i+X^i_\tau\;,
\ea
where $X^i_\tau = X^i_\tau(\x)$ describes the finite displacement of the level surface. The displaced points $y^i_\tau$ span a domain in $\mathbb{R}^3$ with volume ${\cal V}_\tau$, whose boundary $\pd{\cal V}_\tau$  defines a physical realization of the perturbed base manifold, denoted by $\mathfrak{N}_\tau=\{y^i_\tau\in{\cal V}_\tau\cup\pd{\cal V}_\tau\}$. 

In general, for a compressible fluid, the transformations \eqref{k9x} and \eqref{f3v} differ, implying that $\mathfrak{N}_\tau\neq\MM_\tau$. The two representations of the perturbed base manifold coincide only in the special case of an incompressible fluid -- a scenario that is typically inadequate for modeling stars or giant planets in the solar system. 

The vector field $X^i_\tau = X^i_\tau(\x)$, characterizes the finite displacement of level surfaces within space $\mathbb{R}^3$, accompanied by a physical deformation of the fluid body's initially spherical configuration. Mathematically, $X^i_\tau$ is interpreted as an element of the Lie group $\mathfrak{G}$ associated with the base manifold $\MM$. Accordingly, we postulate that this displacement field is defined via the exponential map with generator ${\bm\zeta}$ applied to the identity element $\x$, yielding the transformed position $\y_\tau$. This leads to the expression:
\bal{k5s}
X^i_\tau&=&\bigl[\exp\left(\tau L_{\bm\zeta}\right)-1\bigr]x^i=\tau\zeta^i+\frac{\tau^2}{2!}\zeta^j\pd_j\zeta^i+\frac{\tau^3}{3!}\zeta^k\pd_k\left(\zeta^j\pd_j\zeta^i\right)+\ldots\;. 
\ea
Here, the linear operator $L_{\bm\zeta}$ is defined as $L_{\bm\zeta}:=\zeta^i\pd_i$. The partial derivative of the generator is given by: $\pd_j \zeta^i = \pd_j (\zeta n^i) = n^i \pd_j \zeta + (\zeta/r) {\cal P}_{ij}$, where ${\cal P}_{ij} = \delta_{ij} - n_i n_j$ denotes the projection operator onto the plane orthogonal to the unit radial vector $n^i$. Substituting this expression into the expansion above demonstrates that if the generator $\zeta^i$ is purely radial, then the resulting vector field $X_\tau^i$  is also radial, i.e., $X_\tau^i = X_\tau n^i$, with $X_\tau = X_\tau(\x)$ being a scalar function of position.

We refer to the radial displacement of the level surface, $X_\tau = X_\tau(\x)$, as the {\it height function}. This function is related to the infinitesimal generator of radial translation of level surfaces, $\zeta = \zeta(\x)$, through the push-forward exponential map, as a direct consequence of Eq. \eqref{k5s}:
\bal{wx4}
X_\tau&=&\bigl[\exp\left(\tau L_{\bm\zeta}\right)-1\bigr]r=\tau\zeta+\frac{\tau^2}{2!}\zeta\pd_r\zeta+\frac{\tau^3}{3!}\zeta\pd_r\left(\zeta\pd_r\zeta\right)+\ldots\;. 
\ea
In the general case of a compressible fluid, $\zeta(\x) \neq \xi(\x)$, and the displacement $X^i_\tau \neq {\mathscr X}^i_\tau$. These quantities coincide only in the incompressible fluid case. At this stage, explicit expressions for $\zeta$ and $X_\tau$ are not required for the discussion of subsequent transformations of the gravitational potential. The generator $\zeta$ is obtained by solving the Clairaut equation \eqref{mn5c32}, while the differential equation governing the height function $X_\tau$ will be derived in Section \ref{dmeq43}.

A particular solution to Eq. \eqref{pois5}, expressed in Cartesian coordinates and regular at infinity, is given by:
\bal{vg6f}
\U_\tau(\x)&=&G\int\limits_{\V_\tau}\frac{\rho_\tau({\bm y}')d^3y'}{|\x-{\bm y}'|}\;,
\ea
where $\V_\tau$ denotes the volume occupied by the fluid body deformed under the influence of the rotational potential $\tau W_Q$. Here, $\x \in \mathbb{R}^3$ represents the Cartesian coordinates of the field point, and ${\bm y}' \in \mathfrak{N}_\tau$ are the Cartesian coordinates of the integration points.

Evaluating the integral in Eq. \eqref{vg6f} constitutes a central task in Newtonian gravitational theory. This computation becomes tractable when both the perturbed density $\rho_\tau$ and the deformed volume $\V_\tau$ are known. These quantities can be determined if the diffeomorphism $X_\tau$ is known -- a determination that lies at the heart of the theory of rotating figures (TOF) \citep{Zharkov_1978book, Hubbard-book}.

To solve this problem, it is necessary to decompose the perturbed Newtonian potential $\U_\tau(\x)$ into two components: a volume integral over the unperturbed domain $\V$ of the reference configuration, involving the perturbed density $\rho_\tau$, and a surface integral over the boundary $\partial\V$, accounting for the density of surface layers. For infinitesimally small perturbations, this decomposition -- $\U_\tau = \U^\dagger + \mathcal{O}(\tau^2)$ -- was discussed in Section \ref{uuu23} and is presented in Eq. \eqref{a8juy}.

A similar decomposition can be extended to cases involving strong gravitational perturbations and significant deformations of the spherical reference volume $\V$. This requires isolating the finite Eulerian perturbation of the density, $\varrho_\tau$ (as defined in Eq. \eqref{b6vc}), within the integrand of Eq. \eqref{vg6f}, in a manner consistent with the finite Eulerian variation of the Poisson equation \eqref{pois5}.

However, the integration domain ${\cal V}_\tau$ in Eq. \eqref{vg6f} corresponds to the perturbed configuration, mapped from the reference volume $\V$ via the diffeomorphism  $\V \to {\cal V}_\tau$, as defined by Eq. \eqref{k9x}. Consequently, the Eulerian variation of the gravitational potential cannot be obtained by directly subtracting the unperturbed potential $\U(\x)$ (see Eq. \eqref{vt3s}) from $\U_\tau(\x)$, since the respective integrals are defined over different manifolds: the base manifold $\mathfrak{M}$ for $\U(\x)$ and the perturbed manifold $\mathfrak{N}_\tau$ for $\U_\tau(\x)$.

To address this, and following the approach used in the linearized perturbation theory of Section \ref{uuu23}, we apply a pullback diffeomorphism to transform the integration in Eq. \eqref{vg6f} from manifold $\mathfrak{N}_\tau$ back to the base manifold $\mathfrak{M}$. This reduces the integration domain to the spherical reference volume $\V$, enabling the decomposition of $\U_\tau$ into volume and surface integrals.

\subsection{Pull Back Transformation of Gravitational Potential}\label{ee31}

The volume of integration ${\cal V}_\tau$ in Eq. \eqref{vg6f} consists of a continuous set of stratified level surfaces which do not overlap and can be naturally considered as representing three-dimensional coordinate system on the perturbed based manifold $\mathfrak{N}_\tau$ with Cartesian coordinates ${\bm y}_\tau$. On the other hand, diffeomorphism \eqref{k9x} can be considered as a coordinate transformation between the coordinates of the base $\MM$ and perturbed $\mathfrak{N}_\tau$ manifolds. We identify the coordinates ${\bm y}'$ in the integral \eqref{vg6f} with the coordinates ${\bm y}'_\tau$ defined by the set of the level surfaces
\bal{nb5c}
y'^i:=y'^i_\tau=x'^i+X'^i_\tau\;,
\ea
where $X'^i_{\tau}=X^i_{\tau}(\x')$ is given by Eq. \eqref{k5s} after replacing the generator ${\bm\zeta}\to{\bm\zeta}':={\bm\zeta}(\x')$ and $L_{\bm\zeta}\to L_{{\bm\zeta}'}:=\zeta^i(\x')\pd/\pd x'^i$. The transformation \eqref{nb5c} brings the deformed volume of integration ${\cal V}_\tau$ back to the spherical volume $\V$ of the base manifold $\MM$ and transform the integral \eqref{vg6f} to the following form, explicitly depending on the coordinates $\x'$ of the unperturbed base manifold $\MM$:
\bal{vgy6v}
\U_\tau(\x)&=&G\int\limits_{\cal V}\frac{\rho_\tau({\bm x}'+\X'_{\tau})}{|{\bm x}-\left({\bm x}'+\X'_{\tau}\right)|}{\rm det}\left[\frac{\pd y'^i}{\pd x'^j}\right]d^3x'\;,
\ea
where the integration is over the coordinates $\x'\in\MM$.

Now, using Appendix Eq. \eqref{nbv5t}, we write the ratio of two functions in the above integral in the form of the exponential mapping:
\bal{pnb6v}
\frac{\rho_\tau({\bm x}'+\X'_{\tau})}{|{\bm x}-\left({\bm x}'+\X'_{\tau}\right)|}&=&\exp\left(\tau L_{{\bm\zeta}'}\right)\left[\frac{\rho_\tau({\bm x}')}{|{\bm x}-{\bm x}'|}\right]\;,
\ea
where the operator $L_{{\bm\zeta}'}=\zeta'^i\pd/\pd x'^i$. The Jacobian of the coordinate transformation in the integral of Eq. \eqref{vgy6v} is
\bal{xcre4}
{\rm det}\left[\frac{\pd y'^i}{\pd x'^j}\right]&=&{\rm det}\left[\delta_{ij}+A_{ij}\right]=\exp\left[-\sum_{n=1}^\infty\frac{(-1)^n}{n}A_{i_1i_2}A_{i_2i_3}\ldots A_{i_ni_1}\right]\;,
\ea
where the repeated indices imply Einstein's summation over three coordinates, and the matrix
\bal{srv4}
A_{ij}&=&\frac{\pd X'^i_\tau}{\pd x'^j}=\sum_{n=1}^\infty\frac{\tau^n}{n!}\frac{\pd}{\pd x'^j}\left[L_{{\bm\zeta}'}^{n-1}\zeta'^i(\x')\right]\;.
\ea
Directly computing the matrix $A_{ij}$ in the coordinates $x'^i$ requires the introduction of multivariate tensorial Bell polynomials \citep{evers2023}, extending those introduced in Appendix Eq. \eqref{het65}. However, calculating the determinant \eqref{xcre4} with these multivariate polynomials is a formidable task. Fortunately, there is an alternative, more elegant approach utilizing the invariance of the determinant under coordinate transformations. We choose the orientation of the coordinates such that at the tangent space of the point $\x'$, the generator ${\bm\zeta}'$ of the Lie group is aligned along, say, the $x'^1$ axis, that is $\zeta'^i=(\zeta'^1,0,0)$, where $\zeta'^1=\zeta'^1(\x')$. In such a coordinate system, the matrix $A_{ij}$ has only three components different from zero, which significantly simplifies the calculation of the determinant \eqref{xcre4}. The result is: 
\bal{xv5d}
{\rm det}\left[\frac{\pd y'^i}{\pd x'^j}\right]&=&\left| \begin{array}{ccc}
1+A_{11} & A_{12} & A_{13} \\
0 & 1 & 0 \\
0 & 0 & 1
\end{array} \right|=1+A_{11}\;.
\ea
Thus, the calculation of the determinant is reduced to the calculation of a single element $A_{11}$ of the matrix $A_{ij}$. Direct calculation of $A_{11}$ relies upon the application of the Fa\`a di Bruno formula \citep{deBruno} for the derivatives of a composite function and results in: 
\bal{iko87}
A_{11}&=&\sum_{n=1}^\infty\frac{\tau^n}{n!}\frac{\pd}{\pd x'^1}\left[L_{{\bm\zeta}'}^{n-1}\zeta'^1(\x')\right]=\sum_{n=1}^\infty\frac{\tau^n}{n!}{\bf B}_n\left(\theta',L_{{\bm\zeta}'}\theta',L_{{\bm\zeta}'}^2\theta',...,L_{{\bm\zeta}'}^{n-1}\theta'\right)\;,
\ea
where $\theta':=\pd\zeta'^1/\pd x'^1$ is the divergence of the vector field $\zeta'^i$, and the multi-argument function
\bal{pm5c}
{\bf B}_n\left(x_1,x_2,\ldots,x_{n}\right)&=&\sum_{p=1}^n{\bf B}_{n,p}\left(x_1,x_2,...,x_{n-p+1}\right)\;,
\ea
is the complete Bell polynomial \citep{bellpol}. Using the generating function of the complete Bell polynomials \citep{bellpol}, the final result for the determinant \eqref{xv5d} is:  
\bal{pon94}
{\rm det}\left[\frac{\pd y'^i}{\pd x'^j}\right]&
=&\exp\left[\frac{\exp\bigl(\tau L_{{\bm\zeta}'}\bigr)-1}{L_{{\bm\zeta}'}}\theta'\right]\;.
\ea 
The value of the determinant does not depend on the choice of the coordinate system. Therefore, the formula \eqref{pon94} for the determinant remains exactly the same in the original coordinates, with $\theta'=\pd\zeta'^i/\pd x'^i$ representing the divergence of the vector flow with the generator ${\bm\zeta}'={\bm\zeta}(\x')$ at the point $\x'$ on the base manifold $\MM$.

Now, we substitute Eqs. \eqref{pnb6v} and \eqref{pon94} into integral in the right hand-side of Eq. \eqref{vgy6v}. It yields
\bal{t12}
\U_\tau(\x)&=&G\int\limits_{\cal V}\exp\left[\frac{\exp\bigl(\tau L_{{\bm\zeta}'}\bigr)-1}{L_{{\bm\zeta}'}}\theta'\right]\exp\bigl(\tau L_{{\bm\zeta}'}\bigr)\Biggl[\frac{\rho_\tau({\bm x}')}{|{\bm x}-{\bm x}'|}\Biggr] d^3x'\;.
\ea
The product of two exponential functions in the integrand of \eqref{t12} cannot be immediately simplified because the arguments of the exponents are elements of the Lie algebroid $\g\mapsto\MM$, and they do not commute. Therefore, we need to employ the BCH formula \citep{enwiki_BCH}, which provides a method to express the product of two non-commuting exponential operators. We have found that the most optimal approach is to use the BCH formula in its Zassenhaus form \citep{Casas_2012}. By computing the commutators of the arguments of the exponentials in Eq. \eqref{t12} as prescribed by the Zassenhaus formula \citep{Casas_2012}, we find out that the product of the two exponential operators in Eq. \eqref{t12} reduces to the exponential Lie derivative:
\bal{hjn67}
\exp\left[\frac{\exp\bigl(\tau L_{{\bm\zeta}'}\bigr)-1}{L_{{\bm\zeta}'}}\theta'\right]\exp\bigl(\tau L_{{\bm\zeta}'}\bigr)&=&\exp\bigl(-\tau\pounds_{{\bm\zeta}'}\bigr)\;,
\ea
where the action of the operator of the Lie derivative is understood as $\pounds_{{\bm\zeta}'}f(\x')=-\pd(\zeta'^if(\x'))/\pd x'^i=-\left(L_{{\bm\zeta}'}+\theta'(\x')\right)f(\x')$ for any scalar density $f(\x')$ of weight +1 \citep{dfn,Petrov_2017book}.
Using Eq. \eqref{hjn67} we write Eq. \eqref{t12} for the perturbed potential in a more simple form: 
\bal{t14}
\U_\tau(\x)&=&G\int\limits_{\cal V}\exp\bigl(-\tau\pounds_{{\bm\zeta}'}\bigr)\Biggl[\frac{\rho_\tau({\bm x}')}{|{\bm x}-{\bm x}'|}\Biggr] d^3x'\\\nonumber
&=&G\int\limits_{\cal V}\frac{\rho_\tau({\bm x}')}{|{\bm x}-{\bm x}'|} d^3x'+
G\int\limits_{\cal V}\Bigl[\exp\bigl(-\tau\pounds_{{\bm\zeta}'}\bigr)-1\Bigr]\Biggl[\frac{\rho_\tau({\bm x}')}{|{\bm x}-{\bm x}'|}\Biggr] d^3x'\;.
\ea 
 
Subsequent calculations are conducted in the radial gauge in which ${\bm\zeta}'=\zeta'{\bm n}'$, $X'_\tau=X'_\tau {\bm n}'$, and $X'_\tau=X_\tau(\x')$, where ${\bm n}'$ is a unit vector in the direction of vector $\x'=r'{\bm n}'$. We shall also use spherical coordinates for calculations. It turns out that the integrand with the exponential map on the right-hand side of formula \eqref{t14} can be expressed in terms of the finite translation $X_\tau(\x')$ as follows:
\bal{hu7v5}
\Bigl[\exp\bigl(-\tau\pounds_{{\bm\zeta}'}\bigr)-1\Bigr]\Biggl[\frac{\rho_\tau({\bm x}')}{|{\bm x}-{\bm x}'|}\Biggr]&=&\frac{1}{r'^2}\frac{\pd}{\pd r'}\sum_{n=0}^\infty\frac{X^{n+1}_\tau(\x')}{(n+1)!}\frac{\pd^n}{\pd r'^n}\Biggl[\frac{r'^2\rho_\tau({\bm x}')}{|{\bm x}-{\bm x}'|}\Biggr]\;,
\ea
where we have used the equivariance formula \eqref{nbv5t} and its Taylor expansion analogue in Eq. \eqref{bvtc5a}.

Substituting Eq. \eqref{hu7v5} into the right-hand side of Eq. \eqref{t14} and integrating the term with the total divergence in spherical coordinates yields: 
\bal{xsa2} 
\U_\tau(\x)&=&G\int\limits_{\cal V}\frac{\rho_\tau({\bm x}')d^3x'}{|{\bm x}-{\bm x}'|}+G\oint\limits_{{\mathbb S}^2}d^2\Omega({\bm n}_a) \sum_{n=0}^\infty\frac{X^{n+1}_\tau({\bm a})}{(n+1)!}\frac{\pd^n}{\pd a^n} \left[\frac{a^2\rho_\tau({\bm a})}{|{\bm x}-{\bm a}|}\right]\;, 
\ea 
where functions $X_\tau({\bm a})$ and $\rho_\tau({\bm a})$ are evaluated on the surface of the reference configuration, a sphere of radius $r=a$. Here, $X_\tau({\bm a})\equiv X_\tau(a,\theta_a,\varphi_a)$, $\rho_\tau({\bm a})\equiv \rho_\tau(a,\theta_a,\varphi_a)$, ${\bm a}=a{\bm n}_a$, and ${\bm n}_a=(\sin\theta_a\cos\varphi_a,\sin\theta_a\sin\varphi_a,\cos\theta_a)$ is the unit vector in the direction of the integration point on the unit sphere. The solid angle element is $d^2\Omega({\bm n}_a)=\sin\theta_a d\theta_a d\varphi_a$, and the surface integral is over the unit sphere ${\mathbb S}^2: \{\x\in\mathbb{R}^3\;,\;|\x|=1\}$. The correctness of Eq.~\eqref{xsa2} can alternatively be verified using only the Taylor series, without invoking the exponential operator or the Lie series. This verification of Eq.~\eqref{xsa2} is presented in Appendix~\ref{pulBB}.

Equations \eqref{vg6f} and \eqref{xsa2} yield the same result for the perturbed gravitational potential $\U_\tau(\x)$ both inside and outside the fluid volume. However, Eq. \eqref{xsa2} offers a distinct mathematical advantage for calculations, as it involves integration over a known, spherically symmetric domain $\V$ of the fluid's reference configuration. The surface integral in Eq. \eqref{xsa2} naturally accounts for the gravitational field contribution from the fluid contained within the domain between the deformed volume ${\cal V}_\tau$ and the spherical volume $\V$. From the perspective of differential equations, the surface integral in Eq. \eqref{xsa2} solves the homogeneous Laplace equation and is crucial for formulating boundary conditions to match solutions for the gravitational potential inside and outside the spherical surface of the base manifold $\MM$.

Hubbard \citep{Hubbard_1975SvA} observed that the perturbed potential could be presented as a superposition of contributions from volume and surface integrals in the specific case of a polytrope with a unit index. Eq. \eqref{xsa2} extends this observation to rotating fluid bodies with an arbitrary barotropic equation of state.

\subsection{Poisson Equation for Total Gravitational Perturbation}

Let's revisit the effective potential $U$ defined in Eq. \eqref{bs4x} and examine its perturbation. We define the perturbed effective potential $U_\tau$ as a map $\phi_\tau: \U\to U_\tau$, given by:  
\bal{b5z3}
U_\tau&=&\U_\tau+W_R\;.
\ea
According to Eq. \eqref{nbvrt7}, the Eulerian variation of the potential $W_R$ is zero. Thus, the perturbed effective potential $U_\tau$ is expressed by the exponential mapping:
\bal{cc87} 
U_\tau(\x)&:=&\exp\left(\tau\d_{\bm\xi}\right)U(\x)\;, 
\ea
where $U(\x)$ is given by Eq. \eqref{bs4x}. 
This mapping can be represented in the form of the integral accounting for the Eulerian variation of the fluid density $\d_{\bm\xi}\rho$ within the volume $\V$ and the surface integral accounting for the variation caused by the deformation of the volume $\phi_\tau: \V\mapsto\V_\tau$. More explicitly, we use Eq. \eqref{vt3s} for $\U$ and Eq. \eqref{xsa2} for $\U_\tau$, to represent the finite Eulerian variation of the effective potential $U$ as follows:  
\bal{mi4v} 
\left[\exp\left(\tau\delta_{\bm\xi}\right)-1\right]U(\x)&=&G\int\limits_{\cal V}\frac{\varrho_\tau({\bm x}')d^3x'}{|{\bm x}-{\bm x}'|}+ G\oint\limits_{{\mathbb S}^2}d^2\Omega({\bm n}_a) \sum_{n=0}^\infty\frac{X^{n+1}_ \tau({\bm a})}{(n+1)!}\frac{\pd^n}{\pd a^n} \left[\frac{a^2\rho_\tau({\bm a})}{|{\bm x}-{\bm a}|}\right]\;, 
\ea 
where the perturbation of the density $\varrho_\tau(\x)$ is given by the exponential mapping \eqref{ty74x}.

To denote the overall value of the gravitational field at point $\x\in{\mathbb R}^3$, we use: 
\bal{un7a5} 
V_\tau(\x)&:=&U(\x)+K_\tau(\x)\;. 
\ea 
Here, $K_\tau$ represents the perturbation of the gravitational field, comprising the perturbation of the effective potential $U$ of the body and the external quadrupole perturbation $W_Q$. It is defined as: 
\bal{hj84ss} 
K_\tau(\x)&:=&\left[\exp\left(\tau\delta_{\bm\xi}\right)-1\right]U(\x)+\tau W_Q(\x)=\tau\delta_{\bm\xi}\U(\x)+\frac{\tau^2}{2}\delta^2_{\bm\xi}\U(\x)+\ldots+\tau W_Q(\x)\;. 
\ea 
Applying the Laplace operator to both sides of Eq. \eqref{hj84ss} and considering Eqs. \eqref{vt3s}, \eqref{3afk}, \eqref{xsa2}, along with the Green's function of the Laplace equation, we find that the perturbation $K_\tau$ satisfies the following Poisson equation:
\ba \label{41cs} 
\Delta K_\tau&=&-4\pi G\left(\varrho_\tau+\nu_\tau\right)\;, 
\ea 
where $\nu_\tau=\nu_\tau(\x)$ denotes the density of the surface layer localized on the spherical shell of radius $a$. This surface density is given by:
\bal{nnn6x} 
\nu_\tau(\x)&=&\sum_{n=0}^\infty\frac{X^{n+1}_\tau({\bm a})}{(n+1)!}\frac{\pd^n}{\pd a^n} \left[\frac{a^2}{r^2}\rho_\tau({\bm a})\delta(r-a)\right]\;. \ea 
where $\delta(r-a)$ denotes the Dirac delta function. We emphasize that functions standing in the right-hand side of the Poisson equation \eqref{41cs} are defined on the base manifold $\MM$ corresponding to the reference configuration.

The emergence of a surface layer density in the Poisson equation \eqref{41cs} is a purely mathematical artifact resulting from the transformation of the integration domain from the perturbed physical volume ${\cal V}_\tau \in \mathfrak{N}_\tau$ to the reference volume $\V \in \MM$. Specifically, this artifact arises when the integral in Eq. \eqref{vg6f} is reformulated by changing coordinates from $\y$ on the perturbed manifold $\mathfrak{N}_\tau$ back to $\x$ on the base manifold $\MM$. Such a surface term does not appear in the original formulation of the Poisson equation, where the density is supported over the actual perturbed volume ${\cal V}_\tau$ of the fluid body. Crucially, the contribution of the surface layer originates primarily from the radial derivatives of the fluid density at the boundary of the fluid body. These contributions are particularly significant in determining the figure of a rotating fluid body, where surface gradients play a central role in shaping the equilibrium configuration.

To determine the perturbation of the gravitational field $K_\tau$ by solving the Poisson equation \eqref{41cs}, we utilize the Green function of the Laplace equation convoluted with the volume density perturbation $\varrho_\tau$, integrating it over the unperturbed, spherically symmetric volume $\V$ of the fluid body's reference configuration (calculated in the previous step). The contribution of the surface-layer density \eqref{nnn6x} is derived from a solution of the Laplace equation and must be incorporated into the boundary conditions imposed on the perturbation $K_\tau$ in both the internal and external regions of the spherical volume $\V$. The boundary value problem is detailed in Section \ref{sec6}.

The results in this section clarify and reconcile various methods used by researchers to solve for the gravitational field perturbation $K_\tau$. Our approach, utilizing the Lie group of diffeomorphisms, shows that the integration domain in the Poisson equation \eqref{41cs} for $K_\tau$ can be reduced to the spherically symmetric volume $\V$ of the base manifold $\MM$ using the pullback transformation of coordinates of the level surfaces. This method is valid as long as the infinite power series defining the perturbations are convergent. Determining the precise convergence domain of these series is a separate mathematical problem not addressed in this paper.

\subsection{Nonlinear Interaction between Fluid Density and Gravitational Field Perturbations}\label{nco51}

In Section \ref{y1ub23}, we demonstrated that, under the linear approximation, the Poisson equation \eqref{him4x3} for the gravitational field perturbation ${\cal K}$ can be transformed into the Helmholtz equation \eqref{za45} for the same perturbation. This transformation is made possible by the linear coupling between the infinitesimal perturbations of the density and the gravitational field, as expressed in Eq. \eqref{za44}. Extending this result to the nonlinear regime of perturbation theory, the underlying coupling mechanism remains valid; however, the corresponding coupling equation becomes highly nonlinear. In this section, we derive the nonlinear coupling equation that establishes the mathematical relationship between the finite gravitational perturbation $K_\tau$ and the finite density perturbation $\varrho_\tau$.

To begin, we restate Eq. \eqref{za44} in a form more suitable for further calculations. Using Eqs. \eqref{t5} and \eqref{ko9h}, we express the ratio $U'/\rho'$ of the radial derivatives of the effective potential $U$ and the fluid density $\rho$ as:  
\bal{jkin09}
\frac{U'}{\rho'}&=&\frac{1}{\rho}\frac{\pd p}{\pd\rho}\;.
\ea
Here, the fluid density $\rho$ is treated as an independent radial variable, with $U$ and $p$ considered as functions of $\rho$, i.e., $U = U(\rho)$ and $p = p(\rho)$. Substituting Eq. \eqref{jkin09} into Eq. \eqref{za44} transforms it into:
\bal{41c}
\delta_{\bm\xi}U+W_Q&=&{\cal A}\delta_{\bm\xi}\rho\;,
\ea
where we use a definition of ${\cal K}$ from Eq. \eqref{za46} and introduce the shorthand notation: 
\bal{pnv34} {\cal A}&:=&\frac{1}{\rho}\frac{\pd p}{\pd\rho}\;, 
\ea 
for the pressure derivative with respect to the fluid density $\rho$. Naturally, ${\cal A} = {\cal A}(\rho)$ is a function of density, and its explicit form is determined by the barotropic equation of state, $p = p(\rho)$.

Next, we apply the exponential operator to both sides of Eq. \eqref{41c} and account for the vanishing Eulerian variation of the external perturbation $W_Q$ (see Eq. \eqref{nbvrt7}). Consequently, applying the exponential operator to both sides of Eq. \eqref{41c} yields: 
\bal{xx1}
K_\tau&=&\frac{e^{\tau\delta_{\bm\xi}}-1}{\delta_{\bm\xi}}\left[{\cal A}(\rho)\delta_{\bm\xi}\rho\right]\;,
\ea
where the gravitational perturbation $K_\tau$ is defined in Eq. \eqref{hj84ss}. Equation \eqref{xx1} establishes a functional relationship between $K_\tau$ and the infinitesimal variation of the fluid density $\delta_{\bm\xi}\rho$.

We can further develop the right-hand side of Eq. \eqref{xx1} using the Taylor series expansion of the exponential operator:
\bal{xx2}
\frac{e^{\tau\delta_{\bm\xi}}-1}{\delta_{\bm\xi}}\left[{\cal A}(\rho)\delta_{\bm\xi}\rho\right]=\sum_{n=0}^\infty\frac{\tau^{n+1}}{(n+1)!}\,\delta^n_{\bm\xi}\left[{\cal A}(\rho)\delta_{\bm\xi}\rho\right]\;,
\ea
where $\delta^n_{\bm\xi}\left[{\cal A}(\rho)\delta_{\bm\xi}\rho\right]$ indicates that the Eulerian variation operator $\delta_{\bm\xi}$ is applied $n$ times to the product ${\cal A}(\rho)\delta_{\bm\xi}\rho$. This expansion allows us to systematically account for higher-order perturbations, facilitating further transformations and analysis of the coupling equation.

In the next step, we use the fact that the Eulerian variation of fluid density is a linear operator of the variational derivative \citep{Petrov_2017book}. This allows us to employ the binomial Leibniz rule for derivatives \citep{gradryzh} to compute the $n$-th order variation of the product of two functions: 
\bal{njui0}
\delta^n_{\bm\xi}\left[{\cal A}(\rho)\delta_{\bm\xi}\rho\right]=\sum_{k=0}^n\frac{n!}{k!(n-k)!}\left[\delta_{\bm\xi}^{n-k}{\cal A}(\rho)\right]\delta^{k+1}_{\bm\xi}\rho\;.
\ea
Substituting Eq. \eqref{njui0} into Eq. \eqref{xx2} yields:
\bal{bvc45m}
\frac{e^{\tau\delta_{\bm\xi}}-1}{\delta_{\bm\xi}}\left[{\cal A}(\rho)\delta_{\bm\xi}\rho\right]=\sum_{n=0}^\infty\sum_{k=0}^\infty\frac{\tau^{n+k+1}}{n+k+1}\frac{\delta_{\bm\xi}^n{\cal A}(\rho)}{n!}\frac{\delta^{k+1}_{\bm\xi}\rho}{k!}\;.
\ea

The function ${\cal A}(\rho)$ defined in \eqref{pnv34} is composite, and computing its $n$-th order variation is laborious. This variation can be computed using the Fa{\`a} di Bruno formula in Riordan's form: \citep{deBruno},
\bal{byt67}
\delta^n_{\bm\xi}{\cal A}(\rho)&=&\sum_{p=0}^n  {\bf B}_{n,p}\left(\delta_{\bm\xi}\rho,\delta^2_{\bm\xi}\rho,...,\delta^{n-p+1}_{\bm\xi}\rho\right)\,\pd^p_\rho {\cal A}(\rho)\;,
\ea
where $\pd_\rho=\pd/\pd\rho$, and ${\bf B}_{n,p}\left(x_1,x_2,...,x_{n-p+1}\right)$ is the incomplete Bell polynomial \citep{bellpol} with arguments $x_j\equiv \delta^j_{\bm\xi}\rho$, similar to that defined in Eq. \eqref{het65} for the scalar case. Substituting Eq. \eqref{byt67} into Eq. \eqref{bvc45m}, we obtain:
\bal{bvg9}
\frac{e^{\tau\delta_{\bm\xi}}-1}{\delta_{\bm\xi}}\left[{\cal A}(\rho)\delta_{\bm\xi}\rho\right]=\sum_{n=0}^\infty\sum_{k=0}^\infty\sum_{p=0}^n\frac{\tau^{n+k+1}}{n+k+1}\frac{\pd^p_\rho {\cal A}(\rho)}{n!}\frac{\delta^{k+1}_{\bm\xi}\rho}{k!}{\bf B}_{n,p}\left(\delta_{\bm\xi}\rho,\delta^2_{\bm\xi}\rho,...,\delta^{n-p+1}_{\bm\xi}\rho\right)\;.
\ea
This formula, along with Eq. \eqref{xx1}, defines the overall perturbation of the gravitational potential as:
\bal{hgc6}
K_\tau&=&\sum_{p=0}^\infty\sum_{n=0}^\infty\sum_{k=0}^\infty\frac{\tau^{n+p+k+1}}{n+k+p+1}\frac{\pd^p_\rho {\cal A}(\rho)}{(n+p)!}\frac{\delta^{k+1}_{\bm\xi}\rho}{k!}{\bf B}_{n+p,p}\left(\delta_{\bm\xi}\rho,\delta^2_{\bm\xi}\rho,...,\delta^{n+1}_{\bm\xi}\rho\right)\;,
\ea
where we have exchanged the order of summation compared to \eqref{bvg9}.

The right-hand side of Eq. \eqref{hgc6} can be simplified and expressed solely in terms of the total variation \eqref{ty74x} of density $\varrho_\tau$. To achieve this, consider the sum in \eqref{hgc6} for a fixed value of the index $p$:
\bal{s1}
&&\sum_{n=0}^\infty\sum_{k=0}^\infty\frac{\tau^{n+p+k+1}}{(n+k+p+1)}\frac{{\bf B}_{n+p,p}\left(\delta_{\bm\xi}\rho,\delta^2_{\bm\xi}\rho,...,\delta^{n+1}_{\bm\xi}\rho\right)}{(n+p)!}\frac{\delta^{k+1}_{\bm\xi}\rho}{k!}=\\\nonumber&&\hspace{6cm}\Biggl[\sum_{n=0}^\infty\frac{{\bf B}_{n+p,p}\left(\delta_{\bm\xi}\rho,\delta^2_{\bm\xi}\rho,...,\delta^{n+1}_{\bm\xi}\rho\right)}{(n+p)!}\Biggr]
\Biggl[\sum_{k=0}^\infty\frac{\tau^{n+p+k+1}}{n+k+p+1}\frac{\delta^{k+1}_{\bm\xi}\rho}{k!}\Biggr]\;.
\ea
To facilitate the calculation of the product of two sums on the right-hand side of Eq. \eqref{s1}, we express the sum with respect to the index $k$ as an integral:
\bal{s2}
\sum_{k=0}^\infty\frac{\tau^{n+p+k+1}}{n+k+p+1}\frac{\delta^{k+1}_{\bm\xi}\rho}{k!}=\Biggl[\int_{-\tau\delta_{\bm\xi}}^0\left(-\frac{t}{\delta_{\bm\xi}}\right)^{n+p}e^{-t}dt\Biggr]\rho\;.
\ea
Assuming that the operations of summation and integration commute (because the series are assumed to be convergent), we use the definition of the generating function for the Bell polynomials \citep{bellpol} to obtain:
\bal{s3}
&&\sum_{n=0}^\infty\frac{{\bf B}_{n+p,p}\left(\delta_{\bm\xi}\rho,\delta^2_{\bm\xi}\rho,...,\delta^{n+1}_{\bm\xi}\rho\right)}{(n+p)!}\left(-\frac{t}{\delta_{\bm\xi}}\right)^{n+p}\,=\,\sum_{n=p}^\infty\frac{{\bf B}_{n,p}\left(\delta_{\bm\xi}\rho,\delta^2_{\bm\xi}\rho,...,\delta^{n-p+1}_{\bm\xi}\rho\right)}{n!}\left(-\frac{t}{\delta_{\bm\xi}}\right)^{n}\,=\\\nonumber
&&\hspace{5cm}\frac{1}{p!}\left[\sum_{j=1}^\infty\left(-\frac{t}{\delta_{\bm\xi}}\right)^j\frac{\delta^j_{\bm\xi}\rho}{j!}\right]^p=\frac{1}{p!}\left[\sum_{j=1}^\infty\frac{\left(-t\right)^{j}}{j!}\rho\right]^p
=\frac{1}{p!}\left[\left(e^{-t}-1\right)\rho\right]^p\;.
\ea
What remains is to perform the integral with respect to $t$, which yields the sum on the left-hand side of Eq. \eqref{s1} in the following form:
\bal{s4}\hspace{-0.5cm}
\sum_{n=0}^\infty\sum_{k=0}^\infty\frac{\tau^{n+p+k+1}}{(n+k+p+1)}\frac{{\bf B}_{n+p,p}\left(\delta_{\bm\xi}\rho,\delta^2_{\bm\xi}\rho,...,\delta^{n+1}_{\bm\xi}\rho\right)}{(n+p)!}\frac{\delta^{k+1}_{\bm\xi}\rho}{k!}&=&\frac{1}{p!}\left\{\int_{-\tau\delta_{\bm\xi}}^0\left[\left(e^{-t}-1\right)\rho\right]^pe^{-t}dt\right\}\rho\\\nonumber
&=&\frac{1}{(p+1)!}\Bigl[\left(e^{\tau\delta_{\bm\xi}}-1\right)\rho\Bigr]^{p+1}=\frac{\varrho_\tau^{p+1}}{(p+1)!}\;,
\ea
where we have used the definition of $\varrho_\tau$ given in Eq. \eqref{ty74x} to obtain the final term on the right-hand side of Eq. \eqref{s4}.

Finally, we perform the summation with respect to the index $p$ in the triple sum in \eqref{hgc6} and obtain:
\bal{s5}
\sum_{p=0}^\infty\sum_{n=0}^\infty\sum_{k=0}^\infty\frac{\tau^{n+p+k+1}}{n+k+p+1}\frac{\pd^p_\rho {\cal A}(\rho)}{(n+p)!}\frac{\delta^{k+1}_{\bm\xi}\rho}{k!}{\bf B}_{n+p,p}\left(\delta_{\bm\xi}\rho,\delta^2_{\bm\xi}\rho,...,\delta^{n+1}_{\bm\xi}\rho\right)&=&\sum_{p=0}^\infty\frac{\varrho_\tau^{p+1}}{(p+1)!}\pd^p_\rho {\cal A}(\rho)\;.
\ea
Thus, the functional equation \eqref{xx1} for the gravitational perturbation $K_\tau$ takes its final form as a Taylor series with respect to the total density perturbation $\varrho_\tau$. Specifically,
\bal{nber5}
K_\tau=\sum_{n=0}^\infty\frac{\varrho_\tau^{n+1}}{(n+1)!}\pd_\rho^n {\cal A}(\rho)=\varrho_\tau\left[ {\cal A}(\rho)+\frac{\varrho_\tau}{2!}\pd_\rho {\cal A}(\rho)+\frac{\varrho_\tau^2}{3!}\pd_\rho^2 {\cal A}(\rho)+...\right] \;,
\ea
where the function ${\cal A}={\cal A}(\rho)$ is defined in Eq. \eqref{pnv34} as the derivative of pressure.

Although calculating the gravitational field perturbation $K_\tau$ in Eq. \eqref{nber5} might seem to require explicit knowledge of the equation of state $p=p(\rho)$, we can bypass this by using the hydrostatic equilibrium equation \eqref{jkin09}:  
\bal{zvsf}
{\cal A}&=&\frac{U'}{\rho'}\;.
\ea
This indicates that the function ${\cal A}$ can be determined if the unperturbed gravitational potential $U=U(r)$ and the density $\rho=\rho(r)$ are known.

In the unperturbed fluid configuration, density $\rho$ and radial coordinate $r$ are inverse functions, such that $\rho^{-1}(\rho(r))=r$. Consequently, partial derivatives of the function ${\cal A}$ with respect to density can be expressed in terms of radial coordinate derivatives using the operator equation:
\bal{as7}
\frac{\pd}{\pd\rho}=\frac{1}{\rho'}\frac{\pd}{\pd r}\;,
\ea
where $\rho' = d\rho/dr$ represents the radial derivative of density. This method enables direct calculation of $K_\tau$ in terms of the total density perturbation $\varrho_\tau$, along with the radial derivatives of the unperturbed density $\rho$ and the effective gravitational potential $U=\U+W_R$.

\subsection{Nonlinear Molodensky Equation for Gravitational Perturbations}\label{bubu54}

The field equation \eqref{41cs} for the strongly perturbed gravitational potential $K_\tau$ is determined by the total variation of density $\varrho_\tau$. Previously, we showed that applying the equation of hydrostatic equilibrium to perturbations allows the gravitational field perturbation $K_\tau$ to be expressed as an infinite series \eqref{nber5} of the density perturbation $\varrho_\tau$. Substituting this series into equation \eqref{41cs} results in a non-linear differential equation for the density perturbation $\varrho_\tau$, which can be solved iteratively. More effectively, however, is to work with the gravitational field perturbation $K_\tau$.

To derive a differential equation for $K_\tau$, we need to invert Eq. \eqref{nber5} using the Lagrange method of inversion of power series \citep{lagrangeinv}. First, we rewrite \eqref{nber5} as follows: 
\bal{s8}
K_\tau=\sum_{{n}=1}^\infty\frac{h_{n}}{{n}!}\varrho_\tau^{n}\;,\qquad\qquad h_{n}=\pd_\rho^{{n}-1} {\cal A}\;.
\ea
Applying the Lagrange method to invert Eq. \eqref{s8}, we express $\varrho_\tau$ as a power series of the gravitational potential perturbation $K_\tau$:  
\bal{s9}
\varrho_\tau&=&\sum_{n=1}^\infty\frac{g_{n}}{{n}!}\left(\frac{K_\tau}{h_1}\right)^{n}=\frac{K_\tau}{{\cal A}}+\sum_{n=2}^\infty\frac{g_{n}}{{n}!}\left(\frac{K_\tau}{{\cal A}}\right)^{n}\;,
\ea
where $g_1=1$ and the other coefficients
\bal{s10}
g_{n}&=&\sum_{k=1}^{{n}-1}(-1)^k({n})_{k}{\bf B}_{{n}-1,k}\left(c_1,c_2,...,c_{{n}-k}\right)\;,\qquad\qquad ({n}\ge 2)
\ea
are given in terms of the incomplete Bell polynomials ${\bf B}_{{n}-1,k}\left(c_1,...,c_{{n}-k}\right)$ with arguments
\bal{s11}
c_k:&=&\frac{1}{k+1}\frac{h_{k+1}}{h_1}=\frac{1}{k+1}\frac{\pd_\rho^k {\cal A}}{{\cal A}}\;,
\ea
and 
\bal{s12}
({n})_k:&=&{n}({n}+1)...({n}+k-1)=\frac{\Gamma({n}+k)}{\Gamma({n})}\;,
\ea
is the Pochhammer symbol, also known as the {\it rising} factorial \citep{risingf}, and $\Gamma(n)$ is the gamma function \citep{Arfken-Weber}. 

The coefficients $g_n$ depend solely on the unperturbed (spherically-symmetric) fluid density $\rho=\rho(r)$ and its radial derivatives, which are determined by solving the equations in Section \ref{basman1} that define the body's reference configuration. The first few coefficients $g_n$ $(n\ge 2)$ are given by: 
\bal{s13}
g_2&=&-\frac{\pd_\rho A}{{\cal A}}\;,\\
g_3&=&\frac{3(\pd_\rho {\cal A})^2-{\cal A}\pd_\rho^2{\cal A}}{{\cal A}^2}\;,\\
g_4&=&-\frac{15(\pd_\rho {\cal A})^3-10{\cal A}\pd_\rho {\cal A}\pd_\rho^2{\cal A}+{\cal A}^2\pd_\rho^3{\cal A}}{{\cal A}^3}\;,
\ea
and so forth.

The partial derivatives of the function ${\cal A}={\cal A}(\rho)$ with respect to the density $\rho$ are computed using formula \eqref{as7}, for example:
\bal{s14}
{\cal {\cal A}}&=&\frac{U'}{\rho'}\;,\\
\pd_\rho {\cal {\cal A}}&=&\frac{1}{\rho'}\frac{d}{dr}\left(\frac{U'}{\rho'}\right)=\frac{\rho'U''-U'\rho''}{\rho'^3}\;,\\
\pd^2_\rho {\cal A}&=&\frac{1}{\rho'}\frac{d}{dr}\left[\frac{1}{\rho'}\frac{d}{dr}\left(\frac{U'}{\rho'}\right)\right]=\frac{3\rho''^2U'+\rho'^2U'''-\rho'\left(3\rho''U''+U'\rho'''\right)}{\rho'^5}\;,
\ea
and so on.

For the polytropic equation of state, $p\sim\rho^{1+1/n}$ (where $n$ is the polytropic index), the function ${\cal A}$, its partial derivatives, and the coefficients $g_n$ in Eq. \eqref{s9} can be directly computed from the thermodynamic equation \eqref{pnv34}. This case is discussed in Appendix \ref{secD}.

For small deformations of the fluid body, the density perturbation varies linearly with the perturbation in the gravitational field, as described by Eq. \eqref{41c}. However, as the deformation increases, the density perturbation $\varrho_\tau$ enters a nonlinear regime, becoming increasingly influenced by higher-order gravitational field perturbations, denoted by $K_\tau$. This relationship is captured by an infinite power series expansion, given in Eq. \eqref{s9}.

By substituting Eq. \eqref{s9} into Eq. \eqref{41cs}, the explicit dependence on the volume density perturbation $\varrho_\tau$ is eliminated, yielding a fully nonlinear equation governing the gravitational field perturbation within the interior of the body:
\bal{fe87}
\Delta K_\tau+4\pi G\frac{K_\tau}{{\cal A}}+4\pi G\sum_{n=2}^\infty\frac{g_{n}}{{n}!}\left(\frac{K_\tau}{{\cal A}}\right)^{n}&=&0\;,\qquad\qquad\qquad (r < a)\;.
\ea
In this formulation, the surface density $\nu(x)$ is omitted, as the analysis is restricted to the region strictly within the body's boundary, where $r<a$.

In the vacuum region outside the body, the general equation \eqref{41cs} reduces to the Laplace equation:
\bal{jk6v7}
\Delta K_\tau&=&0\;,\qquad\qquad\qquad (r > a)\;.
\ea
This linear form is consistent with the superposition principle in Newtonian gravity, which states that the total gravitational potential in a vacuum is the sum of the potentials due to individual sources.

Equation \eqref{fe87} generalizes the linearized Molodensky equation \eqref{za45} to account for finite, non-linear perturbations in both density and the gravitational field. Notably, it reveals a compelling phenomenon: the gravitational field perturbation $K_\tau$ exhibits self-interaction through non-linear terms. At first glance, this may appear to contradict the linear nature of Newtonian gravity, which is governed by the superposition principle. However, this principle strictly applies only in vacuum, where the gravitational potential satisfies the Laplace equation \eqref{jk6v7}. Within matter, the gravitational field obeys the non-linear equation \eqref{fe87}.

When the system is subjected to a sufficiently strong external perturbation $W_Q$ -- such as that induced by rapid rotation -- it generates a significant density perturbation $\varrho_\tau$, which is non-linearly coupled to $K_\tau$ via Eq. \eqref{s9}. As a result, the fluid's response to gravitational perturbations becomes inherently non-linear. This behavior is analogous to non-linear electrodynamics, where strong electromagnetic fields interacting with matter lead to non-linear corrections to Maxwell's equations \citep{Ida_1997}.

The non-linear terms in Eq. \eqref{fe87} may be empirically tested by analyzing the gravitational field of rapidly rotating bodies such as Jupiter, whose gravitational inhomogeneities are now measured with exceptional precision \citep{J2-Jupiter_2018}. These investigations are closely tied to the study of rotational deformation, Love numbers, and the multipole structure of the external gravitational field -- topics that are explored in the subsequent sections.

\section{Nonlinear Analysis of Rotational Deformations in Fluid Bodies}\label{sec4}
\stepcounter{equationschapter}
\renewcommand{\theequation}{6.\arabic{equation}}

This section develops an exact theoretical framework for determining the shape of rotating fluid bodies, grounded in the mathematical structure of the Lie group of diffeomorphisms and the exponential mapping of gravitational and hydrostatic perturbations. Within this formalism, the classical hydrostatic equilibrium equations are reformulated as functional equations for the fluid density and gravitational field. This approach establishes a direct and intrinsic link between the geometry of equipotential surfaces and the underlying gravitational perturbations.

By substituting the solutions of these functional equations into the Poisson equations governing gravitational perturbations, one derives a fundamental nonlinear partial differential equation for the height function generalizing the Molodensky equation \eqref{mn5c32}. This equation enables, in principle, the exact determination of the deformation of level surfaces and the body's shape as functions of angular velocity, internal density distribution, and the equation of state -- without relying on perturbative or approximation methods.

This exact theory represents a substantial advancement over traditional approximation-based models. It provides mathematically rigorous insights into the relationship between the body's internal structure -- quantified by Love numbers -- and the multipole moments of its external gravitational field. Furthermore, it supports high-precision predictions across a broad spectrum of physical conditions and offers a robust framework for assessing the accuracy of approximate computational models derived from observational data, by enabling the explicit computation of residual terms.

\subsection{Functional Equation Approach to the Geometry of Level Surfaces}\label{pon5c}
\subsubsection{Functional Equation for Level Surfaces in Rotating Fluid Bodies}\label{fehf2}

The base manifold $\MM$ of the fluid reference configuration initially features spherical level surfaces parameterized by the radial coordinate $r \in [0,a]$. Applying an external quadrupole potential $W_Q$ induces rotational perturbations, deforming the manifold into a new configuration $\MM \mapsto \mathfrak{N}_\tau$ with non-spherical level surfaces. This deformation is described by the diffeomorphism \eqref{k9x}, mapping each point $x^i \in \MM$ to $x^i + X^i_\tau \in \mathfrak{N}_\tau$. The finite displacement vector $X^i_\tau(\x) \in {\mathfrak G}$ arises from the transformation generated by an infinitesimal element $\zeta^i \in \g$ of the Lie algebra via the exponential map \eqref{k5s}.

This section derives an exact functional relationship between $X^i_\tau$ and the gravitational perturbation $K_\tau$, assuming the latter is known from the solution of the nonlinear Molodensky equation \eqref{fe87}. The resulting equation captures the correspondence between the deformation of the level surfaces and the magnitude of the gravitational perturbation.

Calculations are performed in the radial gauge, where both the fluid diffeomorphism generator ${\bm\xi} = \xi{\bm n}$ and the level surface generator ${\bm\zeta} = \zeta{\bm n}$ possess only radial components, with $\xi = \xi(\x)$ and $\zeta = \zeta(\x)$. This gauge significantly simplifies the derivation of functional equations linking Eulerian variations in matter and the gravitational field to the geometric deformation of level surfaces.

We begin by establishing the relationship between the fluid and level surface diffeomorphism generators, ${\bm\xi}$ and ${\bm\zeta}$. To facilitate this, we adopt the unperturbed density $\rho = \rho(r)$ as a proxy for the radial coordinate $r$. This choice is motivated by the fact that, in hydrostatic equilibrium, the level surfaces of the body correspond to isodensity shells, naturally parameterized by the fluid density. Recasting the radial coordinate in terms of $\rho$ amounts to transitioning from spherical coordinates $(r, \theta, \varphi)$ to isopycnal coordinates $(\rho, \theta, \varphi)$, a common framework in fluid dynamics \citep{Vallis_2017,Huang2020}.

We introduce the notation for the effective gravitational potential $U(r(\rho)):=\mathsf{U}(\rho)$ and express Eq. \eqref{zvsf} in terms of $\mathsf{U}$:
\bal{sc54x}
{\cal A}(\rho)&=&\frac{U'}{\rho'}=\frac{\pd{\mathsf U}}{\pd \rho}\;.
\ea
Substituting this expression for ${\cal A}(\rho)$ into Eq. \eqref{nber5} yields: 
\bal{tt34}
K_\tau&=&\sum_{n=1}^\infty\frac{\varrho_\tau^{n}}{n!}\frac{\pd^n {\mathsf U}}{\pd \rho^n}\;,
\ea
which represents a Taylor expansion of the gravitational perturbation $K_\tau = K_\tau(\x)$ about the point $(\rho, \theta, \varphi)$ on a level surface of constant density $\rho$, with respect to the finite Eulerian density variation $\varrho_\tau = \varrho_\tau(\x)$.
This expansion is generated by the following functional expression:
\bal{tt83}
K_\tau&=&{\mathsf U}(\rho+\varrho_\tau)-{\mathsf U}(\rho)\;.
\ea
According to the equivariance theorem established in Appendix \ref{appA1}, the function $\mathsf{U}(\rho + \varrho_\tau)$ can be expressed via the exponential map: 
\bal{tt39}
{\mathsf U}(\rho+\varrho_\tau)&=&\exp\left(\tau\chi\pd_\rho\right){\mathsf U}(\rho)\;,
\ea
where the generator $\chi = \chi(\x)$ is related to the variation $\varrho_\tau=\varrho_\tau(\x)$ through:
\bal{dif3a}
\varrho_\tau&=&\left[\exp\left(\tau\chi\pd_\rho\right)-1\right]\rho\;.
\ea

The generator $\chi = \chi(\x)$ can now be identified by noting that the density variation $\varrho_\tau$ is also expressed via the exponential operator \eqref{ty74x}, involving the fluid diffeomorphism generator $\delta_{\bm\xi}$. Equating this with the exponential form in Eq. \eqref{dif3a} yields the operator identity: $\chi \partial_\rho = \delta_{\bm\xi}$, valid in the radial gauge. Applying this identity to any smooth function of $\rho$, such as the pressure $p = p(\rho)$ or the effective gravitational potential $\mathsf{U} = \mathsf{U}(\rho)$, gives the explicit form of the generator: $\chi = \delta_{\bm{\xi}} \rho = \pounds_{\bm{\xi}} \rho$.

We now return from the isopycnal density variable $\rho = \rho(r)$ to the spherical radial coordinate $r = r(\rho)$. To describe the radial displacement of level surfaces, we introduce the generator of infinitesimal radial translations, $\zeta^i = \zeta n^i$, where $\zeta = \zeta(\x)$ is defined via the correspondence $\chi \partial_\rho = -\zeta \partial_r$. Using the identity $\partial_\rho = (1/\rho') \partial_r$, this yields $\zeta = -\chi / \rho'$, where the prime denotes differentiation with respect to $r$. Recalling that $\chi = \pounds_{\bm{\xi}} \rho$, we obtain the explicit form of the generator of the infinitesimal radial translation of the level surface: 
\bal{qq7}
\zeta&=&-\frac{\pounds_{\bm\xi}\rho}{\rho'}=\xi+\frac{\rho}{\rho'}\left(\xi'+\frac{2\xi}{r}\right)\;.
\ea
This expression coincides with the generator $\zeta$ appearing in Eq. \eqref{wcx41} in the linearized perturbation regime discussed in Section \ref{popo9}. It shows that, in general, for a compressible fluid, the generator ${\bm\zeta}$ of radial level surface displacements differs from the fluid diffeomorphism generator ${\bm\xi}$ responsible for density variations. This discrepancy arises due to fluid compressibility, which introduces a non-zero divergence $\partial_i \xi^i = \xi' + 2\xi/r$ of the vector field ${\bm\xi}$. In the incompressible fluid limit, where $\partial_i \xi^i = 0$, the two generators are equal: ${\bm\xi} = {\bm\zeta}$.

Our analysis shows that, under large deformations, the radial displacement $X_\tau$ of a level surface -- initially spherical with radius $r$ -- is governed by an exponential map generated by $\zeta = \zeta(\x)$. In hydrostatic equilibrium, where the level surfaces of density and gravitational potential coincide, this same exponential map describes variations in both fields: 
\bal{dif3}
\varrho_\tau(\x)&=&\left[\exp\left(-\tau L_{\bm\zeta}\right)-1\right]\rho(\x)\;,
\\
\label{bb7}
K_\tau(\x)&=&\bigl[\exp\left(-\tau L_{\bm\zeta}\right)-1\bigr]U(\x)\;,
\ea
where the linear operator $L_{\bm\zeta} = \zeta^i \partial_i = \zeta \partial_r$ was introduced in Eq. \eqref{k5s}. 

Equations \eqref{dif3} and \eqref{bb7} represent pullback diffeomorphisms. It is more convenient to work with their pushforward counterparts, expressed as:
\bal{hh67}
\exp\left(\tau L_{\bm\zeta}\right)\left[\rho(\x)+\varrho_\tau(\x)\right]&=&\rho(\x)\;,\\
\label{hh68}
\exp\left(\tau L_{\bm\zeta}\right)\left[U(\x)+K_\tau(\x)\right]&=&U(\x)\;.
\ea
These functional equations describe finite deformations of level surfaces through the exponential map. However, it is more practical to express them in terms of the finite radial displacement of a level surface, represented by the height function $X_\tau = X_\tau(\x)$. This function is related to the infinitesimal generator of radial translations, $\zeta = \zeta(\x)$, via the exponential map defined in Eq. \eqref{wx4}. By invoking the equivariance property of Lie group actions, as detailed in Appendix \ref{appA1}, Eqs. \eqref{hh67} and \eqref{hh68} can be reformulated in terms of the height function, yielding the following functional equations: 
\ba\label{pic7}
\rho(\x+\X_\tau)+\varrho_\tau(\x+\X_\tau)&=&\rho(\x)\;,\\
\label{pi6c}
U(\x+\X_\tau)+K_\tau(\x+\X_\tau)&=&U(\x)\;.
\ea
These equations determine the height function $X_\tau$ in terms of either the density perturbation $\varrho_\tau$ or the gravitational field perturbation $K_\tau$ within the body. They form the foundation for deriving the differential equation that governs the shape of the deformed level surfaces.

Functional equations \eqref{pic7} and \eqref{pi6c} describe a homotopic, continuous transformation of the density and gravitational potential, parameterized by $\tau\in[0,1]$. These equations assert that the values of the density and gravitational potential at a given level surface and point $\x$ remain invariant as the surface undergoes a radial displacement to the point $\x+\X_\tau$, driven by the external perturbation $\tau W_Q(\x)$. A similar conservation principle applies to the fluid pressure. Equation \eqref{pi6c} generalizes the fundamental relation \eqref{xa1} that defines level surfaces in the Zharkov-Trubitsyn theory \citep{Zharkov_1978book}.

From this point forward, we consider finite perturbations of the fluid body, characterized by the exponential mappings $\phi_\tau(\bm{\xi}) = \exp(\tau \delta_{\bm{\xi}})$ and $\phi_\tau(\bm{\zeta}) = \exp(\tau \delta_{\bm{\zeta}})$, evaluated at the homotopy parameter $\tau = 1$. For notational simplicity, we omit the subscript $\tau = 1$ and adopt the following conventions: $\varrho \equiv \varrho_1(\x)$, $K \equiv K_1(\x)$, and $\bm{X} \equiv \bm{X}_1(\x)$. The perturbed fluid density and gravitational potential at $\tau = 1$ are denoted by $\mu \equiv \rho_1(\x)$ and $\Phi \equiv \U_1(\x)$, respectively, in accordance with the notation introduced in Eqs. \eqref{001}--\eqref{003}. The volume of the fluid body deformed by rotation is denoted by ${\cal D}\equiv\V_1$. 

\subsubsection{Solving Functional Equations for the Height Function with the Shift Operator}\label{sfeu45}

We now turn to solving the functional equations \eqref{pic7} and \eqref{pi6c} for the height function $X$. In both cases, the solution can be constructed using a Neumann series \citep{neumann}, generated by the translation operator defined as $\hat{\mathsf S}_\X:=1+\hat{\mathsf T}_\X$, where $\hat{\mathsf T}_\X$ is the {\it shift} operator. This operator acts on any analytic function $F(\x)$ via the rule: $\hat{\mathsf S}_{\X}F(\x)=F(\x+{\bm X})$. In this work, we adopt the shift operator $\hat{\mathsf T}_{\bm X}$ as the primary tool for expressing translations. All calculations are performed in the {\it radial gauge}, where the displacement vector is purely radial: $\X=X(\x){\bm n}$.

The unperturbed reference configuration of the fluid is described by the functions $\rho = \rho(r)$, $p = p(r)$, and $U = U(r)$, which are assumed to be known either analytically or through numerical computation to any desired degree of accuracy. These functions depend solely on the radial coordinate $r$ and are fully determined by the solutions to the hydrostatic equilibrium equations presented in Section \ref{basman1}.
 
The total gravitational potential $V$, defined as the sum of the unperturbed potential $U$ and the perturbation $K$, represents the complete gravitational field (cf. Eq. \eqref{un7a5}):
\bal{gt6vc}
V(\x)=U(\x)+K(\x)\;,
\ea
where the perturbation term $K = (\Phi - \U) + W_Q$ encompasses both the deviation $\Phi - \U$ arising from the rotating fluid and the rotational quadrupole contribution $W_Q$ (refer to Eq. \eqref{hj84ss} for $\tau = 1$).
The functional equation \eqref{pi6c}, which governs the deformation of level surfaces for $\tau = 1$, can thus be reformulated as:
\bal{bbb75v}
V(\y):=\Phi(\y)+W(\y)&=&U(\x)\;,
\ea
where $\y=\x+\X$, and ${\bm X}={\bm X}({\bm x})$ denotes the displacement vector of the level surface. This equation may be directly compared with the condition \eqref{xa1} from the Zharkov-Trubitsyn theory \citep{Zharkov_1978book}, which defines the level surfaces of the total potential. The comparison reveals that the Lie group approach identifies $(4\pi G/3)r^2A_0(r)\equiv U(r)$, where $U(r)$ is the effective gravitational potential of the reference fluid configuration, assumed to be known. This identification enables a significantly more efficient methodology for determining the shape of equipotential surfaces and the figure of a rotating fluid body. By employing the shift operator technique, this approach offers substantial advantages over the classical Zharkov-Trubitsyn framework \citep{Zharkov_1978book}. The details of this method are presented in the following sections.

To support the application of the shift operator method, it is beneficial to express the perturbed gravitational potential $K$ in terms of the total potential $V$:
\bal{c3q}
K({\bm x})=V({\bm x})-V\left({\bm x}+{\bm X}\right)\;,
\ea
where ${\bm X}={\bm X}({\bm x})$ denotes the displacement vector of the level surface. As demonstrated in the following section, this formulation serves as the foundation for determining the displacement vector ${\bm X}$, which characterizes the geometry of the level surfaces within a rotating fluid body. Moreover, it enables the determination of the body's external gravitational field and boundary shape, given a prescribed density distribution and equation of state for the reference configuration.

To solve Eq. \eqref{c3q}, we expand $V\left({\bm x}+{\bm X}\right)$ in a Taylor series with respect to the displacement vector ${\bm X}$, assuming the series converges throughout the interior of the fluid body. This yields:
\bal{nb4x}
K({\bm x})&=&-\hat{\mathsf T}_{\bm X}V({\bm x})=-\hat{\mathsf T}_{\bm X}\bigl[U({\bm x})+K({\bm x})\bigr]\;,
\ea
where $\hat{\mathsf T}_{\bm X}$ is the shift operator defined via the infinite Taylor series as follows: 
\bal{asrx5}
\hat{\mathsf T}_{\bm X}&:=&\sum_{n=1}^\infty\frac{1}{n!} X^{i_1}X^{i_2}...X^{i_n}\pd_{i_1i_2...i_n}=X^{i}\pd_{i}+\frac1{2!}X^{i}X^{j}\pd_{ij}+\frac1{3!}X^{i}X^{j}X^k\pd_{ijk}+...\;,
\ea 
using multi-index notation and Einstein summation over repeated spatial indices.

Equation \eqref{nb4x} can be solved iteratively for $K(\x)$, assuming convergence. This leads to a nested expression:
\bal{ft6d3}
K({\bm x})&=&-\hat{\mathsf T}_{\bm X}\left[U({\bm x})-\hat{\mathsf T}_{\bm X}\left[U({\bm x})-\hat{\mathsf T}_{\bm X}\Bigl[U({\bm x})-...\Bigr]\right]\right]\;.
\ea
This is equivalent to expressing $K(\x)$ as a Neumann series \citep{neumann} in powers of the shift operator $\hat{\mathsf T}_{\bm X}$
\bal{bv4rd6}
K&=&\sum_{n=1}^\infty\left(-\hat{\mathsf T}_{\bm X}\right)^nU({\bm x})\;.
\ea

The same solution for $K(\x)$ can also be derived using the symbolic operator method. Rewriting Eq. \eqref{nb4x} gives:
\bal{oper21}
\left(1+\hat{\mathsf T}_{\bm X}\right)K({\bm x})&=&-\hat{\mathsf T}_{\bm X}U({\bm x})\;.
\ea
The formal operator solution of this equation is: 
\bal{iy4d3}
K({\bm x})&=&-\frac{\hat{\mathsf T}_{\bm X}U({\bm x})}{1+\hat{\mathsf T}_{\bm X}}\;.
\ea
Expanding the denominator in a power series confirms the equivalence of Eqs. \eqref{bv4rd6} and \eqref{iy4d3}.
 
It is important to note that the displacement $X^i = X^i({\bm x})$ is a spatially varying vector field. Thus, repeated applications of $\hat{\mathsf T}_{\bm X}$ in Eq. \eqref{bv4rd6} for $n=2,3,4,...$, involve differentiating both the potential $U({\bm x})$ and the components of $X^i({\bm x})$, as defined in the Taylor expansion \eqref{asrx5}. For instance, the quadratic approximation of Eq. \eqref{iy4d3} yields: 
\bal{ioy6}
K(\x)&=&-X^i\pd_iU(\x)+\frac12 X^iX^j\pd_{ij}U(\x)+X^i\pd_iX^j\pd_jU(\x)+\ldots\;,
\ea

The level surfaces of the gravitational potential coincide with surfaces of constant pressure and density \citep{Tassoul_1978book,Zharkov_1978book,horedt_2004book}. As a result, the perturbed density $\rho + \varrho_\tau$ at ${\bm y} = {\bm x} + {\bm X}$ must equal the unperturbed density $\rho$ at the original position $\x$, in accordance with Eq. \eqref{pic7}. For the case $\tau = 1$, this condition can be equivalently expressed as:
\bal{ikr9}
\sigma({\bm x}+{\bm X})+\varrho({\bm x}+{\bm X})&=&\sigma({\bm x})\;.
\ea
where $\sigma$ denotes the reference configuration density, as defined in Eq. \eqref{hunb6}. The substitution $\rho \to \sigma$ is justified here, since $\rho$ and $\sigma$ differ only by a constant offset that cancels out in Eq. \eqref{ikr9}, thereby preserving consistency with Eq. \eqref{pic7}.

Equation \eqref{ikr9} can be reformulated using the shift operator technique, yielding:
\bal{fdr5}
\left(1+\hat{\mathsf T}_{\bm X}\right)\varrho({\bm x})=-\hat{\mathsf T}_{\bm X}\sigma({\bm x})\;.
\ea
This indicates that the density perturbation $\varrho({\bm x})$ satisfies an equation structurally analogous to Eq. \eqref{oper21} for the gravitational perturbation $K(\x)$, involving the same displacement vector ${\bm X}$. Accordingly, its solution takes the form of a Neumann series similar to Eq. \eqref{bv4rd6}:
\bal{jht6a}
\varrho({\bm x})&=&\sum_{n=1}^\infty\left(-\hat{\mathsf T}_{\bm X}\right)^n\sigma({\bm x})\;.
\ea

Equations \eqref{bv4rd6} and \eqref{jht6a} express the gravitational perturbation $K(\x)$ and the density perturbation $\varrho(\x)$, respectively, as power series in the shift operator $\hat{\mathsf T}_{\bm X}$ acting on the unperturbed potential $U({\bm x}) = U(r)$ and density $\rho({\bm x}) = \rho(r)$. The displacement vector ${\bm X} = {\bm X}({\bm x})$ represents the finite radial deformation of the level surfaces in the reference configuration due to the rotational quadrupole perturbation $W_Q$.

As in the linearized theory, the height function $X = X(\x)$ can be determined by two methods. The first method generalizes Bruns' theorem (Eq. \eqref{bruns}) and involves solving Eq. \eqref{oper21}, which relates $K(\x)$ and $X(\x)$, in the form of a power series. This approach requires the explicit form of the gravitational perturbation $K(\x)$, which is obtained by solving the nonlinear Molodensky equation \eqref{fe87}.

The second method for determining the height function $X$ generalizes the Molodensky equation \eqref{mn5c32} to the nonlinear regime, accounting for finite deformations of the level surfaces. This approach involves a detailed manipulation of the shift operator $\hat{\mathsf T}_{\bm X}$ to derive a fundamental differential equation for $X(\x)$ that eliminates the explicit dependence on the gravitational perturbation $K(\x)$. A comprehensive discussion of both methods is presented in the following two sections.

\subsection{Developing Power Series Solution for the Height Function}\label{ssec5}

In this section, we assume that the perturbation field $K=K(\x)$ is known as a solution to the Molodensky equation \eqref{fe87}. To determine the height function $X = X(\x)$ of the perturbed level surface above a reference sphere of radius $r$, we apply the power series method to Eq. \eqref{oper21}, utilizing the expansion of the shift operator $\hat{\mathsf T}_{\bm X}$ given in Eq. \eqref{asrx5}. Since the displacement vector $X^i = X n^i$ is purely radial, Eq. \eqref{oper21} simplifies to a scalar relation involving only the height function $X$ and the radial derivatives of the gravitational potential $U$ and its perturbation $K$: 
\bal{oper22}
\sum_{n=1}^\infty \frac{\a_n}{n!}X^n&=&-K\;,
\ea
where the coefficients are defined as $\a_n = U^{(n)} + K^{(n)}$. Here, $U = U(r)$ is the known gravitational potential of the unperturbed configuration, and $K = K(\bm{x})$ is the perturbation field obtained from Eq. \eqref{fe87}. The terms $U^{(n)} := \partial_r^n U(r)$ and $K^{(n)} := \partial_r^n K(\x)$ denote the $n$-th order radial derivatives. The objective is to invert Eq. \eqref{oper22} to express the height function $X$ explicitly in terms of the known functions $U$ and $K$.

To invert Eq. \eqref{oper22}, we apply the Lagrange inversion theorem \citep{lagrangeinv}, following the approach outlined in Section \ref{bubu54}. This yields a power series representation for $X$:
\bal{oper23}
X&=&\sum_{n=1}^\infty(-1)^n\frac{\b_n}{n!}\left(\frac{K}{\a_1}\right)^n\;,
\ea
where $\b_1 = 1$, and for $n \ge 2$,
\bal{oper24}
\b_{n}&=&\sum_{k=1}^{{n}-1}(-1)^k({n})_{k}{\bf B}_{{n}-1,k}\left(\gamma_1,\gamma_2,...,\gamma_{n-k}\right)\;,\
\ea
with ${\bf B}_{n-1,k}$ denoting the incomplete Bell polynomials -- see Eq. \eqref{het65}. The arguments $\gamma_k$ are defined as
\bal{s11aa}
\gamma_k:&=&\frac{1}{k+1}\frac{\a_{k+1}}{\a_1}=\frac{1}{k+1}\frac{U^{(k+1)}+K^{(k+1)}}{U'+K'}\;,
\ea
with the prime denoting differentiation with respect to the radial coordinate $r$.

Since the perturbation magnitude satisfies $K \sim \m$, the displacement vector $X$ can be expressed as an infinite power series in the small parameter $\m$. This expansion involves the unperturbed gravitational potential $U$, its perturbation $K$, and their radial derivatives. Such a formulation extends the classical linearized Bruns' theorem (cf. Eq. \eqref{bruns}) to higher-order perturbation theory, providing a systematic framework for analyzing nonlinear deformations of level surfaces in rotating fluid bodies.

In practical applications, only a finite number of terms from the series expansion in Eq. \eqref{oper23} are required for computation. Higher-order derivatives of the unperturbed potential $U$ can be analytically reduced to lower-order terms using the field equation \eqref{mku6}, as demonstrated in Eqs. \eqref{klm8h}-\eqref{jn7bv}.

The angular dependence of the solution $X = X(r, \theta, \varphi)$ arises through the perturbation $K = K(r, \theta, \varphi)$, which is assumed to be known from the solution of Eq. \eqref{fe87}.

When both $U$ and $K$ are finite-order polynomials in the radial coordinate $r$, the number of arguments in the Bell polynomial appearing in Eq. \eqref{oper24} is bounded by the polynomial degree. Consequently, only a finite number of coefficients $\b_n$ contribute to the expansion in Eq. \eqref{oper23}, reducing the solution for $X$ to a finite series. However, since $K$ also appears in the denominator of the coefficients $\gamma_k$, its expansion in powers of $K \sim \m$ introduces an infinite series. Thus, even with finitely many $\b_n$, the full solution \eqref{oper23} remains an infinite series in the perturbation parameter $\m$.

A detailed application of this power series method for computing the height function $X$ in case of a polytropic model with index $n=1$ is presented in Section \ref{subD2}.

\subsection{Differential Equations for Fluid Body Deformations}\label{ref55}

\subsubsection{Master Equation for the Deformation Gradient}\label{hhh32}

Traditionally, the structure of level surfaces and the shape of a rotating fluid body -- such as a planet or star -- are analyzed using perturbation theory, with a small parameter $\m$, defined in Eq. \eqref{tttr4}. Within this framework, the perturbed gravitational potential $V$ is expanded in a Legendre polynomial series, and the height function is expressed as $X = \sum_{l=0}^\infty X_l(r) P_l(\cos\theta)$, where each spectral harmonic scales as $X_l \sim \m^{l/2}$. At each order $l$, the corresponding equations for $X_l$ are derived using various perturbative techniques, as outlined in the introduction of this manuscript. Regardless of the specific scheme employed, the computational complexity of formulating and solving these equations increases substantially with higher-order harmonics.

The most advanced developments in this field originate from the Zharkov-Trubitsyn theory of figures \citep{Zharkov_1978book}, which has been extended up to seventh order ($l = 14$, corresponding to $\m^7$) \citep{Nettelmann_2021}. However, the feasibility of further extending perturbative methods remains uncertain. Deriving higher-order equations for $X_l$ leads to increasingly complex, nonlinear integro-differential equations with a rapidly growing number of terms, necessitating extensive symbolic computation. Moreover, the numerical evaluation of higher-order contributions is prone to instability, as small inaccuracies in lower-order terms can propagate and significantly distort the final results. These challenges underscore the need for carefully designed algorithms and robust numerical strategies \citep{jorba_2022}.

The challenges associated with computing higher-order perturbations can be effectively addressed by adopting a non-perturbative approach to deriving the differential equation for the height function $X$, thereby bypassing the traditional Legendre polynomial decomposition of the gravitational potential $V$. The method presented in this section avoids reliance on successive approximations and instead exploits the mathematical structure of the shift operator $\hat{\mathsf T}_{\bm X}$, analytic summation of infinite series, and tools from linear matrix algebra. This framework enables the derivation of a general, exact master differential equation for $X$. Once this equation is obtained, the spectral harmonics $X_l$ can be extracted by expanding the master equation in Legendre polynomials using Wigner's addition theorem for spherical harmonics -- a straightforward algebraic procedure \citep{gelfand_1963}.

The derivation begins with the formulation of a master equation for the {\it deformation gradient matrix} $A_{ij} := \pd_i X^j$, where the displacement vector is given by $X^i = X n^i$ representing a purely radial field. The starting point is Eq. \eqref{nb4x}, which, using the definition of the total potential $V$ from Eq. \eqref{gt6vc}, can be rewritten as:
\bal{xc1}
\left(1+\hat{\mathsf T}_{\bm X}\right)V({\bm x})=U({\bm x})\;.
\ea
Differentiating Eq. \eqref{xc1} and applying the Leibniz rule to the product $\hat{\mathsf T}_{\bm X} V({\bm x})$ yields:
\bal{xc2}
\left(\pd_i\hat{\mathsf T}_{\bm X}\right)V({\bm x})+\left(1+\hat{\mathsf T}_{\bm X}\right)\pd_iV({\bm x})=\pd_i U({\bm x})\;.
\ea
The derivative of the shift operator, based on its definition in Eq. \eqref{asrx5}, is given by:
\bal{xc3}
\pd_i\hat{\mathsf T}_{\bm X}&=&\pd_iX^p\left(1+\hat{\mathsf T}_{\bm X}\right)\pd_p\;.
\ea
Substituting this into Eq. \eqref{xc2} leads to the matrix equation:
\bal{xc4}
M_{ij}({\bm X})\left(1+\hat{\mathsf T}_{\bm X}\right)\pd_j V({\bm x})&=&\pd_i U({\bm x})\;,
\ea
where the matrix 
\bal{xc5}
M_{ij}({\bm X})&=&\frac{\pd y^j}{\pd x^i}=\delta_{ij}+A_{ij}\;,
\ea
and $A_{ij} := \pd_i X^j$ is the deformation gradient \citep{Vanicek_book}. Notably, $A_{ij}$ is generally non-symmetric ($A_{ij} \ne A_{ji}$), and therefore $M_{ij}$ is also non-symmetric -- even in the case of an ideal fluid.

The inverse matrix $M^{-1}_{ij}({\bm X})$ is defined by the standard relations:
\bal{xc5a}
M^{-1}_{ik}({\bm X})M_{kj}({\bm X})=\delta_{ij}\qquad&\;\mbox{and}&\qquad M_{ik}({\bm X})M^{-1}_{kj}({\bm X})=\delta_{ij}\;,
\ea
with summation implied over repeated indices from 1 to 3. Substituting the definition of $M_{ij}$ from Eq. \eqref{xc5} into these identities yields the explicit form:
\bal{xc6}
M^{-1}_{ij}({\bm X})&=&\frac{\pd x^j}{\pd y^i}=\delta_{ij}+B_{ij}({\bm X})\;,
\ea
where $B_{ij}$ is given by the Neumann series expansion in powers of the deformation gradient:
\bal{xc6a}
B_{ij}({\bm X})=\sum_{n=1}^\infty(-1)^n({\bf A}^n)_{ij}=\sum_{n=1}^\infty(-1)^nA_{ip_1}A_{p_1p_2}...A_{p_{n-1}j}\;.
\ea
As before, repeated indices imply summation over the range 1 to 3.

Using the inverse matrix \eqref{xc6}, Eq. \eqref{xc4} can be rewritten as:
\bal{xc7}
\left(1+\hat{\mathsf T}_{\bm X}\right)\pd_i V({\bm x})&=&M^{-1}_{ij}({\bm X})\pd_j U({\bm x})\;.
\ea
This equation mirrors the structure of Eq. \eqref{xc1}, which becomes evident by substituting $V({\bm x}) \rightarrow \pd_i V({\bm x})$ and $U({\bm x}) \rightarrow M^{-1}_{ij}({\bm X}) \pd_j U({\bm x})$. Following the same reasoning that led from \eqref{xc1} to \eqref{xc7}, we obtain the corresponding equation for the second partial derivative:
\bal{xc8}
\left(1+\hat{\mathsf T}_{\bm X}\right)\pd_{ij} V({\bm x})&=&M^{-1}_{ip}({\bm X})\pd_p\left[M^{-1}_{jq}({\bm X})\pd_q U({\bm x})\right]\;.
\ea
The left-hand side of Eq. \eqref{xc8} is symmetric in the indices $i$ and $j$. Therefore, the compatibility condition requires that the antisymmetric part of the right-hand side vanishes identically:
\bal{ii1aa}
M^{-1}_{[ip}({\bm X})\pd_p\left[M^{-1}_{j]q}({\bm X})\pd_q U({\bm x})\right]\equiv 0\;.
\ea
Although this identity may not be immediately evident, it can be rigorously proven (see Appendix \ref{antisym}). 

Contracting the free indices in Eq. \eqref{xc8} yields:
\bal{xc9}
\left(1+\hat{\mathsf T}_{\bm X}\right)\Delta V({\bm x})&=&M^{-1}_{ip}({\bm X})\pd_p\left[M^{-1}_{iq}({\bm X})\pd_q U({\bm x})\right]\;.
\ea
This expression can be simplified using the definition of $V$ from Eq. \eqref{gt6vc}. The Laplacian of $V$ is obtained via the Poisson equations for $U$ and $K$ (Eqs. \eqref{tyc438} and \eqref{41cs}), omitting the surface density term $\nu\equiv\nu_1$ since we are working within the domain $r < a$:
\bal{bxg655}
\Delta V({\bm x})&=&-4\pi G\left[\sigma({\bm x})+\varrho({\bm x})\right]\;,
\ea
where $\varrho$ is the density perturbation defined in Eq. \eqref{jht6a}. Applying Eq. \eqref{fdr5}, which can be equivalently written as:
\bal{nb5f2}
\left(1+\hat{\mathsf T}_{\bm X}\right)\left[\sigma({\bm x})+\varrho({\bm x})\right]&=&\sigma({\bm x})\;,
\ea
we find that the left-hand side of Eq. \eqref{xc9} simplifies to:
\bal{gt4x3}
\left(1+\hat{\mathsf T}_{\bm X}\right)\Delta V({\bm x})&=&-4\pi G\sigma({\bm x})\;.
\ea
Substituting this result into Eq. \eqref{xc9} yields the {\it master equation for the deformation gradient}:
\bal{xc9a}
\delta^{ij}M^{-1}_{ip}({\bm X})\pd_p\left[M^{-1}_{jq}({\bm X})\pd_q U({\bm x})\right]=-4\pi G\sigma({\bm x})\;.
\ea
This equation represents a central result in the theory of figures of rotating fluid bodies. Although exact, it is highly nonlinear due to the dependence of the inverse matrix $M^{-1}_{ij}$ on the deformation gradient $A_{ij}$, which itself is expressed as an infinite Neumann series -- see Eqs. \eqref{xc6} and \eqref{xc6a}.

At first glance, Eq. \eqref{xc9a} may resemble a Poisson equation for the effective gravitational potential $U = U(\x)$ in the reference configuration, but expressed in the deformed coordinates $y^i = x^i + X^i(\x)$. Indeed, by using the definitions of the matrix $M_{ij}$ and its inverse, Eq. \eqref{xc9a} can be rewritten as:
\bal{cax5}
\delta^{ij}\frac{\pd^2 U(\x)}{\pd y^i\pd y^j}&=&-4\pi G\sigma(\x)\;.
\ea
This form might be naively interpreted as a straightforward coordinate transformation of the Poisson equation \eqref{mku6} from the original coordinates $\x$ to the deformed coordinates $\y=\x+\X$. However, this interpretation is incorrect. When Eq. \eqref{mku6} is properly expressed in terms of the deformed coordinates $\y$, it takes the form:
\bal{nz5c4}
\delta^{ij}M_{ip}(\X)\frac{\pd}{\pd y^p}\left[M_{jq}(\X)\frac{\pd U(\x)}{\pd y^q}\right]&=&-4\pi G\sigma(\x)\;.
\ea 
This equation is not equivalent to either Eq. \eqref{cax5} or Eq. \eqref{xc9a}. The original Poisson equation \eqref{mku6} is used to determine the reference potential $U=U(\x)$ as a function of the radial coordinate $r=|\x|$, as shown in Eq. \eqref{om5}. In contrast, Eq. \eqref{cax5} (or Eq. \eqref{xc9a}) is employed to determine the height function $\X = \y - \x$, under the assumption that the unperturbed fluid density $\sigma(\x)$ and the potential $U(\x)$ are already known from the solution describing the unperturbed fluid configuration.

Equation \eqref{xc9a} is preferable to Eq. \eqref{cax5} for further theoretical development. The primary advantage lies in its matrix formulation, which features the inverse matrix $M^{-1}_{ij}$ expressed as an infinite Neumann series. Remarkably, this series can be summed analytically, reducing Eq. \eqref{xc9a} to a single second-order partial differential equation for the height function $X$. This leads to a desirable nonlinear generalization of the classical Clairaut equation. 

This result is novel: earlier perturbative methods were unable to manage the infinite series of nonlinear terms that arise when attempting a closed-form expansion of the perturbed gravitational potential $\Phi$. In contrast, our method -- based on the summable Lie-Neumann series -- offers a more transparent and tractable framework for studying the stratification of level surfaces in rotating fluid bodies. Unlike traditional techniques, it avoids expanding the gravitational potential in terms of Legendre polynomials, thereby sidestepping the divergence issues that commonly afflict such expansions.

\subsubsection{Master Equation for the Height Function}\label{dmeq43}

The master equation \eqref{xc9a} governing the deformation gradient is formulated in terms of the inverse matrix $M^{-1}_{ij}$, which is represented as a Neumann series comprising an infinite sum of terms involving the deformation gradient matrix $A_{ij}$, itself dependent on the derivative of the radial displacement vector $X^i$. At first glance, this formulation may appear to offer no immediate simplification in deriving the differential equation for the height function. However, this is not the case. Notably, the Neumann series in Eq. \eqref{xc9a} can be summed analytically, allowing the original matrix-based formulation to be recast as a scalar partial differential equation involving only the height function $X$.

This transformation is of considerable significance: it reduces a complex, nonlinear matrix equation to a more tractable scalar form, while fully preserving the nonlinearity inherent in the original formulation. The key steps of this transformation are outlined below:
\begin{enumerate}
\item The inverse matrix $M^{-1}_{ij}$ is expressed in closed form by leveraging the analytic structure of the deformation gradient.
\item This closed-form expression is substituted into the master equation \eqref{xc9a}, reformulating it entirely in terms of the height function $X=X(\x)$ and its derivatives.
\item The resulting expression is contracted and simplified, yielding a second-order partial differential equation for the scalar field $X=X(\x)$.
\end{enumerate}

Taking the partial derivative on the left-hand side of the master equation \eqref{xc9a} and dividing both sides by the radial derivative of the unperturbed gravitational potential, $U'$, yields the following expression:
\bal{oo2}
n^qM^{-1}_{ip}({\bm X})\pd_pM^{-1}_{iq}({\bm X})+M^{-1}_{ip}({\bm X})M^{-1}_{iq}({\bm X})\left[\frac{{\cal P}^{pq}}{r}-n^pn^q\left(\frac{4\pi G\sigma}{U'}+\frac{2}{r}\right)\right]=-\frac{4\pi G\sigma}{U'}\;,
\ea
where repeated indices imply summation over the range 1 to 3, and the tensor ${\cal P}^{pq}:=\delta^{pq}-n^pn^q$ denotes the projection operator onto the plane orthogonal to the radial unit vector ${\bm n}=(n^p)$.
Introducing the auxiliary vector $N^i := M^{-1}_{ip} n^p$, the equation simplifies to:
\bal{oo4}
M^{-1}_{ip}\pd_pN^i+\frac{3{\upbeta}}{r}\left(N^2-1\right)-\frac{2}{r}N^2&=&0\;,
\ea
where $N^2 := \delta_{ij} N^i N^j$, and the substitution $4\pi G \sigma / U' \to -3\upbeta / r$ follows from Eq. \eqref{bni5}.

To further simplify the expression, we evaluate the divergence term as follows:
\bal{t19}
M^{-1}_{ip}\pd_pN^i&=&M^{-1}_{ip}\delta^{pq}\pd_qN^i=M^{-1}_{ip}\left({\cal P}^{pq}+n^pn^q\right)\pd_qN^i=M^{-1}_{ip}{\cal P}^{pq}\pd_qN^i+ \frac12\pd_rN^2\;.
\ea
Applying Eq. \eqref{ek3}, this expression can be rewritten as:
\bal{tty1}
M^{-1}_{ip}\pd_pN^i&=&\frac{r}{R}\left[\pd_iN^i-\pd_r\left(\frac{1}{R'}\right)\right]+ \frac12\pd_rN^2\;,
\ea
where we introduce the auxiliary variable $R \equiv r + X$, with $R' = 1 + X'$, and the prime denotes partial differentiation with respect to the radial coordinate $r$. Substituting Eq. \eqref{tty1} into Eq. \eqref{oo4} yields the following scalar partial differential equation for the height function $X$:
\bal{tty2}
\pd_iN^i+\frac{R''}{R'^2}+ \frac{R}{r}\left(\frac12\pd_rN^2-\frac{2N^2}{r}+3\frac{N^2-1}{r}{\upbeta}\right)&=&0\;.
\ea

In terms of the variable $R\equiv r+X$, the vector $N^i$, as defined explicitly in Eq. \eqref{ek10}, can be expressed using the radial displacement $X$ and its derivative $X'=\pd_r X$, as follows:
\bal{vv1}
N^i&=&\frac{n^i}{R'}-\frac{r{\cal P}^{ij}{R_j}}{R'R}=\frac{n^i}{R'}+\frac{rn^i}{R}-\frac{r{R_i}}{R'R}\;,\\
\label{ek11}
N^2&=&\frac{1}{R'^2}+\frac{r^2{\cal P}^{ij}{R_i}{R_j}}{R'^2R^2}=\frac{1}{R'^2}-\frac{r^2}{R^2}+\frac{r^2{R_i}{R_i}}{R'^2R^2}\;,
\ea
where $R_i\equiv\pd_i R=\pd R/\pd x^i$ denotes the partial derivative of $R$. The divergence of $N^i$ is given by:
\bal{tty4}
\pd_i N^i&=&\frac{2}{rR'}+\frac{1}{R}-\frac{r R'}{R^2}-\frac{R''}{R'^2}+\frac{rn^p{R_i} \pd_{ip}R}{R'^2R}+\frac{R+rR'}{R'^2R^2}{R_i}{R_i}-\frac{r\Delta R}{R'R}\;,
\ea
where $\Delta=\delta^{ij}\pd^2/\pd x^i\pd x^j$ is the Laplace operator.
The radial derivative of $N^2$, as derived from Eq. \eqref{ek11}, is given by:
\bal{bgte5}
\frac12\pd_rN^2&=&\frac12\pd_r\left(\frac{1}{R'^2}-\frac{r^2}{R^2}+\frac{r^2{R_i}{R_i}}{R'^2R^2}\right)\\\nonumber
&=&-\frac{R''}{R'^3}-\frac{r}{R^2}+\frac{r^2 R'}{R^3}+ \frac{r{R_i}{R_i}}{R'^2R^2}+\frac{r^2n^p\pd_{ip}R{R_i}}{R'^2R^2}-\frac{r^2R''{R_i}{R_i} }{R'^3R^2}-\frac{r^2{R_i}{R_i} }{R'R^3}\;.
\ea

By adding Eqs. \eqref{tty4} and \eqref{bgte5}, we arrive at the following result:
\bal{tys21}
&&\pd_iN^i+\frac{R''}{R'^2}+ \frac{R}{r}\left(\frac12\pd_rN^2-\frac{2N^2}{r}\right)\\\nonumber
&&\hspace{2cm}=\frac{r}{R'R}\left[-\Delta{R}+\frac{2R'}{r}+\frac{2R}{r^2}-\frac{R''R^2}{r^2R'^2}-\frac{2R^2}{r^3R'}+\frac{2n^p{R_i}\pd_{ip}R}{R'}-\frac{R''{R_i}{R_i}}{R'^2}\right]\;,
\ea
which simplifies to:
\bal{ttys2}\pd_i N^i + \frac{R''}{R'^2} + \frac{R}{r} \left( \frac{1}{2} \pd_r N^2 - \frac{2 N^2}{r} \right)&=&\frac{r}{R'R}\left[-\Delta{R}+\frac{2R'}{r}+\pd_r\left(\frac{R^2}{r^2R'}+\frac{{R_i}{R_i}}{R'}\right)\right]
\;.
\ea
Substituting this result into Eq. \eqref{tty2}, and incorporating the relation from Eq. \eqref{ek11}, we obtain the final expression:
\bal{aaa234}
\Delta{R}-\frac{2R'}{r}-\pd_r\left(\frac{R^2}{r^2R'}+\frac{{R_i}{R_i}}{R'}\right)+\frac{3{\upbeta}}{r}\left[\frac{R^2}{r^2}R'
-\frac{1}{R'}\left(\frac{R^2}{r^2}+{\cal P}^{ij}{R_i}{R_j}\right)\right]&=&0\;.
\ea
This nonlinear second-order partial differential equation governs the total radial displacement $R=r+X$ of the level surface from its unperturbed spherical configuration with radius $r$. It encapsulates the influence of the quadrupole rotational perturbation $W_Q$ on the equilibrium shape of a rotating fluid body. 

\subsubsection{Master Equation for the Shape Function}

The height function $X=X(\x)$ describes the radial displacement of a level surface above a reference sphere of radius $r$. It plays a fundamental role in analyzing and computing perturbations in the density, pressure, and gravitational field of a rotating fluid body. Moreover, it is essential for determining the multipole moments of the external gravitational field of planets and stars, thereby linking local deformations to global gravitational signatures.

For global analyses of the shape of a rotating astronomical body, it is often more practical to introduce a dimensionless measure of deformation -- the shape function $f=f(\x)$ -- defined as the normalized radial displacement:
\bal{oi831}
f(\x)\equiv\frac{X(\x)}{r} \;.
\ea
The shape function offers a concise, scale-invariant characterization of the global deviation of the perturbed level surface from a perfect sphere of radius $r$, making it particularly useful for comparative studies across different planetary bodies or rotational regimes.

The master equation for the shape function $f$ is derived from Eq. \eqref{aaa234} through a sequence of variable transformations. As a first step, we introduce the logarithmic variable:
\bal{nui86}
Z&\equiv&\log\left(\frac{R}{r}\right)=\log\left(1+f\right)\;.
\ea
Rewriting Eq. \eqref{aaa234} in terms of $Z$ yields the following nonlinear partial differential equation: 
\bal{ecty6}
\Delta Z-\frac{2}{r}Z'-\frac{\pd}{\pd r}\left(\frac{r{\pd_iZ}{\pd_iZ}}{1+rZ'}\right)-3{\upbeta}\left[\frac{1-e^{2Z}}{r^2}-\frac{1+e^{2Z}}{r}Z'+\frac{{\pd_iZ}{\pd_iZ}}{1+rZ'}\right]&=&0\;,
\ea
where $\pd_iZ \equiv \pd Z/\pd x^i$.
This equation is transcendental and highly nonlinear. Once a solution for $Z$ is obtained, the shape function $f$ can be recovered by inverting Eq. \eqref{nui86}. Alternatively, one may formulate and solve a differential equation directly in terms of $f$.

To proceed, we invert Eq. \eqref{nui86} to express the shape function $f$ in terms of variable $Z$:
\bal{mnu9}
e^Z&=&1+f\;.
\ea
Differentiating both sides and introducing the notations $\pd_if \equiv \pd f/\pd x^i$, we obtain:
\bal{mn10}
e^Z\pd_iZ=\pd_if\;\quad, \qquad\quad e^Z Z'=f'\;\quad,\qquad\quad e^Z\Delta Z=\Delta f-\frac{\pd_if\pd_if}{1+f}\;,
\ea
and
\bal{mn51}
e^Z\frac{{\pd_iZ}{\pd_iZ}}{1+rZ'}=\frac{\pd_if\pd_if}{1+(rf)'}\;\quad,\qquad\qquad e^Z\frac{\pd}{\pd r}\left(r\frac{{\pd_iZ}{\pd_iZ}}{1+rZ'}\right)=\frac{\pd}{\pd r}\left[\frac{r\pd_if\pd_if}{1+(rf)'}\right]-\frac{\pd_if\pd_if}{1+f}\frac{rf'}{1+(rf)'}\;.
\ea
Multiplying Eq. \eqref{ecty6} by $e^Z$ and applying the relations \eqref{mnu9}--\eqref{mn51}, we arrive at:
\bal{cre5}
\Delta f-\frac{2f'}{r}+\frac{6{\upbeta}}{r}\left(f'+\frac{f}{r}\right)-r\frac{\pd}{\pd r}\left[\frac{\pd_if\pd_if}{1+X'}\right]-\frac{2\pd_if\pd_if}{1+X'}
+3{\upbeta}\left[\frac{3f^2+f^3}{r^2}+\frac{(2f+f^2)f'}{r}-\frac{\pd_if\pd_if}{1+X'}\right]&=&0\;.
\ea
This equation contains both linear and nonlinear terms. By isolating the linear terms on the left-hand side and moving the nonlinear terms to the right-hand side, we obtain the master equation for the shape function $f=f(\x)$ in the form:
\bal{cre5a}
\hat{\mathsf D}f&=&\frac{S(f)}{r}\;,
\ea
where the second-order differential (Clairaut) operator is defined as:
\bal{jx4}
\hat{\mathsf D}&\equiv&\Delta-\frac{2}{r}\frac{\pd}{\pd r}+\frac{6\upbeta}{r}\left(\frac{\pd}{\pd r}+\frac{1}{r}\right)\;.
\ea
The nonlinear source term on the right-hand side of Eq. \eqref{cre5a} is defined as:
\bal{u4iw}
S(f)&\equiv& A'-\upbeta B'+\frac{3\upbeta}{r}(A-B)\;.
\ea
where we introduce the shorthand notations:
\bal{t5cx}
A&\equiv&\frac{Y}{1+X'}\qquad\quad,\quad\qquad B\equiv 3f^2+f^3\;,
\ea
with $X'=(rf)'=rf'+f$ and $Y\equiv r^2 \pd_if\pd_if$. Here, the prime denotes partial differentiation with respect to the radial coordinate $r$.

The master equation \eqref{cre5a} for the shape function $f=f(\x)$ is a new result not previously obtained in the literature. It is exact, depends on all three spatial coordinates, and is nonlinear. Remarkably, in the case of constant density, where the parameter ${\upbeta} = 1$, Eq. \eqref{cre5a} admits several exact solutions. Among these are the classical Maclaurin spheroid and the Jacobi triaxial ellipsoid, both of which are discussed in detail in Section \ref{appmac}. The explicit formulation of the master equation for the shape function provides a rigorous foundation for establishing the asymptotic nature of the perturbation series previously employed by various researchers in the study of rotating fluid bodies in both geophysical and astrophysical contexts \citep{Lebovitz_1970}.

Unfortunately, Eq. \eqref{cre5a} cannot be solved analytically in the general case, necessitating the use of approximation methods. The following section outlines an approach for constructing approximate solutions to Eq. \eqref{cre5a} under the assumption of axisymmetric level surfaces, where the shape function $f(\x)=f(r,\theta)$ depends only on the spherical coordinates $r$ and $\theta$. The approximations are developed using the method of separation of variables, based on the spectral decomposition of the shape function into zonal spherical harmonics, which is a standard technique in planetary and stellar modeling \citep{Zharkov_1978book,horedt_2004book}. 

A key advantage of our approach is that it naturally leads to ordinary differential equations governing the spectral radial harmonics of the shape function. This represents a significant improvement over the Zharkov-Trubitsyn theory, in which deriving such equations from the underlying integro-differential framework is both complex and cumbersome \citep{Lanzano1962, Lanzano_1974, Lanzano_1982}. Moreover, the traditional method does not provide a general analytic form for these equations at arbitrary orders of approximation. In contrast, our formulation offers a systematic and transparent pathway for obtaining such results, enabling more efficient and scalable analyses of rotating fluid bodies.

\section{Differential Equation Framework for Radial Spectral Harmonics}\label{sec5}
\stepcounter{equationschapter}
\renewcommand{\theequation}{7.\arabic{equation}}

In this section, we employ the Legendre polynomial series method to systematically reduce the nonlinear partial differential equation \eqref{cre5a} for the shape function $f({\bm x}) = f(r, \theta)$ to a set of ordinary differential equations. These equations govern the radial dependence of the Legendre coefficients -- referred to as radial harmonics -- which encapsulate the angular structure of the solution through a spectral expansion in terms of Legendre polynomials.

We assume that the dimensionless parameter $\m$, introduced in Eq. \eqref{tttr4}, is small ($\m \ll 1$), and perform a perturbative expansion of the master equation \eqref{cre5a} in powers of $\m$. This expansion linearizes the problem order by order, yielding a hierarchy of inhomogeneous linear ordinary differential equations for the radial harmonics. At each order, the source terms on the right-hand side are explicitly determined by the solutions obtained at lower orders. This structure enables an iterative solution procedure, where the system is solved sequentially, starting from the lowest-order approximation.

Our approach to deriving the system of approximate equations offers a significant improvement in efficiency over traditional methods, which typically involve expanding the gravitational potential into spherical harmonics. Such methods lead to a system of coupled integro-differential equations that are often complex and computationally demanding \citep{Zharkov_1978book, Zharkov-book-1986, Hubbard-book}. In contrast, the present work introduces a direct procedure for obtaining differential equations for the radial harmonics of the shape function from the master equation \eqref{cre5a}, thereby eliminating the need to handle integro-differential formulations altogether.

\subsection{Spectral Decomposition of the Shape Function}\label{onuv5}

In the present paper, we assume that the uniformly rotating body is axially symmetric. This implies that the height function $X$, which characterizes the perturbation of the level surface of a uniformly rotating fluid body, depends only on two coordinates: $X = X(r, \theta)$, where $r$ and $\theta$ denote the radial and polar angular variables, respectively. Under this assumption, the shape function $f$ also depends solely on $r$ and $\theta$, i.e., $f \equiv f(r, \theta)$. It can be expanded in a series of Legendre polynomials (zonal harmonics) as follows:
\bal{zx2}
f(r,\theta)=\sum_{k=0}^\infty f_{k}(r)P_{k}(\cos\theta)\;,
\ea
where $P_k(\cos\theta)$ are the Legendre polynomials of degree $k$, and $f_k(r)$ are the corresponding radial harmonics. We assume that the series \eqref{zx2} is convergent and that each term satisfies the scaling $f_k \sim {\mathsf{m}}^{k/2}$ for $k \ge 1$. The monopole term $f_0$ is of higher order, scaling as $f_0 \sim {\mathsf{m}}^2$, as will be discussed in detail below.

It is also useful to express the spectral decomposition of the height function $X(r, \theta) = r f(r, \theta)$ and its radial derivative, both of which appear in the master equation \eqref{cre5a}. The height function can be expanded as:
\bal{zx2ss}
X(r,\theta)=\sum_{k=0}^\infty X_{k}(r)P_{k}(\cos\theta)\;,
\ea
and its radial derivative as: 
\bal{zx2ab}
X'(r,\theta)=\sum_{k=0}^\infty X'_{k}(r)P_{k}(\cos\theta)\;,
\ea
where the radial harmonics of the height function are given by $X_k(r) = r f_k(r)$, and their radial derivatives are: 
\bal{u7vqw}
X'_{k}&=&rf'_{k}+f_{k}\;.
\ea

It is important to emphasize that the expansions in Eqs. \eqref{zx2} and \eqref{zx2ss} include the monopole harmonic $f_0$, corresponding to the index $k = 0$. This harmonic is purely radial and, as demonstrated below, arises from the nonlinear coupling of higher-order harmonics $f_k$ with $k \ge 2$ in the nonlinear master equation \eqref{cre5a}. The monopole term $f_0$ can be interpreted as an infinitesimal gauge transformation of the radial coordinate $r$. Its presence reflects the inherent freedom in parametrizing the level surfaces, which stems from the arbitrariness in choosing the reference unit for measuring the radial coordinate \citep{Zharkov_1978book}. This gauge freedom may manifest differently depending on the computational framework adopted to solve the system of equations governing the radial harmonics of the shape function via successive approximations.

For example, within the Zharkov-Trubitsyn framework \citep{Zharkov_1978book}, the governing system consists of a set of equations of the form $A_{2n}=0$, where each functional $A_{2n}=A_{2n}\left(\rho, f_0,f_2,...\right)$, for all $n\ge 1$ depends on the density profile $\rho=\rho(r)$, the shape functions $f_{2n}=f_{2n}(r)$ with $n=0,1,2,\ldots$, and their integrals. These equations are supplemented by the hydrodynamic equilibrium equation:  
\bal{hz3a}
\rho\frac{dU}{dr}-\frac{dp}{dr}&=&-\frac{4\pi G\rho}{3}\frac{d}{dr}\left(r^2A_0-\frac{3U}{4\pi G}\right)\;,
\ea
where the right-hand side represents the radial force arising from rotational perturbations, with the potential $U=U(r)$ defined in Eq. \eqref{om5}. The functional $A_0=A_0\left(\rho, f_0,f_2,...\right)$ is constructed from the same density profile and shape functions as the $A_{2n}$ functionals, and contributes to the description of the body's internal structure and equilibrium shape.

A count of the unknown functions and equations reveals that the system contains one more unknown function in the set $\{\rho, f_0, f_2,f_4,\ldots\}$ than the number of equations. This implies the presence of one degree of freedom, which originates from the arbitrariness in defining the reference unit for radial distance. The overdeterminacy of the system can be resolved in various ways, depending on the computational scheme employed.

Zharkov and Trubitsyn \citep{Zharkov_1978book} treated the monopole (purely radial) perturbation of the gravitational field and the level surfaces separately from the higher-order harmonics. In their formulation, the choice of the radial coordinate is fixed by requiring that the monopole term $f_0$ ensures volume conservation: the volume enclosed by the perturbed level surface, described by the function $R(r, \theta) = r + X(r, \theta)$, remains equal to that of the unperturbed sphere of radius $r$. This condition imposes the identity: 
\bal{gt4}
\frac{4\pi r^3}{3}=\int_0^{R(r,\theta)}\int_0^{\pi}\int_0^{2\pi}\xi^2\sin\theta d\xi d\theta d\varphi\;,
\ea
which guarantees volume conservation under the perturbation. The integral on the right-hand side can be evaluated using the Wigner decomposition of products of Legendre polynomials, as detailed in Appendix~\ref{wignerdecomp}. Equation~\eqref{gt4} can then be solved iteratively for $f_0$ to any desired order of accuracy. The first few terms of the expansion are
\bal{nuib6}
f_0&=&-\frac{1}{5}f_2^2-\frac{2}{105}f_2^3-\frac{1}{9}f_4^2-\frac{2}{35}f_2^2f_4-\ldots\;,
\ea
indicating that, within the Zharkov-Trubitsyn framework, the monopole term $f_0$ emerges at second order in the small parameter (i.e., $\sim \m^2$) as a predetermined function of products of the higher-order shape harmonics $f_{2n}$ with $n\ge 1$. Fixing $f_0$ in this way closes the system of equations for the remaining set of unknown functions $\{\rho, f_2, f_4,\ldots\}$, thereby rendering it solvable. This approach eliminates the need to solve an equation for $f_0$ but it incorporates the full hydrodynamic equilibrium equation~\eqref{hz3a} into the computational scheme, which necessitates recalculating the density profile $\rho$ along with the shape functions $f_2, f_4,\ldots$ at each successive order of perturbation theory.

In contrast to the Zharkov-Trubitsyn approach, we do not impose the constraint given by Eq.~\eqref{nuib6} on the monopole harmonic $f_0$, as doing so would decouple its determination from the general computational framework adopted in this manuscript. Instead, the theory presented here treats the monopole harmonic $f_0$ on equal footing with all other harmonics $f_{2n}$ for all $n\ge 1$. The radial gauge freedom is fixed by postulating that the left-hand side of the hydrodynamic equilibrium equation~\eqref{hz3a}, which represents the equilibrium condition for the reference fluid configuration as defined in Eq.~\eqref{w9ab}, vanishes. This convention causes the right-hand side of Eq.~\eqref{hz3a} to vanish as well, thereby yielding a differential equation for determining the radial function $f_0$. As a result, the system of equations for the shape functions $f_0,f_2,f_4,\ldots$ becomes closed and solvable.  

In our approach, the monopole term $f_0$ remains of second order in the small parameter $\m$, but its relationship to the higher-order harmonics is not prescribed a priori, unlike in the Zharkov-Trubitsyn theory. Rather, it emerges naturally from the differential equation governing $f_0$, together with the boundary conditions imposed on the gravitational field perturbations at the surface of the body. Within our computational scheme, the density profile $\rho$ is determined by solving the unperturbed hydrodynamic equilibrium equation~\eqref{w9ab}, and it remains fixed across all orders of perturbation theory. Consequently, the shape functions of lower order do not require recalculation when higher-order effects are considered. This renders our computational approach more efficient than the Zharkov-Trubitsyn formalism, in which both the density profile and all shape functions must be recalculated as the perturbative order increases.

\subsection{Spectral Decomposition for the Shape Function Products}

The equations from the previous section suffice for the spectral decomposition of the left-hand side of the master equation \eqref{cre5a}, as it contains only linear terms. However, the spectral decomposition of the right-hand side is more involved due to the presence of non-linear terms.

We begin by decomposing the function $Y =Y(r,\theta) = r^2 \pd_if \pd_if$ into spherical harmonics. The gradient $\pd_i f$ of the shape function $f = f(r, \theta)$ is given by: 
\bal{b6v}
\pd_if&=&f'n^i+\dot f\pd_i\theta\;,
\ea
where the derivatives are defined as $f' \equiv \partial f / \partial r$ and $\dot{f} \equiv \partial f / \partial \theta$. Using Eq. \eqref{xc15}, we have: 
\bal{vrt6}
\pd_i\cos\theta&=&\frac{{\cal P}^{i3}}{r}\;,
\ea
where ${\cal P}^{ij}=\delta^{ij}-n^in^j$. 
Therefore,
\bal{c6d}
\pd_if&=&f'n^i-\frac{\dot f\,{\cal P}^{i3}}{r\sin\theta}\;,
\ea
which leads to:
\bal{g6x}
Y&=&\left(rf'\right)^2+{\dot f}^2\;.
\ea
By computing the derivatives from $f(r, \theta)$ given in Eq. \eqref{zx2} and substituting into Eq. \eqref{g6x}, we obtain:
\bal{6g3}
Y&=&\sum_{n=0}^\infty\sum_{m=0}^\infty\left[r^2f'_{n}(r)f'_{m}(r)P_{n}(\cos\theta)P_{m}(\cos\theta)+f_{n}(r)f_{m}(r)\sin^2\theta \frac{dP_{n}(\cos\theta)}{d(\cos\theta)}\frac{dP_{m}(\cos\theta)}{d(\cos\theta)}\right]\;.
\ea
The products of Legendre polynomials and their derivatives are decomposed using Eqs. \eqref{k1s} and \eqref{xc20ss}, yielding the spectral decomposition:
\bal{ny6g}
Y&=&\sum_{k=0}^\infty Y_k(r)P_{k}(\cos\theta)\;,
\ea
where the coefficients $Y_k(r)$ are:
\bal{q3x}
Y_{k}(r)&=&\sum_{n=0}^\infty\sum_{m=|n-k|}^{n+k}\left\{r^2f'_{n}f'_{m}+\frac12f_{n}f_{m}\left[n(n+1)+m(m+1)-k(k+1)\right]\right\}T^{nmk}\;.
 \ea
Here, the symbol $T^{nmk}$ denotes the Wigner matrix, as defined in Appendix \ref{wignerdecomp} by Eqs. \eqref{wig1}--\eqref{k2s}.

The function $A=A(r,\theta)$, defined in Eq. \eqref{t5cx}, admits a spectral decomposition in terms of Legendre harmonics: 
\bal{v5d}
A&=&\sum_{l=0}^\infty A_{l}(r)P_{l}(\cos\theta)\;,
\ea
where the harmonic amplitudes $A_l$ are obtained by expanding the denominator of $A$ into a Taylor series: 
\bal{h6ce}
A&=&Y\sum_{k=0}^\infty(-1)^kX'^k\;,
\ea 
and substituting the Legendre expansions for each function in Eq. \eqref{h6ce}, followed by reducing the resulting products using the Wigner decomposition \eqref{k1hv}. Inserting expansions \eqref{zx2ab} and \eqref{q3x} into the right-hand side of Eq. \eqref{h6ce} yields:
\bal{j7v}
A&=&\sum_{k=0}^\infty
\sum_{n_0=0}^\infty\sum_{n_1=0}^\infty...\sum_{n_k=0}^\infty (-1)^k Y_{n_0}(r)X'_{n_1}(r)...X'_{n_k}(r)P_{n_0}(\cos\theta)P_{n_1}(\cos\theta)...P_{n_k}(\cos\theta)\;,
\ea
which corresponds to the left-hand side of Eq. \eqref{v5d}. Multiplying Eq. \eqref{j7v} by $P_l(\cos\theta)$, integrating over $\theta$, and applying the orthogonality relation \eqref{pm64}, we obtain:
\bal{g3c}
\hspace{-1cm}A_l&=&\frac{2l+1}{2} \sum_{k=0}^\infty
\sum_{n_0=0}^\infty\sum_{n_1=0}^\infty...\sum_{n_k=0}^\infty (-1)^k Y_{n_0}(r)X'_{n_1}(r)...X'_{n_k}(r)\\\nonumber
&&\hspace{5cm}\times\int\limits_0^\pi P_{n_0}(\cos\theta)P_{n_1}(\cos\theta)...P_{n_k}(\cos\theta)P_{l}(\cos\theta)\sin\theta d\theta\;.
\ea

The integral on the right-hand side of Eq. \eqref{g3c} is evaluated using Wigner's formula \eqref{yv65}, resulting in the expansion coefficients $A_l$:
\bal{c3e2}
A_l(r)&=&\sum_{k=0}^\infty
\sum_{n_0=0}^\infty\ldots\sum_{n_{k-1}=0}^\infty\sum_{n_k=|l-a_{k-1}|}^{l+a_{k-1}} \sum_{a_1=|n_1-n_0|}^{n_1+n_0}...\sum_{a_{k-1}=|n_{k-1}-a_{k-2}|}^{n_{k-1}+a_{k-2}}(-1)^k Q^{n_0...n_{k-1}n_kl}_{a_1...a_{k-1}}Y_{(n_0}X'_{n_1}...X'_{n_k)}\;.
\ea
Here, $a_0 = n_0$, and the parentheses around indices $(n_0, n_1, \ldots, n_k)$ denote full symmetrization. The coefficients
\bal{juy6}
Q^{n_0...n_kl}_{a_1...a_{k-1}}&\equiv&\Biggl\{\begin{array}{ll}
  1&\;,\qquad \; k=0\;,\\
  T^{n_0n_1l}&\;,\qquad \; k=1\;,\\
  T^{n_0n_1a_1}T^{a_1n_2a_2}...T^{a_{k-2}n_{k-1}a_{k-1}}T^{a_{k-1}n_kl}&\;,\qquad\; k\ge 2\;,
\end{array} 
\ea
represent products of $k$ Wigner matrices.

The spectral decomposition of the function $B=B(r,\theta)$ is given by: 
\bal{c5z}
B&=&\sum_{l=0}^\infty B_l(r)P_l(\cos\theta)\;,
\ea
where the expansion coefficients $B_l(r)$ depend on the radial coordinate $r$ and are defined as:
\bal{h7e}
B_l&=&3(f^2)_l+(f^3)_l\;,
\ea
with the spectral components $(f^2)_l$ and $(f^3)_l$ expressed as: 
\bal{e4c}
(f^2)_l&=&\sum_{n=0}^\infty\sum_{m=|n-l|}^{n+l} f_{n}(r)f_{m}(r)T^{nml}\;,\\\label{ib6v}
(f^3)_l&=&\sum_{n=0}^\infty\sum_{m=0}^\infty\sum_{k=|l-a|}^{l+a}\sum_{a=|m-n|}^{m+n} f_{n}(r)f_{m}(r)f_{k}(r)T^{nma}T^{akl}\;.
\ea

\subsection{Differential Equations for Radial Harmonics of the Shape Function}

The equations from the previous section are employed to derive ordinary differential equations (ODEs) for the radial spectral harmonics $f_l$ of the shape function. The spectral decomposition of the Laplacian of $f$ is given by: 
\bal{p9n}
\Delta f&=&\sum_{l=0}^\infty\left[f_l''+\frac{2f_l'}{r}-\frac{l(l+1)f_l}{r^2}\right]P_l(\cos\theta)\;.
\ea
Using this expression, together with the spectral decomposition of the non-linear terms in the master equation \eqref{cre5a}, we obtain the ODE governing each harmonic $f_l$: 
\bal{v3cd}
f_l''+\frac{6\upbeta}{r}\left(f_l'+\frac{f_l}{r}\right)-\frac{l(l+1)f_l}{r^2}&=&\frac{S_l}{r}\;,
\ea
where
\bal{o86g}
S_l&=&A_l'-\upbeta B_l'+\frac{3\upbeta}{r}\left(A_l-B_l\right)\;.
\ea
The spectral harmonics $A_l$ and $B_l$ of the non-linear terms are given in Eqs. \eqref{c3e2} and \eqref{h7e}, respectively. Due to the non-linear nature of these terms, Eq. \eqref{v3cd} is solved iteratively. Analyzing the first few iterations provides insight into the solution structure and facilitates comparison with results obtained by other methods. In this context, we focus on the first and second iterations. Higher-order iterations will be addressed in future work.

\subsubsection{First Iteration: The Clairaut Equation}\label{clairaut}

In the first (linear) iteration, all non-linear terms are neglected, reducing Eq. \eqref{v3cd} to the classical Clairaut equation: 
\bal{v1cd}
f_l''+\frac{6\upbeta}{r}\left(f_l'+\frac{f_l}{r}\right)-\frac{l(l+1)f_l}{r^2}&=&0\;,
\ea
which describes axisymmetric rotational perturbations. This equation is traditionally derived by expanding the kernel of the gravitational potential in Legendre polynomials and transforming the resulting integro-differential equation for the radial harmonics into the differential Clairaut form \citep{Clairaut-book-1743, Kopal_1960book, Zharkov_1978book, Tassoul_1978book, horedt_2004book}.

Milne \citep{Milne_1923MNRAS} and Chandrasekhar \citep{Chandrasekhar_1933MNRAS} derived the Clairaut equation by expanding the density and gravitational potential in powers of the small rotational parameter $\m$. Hubbard, Slattery, and Devito \citep{Hubbard_1975ApJ} obtained the same result by decomposing the perturbed density $\mu$ of a rotating body into Legendre polynomials. More recently, Chao and Shih \citep{ChaoBF} demonstrated an alternative derivation using the gravitational multipole formalism.

In contrast, we have derived the Clairaut equation \eqref{v1cd} by applying a perturbative approach grounded in Lie group theory. 

\subsubsection{Second Iteration: Darwin -- de Sitter Theory}\label{darwin}

In the second iteration, the spectral components of the non-linear term $S_l$ on the right-hand side of Eq. \eqref{v3cd} are given by $A_l = Y_l$ and $B_l = 3(f^2)_l$, as defined in Eqs. \eqref{q3x} and \eqref{e4c}, respectively. To evaluate $S_l$, we also require the radial derivatives $A_l'$ and $B_l'$.

The derivative $A_l'$ is obtained by differentiating Eq. \eqref{q3x} and substituting the second derivative $f_l''$ using the Clairaut equation \eqref{v1cd}, yielding: 
\bal{n5c3}
A_l'&=&\sum_{n=0}^\infty\sum_{m=|n-l|}^{n+l}\left[2r\left(1-6\upbeta\right)f_n'f_m'+n(n+1)f_nf_m'+m(m+1)f_mf_n'
\right]T^{nml}\\\nonumber
&+&\sum_{n=0}^\infty\sum_{m=|n-l|}^{n+l}\left[n(n+1)+m(m+1)-l(l+1)-12\upbeta\right]f_{(n}^{\phantom'}f_{m)}'T^{nml}
\ea
Differentiating \eqref{e4c} gives:
\bal{e4x}
B_l'&=&6\sum_{n=0}^\infty\sum_{m=|n-l|}^{n+l} f_{(n}f_{m)}'T^{nml}=3\sum_{n=0}^\infty\sum_{m=|n-l|}^{n+l} \left(f_{n}f_{m}'+f_{m}f_{n}'\right)T^{nml}\;.
\ea
Substituting these expressions into Eq. \eqref{o86g}, the non-linear source term $S_l$ in the quadratic approximation becomes: 
\bal{z2a}
S_l
&=&\sum_{n=0}^\infty\sum_{m=|n-l|}^{n+l}\left[r\left(2-9\upbeta\right)f_n'f_m'+n(n+1)f_nf_m'+m(m+1)f_mf_n'
\right]T^{nml}\\\nonumber
&+&\sum_{n=0}^\infty\sum_{m=|n-l|}^{n+l}\left[n(n+1)+m(m+1)-l(l+1)-18\upbeta\right]f_{(n}^{\phantom'}f_{m)}'T^{nml}\\\nonumber
&+&\frac{3\upbeta}{2r}\sum_{n=0}^\infty\sum_{m=|n-l|}^{n+l}\left[n(n+1)+m(m+1)-l(l+1)-6\right]f_{n}f_{m}T^{nml}\;.
\ea

Equations \eqref{z2a} include all terms of quadratic order in the radial spectral harmonics $f_l$ of the shape function. However, since each harmonic scales as $f_0\sim\m^2$ for $l=0$ and $f_l \sim \m^{l/2}$ for $l\ge2$, their magnitudes decrease rapidly with increasing $l$. Consequently, in the second-order approximation with respect to the rotation parameter $\m$, only the lowest harmonics -- specifically $f_0$, $f_2$, and $f_4$ -- contribute significantly to the solution.

In this approximation, the non-linear terms contributing to the source function $S_l$ involve only products of the harmonic $f_2 \sim \mathsf{m}$ and its derivatives. The harmonic $f_0 \sim \mathsf{m}^2$ is of second order, as noted in the final paragraph of Section \ref{onuv5}, and its contribution to non-linear terms can be neglected. A direct evaluation of $S_l$ for $l = 0, 2, 4$ using Eq. \eqref{z2a} with $n = m = 2$ yields the following system:
\bal{o7b5}
f_0''+\frac{6\upbeta}{r}\left(f_0'+\frac{f_0}{r}\right)&=&\frac15\left(2-9\upbeta\right)f_2'f_2'+\frac{6}{5r}\left(4-3\upbeta\right)f_2f_2'+\frac{9\upbeta}{5r^2}f_2^2\;,\\\label{kt5v}
f_2''+\frac{6\upbeta}{r}\left(f_2'+\frac{f_2}{r}\right)-\frac{6f_2}{r^2}&=&\frac{2}{7}\left(2-9\upbeta\right)f_2'f_2'+\frac{36}{7r}\left(1-\upbeta\right)f_2f_2'\;,\\\label{on6}
f_4''+\frac{6\upbeta}{r}\left(f_4'+\frac{f_4}{r}\right)-\frac{20f_4}{r^2}&=&\frac{18}{35}\left(2-9\upbeta\right)f_2'f_2'+\frac{36}{35r}\left(2-9\upbeta\right)f_2f_2'-\frac{54\upbeta}{5r^2}f_2^2\;,
\ea
where the function $\upbeta$ is defined by (c.f. Eq. \eqref{uu2}):
\bal{kiu7}
\upbeta&=&\upalpha+\frac{2{\mathsf{m}}}{3}\frac{\bar\rho(a)}{\bar\rho(r)}\left(\upalpha-1\right)\;.
\ea
Equations \eqref{kt5v} and \eqref{on6}, commonly referred to as the Darwin-de Sitter and Darwin equations, respectively, represent important generalizations of Airy's classical work on second-order corrections to the gravitational potential of rotating celestial bodies \citep{Airy_1826, Darwin_1899, Zharkov_1978book}. These equations have also been independently derived by Kopal \citep{Kopal_1960book} and Lanzano \citep{Lanzano_1982}, further underscoring their foundational role in the theory of planetary figures.

As noted by Chambat et al. \citep{Chambat_2010GeoJI}, there is a typographical error in Kopal's differential equation for the harmonic component $f_4$. Specifically, the coefficient of the term proportional to $f_2 f'_2$ is incorrectly given as $1 - 9\upbeta$ in Kopal's formulation, whereas the correct expression should be $2 - 9\upbeta$. This discrepancy, if uncorrected, may lead to inaccuracies in subsequent theoretical or numerical analyses.

The derivation of the Darwin-de Sitter equations presented in this section offers a novel perspective. Unlike traditional approaches that rely on gravitational potential expansions or multipole formalism, this formulation emerges naturally from a perturbative framework grounded in Lie group theory. By systematically incorporating non-linear corrections up to second order in the rotation parameter $\mathsf{m}$, the resulting equations not only recover classical results but also provide deeper insight into the structure and coupling of spectral harmonics in rotating celestial bodies.

\subsection{Spectral Analysis of Gravitational Perturbations in the Body Interior}
The gravitational field of a rotating fluid body lacks spherical symmetry and exhibits a complex spatial structure. This structure can be analyzed by expanding the gravitational field in spherical harmonics, both outside and inside the body. While the external expansion is well established -- its coefficients known as multipole moments -- this section focuses on the internal spectral expansion, where the harmonics are functions of the radial coordinate $r$ and are referred to as radial spectral harmonics. The objective is to derive differential equations governing these harmonics.

We consider an axially symmetric perturbation of the gravitational field, described by the function $K$ introduced in Eq. \eqref{c3q}. Inside the body, $K$ is expanded in Legendre polynomials:
\bal{h1}
K=\sum_{l=0}^\infty K_l(r)P_l(\cos\theta)\;,
\ea
where $K_l(r)$ are the radial harmonics. The finite density perturbation $\varrho\equiv\varrho_1$, defined in Eq. \eqref{ty74x}, is a nonlinear function of $K$, as given by Eq. \eqref{s9} for $\tau = 1$. It is similarly expanded:
\bal{h2}
\varrho=\sum_{l=0}^\infty \varrho_l(r)P_l(\cos\theta)\;,
\ea
where $\varrho_l(r)$ denotes the $l$-th radial harmonic of the density perturbation.

Substituting Eq. \eqref{h1} into Eq. \eqref{s9} yields the spectral expansion:
\bal{h2aa}
\varrho=\sum_{j=1}^\infty\frac{g_j}{{\cal A}^jj!}\sum_{n_1=0}^\infty...\sum_{n_j=0}^\infty K_{n_1}(r)...K_{n_j}(r)P_{n_1}(\cos\theta)...P_{n_j}(\cos\theta)\;,
\ea
where $g_j = g_j(r)$ are defined in Eqs. \eqref{s10}--\eqref{s11}, and ${\cal A}$ is given in Eq. \eqref{s14}. Applying the Wigner decomposition \eqref{k1hv} to the product of Legendre polynomials, we obtain:
\bal{mop78}
\varrho_l(r)=\sum_{j=1}^\infty\frac{g_j}{j!}\frac{k_{jl}}{{\cal A}^j}\;,
\ea
where the $k_{jl}\equiv\left(K^j\right)_l$ is the $l$-th spectral harmonic of $K^j$:
\bal{omcx}
k_{jl}&=&\sum_{n_1=0}^\infty...\sum_{n_j=|a_{j-1}-l|}^{a_{j-1}+l}\sum_{a_2=|n_2-n_1|}^{n_2+n_1}...\sum_{a_{j-1}=|n_{j-1}-a_{j-2}|}^{n_{j-1}+a_{j-1}}T^{n_1n_2a_2}...T^{a_{j-1}n_jl}K_{n_1}(r)...K_{n_j}(r)\;.
\ea

The first few harmonics are: 
\bal{m78qw}
k_{1l}&=&K_l(r)\;,\\
k_{2l}&=&\sum_{n=0}^\infty\sum_{m=|n-l|}^{n+l}T^{nml}K_{n}(r)K_{m}(r)\;,\\
k_{3l}&=&\sum_{n=0}^\infty\sum_{m=0}^{\infty}\sum_{p=|n-m|}^{n+m}\sum_{q=|l-p|}^{l+p}T^{nmp}T^{pql}K_{n}(r)K_{m}(r)K_{q}(r)\;.
\ea 

Substituting expansions \eqref{h1} and \eqref{h2} into the field equation \eqref{fe87} for $\tau = 1$, we derive the second-order ODE for $K_l(r)$: 
\bal{h3}
K''_l+\frac{2}{r}K'_l-\frac{l(l+1)}{r^2}K_l+4\pi G\frac{K_l}{A}+4\pi G\sum_{j=2}^\infty\frac{g_j}{j!}\frac{k_{jl}}{A^j}=0\;.
\ea
This nonlinear equation can be solved iteratively. For the first three harmonics, we obtain: 
\bal{c8}
&&K''_0+\frac{2K'_0}{r}+4\pi G\frac{\sigma'}{U'}K_0+\frac{2\pi G}{5}\left(\sigma''+\frac{2\sigma'}{r}+4\pi G\frac{\sigma\sigma'}{U'}\right)\left(\frac{K_2}{U'}\right)^2=0\;,\\\label{r2s}
&&K''_2+\frac{2K'_2}{r}-\frac{6K_2}{r^2}+4\pi G\frac{\sigma'}{U'}K_2+\frac{4\pi G}{7}\left(\sigma''+\frac{2\sigma'}{r}+4\pi G\frac{\sigma\sigma'}{U'}\right)\left(\frac{K_2}{U'}\right)^2=0\;,\\\label{ob32}
&&K''_4+\frac{2K'_4}{r}-\frac{20K_4}{r^2}+4\pi G\frac{\sigma'}{U'}K_4+\frac{36\pi G}{35}\left(\sigma''+\frac{2\sigma'}{r}+4\pi G\frac{\sigma\sigma'}{U'}\right)\left(\frac{K_2}{U'}\right)^2=0\;.
\ea

These equations are fully consistent with Eqs. \eqref{o7b5}--\eqref{on6}, which govern the spectral harmonics $f_0$, $f_2$, and $f_4$ of the shape function $f$. Their equivalence follows from Eq. \eqref{bv4rd6}, which relates $K$ to $f$. For the first three harmonics: 
\bal{yt209}
K_0&=&-U'X_0+\frac15 U'X_2X'_2+\frac{1}{10}U''X_2^2\;\\
K_2&=&-U'X_2+\frac27 U'X_2X'_2+\frac{1}{7} U''X_2^2\;\\\label{c4x3}
K_4&=&-U'X_4+\frac{18}{35} U'X_2X'_2+\frac{9}{35} U''X_2^2\;,
\ea
where $X_l = r f_l$ are the radial spectral harmonics of the height function $X$. These expressions are truncated at second order in $\mathsf{m}$. Substituting Eqs. \eqref{yt209}--\eqref{c4x3} into Eqs. \eqref{c8}--\eqref{ob32}, and using Eqs. \eqref{klm8h}--\eqref{jn7bv} to eliminate higher derivatives of $U$, we recover Eqs. \eqref{o7b5}--\eqref{on6}.

This procedure confirms the equivalence between the gravitational harmonics $K_l$ governed by Eq. \eqref{h3} and the shape function harmonics $f_l$ governed by Eq. \eqref{v3cd}, for any order $l \geq 6$.

\subsection{Spectral Analysis of External Gravitational Perturbations Using Multipole Moments}\label{tc43}

Outside the body, the density and its derivatives vanish. Consequently, Eq. \eqref{41cs} reduces to the Laplace equation for $K^+(\x)$, which appears in the matching conditions \eqref{ccc23} and \eqref{ccc24}:
\bal{gh6vc}
\Delta K^+&=&0\;.
\ea
This equation is solved in the external domain, which is the vacuum region complementary to the spherical volume $\V$ occupied by the fluid body. Inside $\V$, the internal solution $K^-$ satisfies the Poisson equation \eqref{fe87}. The external solution $K^+(\x)$ is sought in the form of a series expansion in Legendre polynomials: 
\bal{ko7b5}
K^+(\x)&=&\sum_{l=0}^\infty K^+_l(r)P_l(\cos\theta)\;.
\ea
Substituting this expansion into Eq. \eqref{gh6vc} yields the following ordinary differential equation for each spectral harmonic $K^+_l(r)$: 
\bal{n7v3}
K''^+_l+\frac{2}{r}K'^+_l-\frac{l(l+1)}{r^2}K_l^+&=&0\;.
\ea
The general solution to this equation is a linear combination of two radial power-law terms: 
\bal{oin6}
K_l^+(r)&=&\mathfrak{A}_l\left(\frac{a}{r}\right)^{l+1}+\mathfrak{B}_l\left(\frac{r}{a}\right)^l\;,
\ea
where $\mathfrak{A}_l$ and $\mathfrak{B}_l$ are constants specific to each harmonic degree $l$. Differentiating Eq. \eqref{oin6} with respect to $r$ gives:
\bal{oi4zs}
K'^+_l(r)&=&-(l+1)\frac{\mathfrak{A}_l}{a}\left(\frac{a}{r}\right)^{l+2}+l\frac{\mathfrak{B}_l}{a}\left(\frac{r}{a}\right)^{l-1}\;.
\ea

The coefficients $\mathfrak{A}_l$ characterize the perturbation of the spherically symmetric gravitational field of the fluid body and are defined as: 
\bal{yh7bv}
\mathfrak{A}_l&=&-\frac{G\M}{a}{\cal I}_l\;,
\ea
where $\mathcal{I}_l$ are the gravitational coefficients of the zonal harmonics induced by the external rotational perturbation \citep{Zharkov_1978book}, normalized to the unperturbed radius $a$ of the body. In satellite dynamics, the multipole moments $J_l$, normalized to the equatorial radius $R_{\rm e}$, are typically used. The equatorial radius is obtained from Eq. \eqref{zx2ss} evaluated at $\theta = \pi/2$, and expressed in terms of the shape function $f(a) = X/a$ as: 
\bal{pom76}
\frac{R_{\rm e}}{a}&=&1+\sum_{k=0}^\infty \frac{(-1)^k(2k-1)!!}{2^k k!}\frac{X_{2k}(a)}{a}=1+f_0(a)-\frac{1}{2}f_2(a)+\frac{3}{8}f_4(a)+\ldots\;.
\ea
The relationship between the multipole moments $\mathcal{I}_l$ and $J_l$ is given by:
\bal{evt3}
{\cal I}_l&=&\left(\frac{R_{\rm e}}{a}\right)^lJ_l\;.
\ea 

The coefficients $\mathfrak{A}_l$ and $\mathfrak{B}_l$ are related through the classical definition of the Love numbers $k_l$, widely used in geodesy and geophysics \citep{Love_1909,Vanicek_book,Zharkov_1978book}:
\bal{ion76f}
\mathfrak{A}_l&=&k_l \mathfrak{B}_l\;.
\ea 
The Love numbers $k_l$ quantify the integrated response of the body's material to external rotational perturbations. The coefficients $\mathfrak{B}_l$ in Eq. \eqref{oin6} represent the amplitudes of the spectral harmonics of the external perturbing potential $W_Q$, defined in Eq. \eqref{bw5aa}, and are given by: 
\bal{v5fc}
\mathfrak{B}_l&=&-\frac{\m}{3}\frac{G\M}{a}\delta_{2l}\;,
\ea
where $\delta_{2l}$ is the Kronecker delta. Thus, only the quadrupole term $\mathfrak{B}_2$ is non-zero under purely rotational perturbations. Accordingly, the relation \eqref{ion76f} between $\mathfrak{A}_l$ and $\mathfrak{B}_l$ holds exactly for $l = 2$ in the linear approximation with respect to the perturbation parameter $\m$, implying $\mathfrak{A}_l = 0$ for $l > 2$ at this order. However, at higher orders in $\m$, the coefficients $\mathfrak{A}_l \sim \m^2$ become non-zero and are determined by boundary conditions imposed on the radial harmonics of the height function. This will be elaborated in the subsequent section.

By extending Eq. \eqref{ion76f} to higher-order harmonics, a generalized correspondence between the gravitational multipole moments $J_l$ and the Love numbers $k_l$ is established:
\bal{yhas}
{\cal I}_l&=&\frac{\m}{3}k_l\;.
\ea

It is important to note that the fluid Love number $k_l$ is twice the apsidal motion constant $k^{\rm aps}_l$, as originally introduced by Kopal \citep{kopal-1959book} and further developed by Brooker and Olle \citep{Brooker-1955MNRAS}, i.e., $k_l = 2k^{\rm aps}_l$. In astrophysical literature \citep{Hinderer-2008ApJ,Damour-2009PhRvD,Poisson_2009PhRvD,Yip_2017}, the notation for the apsidal motion constant is often conflated with that of the Love number $k_l$ used in geodesy \citep{Vanicek_book}, geophysics \citep{Zharkov_1978book}, and planetary science \citep{Padovan_2018,Hellard_2019}. This overlap in terminology can lead to confusion, particularly due to the factor-of-two difference. In this work, we adopt the classical definition of the Love number $k_l$ as originally formulated by Love \citep{Love_1909} and consistently used in geophysical and planetary science contexts.

\section{Boundary Conditions and Love Numbers}\label{sec6}
\stepcounter{equationschapter}
\renewcommand{\theequation}{8.\arabic{equation}}
The determination of solutions to the differential equations governing the height function, denoted by $X(\x)=X(r,\theta)$ and the gravitational field perturbation, $K(\x)=K(r,\theta)$, relies critically on the application of appropriate boundary conditions. These functions are not independent; rather, they are coupled through a transformation relationship as indicated by Eq. \eqref{bv4rd6}. This coupling implies that the behavior of one function at the boundaries directly influences the behavior of the other. Consequently, the boundary conditions imposed on $X(\x)$ and $K(\x)$ must be consistent with their interdependence, ensuring that the physical and mathematical constraints of the rotating fluid body are satisfied simultaneously. These boundary conditions typically reflect physical requirements such as regularity at the origin, continuity across interfaces, or decay at infinity, and they play a crucial role in selecting a physically meaningful solution from the general solution space of the differential equations.

\subsection{Boundary Conditions for Gravitational Perturbations}

The gravitational perturbation $K(\x)$ is assumed to be finite and continuous throughout the entire space $\mathbb{R}^3$. The boundary condition at the center of the fluid body is derived from the transformation relation in Eq. \eqref{bv4rd6}, which links $K(\x)$ to the height function $X(\x)$ -- a measure of the fluid surface deformation. For physical consistency, $X(\x)$ must remain finite and continuous across the domain, and we specifically impose the condition $X(0) = 0$ at the center.

This condition is physically motivated. A nonzero value of $X(0)$ would imply a displacement of the central fluid parcel in response to the external perturbing force ${\bm F} = -\rho{\bm\nabla}W_Q$, where $W_Q$ is the rotational perturbation potential defined in Eq. \eqref{iii7}. However, since ${\bm\nabla}W_Q = 0$ at the origin, the force vanishes at the center, indicating that the central fluid element experiences no net acceleration. Therefore, any displacement of the fluid parcel at this point would contradict the absence of force. As a result, the condition $X(0) = 0$, together with Eq. \eqref{bv4rd6}, necessitates that the gravitational perturbation also vanishes at the center: $K(0) = 0$.

Additional boundary conditions for $K(\x)$ are determined by analyzing its behavior at the outer boundary of the fluid body. The physical boundary is axisymmetric and described by the function $R = r + X(r, \theta)$, where $X(r, \theta)$ is the unknown height function. This complicates the direct imposition of boundary conditions on $K(\x)$ at the deformed surface. To address this, we employ the pullback transformation of the integral representation of gravitational potential, as discussed in Section \ref{ee31}. This transformation allows us to treat the boundary of the rotating fluid as a spherical surface of radius $a$, enclosing a volume $\mathcal{V}$ with fluid density $\sigma(\x) + \varrho$, and a surface-layer density $\nu(\x)$ defined on the spherical boundary $\partial\mathcal{V}=\mathbb{S}^2$ of the reference configuration.

Let $K^-(\x) = K^-(r, \theta)$ denote the interior solution valid for $r < a$, and $K^+(\x) = K^+(r, \theta)$ the exterior solution valid for $r > a$. The first boundary condition enforces continuity of the gravitational perturbation across the boundary:
\bal{m5ez}
K^+({\bm a})&=&K^-({\bm a})\;.
\ea
The second boundary condition concerns the radial derivative of $K(\x)$. It is obtained by integrating the field equation \eqref{41cs} over a thin cylindrical shell centered on the boundary and taking the limit as the shell thickness tends to zero. This analysis is performed in the radial gauge $\X = X(\x) \mathbf{n}$, as defined in Section \ref{kji6v5}. The resulting jump condition, which accounts for a surface layer with a Dirac delta-function contribution, is:
\bal{bhyh}
K'^+({\bm a})&=&K'^-({\bm a})-4\pi G \sum_{n=0}^\infty\frac{X^{n+1}({\bm a})}{(n+1)!}\frac{\pd^n}{\pd a^n}\Bigl[\rho(a)+\varrho({\bm a})\Bigr]\;.
\ea
Here, $\rho(a)$ is the unperturbed fluid density at the boundary radius $r = a$, and $\varrho(\mathbf{a})$ denotes the surface-layer density. The latter is defined in terms of the limiting behavior of the height function as:
\bal{dper5}
\varrho({\bm a})&=&\lim_{\x\to{\bm a}}\sum_{k=1}^\infty\left(-\hat{\mathsf T}_{\bm X}\right)^k\sigma(r)\;.
\ea
These conditions fully characterize the behavior of the gravitational perturbation $K(\x)$ at the boundary.

Equation \eqref{bhyh} presents the most general boundary condition for the radial derivative of the gravitational perturbation $K(\x)$, explicitly accounting for a nonzero surface density $\rho(a)$. In many astrophysical contexts, however, the surface density is negligible relative to the mean interior density $\bar{\rho}(a)$ and can be omitted. This simplification is particularly valid for fluid bodies described by a polytropic equation of state with index $0 \leq n < 5$, as given in Eq. \eqref{eqst39} and discussed in \citep{horedt_2004book}.

Nevertheless, even when $\rho(a) = 0$, higher-order derivatives of the density may remain finite at the surface. A notable example is the unit-index polytrope, analyzed in Section \ref{subD2}, where such terms contribute significantly. Therefore, the second term on the right-hand side of Eq. \eqref{bhyh} must be carefully evaluated and retained when these higher-order derivative effects are non-negligible.

\subsection{Boundary Conditions for the Height Function}

The behavior of the height function $X(\x)$ at the center of the body is utilized to impose a constraint on the gravitational field perturbation $K(\x)$ at the origin, $\x = 0$. Conversely, the matching conditions \eqref{m5ez} and \eqref{bhyh}, applied to the gravitational perturbation $K(\x)$ and its normal derivative $K'(\x)$ on the spherical boundary of the base manifold $\MM$, are employed to derive the corresponding boundary conditions for the height function $X(\x)$ and its radial derivative $X'(\x)$ at the surface $r = a$.

Within the volume $\V$ occupied by the base manifold $\MM$, the interior gravitational perturbation $K^-(\x)$ is governed by equation \eqref{bv4rd6}. Substituting the continuity condition \eqref{m5ez} into this expression yields an explicit formula that determines the behavior of the height function $X(\x)$ and its radial derivatives on the boundary surface of the body. 
\bal{bv12}
K^+({\bm a})&=&\sum_{n=1}^\infty\left(-\hat{\mathsf T}_{\bm X}\right)^nU(a)\;,
\ea
where ${\bm X} = X{\bm n}$, $X = X({\bm a})$, and $U(a)$ denotes the unperturbed effective gravitational potential on the reference surface $r = a$, as given in \eqref{oku8}.

The second boundary condition, which involves the height function $X(\x)$ and its radial derivatives, is derived from equation \eqref{xc7}. By incorporating the expression for the potential $V$ from equation \eqref{gt6vc}, this condition can be written in the following form:
\bal{xc7aa}
\left(1+\hat{\mathsf T}_{\bm X}\right)\pd_i K^-({\bm x})&=&M^{-1}_{ij}({\bm X})\pd_j U({\bm x})-\left(1+\hat{\mathsf T}_{\bm X}\right)\pd_i U({\bm x})\;.
\ea
Multiplying both sides of this equation on the left by the unit normal vector $n^i$, and noting that in the radial gauge the vector commutes with the shift operator -- i.e., $n^i \hat{\mathsf T}_{\bm X} = \hat{\mathsf T}_{\bm X} n^i$ -- we obtain:
\bal{gh76v}
\left(1+\hat{\mathsf T}_{\bm X}\right)K'^-({\bm x})&=&n^iM^{-1}_{ij}({\bm X})\pd_j U({\bm x})-\left(1+\hat{\mathsf T}_{\bm X}\right)U'({\bm x})\;,
\ea
where the prime denotes differentiation with respect to the radial coordinate.

The first term on the right-hand side of \eqref{gh76v} is evaluated using equation \eqref{ek8}:
\bal{pop865}
n^iM^{-1}_{ij}({\bm X})\pd_j U({\bm x})&=&\frac{U'(\x)}{1+X'(\x)}\;.
\ea
Substituting this into \eqref{gh76v} yields:
\bal{xc7bb}
\left(1+\hat{\mathsf T}_{\bm X}\right)K'^-(\x)&=&\frac{U'(\x)}{1+X'(\x)}-\left(1+\hat{\mathsf T}_{\bm X}\right)U'(\x)\;.
\ea
The Neumann series solution of this equation is: 
\bal{xc7cc}
K'^-(\x)&=&\sum_{n=0}^\infty\left(-\hat{\mathsf T}_{\bm X}\right)^n\left(\frac{U'(\x)}{1+X'(\x)}\right)-U'(\x)\;.
\ea
Expanding the fraction in a Taylor series with respect to $X'$ and applying the boundary condition \eqref{bhyh} for the radial derivative of the gravitational perturbation, we derive the second boundary condition for the height function:
\bal{h5c2l}\hspace{-0.5cm}
K'^+({\bm a})&=&\sum_{n=1}^\infty\left(-\hat{\mathsf T}_{\bm X}\right)^nU'(a)+\sum_{n=0}^\infty\sum_{k=1}^\infty\left(-\hat{\mathsf T}_{\bm X}\right)^n\biggl[(-X')^kU'(a)\biggr]\\\nonumber
&-&4\pi G \sum_{n=0}^\infty\sum_{k=0}^\infty\frac{X^{n+1}}{(n+1)!}
\frac{\pd^n}{\pd a^n}\left[\left(-\hat{\mathsf T}_{{\bm X}}\right)^k\rho(a)\right]\;,
\ea                                                                                                                                                                                                                                                                                                                                                                                                                                                                                                                                                                                                                                                                                                                                                                                                                                                                                                                                                                                                                                                                                                                                                                                                                                                                                                                                                                                                                                                                                                                                                                                                                                                                                                                                                                                                                                                                                                                                                                                                                                            
where ${\bm X}=X{\bm n}$, $X=X({\bm a})$. 

The boundary conditions \eqref{bv12} and \eqref{h5c2l} are exact and expressed as infinite series in powers of $X$, $X'$, and higher-order radial derivatives. However, due to their complexity, they are not directly solvable and must be approached via successive approximations. In this work, we adopt a quadratic approximation, assuming the height function is small compared to the body's radius, $X \simeq \m a \ll a$. Accordingly, we neglect terms of order ${\cal O}(\m X^2)$ and ${\cal O}(X^3)$.

Expanding equations \eqref{bv12} and \eqref{h5c2l} in powers of $X$ and retaining terms up to second order, we obtain the quadratic boundary conditions:
\bal{c23a}
K^+({\bm a})&=&\left(-X+XX'\right)U'(a)+\frac12 X^2U''(a)\;,\\\label{c24a}
K'^+({\bm a})&=&\left(-X'+X'^2+XX''\right)U'(a)+\left(-X+2XX'\right)U''(a)+\frac12 X^2U'''(a)\\\nonumber
&&\;-4\pi G\left[\rho(a)X-\frac12\rho'(a)X^2\right]\;,
\ea
where all functions and their derivatives are evaluated at $r = a$.

The radial derivatives of the gravitational potential $U$ at the boundary surface $r = a$ are computed using equations \eqref{cok7}--\eqref{cok9}. By substituting the explicit expression for the Newtonian potential, $\U'(a) = -G\M/a^2$, these derivatives are evaluated as follows (cf. Eqs. \eqref{cok7}--\eqref{cok9}): 
\ba\label{cok7a}
U'(a)&=&-\frac{G\M}{a^2}\left(1-\frac{2\m}{3}\right)\;,\\\label{cok8a}
U''(a)&=&\frac{2G\M}{a^3}\left(1+\frac{\m}{3}\right)-4\pi G\rho(a)\;,\\
\label{cok9a}
U'''(a)&=&-\frac{6G\M}{a^4}+\frac{8\pi G\rho(a)}{a}-4\pi G\rho'(a)\;.
\ea
 
For simplicity, we assume that the surface density of the rotating fluid body vanishes, $\rho(a) = 0$, a condition commonly satisfied in astrophysical contexts though not necessarily in geophysical applications. Under this assumption, all terms involving $\rho(a)$ are eliminated. Substituting equations \eqref{cok7a}--\eqref{cok9a} into the boundary conditions \eqref{c23a} and \eqref{c24a}, and noting that terms involving $\rho'(a)$ cancel, we obtain the simplified quadratic boundary conditions: 
\bal{ccc23}\hspace{-2cm}
K^+({\bm a})&=&\left(X-\frac{2\m}{3}X-XX'+\frac{X^2}{a}\right)\frac{G\M}{a^2}\;,\\\label{ccc24}
K'^+({\bm a})&=&\left(X'-2\frac{X}{a}-\frac{2\m}{3}X'-\frac{2\m}{3}\frac{X}{a}-X'^2-XX''+\frac{4XX'}{a}-\frac{3X^2}{a^2}\right)\frac{G\M}{a^2}\;.
\ea  
These boundary conditions are further developed in the subsequent sections, where they are applied to the radial harmonic components of the height and shape functions.

\subsection{Boundary Conditions for Radial Harmonics of the Height Function}

The boundary conditions for the radial spectral harmonics $X_l(\x)$ of the height function $X(\x)$ are obtained by decomposing both sides of equations \eqref{ccc23} and \eqref{ccc24} into series of Legendre polynomials and equating terms of identical spectral order.

The harmonic decomposition of the left-hand sides is straightforward and follows directly from equation \eqref{ko7b5}. The right-hand sides are expanded using equation \eqref{zx2ss}, along with the Wigner decomposition for products of the height function and its derivatives. Matching terms of the same spectral index yields:
\bal{ew30}
K_l^+(a)&=&\left(X_l-\frac{2\m}{3}X_l\right)\frac{G\M}{a^2}-\sum_{n=0}^\infty\sum_{m=|n-l|}^{n+l}T^{nml}\left[X_{(n}X'_{m)}-\frac{X_nX_m}{a}\right]\frac{G\M}{a^2}\;,\\\nonumber\\
\label{yb54}
{K'_l}^+(a)&=&\left[\left(1-\frac{2\m}{3}\right)X'_l-\left(1+\frac{\m}{3}\right)\frac{2X_l}{a}\right]\frac{G\M}{a^2}\\\nonumber
&&-\sum_{n=0}^\infty\sum_{m=|n-l|}^{n+l}T^{nml}\left[X_{(n}''X_{m)}+X'_nX'_m-\frac{4X_{(n}X'_{m)}}{a}+\frac{3X_nX_m}{a^2}\right]\frac{G\M}{a^2}\;,
\ea 
 where all quantities are evaluated at the boundary $r = a$, and parentheses around indices denote full symmetrization.

The second radial derivative $X''_n$ appearing in \eqref{yb54} can be simplified using the Clairaut equation for the height function evaluated at the boundary, where $\upbeta = 0$ due to the assumption $\rho(a) = 0$. This yields: 
\bal{d4es}
X_n''&=&\frac{2X_n'}{a}+\frac{(n+2)(n-1)X_n}{a^2}\;.
\ea
Substituting \eqref{d4es} into \eqref{yb54} gives the final form of the boundary condition: 
\bal{yb5aa}
{K'_l}^+(a)&=&\left[\left(1-\frac{2\m}{3}\right)X'_l-\left(1+\frac{\m}{3}\right)\frac{2X_l}{a}\right]\frac{G\M}{a^2}\\\nonumber
&&-\sum_{n=0}^\infty\sum_{m=|n-l|}^{n+l}T^{nml}\left[X'_nX'_m-\frac{2X_{(n}X'_{m)}}{a}+\left(n^2+n+1\right)\frac{X_nX_m}{a^2}\right]\frac{G\M}{a^2}\;.
\ea

Next, we substitute the expressions for $K_l^+(a)$ and ${K'_l}^+(a)$ from equations \eqref{oin6} and \eqref{oi4zs} into the left-hand sides of \eqref{ew30} and \eqref{yb5aa}, which contain the unknown coefficients $\mathfrak{A}_l$, $\mathfrak{B}_l$, and the boundary values of $X_l(a)$ and $X'_l(a)$.

Using the definitions of the multipole moments $\mathfrak{A}_l$, $\mathfrak{B}_l$ from equations \eqref{ion76f} and \eqref{v5fc}, and canceling like terms, the boundary conditions \eqref{ew30} and \eqref{yb5aa} reduce to: 
\bal{s4rv}\hspace{-1cm}
&&\left(1-\frac{2\m}{3}\right)X_l-\sum_{n=0}^\infty\sum_{m=|n-l|}^{n+l}T^{nml}\left[X_{(n}X'_{m)}-\frac{X_nX_m}{a}\right]=-\frac{\m a}{3}\biggl(k_l+\delta_{2l}\biggr)\;,\\
\label{tf4c}\hspace{-1cm}
&&\left(1-\frac{2\m}{3}\right)X'_l-\left(1+\frac{\m}{3}\right)\frac{2X_l}{a}-\\\nonumber
&&\hspace{2cm}\sum_{n=0}^\infty\sum_{m=|n-l|}^{n+l}T^{nml}\left[X'_nX'_m-\frac{2X_{(n}X'_{m)}}{a}+(n^2+n+1)\frac{X_nX_m}{a^2}\right]=\frac{\m}{3}\biggl[(l+1)k_l-l\delta_{2l}\biggr]\;.
\ea
Solving this system yields the boundary condition for the radial harmonics of the height function: 
\bal{bchf5}\hspace{-1cm}
&&\left(1-\frac{2\m}{3}\right)X'_l+\left[l-1-\frac{2\m}{3}(l+2)\right]\frac{X_l}{a}-\\\nonumber
&&\hspace{2cm}\sum_{n=0}^\infty\sum_{m=|n-l|}^{n+l}T^{nml}\left[X'_nX'_m+(l-1)\frac{X_{(n}X'_{m)}}{a}+(n^2+n-l)\frac{X_nX_m}{a^2}\right]=-\frac{2l+1}{3}\m\delta_{2l}\;.
\ea
An alternative solution to the system \eqref{s4rv}--\eqref{tf4c} expresses the Love number $k_l$ in terms of the radial harmonics of the height function and their derivatives:
\bal{mn7bc4}\hspace{-1cm}
&&\left(1-\frac{2\m}{3}\right)X'_l-\left[l+2-\frac{2\m}{3}(l-1)\right]\frac{X_l}{a}-\\\nonumber
&&\hspace{2cm}\sum_{n=0}^\infty\sum_{m=|n-l|}^{n+l}T^{nml}\left[X'_nX'_m-(l+2)\frac{X_{(n}X'_{m)}}{a}+(n^2+n+l+1)\frac{X_nX_m}{a^2}\right]=\frac{2l+1}{3}\m k_l\;.
\ea

A detailed discussion of the Love numbers and their physical interpretation is presented in Section \ref{lov45}.

\subsection{Boundary Conditions for Radial Harmonics of the Shape Function}

It is instructive to derive the boundary conditions imposed directly on the radial harmonics $f_l = X_l / r$ of the shape function $f = X / r$, which satisfies the nonlinear differential equation \eqref{v3cd}. We focus on the first three harmonics $f_l(r)$ for $l = 0, 2, 4$, within the frameworks of the Clairaut and Darwin-de Sitter approximations, as discussed in Sections \ref{clairaut} and \ref{darwin}.

These boundary conditions follow from the condition \eqref{bchf5} for the height function harmonics $X_l$, by substituting $X_l \rightarrow a f_l$ and $X'_l \rightarrow a f'_l + f_l$ at the body's surface. This yields:
\bal{ik9n}
&&\left(1-\frac{2\m}{3}\right)af'_l+\left[l-\frac{2\m}{3}(l+3)\right]f_l-\\\nonumber
&&\hspace{2cm}\sum_{n=0}^\infty\sum_{m=|n-l|}^{n+l}T^{nml}\left[a^2f'_nf'_m+a(l+1)f_{(n}f'_{m)}+n(n+1)f_nf_m\right]=-\frac{2l+1}{3}\m\delta_{2l}\;.
\ea 
The computation of the Wigner matrix elements $T^{nml}$ for the specific values of the index $l = 0, 2, 4$ yields the boundary conditions governing the radial harmonics of the shape function in a more explicit form. These conditions are expressed as follows:
\bal{nnn4}
af'_0-\frac15\left(a^2{f'_2}^2+af_{2}f'_{2}+6f_2^2\right)&=&0\;,\\
\label{nnn5}
\left(1-\frac{2\m}{3}\right)af'_2+\left(2-\frac{10\m}{3}\right)f_2-\frac27\left(a^2{f'_2}^2+3af_{2}f'_{2}+6f_2^2\right)&=&-\frac{5\m}{3}\;,\\
\label{nnn6}
af'_4+4f_4-\frac{18}{35}\left(a^2{f'_2}^2+5af_{2}f'_{2}+6f_2^2\right)&=&0\;.
\ea
In deriving these expressions, all terms of order ${\cal O}(\m^3)$ and higher have been neglected, under the assumption that parameter $\m$ is sufficiently small for such higher-order contributions to be insignificant.

The boundary conditions given in equations \eqref{nnn5} and \eqref{nnn6} are consistent with those originally derived by Kopal \citep{Kopal_1960book} and later confirmed by Lanzano \citep{Lanzano_1982}. In contrast, Nakiboglu \citep{Nakiboglu_1979GeoJ} derived equation \eqref{nnn5} with an error in the numerical coefficient preceding the term proportional to $f_2 f'_2$, using a value of $1/2$ instead of the correct value of $3$. This error was subsequently propagated in his later work \citep{Nakiboglu_1982PEPI}, and it inadvertently influenced the study by Chambat et al. \citep{Chambat_2010GeoJI}, who performed numerical simulations that appeared to validate Nakiboglu's incorrect coefficient.

Our independent analytical derivation of the boundary condition \eqref{ik9n} fully aligns with the results obtained by Kopal \citep{Kopal_1960book}, Lanzano \citep{Lanzano_1982}, and also with the boundary conditions established by Zharkov and Trubitsyn \citep{Zharkov_1978book}. This agreement provides strong evidence that the boundary conditions proposed by Nakiboglu \citep{Nakiboglu_1979GeoJ} are erroneous and should not be used in geophysical modeling or interpretation.

The boundary condition \eqref{ik9n} can be further simplified by applying iterative approach. In the first iteration, we neglect all terms which are products of the $f_l$, and obtain:
\bal{cl32}
af'_l+lf_l&=&-\frac{2l+1}{3}\m\delta_{2l}\;.
\ea 
This is the boundary condition for the Clairaut equation \eqref{v1cd}. This linear condition implies that only the quadrupole harmonic $f_2$ is excited by the external rotational potential $W_Q$ (see \eqref{bw5aa}), while all others vanish ($f_l = 0$ for $l \neq 2$). Thus, the Clairaut approximation describes a rotating, self-gravitating fluid body as an oblate spheroid, flattened along the axis of rotation.

In the second iteration, we substitute the Clairaut boundary condition \eqref{cl32} into the quadratic terms on the right-hand side of \eqref{ik9n} to eliminate the radial derivatives $f'_l$. This yields the boundary conditions for the harmonics $f_l$ that appear in the Darwin-de Sitter equations \eqref{o7b5}--\eqref{on6}. We restrict our analysis to the harmonics with $l = 0, 2, 4$, as these are the only ones contributing at order $\m^2$ in the Darwin-de Sitter approximation. Higher-order harmonics, which arise at $\mathcal{O}(\m^3)$ and beyond, lie outside the scope of this approximation. Equation \eqref{ik9n} for these harmonics is equivalent to three conditions:
\bal{dr3x}
af_0'-\m f_2-\frac{5}{9}\m^2-\frac{8f_2^2}{5}&=&0\;,\\\label{pom45}
af_2'+2f_2-\frac{52}{21}\m f_2+\frac{20}{63}{{\mathsf{q}}^2}-\frac{8f_2^2}{7}&=&-\frac53{{\mathsf{m}}}\;,\\\label{btv54}
af_4'+4f_4+\frac67 \m f_2-\frac{10}{7}\m^2&=&0\;.
\ea 

The boundary condition \eqref{dr3x} for the monopole harmonic $f_0$ shows that it becomes nonzero at second order in $\m$. This arises from nonlinear coupling with the quadrupole harmonic $f_2$, resulting in an additional, purely radial distortion of the fluid body. This effect supplements the radial distortion caused by the potential $W_R$ (see Eq. \eqref{bw5}) and confirms the earlier estimate of the magnitude of $f_0\sim\m^2$ discussed in the final paragraph of Section \ref{onuv5}.

Equation \eqref{pom45} provides a second-order correction to the boundary condition for the quadrupole harmonic $f_2$ of the shape function. This result is consistent with the boundary condition derived in the Zharkov-Trubitsyn theory, specifically Eq. 33.8 in their monograph \citep{Zharkov_1978book}, which expresses the condition in terms of the variables $\hat{e}$ and $\hat{\eta}$:
\bal{bb59}
\hat{e}\hat{\eta}&=&\frac52\m-2\hat{e}+\frac{10}{21}\m^2+\frac{4}{7}\hat{e}^2-\frac{6}{7}\m\hat{e}\;,
\ea
which are related to the harmonics $f_2$ and $f_4$ of the shape function $f$ (Eqs 30.4 and 33.7 from Zharkov-Trubitsyn's monograph \citep{Zharkov_1978book}):
\bal{bb56}
e&=&-\frac32 f_2 - \frac{69}{56}f_2^2 - \frac{4}{7} k\;,\\\label{bb57}
k&=&\frac{35}{32}f_4-\frac{27}{32} f_2^2\;,\\\label{bb58}
\hat{e}&=&e-\frac{5}{42}e^2+\frac{4}{7}k\;,\\\label{op60}
\hat{\eta}&=&\frac{d\ln\hat{e}}{d\ln r}\;.
\ea
Substituting these into the expression for $\hat{e} \hat{\eta}$ and retaining terms up to second order in $\m$ reproduces Eq. \eqref{pom45}, thereby confirming the equivalence between the two formulations.  

The fourth harmonic $f_4$ of the shape function remains unexcited in the linear Clairaut regime but is induced at second order in the Darwin-de Sitter approximation due to nonlinear self-coupling of the quadrupole harmonic $f_2$. As shown, the boundary condition \eqref{btv54} precisely matches Eq. 33.4 in Zharkov and Trubitsyn's monograph \citep{Zharkov_1978book}, where the function $k$ -- defined in Eq. \eqref{bb57} -- quantifies the second-order deviation of the boundary surface from the Clairaut ellipsoid.

It follows that higher-order harmonics ($l > 4$) will also be excited through similar nonlinear interactions involving $f_2$. However, their contributions arise at order $\m^3$ and beyond, and thus lie outside the scope of the Darwin-de Sitter approximation considered here.

\subsection{The Love Numbers and Multipole Moments of Rotating Fluid Bodies}\label{lov45}

\subsubsection{Love Numbers}

In case of the fluid body, there exist two dimensionless types of Love numbers \citep{Love_1909} for each spectral harmonic $l\ge 0$, denoted correspondingly as $k_l$ and $h_l$. 
The Love numebr $k_l$ was introduced in Eq. \eqref{ion76f}. The second Love number is defined by the following equation:
\bal{24}
h_l:=-\frac{3}{\m}f_l\;,
\ea
where the shape function $f_l$ refers to the surface of the fluid body at $r=a$. The Love numbers $k_l$ characterize the integral response of the body's gravitational potential to a perturbing potential $W_Q$, while the Love numbers $h_l$ characterize the susceptibility of its shape to the perturbing potential. 

The Love numbers $k_l$ and $h_l$ are interconnected. The relashionship between the two Love numbers can be obtained in the most simple way from Eq. \eqref{s4rv} which depends on the height's function harmonic $X_l$ and its first derivative $X'_l$. Replacing in this equation $X_l\rightarrow af_l$ and $X'_l\rightarrow af'_l+f_l$, and using Eq. \eqref{cl32} for replacing $f'_l$, yields:
\bal{ob6tf}
k_l&=&\left(1-\frac{2\m}{3}\right)h_l-\delta_{2l}+\frac{\m}{3}\sum_{n=0}^\infty\sum_{m=|n-l|}^{n+l}T^{nml}\left[\delta_{2(n}h_{m)}+nh_m\delta_{2n}+mh_n\delta_{2m}-\frac12(n+m)h_nh_m\right]\;,
\ea
which clearly demonstrates that the two types of Love numbers are intimately related, and not independent. The shape Love number $h_l$ is more fundamental as it is found by solving equation for the shape function $f_l$. As soon as $h_l$ is known, we can use Eq. \eqref{ob6tf} to calculate the Love number $k_l$.

For the first three non-vanishing harmonics Eq. \eqref{ob6tf} yields
\bal{by5c3}
k_0&=&h_0-\;\frac{\m}{3}\;\left(2h_0-h_2+\frac{2}{5}h_2^2\right)\;,\\\label{gtzx4}
k_2&=&h_2-\frac{4\m}{21}\left(h_2-\frac52 h_4+h_2^2\right)-1\;,\\\label{mmm65v}
k_4&=&h_4+\frac{6\m}{7}\left(h_2-\frac{3}{11}h_4-\frac{2}{5}h_2^2\right)\;.
\ea
The classical theory of Love numbers is typically limited to Clairaut's approximation and does not account for the higher-order terms in the second approximation of the Darwin-de Sitter theory. These nonlinear terms are crucial for a better understanding of the interior structure of rapidly rotating major planets like Jupiter and Saturn.

\subsubsection{Multipole Moments}

The multipole moments of a rotating body are derived from the general expression for the gravitational potential, given by Eq. \eqref{xsa2}, evaluated at the parameter value $\tau = 1$:
\bal{mom1} 
\Phi(\x)&=&G\int_{\cal V}\frac{\mu({\bm x}')d^3x'}{|{\bm x}-{\bm x}'|}+ G\oint_{{\mathbb S}^2}d^2\Omega({\bm n}_a) \sum_{n=0}^\infty\frac{X^{n+1}({\bm a})}{(n+1)!}\frac{\pd^n}{\pd a^n} \left[\frac{a^2\mu({\bm a})}{|{\bm x}-{\bm a}|}\right]\;, 
\ea 
where the perturbed mass density $\mu(\x)$ is expressed as
\bal{mom2}
\mu(\x)&=&\rho(r)+\sum_{n=1}^\infty\left(-\hat{\mathsf T}_{\bm X}\right)^n\rho(r)\;,
\ea
representing a linear combination of the unperturbed density $\rho(r)$ and the perturbation $\varrho(\x)$ defined in Eq. \eqref{jht6a}.

To extract the multipole moments, we expand the gravitational potential in Eq. \eqref{mom1} in terms of Legendre polynomials at a field point $\x$ located at a distance $r > a$: 
\bal{bom1}
\Phi&=&\sum_{l=0}^\infty\frac{G\M_l}{r^{l+1}}P_l(\cos\theta)\;,
\ea
where the coefficients $\mathcal{M}_l$ are, by definition, the gravitational multipole moments:
\bal{mom3}
\M_l&=&\int_\V\mu(\x)r^lP_l(\cos\theta)d^3x+\oint_{{\mathbb S}^2}d^2\Omega({\bm n}_a)P_l\left(\cos\theta_a\right)\sum_{n=0}^\infty\frac{X^{n+1}({\bm a})}{(n+1)!}\frac{\pd^n}{\pd a^n} \biggl[\mu({\bm a})a^{l+2}\biggr]\;.
\ea
These multipole moments are related to the zonal harmonic coefficients $\mathcal{I}_l$ and $J_l$, introduced in Eqs. \eqref{yh7bv} and \eqref{evt3}, via the relations:
\bal{mba2}
\M_l&=&-\M a^l{\cal I}_l=-\M R^l_{\rm e}J_l\;.
\ea
Furthermore, the Love numbers $k_l$ are connected to the multipole moments through the expression (see Eq. \eqref{yhas}):
\bal{bom7}
\M_l&=&-\frac{\m}3\M a^l k_l\;.
\ea
Thus, empirical determination of the multipole moments $\mathcal{M}_l$ provides direct insight into the Love numbers, which characterize the body's elastic response to external gravitational forces. In particular, by combining the results presented in Eqs. \eqref{gtzx4} and \eqref{bom7}, we obtain a generalized form of the classical Clairaut's theorem:
\bal{kim5}
J_2&=&-f_2-\frac{\m}{3}-\frac{\m}{7}f_2-\frac{11}{7}f^2_2\;,
\ea
which expresses the second zonal harmonic $J_2$ in terms of the body's geometric flattening $f_2$ and its angular rotation rate $\sim\m$. This relation extends Clairaut's original formulation by incorporating higher-order corrections and rotational effects.

The dependence of the multipole moments in Eq. \eqref{mom3} on the height function $X(\x)$ is nontrivial. To facilitate further analysis, we adopt the quadratic approximation employed in the preceding section. Under this approximation, the perturbed density becomes:
\bal{bom8}
\mu(\x)&=&\rho+\left(-X+XX'\right)\rho' +\frac12 X^2\rho''+\ldots\;.
\ea
Substituting this into Eq. \eqref{mom3} and retaining terms up to second order yields:
\bal{bom9}
\M_l&=& \int_\V\left[\rho+\left(-X+XX'\right)\rho' +\frac12 X^2\rho''\right]r^{l+2}P_l(\cos\theta)drd^2\Omega({\bm n})\\\nonumber
&+&\oint_{{\mathbb S}^2}d^2\Omega({\bm n}_a)P_l\left(\cos\theta_a\right)Xa^{l+1}\biggl[a\rho+\left(1+\frac{l}{2}\right)\rho X-\frac12aX\rho'\biggr]\;.
\ea
By integrating the volume term by parts, surface contributions cancel, simplifying the expression to:
\bal{bm10}
\M_l&=& \int_\V\rho\frac{\pd}{\pd r}\left\{\left[rX-\left(1+\frac{l}{2}\right)X^2\right]r^{l+1}\right\}P_l(\cos\theta)drd^2\Omega({\bm n})\;.
\ea
The angular integrals in this expression can be evaluated analytically using the Legendre expansion of the shape function $f = X/r$, as given in Eq. \eqref{zx2}. These integrals are computed using the Wigner decomposition technique detailed in Appendix \ref{wignerdecomp}. This yields the final expression:
\bal{cv8}
{\cal M}_l&=&4\pi\int_0^a\rho(r)d\left[r^{l+3}F_l(r)\right]\;,\qquad (l\ge 2)
\ea
where the functions $F_l(r)$ are defined as:
\bal{gtr5}
F_l&=&\frac{1}{2l+1}\left[f_l-\left(1+\frac{l}{2}\right)\sum_{n=0}^\infty\sum_{m=0}^\infty T^{nml}f_nf_m \right]\;,
\ea
with $T^{nml}$ denoting the Wigner matrix elements introduced in Eq. \eqref{wig1}.

This final expression for $\mathcal{M}_l$ is in exact agreement with the formulation of multipole moments in the classical theory of Zharkov and Trubitsyn \citep{Zharkov_1978book} in the quadratic approximation with respect to the parameter $\m$.

\section{Exact Solutions to the Nonlinear Master Equation for the Shape Function}\label{appmac}
\stepcounter{equationschapter}
\renewcommand{\theequation}{9.\arabic{equation}}

The governing differential equation \eqref{cre5a}, which characterizes the equilibrium shape of a rotating fluid body, is inherently nonlinear and poses substantial analytical challenges. To assess the robustness and physical relevance of this equation, it is essential to examine its consistency with established exact solutions derived from classical theories of rotating fluid configurations -- most notably, the Maclaurin spheroids and Jacobi ellipsoids.

These classical solutions serve as critical benchmarks, offering a mean to validate the predictions of the nonlinear formulation. In this section, we present and analyze several exact solutions to the nonlinear equation \eqref{cre5a}, exploring their mathematical structure and physical implications. Through this comparative analysis, we aim to enhance our understanding of the shape function $f$ and its behavior across a range of physically meaningful scenarios, thereby evaluating the applicability and limitations of the nonlinear model in describing rotating fluid bodies.

\subsection{Exact Solutions Leading to Maclaurin and Jacobi Ellipsoids}\label{subD1}
We now turn to the case of a fluid body with uniform (constant) density, denoted by $\rho$, in order to derive an exact solution to the nonlinear master equation \eqref{cre5a} governing the shape function $f$. The reference configuration is assumed to possess a density profile of the form $\rho(r) = \rho H(r)$, where the Heaviside step function $H(r)$ is defined as
\bal{sv4z}
H(r)&=&\begin{cases}
1, & 0\le r < r_0 \\
0, & r \ge r_0\;,
\end{cases}
\ea
indicating that the density remains constant within a spherical region of radius $r_0$ and vanishes identically at its boundary. The effective fluid density $\sigma(r)$, introduced in Eq.~\eqref{hunb6}, is defined analogously:
\bal{hste4}
\sigma(r)&=&\begin{cases}
\sigma, & 0\le r < r_0 \\
0, & r \ge r_0\;,
\end{cases}
\ea
where the constant value of $\sigma$ is given by
\bal{pon4c}
\sigma&=&\rho-\frac{\omega^2}{2\pi G}\;.
\ea

Under the assumption of constant density, it is natural to postulate that the shape function $f = f(\theta, \varphi)$ is independent of the radial coordinate $r$. In this framework, the level surface is defined by the relation
\bal{f3sa}
R&=&r\left[1+f(\theta,\varphi)\right]\;,
\ea
where $r\in\MM$ serves as a continuous parameter labeling the unperturbed level surfaces of the reference configuration.

The assumption of constant density implies $\upbeta = 1$, which simplifies the nonlinear master equation \eqref{cre5a} governing the shape function $f$ to the form
\bal{nz4a}
\Delta f-\frac{2f'}{r}+\frac{6}{r}\left(f'+\frac{f}{r}\right)-\frac{A'-B'}{r}-\frac{3(A-B)}{r^2}&=&0\;,
\ea
where the functions $A$ and $B$ are defined in Eqs.~\eqref{t5cx}. Since $f$ depends solely on the angular coordinates $\theta$ and $\varphi$, all derivatives with respect to the radial coordinate $r$ vanish, reducing Eq.~\eqref{nz4a} to
\bal{hgf5}
\frac{1}{\sin\theta}\frac{\pd}{\pd\theta}\left(\sin\theta\frac{\pd f}{\pd\theta}\right)+\frac{1}{\sin^2\theta}\frac{\pd^2f}{\pd\varphi^2}-\frac{3}{1+f}\left[\left(\frac{\pd f}{\pd\theta}\right)^2+\frac{1}{\sin^2\theta}\left(\frac{\pd f}{\pd\varphi}\right)^2\right]+3f(1+f)(2+f)&=&0\;.
\ea
Although Eq.~\eqref{hgf5} is independent of $r$, it remains highly nonlinear and analytically challenging. To facilitate its solution, we introduce the transformation
\bal{h4c}
f(\theta,\varphi)&=&\frac{1}{\sqrt{1+\chi(\theta,\varphi)}}-1\;,
\ea
where $\chi(\theta, \varphi)$ is an auxiliary function. In terms of $\chi$, the equation defining the level surface becomes
\bal{v3c5d}
R^2\left[1+\chi(\theta,\varphi)\right]&=&r^2\;,
\ea
with $r$ denoting the radius of the unperturbed level surface.

Substituting \eqref{h4c} into \eqref{hgf5} yields:
\bal{f4xs}
\frac{1}{\left(1+\chi\right)^{3/2}}\left(\frac{\pd^2\chi}{\pd\theta^2}+\frac{\cos \theta}{\sin\theta}\frac{\pd \chi}{\pd\theta}+\frac{1}{\sin^2\theta}\frac{\pd^2\chi}{\pd\varphi^2}+6\chi\right)&=&0\;,
\ea
where the prefactor is strictly positive for all physically admissible values of $\chi$. Consequently, the expression in parentheses must vanish identically, leading to the following linear partial differential equation for the auxiliary function $\chi = \chi(\theta, \varphi)$:
\bal{df4s}
\frac{\pd^2\chi}{\pd\theta^2}+\frac{\cos \theta}{\sin\theta}\frac{\pd \chi}{\pd\theta}+\frac{1}{\sin^2\theta}\frac{\pd^2\chi}{\pd\varphi^2}+6\chi&=&0\;.
\ea

Equation~\eqref{df4s} is homogeneous and admits separation of variables. We seek a solution of the form
\bal{m7b}
\chi(\theta,\varphi)&=&\Theta(\theta)\Phi(\varphi)\;,
\ea
where $\Theta(\theta)$ and $\Phi(\varphi)$ are functions of the respective angular coordinates. It is important to note that the symbol $\Phi(\varphi)$ used here denotes an auxiliary function and should not be confused with the gravitational potential, which is represented by the same symbol elsewhere in this manuscript.

Substituting Eq.~\eqref{m7b} into Eq.~\eqref{df4s} yields two ordinary differential equations:
\bal{f5xz}
\frac{\pd^2\Phi}{\pd\varphi^2}+m^2\Phi&=&0\;,\\\label{pa4c}
\frac{\pd^2\Theta}{\pd\theta^2}+\cot\theta\frac{\pd\Theta}{\pd\theta}+\left(6-m^2\csc^2\theta\right)\Theta&=&0\;,
\ea
where $m$ is the separation constant. The general solution to Eq.~\eqref{f5xz}, subject to the periodicity condition $\Phi(0) = \Phi(2\pi)$, is
\bal{z4d}
\Phi(\varphi)&=&a_{1m}\sin m\varphi+a_{2m}\cos m\varphi\;,\qquad\qquad (m=0,\pm 1,\pm 2,...)\;,
\ea
with $a_{1m}$ and $a_{2m}$ being arbitrary constants. The solution to Eq.~\eqref{pa4c} is a linear combination of associated Legendre functions:
\bal{xa5d}
\Theta(\theta)&=&b_{1m} P_2^m(\cos\theta)+b_{2m} Q_2^m(\cos\theta)\;,
\ea
where $b_{1m}$ and $b_{2m}$ are constants. Imposing regularity over the entire domain $0 \leq \theta \leq \pi$ requires the exclusion of the singular Legendre function $Q_2^m$, i.e., $b_{2m} = 0$.

Consequently, the general solution for $\chi(\theta, \varphi)$ is expressed as a linear combination of five terms:
\bal{b6vd}
\chi(\theta,\varphi)&=&2\a_{0} P_2^0(\cos\theta)+(2\a_1\cos \varphi+2\b_1\sin\varphi)P_2^1(\cos\theta)+(\a_2\cos 2\varphi+\b_2\sin 2\varphi)P_2^2(\cos\theta)\;,
\ea
where $\alpha_0$, $\alpha_1$, $\alpha_2$, $\beta_1$, and $\beta_2$ are real constants, and the numerical factor of 2 has been introduced for convenience.

The gravitational potential within a rotating fluid body is governed by Eq.~\eqref{gt6vc}, where the unperturbed potential $U$ is obtained by solving Eq.~\eqref{mku6}, and the perturbation $K$ to $U$ is defined by Eq.~\eqref{bv4rd6}. Direct integration of Eq.~\eqref{mku6} yields a quadratic dependence of the unperturbed potential $U$ on the radial coordinate:
\bal{s5re}
U(r)&=&2\pi G\rho r_0^2-\frac{2\pi G\sigma}{3}r^2\;,
\ea 
where $r_0$ denotes the radius of the unperturbed fluid body. This expression coincides with Eq.~\eqref{om5} under the identification $a = r_0$ and $\rho = \text{const}$.

We now compute the perturbation $K^-=K^-(\x)$ of the gravitational potential inside the fluid body. Applying the shift operator $\hat{\mathsf{T}}_{\bm{X}}$, defined in Eq.~\eqref{asrx5}, to the unperturbed potential $U$ yields a quadratic expression:
\bal{t4x}
\hat{\mathsf T}_{\bm X}U&=&-\frac{2\pi G\sigma}{3}f(2+f)r^2\;,
\ea
where $f = f(\theta, \varphi)$ is given by Eq.~\eqref{h4c}, with the auxiliary function $\chi(\theta, \varphi)$ defined in Eq.~\eqref{b6vd}. Repeated application of the shift operator $n$ times results in
\bal{z4ds}
\left(\hat{\mathsf T}_{\bm X}\right)^nU&=&-\frac{2\pi G\sigma}{3}f^n(2+f)^nr^2\;.
\ea
Substituting these expressions into Eq.~\eqref{bv4rd6} and summing the resulting series yields the perturbation $K^-$ of the interior gravitational potential:
\bal{b6v5}
K^-&=&\sum_{n=1}^\infty\left(-\hat{\mathsf T}_{\bm X}\right)^nU=\frac{2\pi G\sigma r^2}{3}\frac{(2+f)f}{(1+f)^2}=-\frac{2\pi G\sigma r^2}{3}\chi(\theta,\varphi)\;.
\ea
Since the density perturbation $\varrho$ vanishes for a fluid of uniform density, the perturbation $K^-$ must satisfy the Laplace equation. Applying the Laplacian to Eq.~\eqref{b6v5}, with $\chi(\theta, \varphi)$ defined by Eq.~\eqref{b6vd}, confirms that $\Delta K^- = 0$, thereby verifying that $K^-$ is a harmonic function.

A remark is in order regarding the evaluation of radial derivatives appearing on the left-hand side of Eqs.~\eqref{t4x} and \eqref{z4ds}. A more rigorous mathematical treatment requires differentiating the density function $\sigma(r)$, which is defined as a Heaviside step function; see Eq.~\eqref{hste4}. The derivative of the step function yields a Dirac delta distribution, $\sigma'(r) = \sigma(r)\delta(r - r_0)$. However, this term is multiplied by $\sigma(r)$, which vanishes identically at the boundary $r = r_0$. Consequently, $\sigma'(r) = 0$ in the distributional sense \citep{Gelfand_1964,shilov_1968}. For the same reason, all higher-order derivatives of the density function defined in Eq.~\eqref{hste4} vanish and do not contribute to the derivations leading to Eqs.~\eqref{t4x} and \eqref{z4ds}.

The total gravitational potential $V^-$ within the ellipsoid is given by the sum of the unperturbed potential $U$ and its perturbation $K^-$, as defined in Eqs.~\eqref{s5re} and \eqref{b6v5}, respectively:
\bal{g6hb0}
V^-(r,\theta,\varphi)&=&U(r)+K^-(r,\theta,\varphi)=2\pi G\rho r_0^2-\frac{2\pi G\sigma}{3}r^2\left[1+\chi(\theta,\varphi)\right]\;,
\ea
where the angular function $\chi(\theta, \varphi)$ is given by Eq.~\eqref{b6vd}. This expression demonstrates that the interior gravitational potential $V^-$ is a quadratic polynomial in Cartesian coordinates, consistent with the harmonic nature of the solution and the symmetry of the reference configuration.

The level surface of the rotating fluid body is determined by solving the homotopy equation \eqref{pi6c} for the gravitational potential, evaluated at the parameter value $\tau = 1$. This yields the condition
\bal{njec5}
V^-(R,\theta,\varphi)&=&U(r)\;,
\ea
which is equivalent to Eq.~\eqref{v3c5d}, with the function $\chi(\theta, \varphi)$ evaluated on the level surface defined by Eq.~\eqref{v3c5d}. Transforming Eq.~\eqref{b6vd} into Cartesian coordinates for a point $\tilde\x=(\tilde x,\tilde y,\tilde z)$ located on the boundary surface of the body is accomplished via the standard spherical-to-Cartesian transformation:
\bal{yz5k}
\tilde x=R\sin\theta\cos\varphi\qquad\quad,\quad\qquad \tilde y=R\sin\theta\sin\varphi\quad\qquad,\quad\qquad
\tilde z=R\cos\theta\;,
\ea
which recasts $\chi(\theta, \varphi)$ into the Cartesian form
\bal{bvz5}
\chi(\tilde x,\tilde y,\tilde z)&=&-\a_0+3\a_2\frac{\tilde x^2}{R^2}-3\a_2\frac{\tilde y^2}{R^2}+3\a_0\frac{\tilde z^2}{R^2}- 6\a_1 \frac{\tilde x\tilde z}{R^2}-6\b_1\frac{\tilde y\tilde z}{R^2} + 6\b_2\frac{\tilde x\tilde y}{R^2}\;.
\ea
Substituting Eq.~\eqref{bvz5} into Eq.~\eqref{v3c5d} yields the boundary surface equation in Cartesian form:
\bal{suf59}
\left(1-\a_0\right)R^2+3\a_2 {\tilde x}^2-3\a_2{\tilde y}^2+3\a_0{\tilde z}^2- 6\a_1 \tilde x\tilde z-6\b_1 \tilde y\tilde z + 6\b_2\tilde x\tilde y&=&r_0^2\;,
\ea
where $r_0$ denotes the radius of the reference configuration, and $R\equiv R(r_0, \theta, \varphi)$ describes its deformation. This equation can be recast in matrix form:
\bal{bound1}
\tilde\x\cdot{\bm A}\cdot{\tilde\x}^{\rm T}&=&r_0^2\;,
\ea
where $\tilde\x=(\tilde x,\tilde y,\tilde z)$ and ${\tilde\x}^{\rm T}=(\tilde x,\tilde y,\tilde z)^{\rm T}$ is its transpose. The matrix $\mathbf{A}$ is defined as
\bal{matr9}
{\bm A}&=&\begin{bmatrix}
a_{11} & a_{12} & a_{13} \\
a_{21} & a_{22} & a_{23} \\
a_{31} & a_{32} & a_{33}
\end{bmatrix}\;,
\ea 
with components
\bal{pont2}
&&a_{11}=1-\a_0+3\a_2\;,\quad
a_{22}=1-\a_0-3\a_2\;,\quad a_{33}=1+2\a_0\;,\\\nonumber
&&a_{12}=a_{21}=3\b_2\;,\;\qquad a_{13}=a_{31}=-3\a_1\;,\;\;\quad a_{23}=a_{32}=-3\b_1\;.
\ea
Equation~\eqref{bound1} describes a second-order algebraic surface. Depending on the values of the coefficients in Eq.~\eqref{pont2}, this surface can be transformed into one of the 17 canonical forms \citep{encyclop} via rotation of the coordinate system $\tilde\x$ to new system $\x$. The physically relevant case for a bounded rotating fluid body is the {\it oblate}, triaxial Jacobi ellipsoid:
\bal{hj6}
\frac{x^2}{a^2}+\frac{y^2}{b^2}+\frac{z^2}{c^2}&=&1\;,
\ea
where $(x, y, z)$ are the principal axes of the ellipsoid, and the semi-major axes are given by $a = r_0 \lambda_1^{-1/2}$, $b = r_0 \lambda_2^{-1/2}$, $c = r_0 \lambda_3^{-1/2}$, with $\lambda_1 < \lambda_2 < \lambda_3$ being the eigenvalues of the matrix $\mathbf{A}$. Reduction of the quadratic form \eqref{bound1} to the canonical form \eqref{hj6} is possible if and only if all three invariants of $\mathbf{A}$ are positive:
\bal{nj7b}
I_1&=&a_{11}+a_{22}+a_{33}>0\;,\\\label{nj8b}
I_2&=&\begin{vmatrix}
a_{11} & a_{12}  \\
a_{21} & a_{22} \\
\end{vmatrix}
+
\begin{vmatrix}
 a_{22} & a_{23} \\
 a_{32} & a_{33}
\end{vmatrix}
+
\begin{vmatrix}
a_{11} & a_{13} \\
a_{31} & a_{33}
\end{vmatrix} >0\;,\\\label{nj9b}
I_3&=&\begin{vmatrix}
a_{11} & a_{12} & a_{13} \\
a_{21} & a_{22} & a_{23} \\
a_{31} & a_{32} & a_{33}
\end{vmatrix}>0\;.
\ea
These inequalities are satisfied for a broad domain of the real parameters $\alpha_0$, $\alpha_1$, $\alpha_2$, $\beta_1$, and $\beta_2$. A particularly simple realization of the Jacobi ellipsoid arises by setting all off-diagonal components of $\mathbf{A}$ to zero, i.e., $\alpha_1 = \beta_1 = \beta_2 = 0$. In this case, the semi-major axes of the ellipsoid are given by
\bal{xftv6}
a=\frac{r_0}{\left(1-\a_0+3\a_2\right)^{1/2}} \quad,\quad b=\frac{r_0}{\left(1-\a_0-3\a_2\right)^{1/2}}\quad,\quad c=\frac{r_0}{\left(1+2\a_0\right)^{1/2}}\;,
\ea
with the parameters constrained to the domain $0 < \alpha_0 < 1$ and $-1 + \alpha_0 < 3\alpha_2 < 0$. This ensures the ordering $a > b > c$, consistent with the geometry of an oblate Jacobi ellipsoid. In the special case $\alpha_2 = 0$, the ellipsoid becomes axisymmetric with $a = b \ge c$, corresponding to a Maclaurin spheroid. Notably, the expressions in Eq.~\eqref{xftv6} imply the identity
\bal{un4x2}
\frac{1}{a^2}+\frac{1}{b^2}+\frac{1}{c^2}&=&\frac{3}{r_0^2}\;,
\ea
which reflects the trace of the matrix $\mathbf{A}$ and is consistent with Eq.~\eqref{nj7b}.

The internal gravitational potential $V^-$ within the Jacobi ellipsoid, defined by Eq.~\eqref{hj6} with semi-major axes given in Eq.~\eqref{xftv6}, takes the form
\bal{jeop9}
V^-(x,y,z)&=&2\pi G\rho r_0^2-\frac{2\pi G\sigma r_0^2}{3}\left(\frac{x^2}{a^2}+\frac{y^2}{b^2}+\frac{z^2}{c^2}\right)\;,
\ea
where the point $\x=(x,y,z)$ lies within the volume occupied by the ellipsoid. This expression confirms that the gravitational potential inside the body is a quadratic function of Cartesian coordinates, consistent with the harmonic nature of the solution and the ellipsoidal symmetry of the configuration.

It is instructive to compare the internal gravitational potential derived in Eq.~\eqref{jeop9} for a Jacobi ellipsoid with classical results obtained by earlier researchers. MacMillan \citep{MacMillan1930} was among the first to derive the gravitational potential of a triaxial ellipsoid, employing integral representations that are essentially elliptic in nature. His work laid the foundation for subsequent developments in the theory of ellipsoidal figures.

A more explicit identification and systematic use of elliptic integrals in gravitational modeling was later provided by Sretenskii \citep{sret_1946}, whose treatment clarified the mathematical structure of the problem and influenced later formulations. These results were subsequently reproduced and discussed in the context of field theory by Landau and Lifshitz in {\it The Classical Theory of Fields} \citep{Landau1975}.

Chandrasekhar \citep{chandr87} extended these foundational works, focusing on astrophysical applications and the theory of equilibrium figures of rotating fluid masses. In what follows, we adopt the formalism and equations presented in Chandrasekhar's monograph \citep{chandr87} to facilitate a direct comparison with our solution.

The internal gravitational potential $V^-_{\rm C}$ expressed in terms of elliptic integrals, as derived by Chandrasekhar \citep{chandr87}, is given by
\bal{ch123}
V^-_{\rm C}(x,y,z)&=&\pi G\rho\left(I_0-A_1x^2-A_2y^2-A_3z^2\right)+\frac12\omega^2\left(x^2+y^2\right)\;,
\ea
where the constants $I_0$, $A_1$, $A_2$, and $A_3$ are defined via elliptic integrals:
\bal{ji8b}
I_0&=&abc\int\limits_0^\infty\frac{du}{\sqrt{(a^2+u)(b^2+u)(c^2+u)}}\;,\\ 
A_1&=&abc\int\limits_0^\infty\frac{du}{(a^2+u)\sqrt{(a^2+u)(b^2+u)(c^2+u)}}\;,\\
A_2&=&abc\int\limits_0^\infty\frac{du}{(b^2+u)\sqrt{(a^2+u)(b^2+u)(c^2+u)}}\;,\\
A_3&=&abc\int\limits_0^\infty\frac{du}{(c^2+u)\sqrt{(a^2+u)(b^2+u)(c^2+u)}}\;.
\ea
Equating the potentials given by Eqs.~\eqref{jeop9} and \eqref{ch123}, and comparing the coefficients of the quadratic terms in $x$, $y$, and $z$, yields the following relations:
\bal{a9v3x}
I_0=2r_0^2\quad,\quad A_1-\frac{\omega^2}{2\pi G\rho}=\frac{2\sigma}{3\rho}\frac{r_0^2}{a^2}\quad,\quad A_2-\frac{\omega^2}{2\pi G\rho}=\frac{2\sigma}{3\rho}\frac{r_0^2}{b^2}\quad,\quad A_3=\frac{2\sigma}{3\rho}\frac{r_0^2}{c^2}\;.
\ea
These expressions are fully consistent with the identities derived by Chandrasekhar \citep{chandr87}, namely,
\bal{jk4x}
\frac{\omega^2a^2}{\pi G\rho}-2A_1a^2&=&\frac{\omega^2b^2}{\pi G\rho}-2A_2b^2=-2A_3c^2\;,\\
\label{kn6v}
A_1+A_2+A_3&=&2\;,
\ea
where Eq.~\eqref{jk4x} follows directly from Eq.~\eqref{a9v3x}, and Eq.~\eqref{kn6v} is verified by summing the last three equalities in Eq.~\eqref{a9v3x} and invoking Eqs.~\eqref{un4x2} and \eqref{pon4c}.

The first equality in Eq.~\eqref{a9v3x} can be verified by noting that, according to Chandrasekhar \citep{chandr87}, the elliptic integral $I_0$ is equivalent to the surface integral of the squared radius function $R^2(r_0, \theta, \varphi)$ over the boundary of the ellipsoid:
\bal{nr4xc}
I_0&=&\frac{1}{2\pi}\oint\limits_{\mathbb{S}^2}R^2(r_0,\theta,\varphi)\sin\theta d\theta d\varphi\;.
\ea
Evaluating this integral by substituting the right-hand side of Eq.~\eqref{v3c5d} for $R^2(r_0, \theta, \varphi)$ and using Eq.~\eqref{b6vd} for the function $\chi(\theta, \varphi)$ yields the result $I_0 = 2r_0^2$, as claimed.

This confirms the consistency of our theoretical framework with classical results obtained through alternative mathematical techniques, thereby reinforcing the validity of our approach.

\subsection{The Unit Index Polytrope}\label{subD2}
\subsubsection{Unperturbed Configuration of a Polytropic Fluid with Unit Index}
The equation of state for a polytropic fluid with polytropic index $n = 1$ is given by:
\begin{align}
p &= {\mathsf K}_0 \rho^2\;,
\end{align}
where ${\mathsf K}_0$ is a constant. This relation allows for an exact analytic solution for the unperturbed density distribution and gravitational potential of the fluid body \citep{horedt_2004book}.

Solving the system of governing equations \eqref{hunb6}, \eqref{mku6}, and \eqref{w9ab} for the equilibrium configuration yields the following dimensionless density profile: 
\bal{erd4a}
\frac{\rho(\eta)}{\bar\rho(\eta_1)}&=&A\frac{\sin\eta}{\eta}+\frac{2}{3}\m\;,
\ea
where:
\begin{itemize}
\item[-] $A$ is an integration constant,
\item[-] $\eta := \kappa r$ is the dimensionless radial coordinate (Lane-Emden variable),
\item[-] $\kappa = \sqrt{2\pi G / {\mathsf K}_0}$ is a scaling constant,
\item[-] $\bar\rho(\eta_1)$ is the average density of the fluid,
\item[-] $\eta_1 = \kappa a$ is the dimensionless radius of the unperturbed fluid body, incorporating the effect of the centrifugal potential $W_R$.
\end{itemize}
This solution describes the internal structure of a rotating polytropic fluid body with uniform rotation and provides a foundation for analyzing perturbations and stability in such configurations.

\subsubsection{Determination of the Polytropic Radius and Integration Constant}

The fluid density $\rho$ vanishes at the surface of the body, which corresponds to the condition $\rho(\eta_1) = 0$. Substituting this into Eq.~\eqref{erd4a} yields the following expression for the integration constant $A$:
\bal{hg6v}
A&=&-\frac{2\m}{3}\frac{\eta_1}{\sin\eta_1}\;.
\ea
The average density $\bar\rho(\eta_1)$ can be computed using Eq.~\eqref{erd4a} in conjunction with the definition of average density given in Eq.~\eqref{jky7}. This leads to the relation:
\bal{avd5}
\frac{3A}{\eta_1^3}\left(\sin\eta_1-\eta_1\cos\eta_1\right)&=&1-\frac{2\m}{3}\;.
\ea
Solving Eqs.~\eqref{hg6v} and \eqref{avd5} simultaneously provides expressions for both $A = A(\eta_1)$ and $\m = \m(\eta_1)$ in terms of the dimensionless radius $\eta_1$ of the fluid body. Inverting the resulting expression for $\m$ yields an equation for determining $\eta_1$ as a function of $\m$:
\bal{eq222}
\frac{\tan\eta_1}{\eta_1}&=&\frac{2\m}{2\m+\ds\left(1-\frac{2\m}{3}\right)\eta_1^2}\;.
\ea
This transcendental equation can be solved iteratively. Expanding the solution in powers of $\m$ gives:
\bal{eq333}
\eta_1&=&\pi+\frac{2\m}{\pi}+\frac{4\m^2}{3\pi}\left(1-\frac{6}{\pi}\right)+{\cal O}\left(\m^3\right)\;.
\ea
Substituting this expansion into Eq.~\eqref{hg6v} yields the corresponding expression for the integration constant $A$:
\bal{sd97}
\frac{A}{\eta_1}&=&\frac{\pi}{3}\left[1-\frac{2\m}{3}\left(1-\frac{6}{\pi^2}\right)\right]+{\cal O}\left(\m^2\right)\;.
\ea
A comparison of Eqs.~\eqref{eq333} and \eqref{sd97} with the corresponding expressions in Zharkov and Trubitsyn's theory \citep[Eq.~34.11]{Zharkov_1978book} confirms their equivalence, thereby validating the consistency of the analytic approach.

\subsubsection{Density Profile, Gravitational Potential, and Perturbations in a Polytrope with Unit Index}

Using the definition of the density $\sigma$ from Eq.~\eqref{hunb6}, the dimensionless density profile given in Eq.~\eqref{erd4a} can be reformulated as:
\bal{dprof1}
\sigma(\eta)&=&\sigma_0\frac{\sin\eta}{\eta}\;,
\ea
where $\sigma_0 := A \bar\rho(\eta_1)$ is a constant, and $\eta = \kappa r$ is the dimensionless radial coordinate.

This expression allows for an exact evaluation of the function $\upbeta(\eta)$ defined in Eq.~\eqref{v4}, which represents the ratio of the local density to the average density:
\bal{tv654} 
\upbeta(\eta)&=&\frac{\sigma(\eta)}{\bar\sigma(\eta)}=\frac{\eta}{3}\frac{j_0(\eta)}{j_1(\eta)}=\frac{\eta^2}{3(1-\eta \cot\eta)}\;,
\ea
where $\bar\sigma(\eta)$ is the average density computed using Eq.~\eqref{y7}, and $j_0(\eta)$ and $j_1(\eta)$ are spherical Bessel functions of the first kind \citep{Arfken-Weber}.

The unperturbed gravitational potential $U$ of the reference configuration satisfies the Poisson equation \eqref{mku6}, with the density profile given by Eq.~\eqref{dprof1}. Solving this equation yields:
\bal{iop11}
U(\eta)&=&2{\mathsf K}_0 \left[\sigma(\eta)+\frac{2}{3}\eta_1^2\bar\sigma(\eta_1)\right]\;,\qquad\qquad (\eta\le\eta_1)\;.
\ea

The gravitational potential perturbation $K^-$ inside the polytropic fluid satisfies the linear Molodensky equation \eqref{k7a} with a constant coefficient:
\bal{d2a}
\Delta_\eta K^-+K^-&=&0\;,
\ea
where $\Delta_\eta$ denotes the Laplacian in spherical coordinates with $\eta$ as the radial variable. Outside the fluid body, the perturbation $K^+$ satisfies the Laplace equation:
\bal{ggg6}
\Delta_\eta K^+&=&0\;.
\ea

The general solution to Eq.~\eqref{d2a} that remains regular at the center ($\eta = 0$) is given by:
\bal{d2b}
K^-&=&2\sigma_0{\mathsf K}_0\sum_{{l}=0}^\infty c_{l} j_{l}(\eta)P_{l}(\cos\theta)\;,
\ea
where $c_l$ are dimensionless constants, $j_l(\eta)$ are spherical Bessel functions, and $P_l(\cos\theta)$ are Legendre polynomials. The coefficients $c_l$ are determined by enforcing continuity of the gravitational potential and its radial derivative across the boundary $\eta = \eta_1$.

\subsubsection{Rotational Deformation and the Height Function}

The rotational deformation of a self-gravitating fluid body is described by the height function $X = X(\eta, \theta)$, which characterizes the deviation of the fluid surface from spherical symmetry. This function can be obtained using the power series method outlined in Section~\ref{ssec5}, and expressed in its general form by Eq.~\eqref{oper23}. Substituting the specific solutions for the unperturbed gravitational potential $U$ from Eq.~\eqref{iop11} and the perturbation $K$ from Eq.~\eqref{d2b} into this framework yields:
\bal{2dz}
X(\eta,\theta)&=&\sum_{n=1}^\infty\left(-1\right)^n\frac{\b_n}{n!}\left[\frac{\ds\sum_{{l}=0}^\infty c_{l} j_{l}(\eta)P_{l}(\cos\theta)}{j'_0(\eta)+\ds\sum_{{l}=0}^\infty c_{l} j'_{l}(\eta)P_{l}(\cos\theta)}\right]^n\;,
\ea
where $\beta_1 = 1$, and the higher-order coefficients $\beta_n$ are given by:
\bal{2dx}
\b_n&=&\sum_{k=1}^{n-1}(-1)^k(n)_k{\bf B}_{n-1,k}\left(\gamma_1,\gamma_2,...,\gamma_{n-k}\right)\;,
\ea
with ${\bf B}_{n-1,k}\left(\gamma_1,\gamma_2,...,\gamma_{n-k}\right)$ denoting the partial Bell polynomials with arguments:
\bal{2dv}
\gamma_k&=&\frac{1}{k+1}\frac{j^{(k+1)}_0(\eta)+\ds\sum_{{l}=0}^\infty c_{l} j^{(k+1)}_{l}(\eta)P_{l}(\cos\theta)}{j'_0(\eta)+\ds\sum_{{l}=0}^\infty c_{l} j'_{l}(\eta)P_{l}(\cos\theta)}\;.
\ea
Equations \eqref{2dz}--\eqref{2dv} provide an exact power series representation for the height function $X$ describing the equilibrium shape of a uniformly rotating polytrope with polytropic index unity. In the quadratic approximation, Eq.~\eqref{2dz} simplifies to:
\bal{fig17x}
X(\eta,\theta)&=&-\frac{1}{\,j'_0(\eta)}\biggl[c_0 j_0(\eta)+c_2j_2(\eta)P_2(\cos\theta)+c_4j_4(\eta)P_4(\cos\theta)\biggr]\\\nonumber
&&\hspace{0.7cm}+c_2^2\left[j'_2(\eta)-\frac{j_2(\eta)j''_0(\eta)}{2j'_0(\eta)}\right]\frac{j_2(\eta)}{\eta\,{j'_0}^2(\eta)}P_2^2(\cos\theta)\;.
\ea

Expanding the square of the Legendre polynomial $P_2^2(\cos\theta)$ into a sum of harmonics allows the height function to be expressed in spectral form:
\bal{hb6d}
X(\eta,\theta)&=&X_0(\eta)+X_2(\eta)P_2(\cos\theta)+X_4(\eta)P_4(\cos\theta)\;,
\ea
where the spectral components are:
\bal{n6v5}
X_0(\eta)&=&-c_0\frac{j_0(\eta)}{j'_0(\eta)}+\frac{1}{5}c_2^2\left[j'_2(\eta)-\frac{j_2(\eta)j''_0(\eta)}{2j'_0(\eta)}\right]\frac{j_2(\eta)}{{j'_0}^2(\eta)}\;,\\
X_2(\eta)&=&-c_2\frac{ j_2(\eta)}{j'_0(\eta)}+\frac{2}{7}c_2^2\left[j'_2(\eta)-\frac{j_2(\eta)j''_0(\eta)}{2j'_0(\eta)}\right]\frac{j_2(\eta)}{{j'_0}^2(\eta)}\;,\\
X_4(\eta)&=&-c_4\frac{ j_4(\eta)}{j'_0(\eta)}+\frac{18}{35}c_2^2\left[j'_2(\eta)-\frac{j_2(\eta)j''_0(\eta)}{2j'_0(\eta)}\right]\frac{j_2(\eta)}{{j'_0}^2(\eta)}\;.
\ea
The expansion \eqref{hb6d} includes terms up to second order in the small parameter $\m$. Higher-order corrections can be systematically incorporated by extending the power series in Eqs. \eqref{2dz}--\eqref{2dv}.

\subsubsection{Determination of Shape Function Coefficients via Boundary Matching}

The constants entering the solution for the shape function can be determined by matching the interior and exterior solutions for the radial harmonics of the shape function, defined as $f_l(\eta) = X_l(\eta)/\eta$. This matching is performed at the boundary $\eta = \eta_1$, where $\eta_1$ is given by Eq.~\eqref{eq333}. The normalization condition ensures that $f_l(\eta_1) = X_l(\eta_1)/\eta_1$.

The boundary values of the radial harmonics $f_l(\eta)$ at $\eta = \eta_1$ are:
\bal{g555}
f_0(\eta_1)&=&\frac{3c^2_2}{5\pi^2}\left(1-\frac{6}{\pi^2}\right)-\frac{2c_0\m}{\pi^2}\;,\\\label{g556}
f_2(\eta_1)&=&\frac{3c_2}{\pi^2} +\frac{6c^2_2}{7\pi^2}\left(1-\frac{6}{\pi^2}\right)+\frac{2c_2\m}{\pi^2}\left(1-\frac{6}{\pi^2}\right)\;,\\\label{g557}
f_4(\eta_1)&=&\frac{5c_4}{\pi^2}\left(-2+\frac{21}{\pi^2}\right)+\frac{54c^2_2}{35\pi^2}\left(1-\frac{6}{\pi^2}\right)\;.
\ea
The corresponding derivatives at the boundary are:
\bal{guu5}
f'_0(\eta_1)&=&-\frac{c_0}{\pi}+\frac{c^2_2}{5\pi}\left(1-\frac{27}{2\pi^2}+\frac{72}{\pi^4}\right)\;,\\\label{guu6}
f'_2(\eta_1)&=&+\frac{c_2}{\pi}\left(1-\frac{6}{\pi^2}\right)+\frac{2c^2_2}{7\pi}\left(1-\frac{27}{2\pi^2}+\frac{72}{\pi^4}\right)+\frac{36}{\pi^5}c_2\m\;,\\\label{guu7}
f'_4(\eta_1)&=&-\frac{c_4}{\pi}\left(1-\frac{55}{\pi^2}+\frac{420}{\pi^4}\right)+\frac{18c^2_2}{35\pi}\left(1-\frac{27}{2\pi^2}+\frac{72}{\pi^4}\right)\;
\ea 
Substituting Eqs. \eqref{g555}--\eqref{guu7} into the matching conditions \eqref{dr3x}--\eqref{btv54}, and solving the resulting system of equations for the unknown coefficients $c_0$, $c_2$, and $c_4$, we obtain:
\bal{nbt6c}
c_0&=&-\frac{5\m^2}{2\pi^2}\;,\\\label{nbt5c}
c_2&=&-\frac{5\m}{3}-\frac{10\m^2}{9}+\frac{25\m^2}{7\pi^2}\;,\\\label{nbt4c}
c_4&=&\frac{45}{15-\pi^2}\frac{\m^2}{7}\;.
\ea

\subsubsection{Love Numbers and Gravitational Multipole Moments}

The Love numbers $k_l$ quantify the response of a self-gravitating fluid body to rotational deformation and are computed using Eqs.~\eqref{by5c3}--\eqref{mmm65v}. For a polytrope with unit index, the Love numbers up to second order in the small parameter $\m$ are:
\bal{love45}
k_0&=&\frac{20}{21\pi^2}\left(8-\frac{285}{4\pi^2}+\frac{207}{\pi^4}\right)\m^2\;,\\\label{love46}
k_2&=&-1+\frac{15}{\pi^2}+\frac{10}{\pi^2}\left(1-\frac{129}{14\pi^2}\right)\m    \;,\\\label{love47}
k_4&=&-\frac{675}{7\pi^4}\frac{21-2\pi^2}{15-\pi^2}\m\;.
\ea

These Love numbers allow us to compute the gravitational multipole moments ${\cal I}_l$ of the rotating fluid body via the correspondence given in Eq.~\eqref{yhas}. Substituting Eqs.~\eqref{love45}--\eqref{love47} yields:
\bal{ux4a}
{\cal I}_0&=&{\cal O}\left(\m^3\right)\;,\\\label{ux4b}
{\cal I}_2&=&\left(-\frac13+\frac{5}{\pi^2}\right)\m+\frac{10}{\pi^2}\left(\frac{1}{3}-\frac{43}{14\pi^2}\right)\m^2+ {\cal O}\left(\m^3\right)   \;,\\\label{ux4c}
{\cal I}_4&=&-\frac{225}{7\pi^4}\frac{21-2\pi^2}{15-\pi^2}\m^2+{\cal O}\left(\m^3\right)\;.
\ea

To compare with the results of Hubbard \citep{Hubbard_1975SvA}, we express the external gravitational field in terms of the dimensionless multipole moments $J_l$, which are related to ${\cal I}_l$ via Eq.~\eqref{evt3}. Hubbard uses a different parameter $\q$ to characterize the rotational rate, related to $\m$ by Eq.~\eqref{hjuv}.

In terms of $\q$, the multipole moments are expressed as:
\bal{acq3}
J_l&=&\frac{\q}{3}\left(\frac{a}{R_{\rm e}}\right)^{3+l}k_l=\frac{\q}{3}\left[1-(3+l)f_0(\eta_1)+\frac{3+l}{2}f_2(\eta_1)+\ldots\right]k_l\;,
\ea
where $R_{\rm e}$ is the equatorial radius of the body.

Using Eqs.~\eqref{g555}--\eqref{g557}, \eqref{nbt6c}--\eqref{nbt4c}, and \eqref{love45}--\eqref{love47}, we obtain the following expressions for the gravitational moments:
\bal{44a}
J_0&=&{\cal O}\left(\q^3\right)\;,\\\label{44b}
J_2&=&\left(-\frac13+\frac{5}{\pi^2}\right)\q+\frac{5}{2\pi^2}\left(3-\frac{261}{7\pi^2}\right)\q^2+ {\cal O}\left(\q^3\right)   \;,\\\label{44c}
J_4&=&-\frac{225}{7\pi^4}\frac{21-2\pi^2}{15-\pi^2}\q^2+{\cal O}\left(\q^3\right)\;.
\ea 
These results are in exact agreement with Eqs. (23), (28), and (29) from Hubbard's analysis \citep{Hubbard_1975SvA}, thereby validating the consistency of the present formulation with established results.

\section*{Conclusion}

This paper presents a unified, non-perturbative framework for determining the equilibrium shape and gravitational field of uniformly rotating, self-gravitating fluid bodies. The formulation is based on the application of \textit{Lie group theory}, \textit{exponential operator techniques}, and \textit{Neumann series expansions}, which collectively extend Clairaut's classical linear theory into the fully nonlinear regime. The resulting formalism enables accurate modeling of rotational deformations without relying on the assumption of slow rotation.

In contrast to earlier approaches -- such as the \textit{Clairaut-Darwin-de Sitter perturbative expansions}~\citep{Clairaut-book-1743, Darwin_1899, deSitter_1924BAN}, \textit{Lyapunov's integro-differential equations}~\citep{Lyapunov_1906}, and the \textit{Zharkov-Trubitsyn spectral harmonic method}~\citep{Zharkov_1978book,Zharkov-book-1986} -- which are either limited in scope or subject to convergence and computational challenges, the proposed framework offers the following advantages:

\begin{itemize}
    \item[--] It replaces perturbative expansions with exact functional and differential equations for the gravitational potential, fluid density, and shape functions.
    \item[--] It resolves divergence issues associated with Legendre polynomial expansions near the body's surface.
    \item[--] It yields closed-form master equations for the shape and height functions, applicable across a wide range of rotation rates.
    \item[--] It incorporates Wigner's formalism to facilitate the spectral decomposition of radial harmonics, enabling efficient treatment of nonlinear couplings.
    \item[--] It provides rigorously defined boundary conditions and accurate computation of Love numbers, thereby improving the modeling of gravitational multipole moments and enhancing constraints on the internal structure and equation of state of rotating astrophysical bodies.
\end{itemize}

The resulting formalism constitutes a mathematically rigorous and computationally tractable alternative to traditional methods, with broad applicability in planetary science, stellar astrophysics, and gravitational wave astronomy.

\begin{acknowledgments}
The author gratefully acknowledges Claus L\"ammerzahl and J\"urgen M\"uller for their generous hospitality and financial support during his visit to ZARM at the University of Bremen and the Institute of Geodesy at Leibniz University Hannover in the summer of 2023. This support was instrumental in initiating the present research. Valuable discussions with Thibaut Damour (IHES, Paris) and Michael Evers (University of Hohenheim, Stuttgart) are also sincerely appreciated. The author thanks the anonymous referees for their insightful comments and constructive suggestions, which significantly enhanced the clarity and mathematical rigor of the manuscript.
\end{acknowledgments}

\section*{Conflict of Interest Statement}
\noindent
The author (Sergei Kopeikin) has no conflicts to disclose.
\newpage 

\appendix

\section{Relationship between Exponential Flow and Translation Operators}\label{appA1}
\stepcounter{equationschapter}
\renewcommand{\theequation}{A.\arabic{equation}}

In this appendix, we derive a formula that establishes a connection between the shift of the argument $\x$ of a smooth function $V(\x)$ by ${\pmb{\mathscr X}}_\tau$ and the exponential map generated by the vector field ${\bm \xi}$, which induces the shift ${\pmb{\mathscr X}}_\tau$. The key identity is:
\bal{nbv5t}
\exp\left(\tau L_{\bm\xi}\right)V(\x)&=&V(\x+{\pmb{\mathscr X}}_\tau)\;,
\ea
where $L_{\bm\xi} = \xi^i \partial_i$ is the Lie derivative along ${\bm \xi}$, and ${\pmb{\mathscr X}}_\tau = {\pmb{\mathscr X}}_\tau(\x)$ is defined by Eq. \eqref{f3}. This identity can also be expressed in the form:
\bal{nb05}
\exp\left(\tau L_{\bm\xi}\right)V(\x)&=&V\left(\exp\left(\tau L_{\bm\xi}\right)\x\right)\;,
\ea
which illustrates the property of {\it equivariance} in the context of Lie diffeomorphism theory.

To prove Eq. \eqref{nbv5t}, we consider the propagation of the gravitational potential $V(\x)$ from the point $\x$ to the point $\x_\tau$ along the integral curve of the congruence defined by Eq. \eqref{f1}, using the exponential map: 
\bal{vb6c4a}
V_\tau&:=&e^{\tau L_{\bm\xi}}V({\bm x})=\sum_{n=0}^\infty\frac{\tau^n}{n!}L_{\bm\xi}^n V({\bm x})\;.
\ea
The operator $L_{\bm\xi}^n$ can be expanded as:
\bal{jk8ba}
\sum_{n=0}^\infty\frac{\tau^n}{n!}L_{\bm\xi}^n V({\bm x})&=&
\sum_{n=0}^\infty\sum_{p=0}^n\frac{\tau^n}{n!}{\bf B}^{i_1i_2\ldots i_p}_{n,p}\left(L_{\bm\xi}\x,L_{\bm\xi}^2\x,...,L_{\bm\xi}^{n-p+1}\x\right)\pd_{i_1i_2\ldots i_p} V({\bm x})\;,
\ea
where, each of the the arguments $L_{\bm\xi}\x=(L_{\bm\xi}x^i)$, $L^2_{\bm\xi}\x=(L^2_{\bm\xi}x^i)$, and so on, have one free index, and 
\bal{het65}\hspace{-1cm}
{\bf B}^{i_1i_2\ldots i_p}_{n,p}\left(L_{\bm\xi}\x,L_{\bm\xi}^2\x,...,L_{\bm\xi}^{n-p+1}\x\right)&=&\sum_{j_1=0}^n...\sum_{j_n=0}^n\frac{n!}{j_1!j_2! ... j_{n-p+1}!}\left\{\Biggl[\frac{L_{\bm\xi}\x}{1!}\Biggr]^{j_1}\Biggl[\frac{L^2_{\bm\xi}\x}{2!}\Biggr]^{j_2}...\Biggl[\frac{L_{\bm\xi}^{n-p+1}\x}{(n-p+1)!}\Biggr]^{j_n-p+1}\right\}_{(i_1i_2\ldots i_p)}
\ea
denotes the incomplete Bell polynomial \citep{Johnson_2002} of tensor rank $p$. The sum in Eq. \eqref{het65} is taken over all sequences $j_1, j_2, j_3, ..., j_{n-p+1}$ of non-negative integers such that the following two constraints are satisfied:
\bal{nbvrt6}
j_1+j_2+j_3+...+j_{n-p+1}=p\qquad\quad,\quad\qquad j_1+2j_2+3j_3+...+(n-p+1)j_{n-p+1}=n\;.
\ea
In particular, ${\bf B}_{0,0}=1$,  ${\bf B}_{n,0}= 0\;\;\forall n\ge1$, and ${\bf B}_{0,k}= 0\;\;\forall k\ge 1$. The symmetry of ${\bf B}^{i_1i_2\ldots i_p}_{n,p}$ is understood with respect to all free vector indices in its argument, for instance,
\bse
\bal{nytc5}
{\bf B}^{i_1i_2}_{5,2}\left(L_{\bm\xi}\x,L_{\bm\xi}^2\x,L^3_{\bm\xi}\x,L^4_{\bm\xi}\x\right)&=&
10L^2_{\bm\xi}x^{(i_1}L^3_{\bm\xi}x^{i_2)}+5L_{\bm\xi}x^{(i_1}L^4_{\bm\xi}x^{i_2)}\;,\\
{\bf B}^{i_1i_2i_3}_{6,3}\left(L_{\bm\xi}\x,L_{\bm\xi}^2\x,L^3_{\bm\xi}\x,L^4_{\bm\xi}\x\right)&=&
15L^2_{\bm\xi}x^{(i_1}L^2_{\bm\xi}x^{i_2}L^2_{\bm\xi}x^{i_3)}+60L_{\bm\xi}x^{(i_1}L^2_{\bm\xi}x^{i_2}L^2_{\bm\xi}x^{i_3)}+15L_{\bm\xi}x^{(i_1}L_{\bm\xi}x^{i_2}L^4_{\bm\xi}x^{i_3)}\;,
\ea
\ese
and so on.

Reordering the summation in Eq. \eqref{jk8ba}, we obtain: 
\bal{zvtf5a}
&&\sum_{n=0}^\infty\sum_{p=0}^n\frac{\tau^n}{n!}{\bf B}^{i_1i_2\ldots i_p}_{n,p}\left(L_{\bm\xi}\x,L_{\bm\xi}^2\x,...,L_{\bm\xi}^{n-p+1}\x\right)\pd_{i_1i_2\ldots i_p} V({\bm x})\\\nonumber
&&\hspace{6cm}=\,\sum_{p=0}^\infty\pd_{i_1i_2\ldots i_p} V({\bm x})\sum_{n=p}^\infty\frac{\tau^n}{n!}{\bf B}^{i_1i_2\ldots i_p}_{n,p}\left(L_{\bm\xi}\x,L_{\bm\xi}^2\x,...,L_{\bm\xi}^{n-p+1}\x\right)\;.
\ea
Using the generating function for Bell polynomials \citep{bellpol}, we find:
\bal{oin7ga}
\sum_{n=p}^\infty\frac{\tau^n}{n!}{\bf B}^{i_1i_2\ldots i_p}_{n,p}\left(L_{\bm\xi}\x,L_{\bm\xi}^2\x,...,L_{\bm\xi}^{n-p+1}\x\right)&=&\frac{1}{p!}\left(\sum_{n=1}^\infty\frac{\tau^n}{n!}L_{\bm\xi}^nx^{i_1}\right)\left(\sum_{n=1}^\infty\frac{\tau^n}{n!}L_{\bm\xi}^nx^{i_2}\right)\ldots\left(\sum_{n=1}^\infty\frac{\tau^n}{n!}L_{\bm\xi}^nx^{i_p}\right)\;.
\ea
Applying Eq. \eqref{f3}, this becomes:
\bal{jku5c}
\sum_{n=p}^\infty\frac{\tau^n}{n!}{\bf B}^{i_1i_2\ldots i_p}_{n,p}\left(L_{\bm\xi}\x,L_{\bm\xi}^2\x,...,L_{\bm\xi}^{n-p+1}\x\right)&=&
\frac{1}{p!}{\mathscr X}^{i_1}_\tau {\mathscr X}^{i_2}_\tau\ldots {\mathscr X}^{i_p}_\tau\;.
\ea
Substituting this results into Eq. \eqref{zvtf5a}, we obtain:
\bal{bvtc5a}
V_\tau&=&\sum_{p=0}^\infty\frac{1}{p!}{\mathscr X}^{i_1}_\tau {\mathscr X}^{i_2}_\tau\ldots {\mathscr X}^{i_p}_\tau\,\pd_{i_1i_2\ldots i_p} V({\bm x})=V\left({\bm x}+{\pmb{\mathscr X}}_\tau\right)\;,
\ea
which follows directly from the Taylor expansion. Comparing this with Eq. \eqref{vb6c4a}, we confirm the validity of Eq. \eqref{nbv5t}, thus completing the proof.

\section{Independent Proof of Pullback Transformation Formula of Gravitational Potential}\label{pulBB}
\stepcounter{equationschapter}
\renewcommand{\theequation}{B.\arabic{equation}}

To demonstrate the correctness of Eq.~\eqref{xsa2}, we employ the radial gauge and spherical coordinates. In these coordinates, the gravitational potential ${\cal U}_\tau$ from Eq.~\eqref{vg6f} takes the form:
\bal{apb1}
{\cal U}_\tau&=&G\int\limits_{\V_\tau}\frac{\rho_\tau({\bm y}')}{|\x-{\bm y}'|}|{\bm y}'|^2dy'd^2\Omega\;,
\ea
where ${\bm y}':={\bm y}_\tau=\x'+\X'_\tau$ in accordance with Eq. \eqref{nb5c}. In the radial gauge, we have $\x'=r'{\bm n}$ and $\X'_\tau=X'_\tau{\bm n}$, with $X'_\tau=X'_\tau(\x')$. Using spherical coordinates, Eq.~\eqref{apb1} becomes: 
\bal{apb2}
{\cal U}_\tau&=&G\int\limits_{\cal V}\frac{\rho_\tau({\bm x}'+\X'_{\tau})}{|{\bm x}-\left({\bm x}'+\X'_{\tau}\right)|}\left|{\bm x}'+\X'_{\tau}\right|^2\left(1+\frac{\pd X'_\tau}{\pd r'}\right)dr'd^2\Omega\;.
\ea
We now expand the integrand in a Taylor series around the point $\x'$, yielding: 
\bal{apb3}
{\cal U}_\tau&=&G\int\limits_{\cal V}\sum_{n=0}^\infty\frac{{X'_\tau}^n}{n!}\frac{\pd^n}{\pd r'^n}\left[\frac{\rho_\tau(\x')r'^2}{|\x-\x'|}\right]dr'd^2\Omega+G\int\limits_{\cal V}\sum_{n=0}^\infty\frac{{X'_\tau}^n}{n!}\frac{\pd^n}{\pd r'^n}\left[\frac{\rho_\tau(\x')r'^2}{|\x-\x'|}\right]\frac{\pd X'_\tau}{\pd r'}dr'd^2\Omega\;.
\ea
Next, we isolate the $n=0$ term in the first sum on the right-hand side of Eq.~\eqref{apb3} and shift the summation index in the remaining terms via $n \to n+1$. This transforms the first integral into: 
\bal{apb4}
 \int\limits_{\cal V}\sum_{n=0}^\infty\frac{{X'_\tau}^n}{n!}\frac{\pd^n}{\pd r'^n}\left[\frac{\rho_\tau(\x')r'^2}{|\x-\x'|}\right]dr'd^2\Omega&=&\int\limits_{\V}\frac{\rho_\tau(\x')d^3x'}{|\x-\x'|}+
\int\limits_{\cal V}\sum_{n=0}^\infty\frac{{X'_\tau}^{n+1}}{(n+1)!}\frac{\pd^{n+1}}{\pd r'^{n+1}}\left[\frac{\rho_\tau(\x')r'^2}{|\x-\x'|}\right]dr'd^2\Omega \;. 
\ea
The second terms on the right-hand sides of Eqs.~\eqref{apb3} and \eqref{apb4} can be combined into a single expression involving a total derivative with respect to the radial coordinate. Thus, Eq.~\eqref{apb3} simplifies to: 
\bal{apb5}
{\cal U}_\tau&=&G\int\limits_{\V}\frac{\rho_\tau(\x')d^3x'}{|\x-\x'|}+G\int\limits_{\cal V}\sum_{n=0}^\infty\frac{\pd}{\pd r'}\left\{\frac{{X'_\tau}^{n+1}}{(n+1)!}\frac{\pd^{n}}{\pd r'^{n}}\left[\frac{\rho_\tau(\x')r'^2}{|\x-\x'|}\right]\right\}dr'd^2\Omega\;.
\ea
Finally, we observe that the second integral in Eq.~\eqref{apb5} can be evaluated as a surface integral over the boundary of the spherically symmetric unperturbed reference configuration of radius $r = a$. This yields:
\bal{apb6}
\U_\tau(\x)&=&G\int_{\cal V}\frac{\rho_\tau({\bm x}')d^3x'}{|{\bm x}-{\bm x}'|}+ G\oint_{{\mathbb S}^2}d^2\Omega({\bm n}_a) \sum_{n=0}^\infty\frac{X^{n+1}_\tau({\bm a})}{(n+1)!}\frac{\pd^n}{\pd a^n} \left[\frac{a^2\rho_\tau({\bm a})}{|{\bm x}-{\bm a}|}\right]\;. 
\ea 
This final expression for ${\cal U}_\tau$ matches exactly with Eq.~\eqref{xsa2}, thereby confirming their equivalence.

\section{Compatibility Conditions for the Master Equation of the Height Function}\label{antisym}
\stepcounter{equationschapter}
\renewcommand{\theequation}{C.\arabic{equation}}

Due to the symmetry of the second partial derivatives the anti-symmetric part of equation \eqref{xc8} must vanish identically
\bal{ii1}
M^{-1}_{[ip}({\bm X})\pd_p\left[M^{-1}_{j]q}({\bm X})\pd_q U({\bm x})\right]=0\;.
\ea
In order to prove it, let us take the partial derivatives and account for the symmetry of the second partial derivative of the potential $U$ in the above equation. Then, it takes on a simpler form
\bal{ii2}
n^qM^{-1}_{[ip}({\bm X})\pd_pM^{-1}_{j]q}({\bm X})&=&0\qquad\qquad\Longrightarrow \qquad\qquad~M^{-1}_{[ip}\pd_pN_{j]}=0\;, 
\ea
where we have introduced a new vector
\bal{oo3}
N^i&:=&M^{-1}_{iq}n^q\;.
\ea
Equation \eqref{ii2} can be further transformed. We contract it with a direct matrix $M_{ai}$ and get
\bal{pp1}
M_{ai}M^{-1}_{[ip}\pd_pN_{j]}=\frac12\left(\pd_aN_j-M_{ai}M^{-1}_{jp}\pd_pN_{i}\right)&=&0\;.
\ea
Multiply it again with the direct matrix $M_{bj}$ and contract with respect to the index $j$. It yields
\bal{pp2}
M_{bj}\pd_aN_j-M_{aj}\pd_bN_j&=&0\;.
\ea
Integrate this equation by parts and take into account that due to  definition, $A_{ab}=\pd_a X^b$, the following symmetry property is valid, $\pd_a M_{bj}=\pd_b M_{aj}$. Hence, equation \eqref{pp2} is equivalent to
\bal{pp3s}
\pd_a\left(M_{bj}N_j\right)-\pd_b\left(M_{aj}N_j\right)&=&\pd_{[a}n_{b]}=r^{-1}{\cal P}^{[ab]}\equiv 0\;.
\ea
It means that equation {\eqref{ii1} is satisfied for arbitrary vector field $X^i$ and that the anti-symmetric part of the right hand side of equation \eqref{xc8} vanishes identically, q.e.d.

\section{Calculating the Deformation Gradient Matrix}\label{appB}
\stepcounter{equationschapter}
\renewcommand{\theequation}{D.\arabic{equation}}
This appendix examines the properties of the displacement gradient matrix $A_{ij}$ and its inverse $B_{ij}$ in the context of an ideal fluid. In such a case, the displacement vector of a level surface is aligned with the radial direction, and can be expressed as $X^i = n^i X(\bm{x})$, where $n^i$ is the radial unit vector and $X(\bm{x})$ is a scalar function.

By applying the Leibniz rule, the partial derivative of $X^i$ decomposes into the derivative of the unit vector $n^i$ and the derivative of the scalar function $X$. The derivative of $n^i$ is given by:
\bal{xc15}
\pd_i n^j&=&\frac{{\cal P}^{ij}}{r}\;,
\ea
where ${\cal P}^{ij} = \delta^{ij} - n^i n^j$ is the projection operator onto the plane orthogonal to $n^i$. The deformation gradient is then: 
\bal{xc16}
A_{pi}&=&\pd_pX^i={\cal P}^{ip}\frac{X}{r}+n^i\pd_pX={\cal P}^{pq}\left(\frac{X}{r}\delta^{iq}+n^i\pd_qX\right)+n^in^pn^q\pd_qX\;.
\ea
Differentiating again yields:
\bal{uyb6}
\pd_{pq}X^i&=&-\left(n^i{\cal P}^{pq}+n^q{\cal P}^{ip}+n^p{\cal P}^{iq}\right)\frac{X}{r^2}+\frac{1}{r}\left({\cal P}^{ip}\pd_qX+{\cal P}^{iq}\pd_pX\right)+n^i\pd_{pq}X\;.
\ea
Contracting indices in Eq. \eqref{uyb6} gives the Laplacian of the displacement vector:
\bal{pmex5}
\Delta X^i&=&n^i\left(\Delta X-\frac{2X}{r^2}\right)+\frac{2}{r}{\cal P}^{ip}\pd_pX\;.
\ea

From Eq. \eqref{xc16}, we also obtain the identity: 
\bal{ek1}
A_{pi}{\cal P}^{iq}&=&\left({\cal P}^{ip}\frac{X}{r}+n^i\pd_pX\right){\cal P}^{iq}={\cal P}^{pq}\frac{X}{r}\;.
\ea
This identity facilitates the computation of the projection of the inverse matrix $B_{ij}$. Using the definition of $B_{ij}$ and Eq. \eqref{ek1}, we find:
\bal{ek2}
B_{ij}{\cal P}^{jq}&=&\sum_{n=1}^\infty(-1)^nA_{ip_1}A_{p_1p_2}...A_{p_{n-1}j}{\cal P}^{jq}
=\sum_{n=1}^\infty\left(-\frac{X}{r}\right)^n{\cal P}^{iq}=\left[\left(1+\frac{X}{r}\right)^{-1}-1\right]{\cal P}^{iq}\;,
\ea
and thus
\bal{ek3}
M^{-1}_{ij}{\cal P}^{jq}&=&\left(1+\frac{X}{r}\right)^{-1}{\cal P}^{iq}\;.
\ea
In particular, the contraction yields:
\bal{ek4}
M^{-1}_{pq}{\cal P}^{pq}&=&2\left(1+\frac{X}{r}\right)^{-1}\;.
\ea 

Projecting the deformation gradient onto the unit vector $n^i$ gives:
\bal{ek5}
n^pA_{pi}&=&n^iX'\;,
\ea
where $X' = n^p \partial_p X$ is the radial derivative of $X$. Applying this in the definition of $B_{ij}$ leads to:
\bal{ek6}
n^jB_{ji}&=&\sum_{n=1}^\infty(-1)^nn^jA_{jp_1}A_{p_1p_2}...A_{p_{n-1}i}=n^i\sum_{n=1}^\infty(-X')^n=\frac{n^i}{1+X'}-n^i\;,
\ea
and therefore:
\bal{ek7}
n^jM^{-1}_{ji}&=&\frac{n^i}{1+X'}\;.
\ea
From this, we also obtain:
\bal{ek8}
M^{-1}_{pq}n^pn^q&=&\frac{1}{1+X'}\;,\\\nonumber
{\cal P}^{ip}M^{-1}_{pj}&=&M^{-1}_{ij}-n^in^pM^{-1}_{pj}=M^{-1}_{ij}-\frac{n^in^j}{1+X'}\;.
\ea 
Additional identities involving contractions of $A_{ij}$ with $n^i$ include:
\bal{ek9}
A_{pi}n^i&=&\pd_pX={\cal P}^{pq}\pd_q X+X'n^p\;,\\
A_{qp}A_{pi}n^i&=&A_{qp}\pd_pX={\cal P}^{qp}\left(X'+\frac{X}{r}\right)\pd_pX+X'^2n^q\;,\\
A_{bq}A_{qp}A_{pi}n^i&=&A_{bq}A_{qp}\pd_pX=A_{bq}{\cal P}^{qp}\frac{X}{r}\pd_pX+X'A_{bq}\pd_qX\nonumber\\\nonumber
&=&{\cal P}^{bp}\left(X'+\frac{X}{r}\right)\frac{X}{r}\pd_pX+X'^2\pd_bX\\&=&{\cal P}^{bp}\left(X'^2+X'\frac{X}{r}+\frac{X^2}{r^2}\right)\pd_pX+X'^3n^b\;,\\\nonumber
A_{ab}A_{bq}A_{qp}A_{pi}n^i&=&A_{ab}A_{bq}A_{qp}\pd_pX=A_{ab}A_{bq}{\cal P}^{qp}\frac{X}{r}\pd_pX+X'A_{ab}A_{bq}\pd_qX\\\nonumber
&=&A_{ab}{\cal P}^{bp}\left(X'+\frac{X}{r}\right)\frac{X}{r}\pd_pX+X'^2A_{ab}\pd_bX\\
&=&{\cal P}^{ap}\left(X'+\frac{X}{r}\right)\frac{X^2}{r^2}\pd_pX+X'^2\left({\cal P}^{ap}\frac{X}{r}\pd_pX+X'\pd_aX\right)\nonumber\\
&=&{\cal P}^{ap}\left(X'^3+X'^2\frac{X}{r}+X'\frac{X^2}{r^2}+\frac{X^3}{r^3}\right)\pd_pX+X'^4n^a\;.
\ea
Finally, the vector $N^i = M^{-1}_{ij} n^j$ is given by:
\bal{ek10}
N^i=M^{-1}_{ij}n^j&=&-\frac{{\cal P}^{ij}\pd_j X}{\ds~\Biggl(1+X'\Biggr)\left(1+\frac{X}{r}\right)}+\frac{n^i}{1+X'}\;.
\ea

\section{Wigner's Formalism to Decomposing Legendre Polynomial Products}\label{wignerdecomp}
\stepcounter{equationschapter}
\renewcommand{\theequation}{E.\arabic{equation}}

The Wigner decomposition is a mathematical technique that employs Wigner matrices -- objects closely associated with angular momentum theory in quantum mechanics -- to decompose products of Legendre polynomials into linear combinations of Legendre polynomials of various degrees \citep{gelfand_1963}. This decomposition is grounded in the completeness of the Legendre polynomials, which form an orthogonal basis for the space of square-integrable functions on the interval $[-1, 1]$.

Since the product of a finite number of Legendre polynomials is itself a polynomial, it can be uniquely expanded in terms of this orthogonal basis. The Wigner decomposition provides a systematic way to perform this expansion, particularly useful in problems involving spherical symmetry, such as those encountered in gravitational and quantum field theories.

\subsection{Expressing the Wigner Matrix through Clebsch-Gordan Coefficients}
The Wigner matrix is defined as a square of the Clebsch-Gordan coefficients  
\bal{wig1}
T^{nml}\equiv \left(C^{nml}\right)^2\;,
\ea 
where $n, m, l\in \mathbb{N}=\{0,1,2,3,...\}$ are natural numbers and the Clebsch-Gordan coefficient 
\bal{k1e}
C^{nml}&=&(-1)^{n-m}\sqrt{2l+1}\left(\begin{array}{ccc}
  n & m & l \\
  0& 0 & 0 \\
\end{array}\right)\;,
\ea
is proportional to the Wigner $3j$ symbols,
\bal{k2s}
\left(\begin{array}{ccc}
  n & m & l \\
  0& 0 & 0 \\
\end{array}\right)&=&\frac{(-1)^ss!}{(s-n)!(s-m)!(s-l)!}\sqrt{\frac{(2s-2n)!(2s-2m)!(2s-2l)!}{(2s+1)!}}\,\delta_{\mathbb{A}}^n\delta_{\mathbb{B}}^m\delta_{\mathbb{C}}^l\delta^{2s}_{2\mathbb{N}}\;,\ea
with $2s\equiv n+m+l$, and $\delta^{n}_{\mathbb{A}}$, $\delta^{m}_{\mathbb{B}}$, and so on, are the indicator functions. The indicator functions are generalizations of the Kronecker symbol extended to a set of indices, for example,
\bal{knn}
\delta^{n}_{\mathbb{A}}&=&\biggl\{\begin{array}{ll}
  1&\;,\qquad \mbox{if}\; n\in \mathbb{A}\;,\\
  0&\;,\qquad \mbox{if}\; n\not\in \mathbb{A}\;,
\end{array} 
\ea
the set of even integers, $2\mathbb{N}=\{0,2,4,6,...\}$, and the sets $\mathbb{A}, \mathbb{B}, \mathbb{C}$ are defined by the rules
\bal{g5c3}
\mathbb{A}=\{|l-m|,...,l+m\}\qquad,\qquad\mathbb{B}=\{|l-n|,...,l+n\}\qquad,\qquad\mathbb{C}=\{|n-m|,...,n+m\}\;,
\ea
After computing the square of the $3j$ symbols we get the Wigner matrix $T^{nml}$ in explicit form, 
\bal{nxv5}
T^{nml}&=&\frac{2l+1}{2\pi }\frac{(s-n+1)_{-\frac12}(s-m+1)_{-\frac12}(s-l+1)_{-\frac12}}{(s+1)_{\frac12}}\,\delta_{\mathbb{A}}^n\delta_{\mathbb{B}}^m\delta_{\mathbb{C}}^l\delta^{2s}_{2\mathbb{N}}\;,
\ea
where $(a)_p=a(a+1)...(a+n-1)$ is the Pochhammer symbol -- cf. \eqref{s12}. 

We notice that the Wigner $3j$ symbols are fully symmetric with respect to the interchange of a pair of indices, while the Wigner matrix $T^{mnl}$ is symmetric only with respect to the first two indices, 
\bal{iu9}
T^{nml}=T^{(nm)l}\;.
\ea
Moreover, it satisfies the following two identities,
\bal{nb67v}
T^{0ml}=\delta^{ml}\;,\qquad\qquad T^{mn0}=\frac{\delta^{nm}}{2n+1}\;,
\ea
where $\delta^{nm}$ is the Kronecker symbol (the unit matrix). 

\subsection{Wigner's Decomposition of Legendre Polynomial Products}

The Wigner decomposition provides a systematic method for expressing the product of multiple Legendre polynomials as a linear combination of Legendre polynomials of various degrees. This is achieved through an iterative application of the binary product decomposition formula, which relies on Wigner $3j$-symbols or equivalent coefficients $T^{nml}$.

In the first step, the product of two Legendre polynomials is expanded as:
\bal{k1s}
P_{n}(\cos\theta)P_{m}(\cos\theta)&=&\sum_{l=|n-m|}^{n+m}T^{nml}P_{l}(\cos\theta)\;,
\ea
where the coefficients $T^{nml}$ encode the coupling between the angular momenta associated with the degrees $n$ and $m$.

To compute the product of three Legendre polynomials, the decomposition is applied iteratively:
\bal{k1sw}
P_{n_1}(\cos\theta)P_{n_2}(\cos\theta)P_{n_3}(\cos\theta)&=&\sum_{a_2=|n_2-n_1|}^{n_2+n_1}\sum_{a_3=|n_3-a_2|}^{n_3+a_2}T^{n_1n_2a_2}T^{a_2n_3a_3}P_{a_3}(\cos\theta)\;,
\ea
where the intermediate indices $a_1$, $a_2$, and $a_3$ arise from successive pairwise decompositions.

More generally, the Wigner decomposition for the product of $k$ Legendre polynomials takes the form:
\bal{k1hv}
P_{n_1}(\cos\theta)\ldots P_{n_k}(\cos\theta)&=&\sum_{a_2=|n_2-n_1|}^{n_2+n_1}\ldots\sum_{a_{k}=|n_k-a_{k-1}|}^{n_k+a_{k-1}}T^{n_1n_2a_2}T^{a_2n_3a_3}\ldots T^{a_{k-1}n_ka_k}P_{a_k}(\cos\theta)\;.
\ea
This recursive structure is particularly useful in applications involving spherical harmonics, such as quantum mechanics, geophysics, and gravitational potential theory.

\subsection{Wigner's Decomposition of Legendre Polynomial Derivative Products}

To compute the product of derivatives of two Legendre polynomials, we begin by differentiating both sides of Eq. \eqref{k1s} twice with respect to $\cos\theta$, then multiply the result by $\sin^2\theta$. This procedure leverages the differential identity satisfied by Legendre polynomials:
\bal{c45d}
\frac{d}{d(\cos\theta)}\left[\sin^2\theta\frac{dP_n(\cos\theta)}{d(\cos\theta)}\right]&=&-n(n+1)P_n(\cos\theta)\;.
\ea
Applying this identity to the differentiated form of Eq. \eqref{k1s}, we obtain the following expression for the product of the derivatives:
\bal{xc20ss}
\sin^2\theta\frac{dP_{n}(\cos\theta)}{d(\cos\theta)}\frac{dP_{m}(\cos\theta)}{d(\cos\theta)}&=&\frac12\sum_{l=|n-m|}^{n+m}\left[n(n+1)+m(m+1)-l(l+1)\right]T^{nml}P_{l}(\cos\theta)\;.
\ea
This result expresses the product of the derivatives of two Legendre polynomials as a weighted sum of Legendre polynomials, with weights determined by the coupling coefficients $T^{nml}$ and the eigenvalues of the Legendre differential operator.

\subsection{Integrals Involving Products of Legendre Polynomials}

Wigner's decomposition provides a powerful tool for evaluating definite integrals involving products of Legendre polynomials. By expressing such products as linear combinations of Legendre polynomials, integration becomes straightforward due to their orthogonality properties. For instance, using Eq. \eqref{k1s} along with the orthogonality relation:
\bal{pm64}
\int_{0}^{\pi} P_{n}(\cos\theta)P_m(\cos\theta)\sin\theta d\theta&=&\frac{2}{2n+1}\delta_{nm} \;,
\ea
we can directly compute integrals of Legendre polynomial products.

For the triple product, applying Eq. \eqref{k1sw} yields:
\bal{pm87}
\int_{0}^{\pi} P_{n}(\cos\theta)P_{m}(\cos\theta)P_{l}(\cos\theta)\sin\theta d\theta&=&\frac{2}{2l+1}T^{nml}\;,
\ea
where the coefficient $T^{nml}$ is symmetric under permutations of its indices, as follows from its definition via Wigner $3j$-symbols (see Eq. \eqref{wig1}) and their inherent symmetries (cf. Eq. \eqref{k2s}).

This process generalizes to the product of $k$ Legendre polynomials:
\bal{yv65}\hspace{-0.8cm}
\int_{0}^{\pi} P_{n_1}(\cos\theta)...P_{n_k}(\cos\theta)\sin\theta d\theta&=&\frac{2}{2n_k+1}\sum_{a_2=|n_2-n_1|}^{n_2+n_1}\ldots\sum_{a_{k-2}=|n_{k-2}-a_{k-3}|}^{n_{k-2}+a_{k-3}}T^{n_1n_2a_2}T^{a_2n_3a_3}\ldots T^{a_{k-2}n_{k-1}n_k}\;.
\ea
Additionally, the integral of the product of derivatives of Legendre polynomials can be evaluated using Eq. \eqref{xc20ss}, resulting in:
\bal{o6gt}
\int_0^\pi\frac{dP_n(\cos\theta)}{d\theta}\frac{dP_m(\cos\theta)}{d\theta}\sin\theta d\theta&=&\frac{2n(n+1)}{2n+1}\delta^{mn}\;.
\ea 

\section{Differential Equation for Gravitational Field Perturbations in Polytropic Models}\label{secD}
\stepcounter{equationschapter}
\renewcommand{\theequation}{F.\arabic{equation}}
\subsection{General Equation}
The polytropic equation of state is given by
\bal{k1}
p&=&{\mathsf K}_0\rho^{1+\frac{1}{n}}\;,
\ea
where ${\mathsf K}_0$ is a constant and $n$ is the polytropic index. This relation simplifies the computation of the coefficients $h_l$ in the power series expansion \eqref{s8} of the gravitational field perturbation $K$ in terms of the density perturbation $\varrho$.

We begin by evaluating the function ${\cal A}$ defined in Eq. \eqref{pnv34}: 
\bal{k2}
{\cal A}=\frac{1}{\rho}\frac{\pd p}{\pd\rho}=\left(1+\frac{1}{n}\right)p  \;.
\ea
Differentiating ${\cal A}$ yields: 
\bal{k3}
h_{l}&=&\pd_\rho^{{l}-1}A=K_0\frac{\ds n\;\Gamma\left(2+\frac{1}{n}\right)}{\ds\Gamma\left(1+\frac{1}{n}-{l}\right)}\rho^{-{l}+\frac{1}{n}}\;.
\ea 
Substituting these coefficients into Eq. \eqref{s8} allows the summation to be performed explicitly, resulting in an exact analytic relationship between the gravitational and density perturbations: 
\bal{k4}
K=(1+n){\mathsf K}_0\left[\left(\rho+\varrho\right)^{\frac{1}{n}}-\rho^{\frac{1}{n}}\right]\;.
\ea
Solving for $\varrho$ gives: 
\bal{k5}
\varrho&=&\left(\rho^{\frac{1}{n}}+\frac{K}{(1+n){\mathsf K}_0}\right)^n-\rho\;.
\ea
This expression can be used to derive two equivalent forms of the gravitational field perturbation equation.

{\bf First approach}: Expanding Eq. \eqref{k5} using the generalized binomial theorem \citep[Eq. 1.111]{gradryzh} yields:
\bal{k5a}
\varrho&=&\rho\sum_{{l}=1}^n\frac{b_{l}}{{l}!}\left(\frac{K}{{\mathsf K}_0}\right)^{l}\;,
\ea
with coefficients
\bal{k5b}
b_{l}&=&(-1)^{l}(1+n)^{-{l}}(-n)_{l}\rho^{-\frac{{l}}{n}}\;,
\ea
where $(-n)_{l}=(-n)(-n+1)...(-n+{l}-1)$ is the Pochhammer symbol. Substituting Eq. \eqref{k5b} into Eq. \eqref{41cs} leads to a non-linear field equation (c.f. Eq. \eqref{fe87}):  
\bal{k5c}
\Delta K+4\pi G\rho  \sum_{{l}=1}^n\frac{b_{l}}{{l}!}\left(\frac{K}{{\mathsf K}_0}\right)^{l}&=&0\;,
\ea
which can be solved iteratively.

{\bf Second approach}: Define a new function
\bal{k5d}
F&:=&\rho^{\frac{1}{n}}+\frac{K}{(1+n){\mathsf K}_0}\;,
\ea
and add the Laplacian of $\rho^{1/n}$ to both sides of Eq. \eqref{41cs}:
\bal{k5e}
\Delta\rho^{\frac{1}{n}}&=&\frac{\rho^{\frac{1}{n}-1}}{n}\left[\rho''+\frac{2\rho'}{r}+\left(\frac{1}{n}-1\right)\frac{\rho'^2}{\rho}\right]\;.
\ea
This transforms the equation into: 
\bal{k5f}
\Delta F+\kappa^2 F^n&=&\frac{\rho^{\frac{1}{n}-1}}{n}\left[\rho''+\frac{2\rho'}{r}+\left(\frac{1}{n}-1\right)\frac{\rho'^2}{\rho}\right]+\kappa^2\rho\;,
\ea
where 
\bal{k5g}
\kappa&=&\sqrt{\frac{4\pi G}{(1+n){\mathsf K}_0}}\;.
\ea
The right-hand side of Eq. \eqref{k5f} vanishes due to the Lane-Emden equation, yielding the final form:
\bal{unf1}
\Delta F+\kappa^2 F^n&=&0\;.
\ea
This non-linear PDE (except for $n = 1$) is equivalent to Eq. \eqref{k5c} but expressed in a more compact form. It generally requires numerical methods for solution.

\subsection{Special Cases of Polytropes}
Three particular cases of polytropes are of special interest and warrant separate treatment.
\subsubsection{Polytrope of index $n=0$}

For $n = 0$, the density $\rho$ is constant, implying $\varrho = 0$. The gravitational field perturbation satisfies the Laplace equation: 
\bal{bbb21}
\Delta K=0\;,
\ea 
valid both inside and outside the fluid body.

\subsubsection{Polytrope of index $n=1$}

For $n = 1$, Eq. \eqref{k5a} simplifies to: 
\bal{k7}
\varrho=\frac{K}{2{\mathsf K}_0}\;,
\ea
leading to the Helmholtz equation:
\bal{k7a}
\Delta K+\frac{2\pi G}{{\mathsf K}_0}K&=&0\;.
\ea
This equation admits an exact analytic solution, which can be used to determine the shape function $f = X/r$ via Eq. \eqref{oper23}, bypassing the need to solve the non-linear equation \eqref{cre5a}. Further details are provided in Section \ref{subD2}.

\subsubsection{Polytrope of index $n=\infty$} 
The case $n = \infty$ corresponds to an isothermal sphere, relevant for modeling systems like globular clusters \citep{horedt_2004book}. Here, Eq. \eqref{k5a} becomes: 
\bal{k8}
\varrho=\left(e^{K/{\mathsf K}_0}-1\right)\rho\;,
\ea
leading to the field equation:
\bal{k8a}
\Delta K+4\pi G\rho e^{K/{\mathsf K}_0}&=&4\pi G\rho\;.
\ea
The density profile $\rho(r)$ is obtained numerically from the Lane-Emden equation. Since $\rho$ does not vanish at any finite radius, the solution extends to infinity, implying infinite mass. In practice, the model is truncated at a finite radius for astrophysical applications \citep{horedt_2004book}.

\bibliographystyle{apsrev4-2}
\bibliography{Love_numbers_references}
\end{document}